\documentclass[11pt]{article}
\usepackage[utf8]{inputenc}

\usepackage{amsthm, amsfonts, amssymb, amsmath,enumitem, natbib, bbm, mathabx}
\usepackage{cases, amsthm}
\usepackage{fullpage}
\usepackage{mathtools}
\usepackage{mathrsfs} 
\usepackage{etoolbox}

\newcommand{\cosi}[1]{%
  \mathtt{COSI}%
  \ifstrempty{#1}{}{\mbox{-}{#1}}%
}
\usepackage{wasysym} 
\usepackage{verbatim}
\usepackage{fancyhdr}
\usepackage{graphicx}
\usepackage{algorithm}
\usepackage[ruled,linesnumbered, vlined, noend,algo2e]{algorithm2e}
\usepackage{bbm}
\usepackage{caption}
\usepackage{subcaption}
\usepackage{authblk}
\usepackage{setspace}
\let\Algorithm\algorithm
\renewcommand\algorithm[1][]{\Algorithm[#1]\setstretch{1.6}}
\usepackage[colorlinks=true, pdfstartview=FitV,
linkcolor=blue, citecolor=blue, urlcolor=blue]{hyperref}
\usepackage{cleveref}
\newtheorem{definition}{Definition}[section]
\newtheorem{theorem}[definition]{Theorem}
\newtheorem{proposition}[definition]{Proposition}
\newtheorem{lemma}[definition]{Lemma}
\newtheorem{corollary}[definition]{Corollary}
\newtheorem{example}{Example}[section]

\theoremstyle{remark}
\newtheorem{remark}[definition]{Remark}

\usepackage{tocloft}
\usepackage{accents}

\newcommand{\cosici}[3]{\widehat{\mathrm{CI}}_{#1,#2}^{\mathtt{COSI}\mbox{-}#3}}
\newcommand{\cosicit}[3]{\widetilde{\mathrm{CI}}_{#1,#2}^{\mathtt{COSI}\mbox{-}#3}}
\renewcommand{\mp}{\mathfrak{m}+}
\newcommand{\mpp}{\mathfrak{m}{+\!\!+}}
\newcommand{\rmpp}{\mathtt{RM}{+\!\!+}}
\newcommand{\rmp}{\mathtt{RM}{+}}

\DeclareMathOperator*{\argmin}{arg\,min}

\newcommand{\lQ}{\mathrm{lQ}}

\usepackage{xcolor}

\def\SAV{\textrm{COSI}}

\usepackage[top=1in, bottom=1in, left=1in, right=1in]{geometry}
\title{$\cosi{}$: A Framework for Robustly Valid Inference}

\author{Arun Kumar Kuchibhotla}
\author{Sivaraman Balakrishnan}
\author{Larry Wasserman}
\affil{Department of Statistics \& Data Science, Carnegie Mellon University}
\date{}

\begin{document}
\maketitle
\begin{abstract}
We introduce $\cosi{}$ (COnfidence from Scale Invariance), a framework for constructing confidence sets for statistical functionals. $\cosi{}$ builds on the HulC (Hull-based Confidence), which uses median unbiasedness to construct confidence sets from batch estimates. $\cosi{}$ uses scale-invariance properties shared by many common estimators. We divide the data into batches, compute an estimate in each batch, and retain each candidate parameter value for which the batch estimates pass a test of scale invariance. The resulting confidence sets are asymptotically valid under weaker assumptions than classical Wald and bootstrap procedures require, and their coverage can be closer to the nominal level in small samples. Applications include robust inference for the mean, regression coefficients in misspecified linear models, tensor norms, and Fréchet means.
\end{abstract}


\section{Introduction}
Traditional confidence intervals for real-valued 
parameters and functionals are constructed based on estimating the limiting distribution of a ``good'' estimator. These methods can be brittle, in the sense of losing uniform validity, if the limiting distribution changes abruptly. In this paper, our aim is to provide robustly valid confidence sets while relying on weaker assumptions on the estimator used. We provide a host of such methods within a framework we call -- $\cosi{}$: 
COnfidence from Scale Invariance.

Before proceeding further, let us recall a few definitions of the validity of confidence sets and how traditional inference methods work. 
Let $\mathcal{P}$ be a collection of probability distributions. Let $\theta:\mathcal{P}\to\mathbb{R}^d$ be a functional or parameter of interest. (We will allow the functional to lie in an infinite-dimensional space for our results, but assume it to be in a Euclidean space or even the real line for the following review.) For example, $\mathcal{P}$ could be the collection of Normal distributions with mean $\mu\in\mathbb{R}^d$ and a (symmetric, nonnegative definite) covariance matrix $\Sigma\in\mathbb{R}^{d\times d}$. In this example, the functional of interest could be $\theta(P) = \mu$ or $\theta(P) = e_1^{\top}\Sigma e_1$. These are just illustrative examples, and we do not require parametric models for the data generating process in this paper. 

A confidence set $\widehat{\mathrm{CI}}_{n,\alpha}$ based on $n$ IID observations $X_1, \ldots, X_n$ from $P\in\mathcal{P}$ is said to be {\em asymptotically pointwise valid} for a functional $\theta(P)$  at level $\alpha$ if 
\begin{equation}\label{eq:asymptotic-pointwise-valid}
\sup_{P\in\mathcal{P}}\,\limsup_{n\to\infty}\,\mathbb{P}\left(\theta(P)\notin\widehat{\mathrm{CI}}_{n,\alpha}\right) \le \alpha.
\end{equation}
A confidence set $\widehat{\mathrm{CI}}_{n,\alpha}$ is said to be {\em asymptotically uniformly valid} for a functional $\theta(P)$ at level $\alpha$ if
\begin{equation}\label{eq:asymptotic-uniform-valid}
\limsup_{n\to\infty}\,\sup_{P\in\mathcal{P}}\,\mathbb{P}\left(\theta(P)\notin\widehat{\mathrm{CI}}_{n,\alpha}\right) \le \alpha.
\end{equation}
Comparing the left hand sides of~\eqref{eq:asymptotic-pointwise-valid} and~\eqref{eq:asymptotic-uniform-valid}, we note the difference between pointwise and uniform validity lies in the order in which the data generating process and sample size act. In pointwise validity, the distribution is treated as unchanged with sample size and in uniform validity, the distribution is allowed to change with the sample size. It is important to note that the property of pointwise or uniform validity is a function of both the confidence set and the collection of distributions $\mathcal{P}$. If $\mathcal{P}$ is a singleton (i.e., contains only one distribution), then both notions of validity coincide. Ideally, one would and should prefer a confidence set that is uniformly valid over the largest possible collection $\mathcal{P}$, and whenever possible, it would be preferable that the confidence set does not depend on $\mathcal{P}$. In what follows, we construct general purpose confidence sets and {\em implicitly} identify the collection $\mathcal{P}$ over which they remain uniformly valid.   
Traditionally, a confidence set is called $j$-th order accurate if $\mathbb{P}_P(\theta(P)\notin\widehat{\mathrm{CI}}_{n,\alpha}) = \alpha + O(n^{-j/2})$ for all $\alpha\in(0, 1)$, as $n\to\infty$ for all $P\in\mathcal{P}$; it is uniformly $j$-th order accurate if $O(n^{-j/2})$ is uniform over $P\in\mathcal{P}$. 

Another subtle aspect lurking behind the asymptotic notion of validity is the dependence on $\alpha$. In definitions~\eqref{eq:asymptotic-pointwise-valid} and~\eqref{eq:asymptotic-uniform-valid}, we treat $\alpha$ as fixed while the sample size $n$ grows. In some settings, one may also want to vary $\alpha$ with sample size; for example, in multiple testing problems or sequential analysis (with alpha spending), the chosen level $\alpha$ depends on the sample size. It is possible for a confidence set to satisfy uniform validity guarantee as in~\eqref{eq:asymptotic-uniform-valid}, but for a specific choice of $\{\alpha_n\}_{n\ge1}$, $\sup_{P\in\mathcal{P}}\,\mathbb{P}_P(\theta(P)\notin\widehat{\mathrm{CI}}_{n,\alpha_n})/\alpha_n \to \infty$ as $n\to\infty$. Uniformly validity for all sequential levels $\{\alpha_n\}_{n\ge1}$ is equivalent to 
\begin{equation}\label{eq:post-hoc-validity}
\limsup_{n\to\infty}\,\sup_{\alpha\in[0, 1]}\,\sup_{P\in\mathcal{P}}\, \frac{\mathbb{P}_P(\theta(P)\notin\widehat{\mathrm{CI}}_{n,\alpha})}{\alpha} \le 1.
\end{equation}
This goal is related to a recent criterion proposed in the E-values literature called post-hoc validity~\citep[Section 3]{chugg2026post}. We believe this to be an impossible goal to achieve in the general framework we consider. For example, the Wald confidence interval for the population mean of univariate random variables does not satisfy~\eqref{eq:post-hoc-validity}, even when the underlying data is from a Gaussian distribution. More generally, there cannot be a locally bounded confidence interval for the median of a symmetric distribution unless $n \ge \log_2(1/\alpha)$~\citep{lanke1974interval}; also, see~\cite{pmlr-v300-deep26a} for some recent results. However, we would prefer the existence of a ``small'' sequence $\{\tau_n\}_{n\ge1}$ such that
\begin{equation}\label{eq:uniform-validity-beyond-zero}
\limsup_{n\to\infty}\,\sup_{\alpha\in[\tau_n, 1]}\,\sup_{P\in\mathcal{P}}\,\frac{\mathbb{P}_P(\theta(P)\notin\widehat{\mathrm{CI}}_{n,\alpha})}{\alpha} \le 1.
\end{equation}
Such a sequence $\tau_n, n\ge1$ may remain unknown to the analyst. Also, from the ordering of the supremums in~\eqref{eq:uniform-validity-beyond-zero}, it follows that $\{\tau_n\}_{n\ge1}$ can depend only on $\mathcal{P}$, but not on the individual distributions $P\in\mathcal{P}$.

\paragraph{Review of the Wald and Resampling Methods of Inference.} Traditional methods of inference construct confidence intervals in the following way. For any real-valued functional $\theta(P), P\in\mathcal{P}$, let $\widehat{\theta}_n$ be an estimator computed from $n$ observations. Suppose that $\widehat{\theta}_n$ when scaled and centered properly converges in distribution to a random variable $L_P$ for all $P\in\mathcal{P}$, i.e., 
\begin{equation}\label{eq:limiting-distribution}
\limsup_{n\to\infty} \mathbb{P}_P(r_n(\widehat{\theta}_n - \theta(P)) \in [a, b]) \le \mathbb{P}(L_P\in [a, b])\quad\mbox{for all}\quad P\in\mathcal{P}, a,b\in\mathbb{R}.
\end{equation}
Note that assumption~\eqref{eq:limiting-distribution} is a pointwise statement as $P$ is fixed as $n$ changes. A uniform version of~\eqref{eq:limiting-distribution} would be
\begin{equation}\label{eq:uniform-limiting-distribution}
    \limsup_{n\to\infty} \sup_{P\in\mathcal{P}}\left(\mathbb{P}_P(r_n(\widehat{\theta}_n - \theta(P)) \in [a, b]) - \mathbb{P}(L_P\in[a, b])\right)_+ = 0\quad\mbox{for any}\quad a, b\in\mathbb{R}.
\end{equation}
\begin{itemize}
    \item {\bf Wald Inference:} Wald inference assumes that the rate of convergence $r_n$ is known and that the limiting random variable $L_P$ belongs to a known parametric family. By estimating the unknown parameters of the limiting distribution, we can estimate the $\alpha/2$ and $1-\alpha/2$ quantiles of $L_P$ as $\widehat{q}_{\alpha/2}$ and $\widehat{q}_{1-\alpha/2}$. Then the Wald interval is given by
    \[
    \widehat{\mathrm{CI}}_{n,\alpha}^{\mathtt{Wald}} ~:=~ \left[\widehat{\theta}_n - \frac{\widehat{q}_{1-\alpha/2}}{r_n},\, \widehat{\theta}_n -\frac{\widehat{q}_{\alpha/2}}{r_n}\right].
    \]
    For example, if the limiting distribution is $N(0, \sigma^2(P))$, then with an estimate $\widehat{\sigma}_n$ for the unknown variance of the limiting distribution, we get $-\widehat{q}_{\alpha/2} = \widehat{q}_{1-\alpha/2} = \widehat{\sigma}_nz_{\alpha/2}$ where $z_{\alpha/2}$ is the $1-\alpha/2$-th quantile of the standard normal random variable. More generally, if $L_P \sim G_{\eta(P)}$ for a parametric family of univariate distributions $\{G_{\eta}:\,\eta\in\Upsilon\}$, then with a consistent estimator $\widehat{\eta}_n$ for $\eta(P)$, take $\widehat{q}_{\alpha/2}$ and $\widehat{q}_{1-\alpha/2}$ to be the $\alpha/2$ and $1-\alpha/2$ quantiles of $G_{\widehat{\eta}_n}$.
    \item {\bf Resampling Inference:} Resampling inference usually assumes that the rate of convergence is known, but there exist resampling methods that estimate the rate of convergence. Unlike Wald inference, resampling inference does not assume a parametric form for the limiting distribution. Resampling methods draw repeated samples from the observed data with some sampling scheme and estimate the distribution of $L_P$. This, in turn, provides estimates of the $\alpha/2$ and $1-\alpha/2$ quantiles of $L_P$. Hence, the resampling intervals are given by
    \[
    \widehat{\mathrm{CI}}_{n,\alpha}^{\mathtt{Resmp}} ~:=~ \left[\widehat{\theta}_n - \frac{\widehat{q}_{1-\alpha/2}}{\widehat{r}_n},\, \widehat{\theta}_n - \frac{\widehat{q}_{\alpha/2}}{\widehat{r}_n}\right],
    \]
    where $\widehat{r}_n$ is an estimated rate of convergence. There are numerous resampling schemes andwe refer the reader to~\cite{Shao1995},~\cite{Politis1999}, and~\cite{Lahiri2003} for excellent overviews of resampling methods.
\item{\bf The HulC:}
The $\cosi{}$ method in this paper is a generalization of the HulC proposed in~\cite{kuchibhotla2024hulc} (and its generalization in~\cite{paul2024inference}). We briefly review it here to show the differences from the Wald and resampling approaches to inference. Based on $n$ independent observations $X_1, \ldots, X_n$ with common distribution $P$, HulC operates by first partitioning $\{1, 2, \ldots, n\}$ into $B = B_{\alpha}$ disjoint batches $S_1, \ldots, S_B$, with $B_{\alpha} \ge \log_2(2/\alpha)$; this means $S_j\cap S_k = \emptyset$ for $j\neq k$ and $S_1\cup\cdots\cup S_B = \{1, 2, \ldots, n\}$. Then, for $1\le j\le B$, compute the estimator $\widehat{\theta}^{(j)}$ based on data $X_i, i\in S_j$. These estimators can be based on different methods. Then the HulC confidence interval is given by
\[
\widehat{\mathrm{CI}}_{n,\alpha}^{\mathtt{HulC}} ~:=~ \left[\widehat{\theta}_{(\lfloor B/2\rfloor - r_{B,\alpha})},\, \widehat{\theta}_{(\lceil B/2\rceil + r_{B,\alpha})}\right], 
\]
where $r_{B,\alpha}$ is the smallest integer (random variable) such that $\mathbb{P}(\mbox{Binom}(B, 1/2) \ge \lfloor B/2\rfloor - r_{B,\alpha}) = 1 - \alpha/2$; here $\widehat{\theta}_{(1)} \le \ldots \le \widehat{\theta}_{(B)}$ are the order statistics of $\widehat{\theta}^{(j)}, 1\le j\le B$. This formulation is taken from Algorithm 2 of~\cite{paul2025finite}. In~\cite{kuchibhotla2024hulc}, we have $B \in \{\lfloor \log_2(2/\alpha)\rfloor,\, \lceil\log_2(2/\alpha)\rceil\}$, while in the formulation above, we randomize $r_{B,\alpha}$ for exact asymptotic coverage. It is now clear that HulC, unlike Wald/Resampling, does not estimate the limiting distribution. In fact, the validity of the HulC interval relies only on the median bias, which depends only on $\mathbb{P}_P(\widehat{\theta}^{(j)} \ge \theta(P))$ and $\mathbb{P}_P(\widehat{\theta}^{(j)} \le \theta(P))$. 
\end{itemize}
Under assumption~\eqref{eq:limiting-distribution}, Wald and resampling intervals can only be guaranteed to be pointwise valid. Even under assumption~\eqref{eq:uniform-limiting-distribution}, they need not be uniformly valid unless the limiting random variable $L_P$ has adistribution which varies ``continuously'' as $P$ changes. This is related to the classical notion of regularity~\citep[Sec. 8.5]{Vaart_1998}. To distinguish it from median regularity~\citep{kuchibhotla2023median}, we may refer to the classical notation of regularity as distributional regularity. \citet[Theorem 2.1]{beran1997diagnosing} proved that, for regular parametric models, the parametric bootstrap is successful if and only if the estimator is distributionally regular. It is easy to see that distributional regularity can be significantly stronger than median regularity.

In the following section, we present a new framework for inference that yields confidence sets with robust validity guarantees in the sense that the limiting distribution, in its entirety, need not be continuous in $P$. In addition, our proposed methodology completely bypasses the requirement of estimating the limiting distribution or the rate of convergence.

We will also present several examples where either Wald or resampling or both fail to provide uniformly or even pointwise valid confidence intervals. As illustrated by the examples, $\cosi{}$~encompasses several existing methods, including the t-test-based inference of~\cite{ibragimov2010tstatistic}, the randomization tests of~\cite{canay2017randomization}, and the HulC methodology of~\cite{kuchibhotla2024hulc}. $\cosi{}$,~in particular, shares several similarities with the framework of approximate randomization tests of~\cite{canay2017randomization} and we revisit this below. Our framework also goes significantly beyond the univariate setting by providing rate-optimal confidence sets for parameters/functionals taking values in (potentially) infinite-dimensional topological vector spaces (that include Hilbert, Banach or Alexandrov spaces). 

\paragraph{Main Contributions and Reading Guide.} The following are the main contributions of this paper and provide a potential reading guide. 
\begin{enumerate}
    \item HulC already shows that the goal of inference can be achieved without estimating the limiting distribution by relying only on the property of asymptotic median unbiasedness. Section~\ref{sec:SAVARI} provides an abstract generalization of median unbiasedness to general scale-invariance properties and provides a general construction of confidence sets with multiplicative second order miscoverage guarantees. This section works with an arbitrary topological vector space but for first reading it may be beneficial to consider univariate functionals of interest. Section~\ref{sec:applications-univariate} instantiates the results of Section~\ref{sec:SAVARI} with four scale-invariance properties, namely, median unbiasedness (same as the HulC), symmetry around zero, normality with mean zero, and unimodality with mode at zero. For the first reading, one may skip Sections~\ref{subsec:symmetry}--\ref{sec:unimodal}, especially because the development in Sections~\ref{sec:hunt-and-test} and~\ref{sec:illustrations-hunt-and-test} only uses the zero median property.
    \item Section~\ref{sec:hunt-and-test} provides a specific way to use the $\cosi{}$ framework to construct confidence sets for functionals taking values in general topological vector spaces based on the so-called hunt-and-test idea~\citep{moran1973dividing}. For the first reading, one can skip Section~\ref{sec:hunt-and-test}. Section~\ref{sec:illustrations-hunt-and-test} shows the instantiation of the abstract hunt-and-test framework to several interesting problems including high-/infinite-dimensional mean inference, inference under adversarial cellwise contamination, and inference for Fr{\'e}chet means. The main message of Sections~\ref{sec:hunt-and-test} and~\ref{sec:illustrations-hunt-and-test} is that one can maintain coverage and (near) optimal diameter while only requiring marginal CLTs without any distributional assumptions on the joint distribution of the estimator. One can match the diameter of high-dimensional CLT based confidence set (up to constants) even when the high-dimensional CLT does not hold. 
    \item Even in the examples where we assume strong enough regularity for high-dimensional CLTs to hold, our miscoverage guarantees throughout are multiplicative in nature, which implies valid coverages even for $\alpha = \exp(-n^{o(\gamma)})$ for some $\gamma < 1$. This is a significantly stronger guarantee than that offered by CLTs which are additive in most cases. Moreover, our miscoverage guarantees do not depend on the underlying dimension or complexity of the parameter space. 
\end{enumerate}
\section{$\cosi{}$: Intuition and Description}\label{sec:SAVARI}
\begin{figure}[!h]
    \centering
    \includegraphics[width=\linewidth]{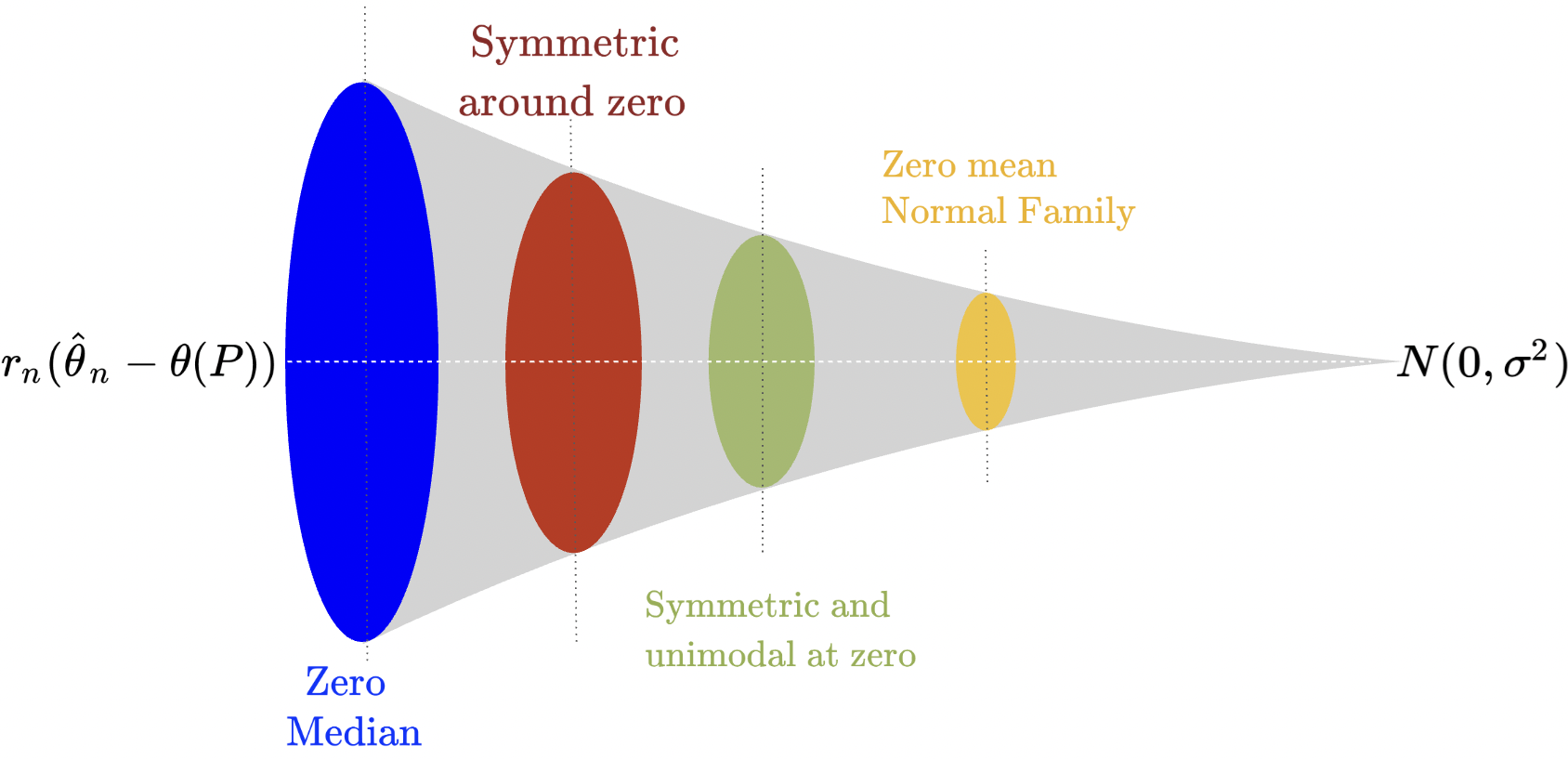}
    \caption{Illustration of some scale-invariant properties satisfied by zero-mean normal distribution. If a properly centered and scaled estimator is converging in distribution to $N(0, \sigma^2)$, then it must first be close to zero median distributions, and then to symmetric distributions, and then to symmetric and unimodal distributions, and then to the family of mean zero normal distributions, and then finally to a specific mean zero normal distribution.}    
    \label{fig:main-figure-savari}
\end{figure}

The basic idea behind $\cosi{}$ is best understood through the illustration in Figure~\ref{fig:main-figure-savari}. 
If we have an estimator $\widehat{\theta}_n$ such that $r_n(\widehat{\theta}_n - \theta(P))$ converges in distribution to $N(0,\sigma^2)$ as $n\to\infty$, then the law of $\widehat{\theta}_n - \theta(P)$ must satisfy several {\em scale invariance} properties, such as zero median, symmetry, unimodality and so on, approximately. Hence, we can construct a confidence set for $\theta(P)$ using these properties instead of relying on the limiting distribution, and relying on these ``soft'' properties of the limiting distribution endows its validity guarantee with additional robustness. Moreover, using stronger properties of the limiting distribution yields better confidence sets while yielding some robustness beyond the classical (Wald/resampling) confidence sets. One other advantage of relying on these ``soft'' properties is that the limiting distributions may remain ``continuous'' (i.e., regular) with respect to these ``soft'' properties, even when distributional regularity fails. 

Before presenting the precise algorithm for $\cosi{}$~confidence sets, we first discuss two examples that highlight the challenges of inference methods based on the limiting distribution.
\begin{example}[Non-negative Mean]
    Suppose $X_1, \ldots, X_n$ are independent random variables with a common distribution $P$ and a mean $\theta(P)$ that is known to be non-negative. A natural estimator for $\theta(P)$ is $\widehat{\theta}_n = \max\{\overline{X}_n, 0\}$, where $\overline{X}_n = \sum_{i=1}^n X_i/n$. Fixing $P$ as $n$ changes (as in pointwise asymptotics), we get
    \[
    n^{1/2}(\widehat{\theta}_n - \theta(P)) \overset{d}{\to} \begin{cases}\max\{Z, 0\}, &\mbox{if }\theta(P) = 0,\\
    Z, &\mbox{if }\theta(P) > 0,\end{cases} 
    \]
    where $Z \sim N(0, \sigma^2(P))$ with $\sigma^2(P) = \mathrm{Var}_P(X_i)$. Observe that the limiting distribution is discontinuous in $P$ at $\theta(P) = 0$. This discontinuity causes Wald and resampling methods to yield invalid confidence intervals in the uniform sense; see, for example,~\cite{bickel1997resampling} and~\cite{andrews2010asymptotic}.

    Observe now that although the limiting distribution itself is discontinuous in $P$, the limiting distribution always has a zero median. Formally, for any $\theta(P) \ge 0$, we have
    \begin{equation}\label{eq:median-bias-non-negative-mean}
        \begin{split}
            \mathbb{P}_P(\widehat{\theta}_n \ge \theta(P)) ~&\ge~ \mathbb{P}_P(\overline{X}_n \ge \theta(P)),\\
            \mathbb{P}_P(\widehat{\theta}_n \le \theta(P)) ~&=~ \mathbb{P}_P(\overline{X}_n \le \theta(P)).
        \end{split}
    \end{equation}
    The probabilities on the right hand side converge to $1/2$ by the central limit theorem (under appropriate conditions). Hence, uniformly over a class of distributions $\mathcal{P}$, the estimator centered at $\theta(P)$ has an approximate median of zero; for example, $\mathcal{P}$ can be taken to be the union of collection of all symmetric distributions and the collection of all distributions $P$ such that $\mathbb{E}_P[|X_i - \theta(P)|^{2 + \delta}] \le L(\mathbb{E}_P[|X_i - \theta(P)|^2])^{1 + \delta/2}$. Therefore, constructing a confidence interval using this approximate zero median property leads to intervals which remain valid even when intervals constructed using the limiting distribution become invalid. In essence, this idea yields the HulC procedure of~\cite{kuchibhotla2024hulc}.
\end{example}

\begin{example}[Population Median]
    Suppose $X_1, \ldots, X_n$ are independent random variables with a common distribution $P$ and a median $\theta(P)$, i.e.,
    \[
    \mathbb{P}_P(X_i \ge \theta(P)) \ge 1/2\quad\mbox{and}\quad \mathbb{P}_P(X_i \le \theta(P)) \ge 1/2.
    \]
    A natural estimator for $\theta(P)$ is the sample median
    \[
    \widehat{\theta}_n = \min\left\{\theta:\, n^{-1}\sum_{i=1}^n \mathbf{1}\{X_i \le \theta\} \ge 1/2\right\}.
    \]
    Assuming local linearity of the distribution function:
    \begin{equation}\label{eq:differentiability-median}
        \mathbb{P}_P(X_i \le \theta(P) + t) - \mathbb{P}_P(X_i \le \theta(P)) = A_Pt + o(t)\quad\mbox{as}\quad t\to0,
    \end{equation}
    we know
    \[
    n^{1/2}(\widehat{\theta}_n - \theta(P)) \overset{d}{\to} A_P^{-1}Z,\quad Z\sim N(0, 1/4).
    \]
    Under~\eqref{eq:differentiability-median}, the Wald and resampling confidence intervals are pointwise valid. Uniform validity also holds for the class of distributions $\mathcal{P}$ satisfying~\eqref{eq:differentiability-median} (uniformly) with $\inf_{P\in\mathcal{P}}\,A_P > 0$. These validity guarantees do not hold if we relax the local linear assumption of~\eqref{eq:differentiability-median} to a local polynomial assumption. For instance, if we assume
    \[
    \mathbb{P}_P(X_i \le \theta(P) + t) - \mathbb{P}_P(X_i \le \theta(P)) = A_P|t|^{\gamma}\mathrm{sign}(t) + o(|t|^{\gamma})\quad\mbox{as}\quad t\to0,
    \]
    then
    \[
    n^{1/(2\gamma)}(\widehat{\theta}_n - \theta(P)) \overset{d}{\to} A_P^{-1}|Z|^{1/\gamma}\mathrm{sign}(Z),\quad Z\sim N(0, 1/4). 
    \]
    See, for example,~\cite{knight2002limiting} for details. From the results of~\cite{knight1999asymptotics}, it follows that the Bootstrap is valid if and only if $\gamma = 1$. Subsampling with an estimated rate of convergence as described in~\cite{bertail1999subsampling} can be applied, but its uniform validity properties are unknown. In any case, its practical performance is unreliable, as illustrated in Figure 5 of~\cite{paul2024inference}. However, no matter what $\gamma \in (0, \infty)$ is, the limiting distribution is symmetric around zero, due to the independence of $\mathrm{sign}(Z)$ and $|Z|$. Hence, in this example, the ``soft'' property of symmetry of the limiting distribution remains true as the underlying distribution changes. Therefore, a method that only uses the symmetry of the Normal distribution to ensure its validity will remain valid for all $\gamma\in(0,\infty)$, but methods that rely on the limiting Normality will only stay valid for $\gamma = 1$.
\end{example}

To describe the generic algorithm for the construction of confidence sets under the $\cosi{}$ framework, we need the following definition. For any collection of distributions $\mathfrak{P}$, we write $W\in\mathfrak{P}$ to say that the distribution of $W$ is an element of $\mathfrak{P}$.
\begin{definition}\label{def:scale-invariance}
    A collection of distributions $\mathfrak{P}$ on any real topological vector space is said to be {\em scale-invariant} if, for any random variable $W\in\mathfrak{P}$, $cW\in\mathfrak{P}$ for any $c > 0$.
\end{definition}
Some examples of scale-invariant classes of distributions are as follows:
\begin{enumerate}
    \item \textbf{Angular symmetry}: Let $W$ be a Borel-measurable random element taking values in a real normed vector space $(V,\|\cdot\|)$. We say $W$ is angularly symmetric if $W/\|W\| \overset{d}{=} -W/\|W\|$, with the convention that $W/\|W\| = 0$ if $W = 0$. In the univariate case, if $\mathbb{P}(W = 0) = 0$, then angular symmetry is the same as the zero median property. 
    \item \textbf{Half-space symmetry}: Let $W$ be a Borel-measurable random element taking values in a real inner-product space $(\mathbb{H}, \langle \cdot, \cdot\rangle)$. We say $W$ is half-space symmetric if $\mathbb{P}(\langle u, W\rangle \ge 0) \ge 1/2$ for all $u\in\mathbb{H}$. Equivalently, 
    \[
    \min\{\mathbb{P}(\langle u, W\rangle \ge 0),\,\mathbb{P}(\langle u, W\rangle \le 0)\} \ge \frac{1}{2}\quad\mbox{for all}\quad u\in\mathbb{H}.
    \]
    \item \textbf{Central symmetry}: Let $W$ be a Borel-measurable random element taking values in a real topological vector space $V$. We say $W$ is centrally symmetric if $W \overset{d}{=} -W$. 
    \item \textbf{Zero rectilinear median}: Let $\mathcal{T}$ be an index set and let $W = (W(t):t\in\mathcal{T})\in\mathbb{R}^{\mathcal{T}}$ be defined by a family of Borel-measurable real-valued random variables $W(t), t\in\mathcal{T}$ on a common probability space. We say $W$ has a zero rectilinear median if $W(t)$ has zero median for all $t\in\mathcal{T}$, i.e., $\min\{\mathbb{P}(W(t) \ge 0),\,\mathbb{P}(W(t) \le 0)\} \ge 1/2$ for all $t\in\mathcal{T}$.
    \item \textbf{$\gamma$-unimodality}: Let $W$ be a Borel-measurable random element taking values in a real topological space $V$. We say $W$ is $\gamma$-unimodal (for some $\gamma > 0$) if for every bounded, non-negative function $g:V\to\mathbb{R}_{\ge0}$, $t\mapsto t^{\gamma}\mathbb{E}[g(tW)], t\in(0,\infty)$ is non-decreasing. In the univariate case, with $\gamma = 1$, this is precisely the notion of unimodality with mode at zero, that is, the Lebesgue density of $W$ is non-decreasing on $(-\infty, 0)$ and decreasing on $(0,\infty)$. See~\cite{Olshen1970} and~\cite{kolyvakis2023multivariate} for more details.
    \item\label{item:normality-scale} \textbf{Mean-zero normality:} Let $W$ be a Borel-measurable random element taking values in a real Banach space $(\mathbb{B},\|\cdot\|)$. We say that $W$ is \emph{mean-zero normal}, or equivalently a \emph{centered Gaussian random element}, if, for every continuous linear functional $\ell\in\mathbb{B}^{*}$, $\ell(W)\sim \mathcal{N}(0,\sigma_{\ell}^{2})$, for some $\sigma_{\ell}\geq 0$.
    \item \textbf{Zero $\gamma$-quantile}: A Borel-measurable real-valued random variable $W$ has a zero $\gamma$-quantile if 
    \[
    \mathbb{P}(W \le 0) \ge \gamma\quad\mbox{and}\quad \mathbb{P}(W \ge 0) \ge 1 - \gamma.
    \]
    We say a random variable $W$ taking values in a real inner-product space $(\mathbb{H}, \langle\cdot, \cdot\rangle)$ has a zero $\gamma$-quantile for $\gamma\in(0, 1/2)$, if $\mathbb{P}(\langle u, W\rangle \le 0) \ge \gamma$ for all $u\in\mathbb{H}$. 
\end{enumerate}
Finally, we mention that if $\mathfrak{P}_1$ and $\mathfrak{P}_2$ are any two scale-invariant classes of distributions, then $\mathfrak{P}_1\cap\mathfrak{P}_2$ is also a scale-invariant property; for example, symmetry and $\gamma$-unimodality is also a scale-invariant class. 
See~\cite{serfling2006multivariate} for the different notions of symmetry mentioned above (in the Euclidean settings) and their characterizations. Usually, symmetry notions are defined around a point, and for our application, we only need symmetry around zero. Example~\ref{item:normality-scale} above shows that mean zero Normality is a scale-invariance property and it is easy to see that mean zero Normality with a specific covariance operator is not a scale-invariant property. All the examples above of scale-invariance classes satisfy the stronger condition that $W\in\mathfrak{P}$ implies $cW\in\mathfrak{P}$ for all $c\in\mathbb{R}$ (not just $c > 0$). The class of all exponential (or folded normal) distributions and zero $\gamma$-quantile for real-valued random variables are examples satisfying the weaker Definition~\ref{def:scale-invariance}. 

The main implication of the scale-invariance property in the context of inference is that if $r_n(\widehat{\theta}_n - \theta(P)) \overset{d}{\to} L_P\in\mathfrak{P}$, then, under appropriate assumptions, we can claim that $\widehat{\theta}_n - \theta(P)$ is close to $\mathfrak{P}$ asymptotically. In other words, we can safely ignore the rate of convergence by the definition of scale-invariance. This would not be possible when the precise limiting distribution is used.

One more ingredient for $\cosi{}$ is a level $\gamma$ test $\phi_{\mathfrak{P},\gamma}(\cdot)\in\{0, 1\}$ for given scale-invariant class $\mathfrak{P}$. 
\begin{enumerate}[label=\textbf{(T\arabic*)}]
    \item $\phi_{\mathfrak{P},\gamma}(\cdot)$ is a finite sample valid test with level $\gamma$ for the class $\mathfrak{P}$, i.e., if $W_j\in\mathfrak{P}$ for $1\le j\le B$ are independent random variables, then $\mathbb{E}[\phi_{\mathfrak{P},\gamma}(\{W_j\}_{j=1}^B)] \le \gamma$ for any $B \ge 1$. \label{assump:finite-sample-validity}
\end{enumerate}
In other words, for independent random variables $W_j\in\mathfrak{P}, 1\le j\le B$, $\phi_{\mathfrak{P},\gamma}(\{W_{j}\}_{j=1}^B)$ is stochastically dominated by Bernoulli$(\gamma)$ for any $\gamma\in(0, 1)$, denoted by $\phi_{\mathfrak{P},\gamma}(\{W_j\}_{j=1}^B)\preceq\mbox{Bernoulli}(\gamma)$.  We require $\phi_{\mathfrak{P},\gamma}(\cdot)\in\{0, 1\}$ and allow it to be a randomized test. (A real-valued random variable $X$ is said to be stochastically dominated by a real-valued random variable $Y$ ($X\preceq Y$) if $\mathbb{P}(X \le t) \ge \mathbb{P}(Y \le t)$ for all $t\in\mathbb{R}$.)
\begin{algorithm2e}[h]
    \caption{$\cosi{}$~framework for confidence sets}
    \label{alg:proposed-conf-int}
    \SetAlgoLined
    \SetEndCharOfAlgoLine{}
    \KwIn{Sample: $X_1,\cdots,X_n\overset{iid}{\sim} P$, confidence level: $1-\alpha$, estimation procedure $\mathcal{A}(\cdot)$ that takes a data set as input and return an estimator for $\theta(P)$, a scale-invariant property $\mathfrak{P}$ along with a test $\phi_{\mathfrak{P},\alpha}(\cdot)$, and integer $B \ge 1$.}
    \KwOut{Confidence set for $\theta(P)$.}
    Randomly split $\{1, 2, \ldots, n\}$ into $B$ disjoint batches $S_1, \ldots, S_B$.\;
    
    Compute estimators $\widehat{\theta}_j = \mathcal{A}(\{X_i: i\in S_j\}),$ for $1\le j\le B$.\;
    
    \Return the confidence sets 
    \begin{equation}\label{eq:confidence-interval-savari}
    \begin{split}
    \cosici{n}{\alpha}{a} ~&:=~ \left\{\theta:\, \phi_{\mathfrak{P},\alpha}(\{\widehat{\theta}_j - \theta\}_{j=1}^B) = 0\right\},\\
    \cosici{n}{\alpha}{m} ~&:=~ \left\{\theta:\, \sum_{j=1}^B \phi_{\mathfrak{P},1/2}(\widehat{\theta}_j - \theta) \le c_{B,\alpha}^*\right\},
    \end{split}
    \end{equation}
    where $c_{B,\alpha}^*$ is any natural number such that $\mathbb{P}(\mbox{Binom}(B, 1/2) \le c^*_{B,\alpha}) = 1 - \alpha$. (We allow $c_{B,\alpha}^*$ to be a random variable.) 
\end{algorithm2e}

Algorithm~\ref{alg:proposed-conf-int} provides a pseudo-code for constructing $\cosi{}$ confidence sets. Informally, we split the data into $B$ disjoint batches to compute $B$ independent estimators $\widehat{\theta}_j$. Then, we test for the scale-invariant property using the data $\{\widehat{\theta}_j - \theta\}_{j=1}^B$; with $\theta = \theta(P)$, the test should not reject with $1-\alpha$ probability. Under~\ref{assump:finite-sample-validity}, $\phi_{\mathfrak{P}, 1/2}(\cdot)$ is a level $1/2$ test with one observation which implies that
\[
\overline{\phi}_{\mathfrak{P},\alpha}(\{W_j\}_{j=1}^B) = \mathbf{1}\left\{\sum_{j=1}^B \phi_{\mathfrak{P}, 1/2}(W_j) > c_{B,\alpha}^*\right\},
\]
is a level $\alpha$ test with $B$ observations. This is the classical ``confidence boosting'' procedure, commonly used in robust/sub-Gaussian mean estimation~\citep{lerasle2011robust}. The designation $\mbox{-}\mathtt{a}$ and $\mbox{-}\mathtt{m}$ for the confidence sets in~\eqref{eq:confidence-interval-savari} stands for additive and multiplicate miscoverage bounds we prove below. Note that for $\cosi{m}$ confidence set to be non-trivial, the acceptance set of the test $\phi_{\mathfrak{P}, 1/2}(\cdot)$ should be non-trivial. 
In implementing Algorithm~\ref{alg:proposed-conf-int}, we would use approximately equal-sized batches, but it is not a requirement for validity. Moreover, our definition of $c^*_{B,\alpha}$ is incomplete because the support of $c^*_{B,\alpha}$ as a random variable is unspecified and can be arbitrary. The validity guarantees do not constrain the specification of $c^*_{B,\alpha}$ but for practical purposes, we recommend choosing $c^*_{B,\alpha}$ to be supported on two consecutive integers to satisfy the probability constraint. In the following subsections, we discuss validity of the confidence sets.

The following sections discuss some finite-sample tests for different scale-invariant properties, including zero median, symmetry, and unimodality. As can be expected, the more powerful the test $\phi_{\mathfrak{P},\alpha}(\cdot)$, the smaller the ``diameter'' of the reported confidence set. The literature on non-parametric optimal tests for several scale-invariant properties is scarce, especially for high-dimensional data. 
\subsection{Validity of~\SAV}\label{subsec:validity-of-savari}
We now provide finite-sample validity results for the $\cosi{}$~confidence sets. For this, we need to define a notion of distance.
For any random variable $W$, a scale-invariant class $\mathfrak{P}$, and a collection $\mathcal{C}$ of sets, define the distance between $W$ and $\mathfrak{P}$ as
\begin{equation}\label{def:dist-scale-invariance}
\mathrm{dist}_{\mathcal{C}}(W, \mathfrak{P}) = \inf_{L\in\mathfrak{P}}\sup_{C\in\mathcal{C}}\,\left(\mathbb{P}(W \in C) - \mathbb{P}(L\in C)\right)_+.
\end{equation}
When this distance is small $W$ approximately satisfies the scale-invariant property $\mathfrak{P}$, as measured by the sets in $\mathcal{C}$. This distance is always upper bounded by the Total Variation (TV) distance between the distribution of $W$ and the collection of distributions $\mathfrak{P}$. Observe that this distance can be strictly smaller than the (corresponding) distance to the limiting distribution even if the limiting distribution belongs to $\mathfrak{P}$. For example, if $X_1, \ldots, X_n$ are independent real-valued random variables from a symmetric location family, i.e., $X_i = \theta + \varepsilon_i$ where $\varepsilon_i$'s are symmetric around 0 random variables. Then $\widehat{\theta}_n = \sum_{i=1}^n X_i/n$ satisfies $\mbox{dist}_{\mathcal{C}}(\widehat{\theta}_n - \theta, \mathfrak{P}^{\text{sym}}) = 0$ for any $n$ and any collection $\mathcal{C}$, but the distance to the limiting distribution, whatever it maybe, can be converging arbitrarily slowly to zero. 

\begin{theorem}[Finite-sample Validity]\label{thm:finite-sample-validity}
    Fix any $B \ge 1$ and $\alpha\in(0, 1)$. For each $j \in \{1, 2, \ldots, B\}$ and $w_1, \ldots, w_{j-1}, w_{j+1}, \ldots, w_B$, set 
    \[
    C_j(w_1, \ldots, w_{j-1}, w_{j+1}, \ldots, w_B) = \{w_j:\, \phi_{\mathfrak{P},\alpha}(w_1, \ldots, w_B) = 1\}.
    \]
    Let $\mathcal{C}_j$ denote the collection of all sets of the form $C_j(w_1, \ldots, w_{j-1}, w_{j+1}, \ldots, w_B)$ over arbitrary choices of $w_1, \ldots, w_{j-1}, w_{j+1}, \ldots, w_B$. If Assumption~\ref{assump:finite-sample-validity} holds,
    then for any $n\ge1$, and any collection of distributions $\mathcal{P}$, we have
    \begin{equation}\label{eq:finite-sample}
    \sup_{P\in\mathcal{P}}\,\mathbb{P}_P\left(\theta(P)\notin\cosici{n}{\alpha}{a}\right) ~\le~ \alpha ~+~ \sup_{P\in\mathcal{P}}\,\sum_{j=1}^B \mathrm{dist}_{\mathcal{C}_j}(\widehat{\theta}_j - \theta(P), \mathfrak{P}),
    \end{equation}
    and
    \begin{equation}\label{eq:finite-sample-multiplicative}
    \sup_{P\in\mathcal{P}}\,\mathbb{P}_P\left(\theta(P)\notin\cosici{n}{\alpha}{m}\right) ~\le~ \alpha~\times~ \exp\left(2\sup_{P\in\mathcal{P}}\,\sum_{j=1}^B (\mathbb{E}_P[\phi_{\mathfrak{P},1/2}(\widehat{\theta}_j - \theta(P))] - 1/2)_+\right).
    \end{equation}
\end{theorem}
A proof of Theorem~\ref{thm:finite-sample-validity} is provided in Section~\ref{appsec:proof-of-finite-sample-validity}.
Taking $\mathcal{P}$ to be a singleton in Theorem~\ref{thm:finite-sample-validity} implies a pointwise validity result as a simple corollary. If $r_n(\widehat{\theta} - \theta(P))$ converges in distribution to $L_P\in\mathfrak{P}$ and $\mathcal{C}_j(\cdot)$'s are uniformity classes sets of $L_P$, then $\mathrm{dist}_{\mathcal{C}_j}(\widehat{\theta}_j - \theta(P), \mathfrak{P})$ converges to zero and hence, validity follows. For the precise definition of uniformity class of sets, see, for example,~\citet[Eqn. (2.2) and Corollary 2.6]{Bhattacharya86}.
If the test $\phi_{\mathfrak{P},\alpha}(\cdot)$ is permutation invariant, then $\mathcal{C}_1 = \mathcal{C}_2 = \ldots = \mathcal{C}_B$. It is worth noting that Theorem~\ref{thm:finite-sample-validity} holds true even if the class $\mathfrak{P}$ does not satisfy Definition~\ref{def:scale-invariance} as long as assumption~\ref{assump:finite-sample-validity} holds. Because of this, the Wald confidence sets can be seen as a special of Theorem~\ref{thm:finite-sample-validity} with $B = 1$, $\mathfrak{P} = \{L_P\}$, and the test being $\phi_{\mathfrak{P},\alpha}(W) = \mathbf{1}\{W \notin Q_{\alpha}\}$ for any set $Q_{\alpha}$ such that $\mathbb{P}(L_P\notin Q_{\alpha}) \le \alpha$; this requires the knowledge of $L_P$.

The fact that the distance used in~\eqref{eq:finite-sample} can be strictly smaller than the distance to the limiting distribution implies that if the $\cosi{}$~confidence sets are unreliable in practice when the limiting distribution belongs to $\mathfrak{P}$, then none of the classical Wald or resampling confidence sets can be reliable. For example, if the (two-sided) Wald confidence interval for $\theta(P)$ based on $\widehat{\theta}_n$ given by
\[
\widehat{\mathrm{CI}}_{n,\alpha}^{\mathrm{Wald}} := \left[\widehat{\theta}_n \pm \frac{\widehat{\sigma}_n}{r_n}z_{\alpha/2}\right],
\]
is valid for all $\alpha\in(0, 1)$ if and only if $r_n|\widehat{\theta}_n - \theta(P)|/\widehat{\sigma}$ converges in distribution to $|N(0, 1)|$. This, in particular, requires $\widehat{\sigma}_n$ to be a consistent variance estimator. On the other hand, such convergence in distribution already implies $|\widehat{\theta}_n - \theta(P)|$ is close to the scale-invariance class of all folded normal distributions (with mode at zero). Hence, $\cosi{}$ with this scale-invariance class is asymptotically valid as long as the Wald interval is asymptotically exactly valid. In other words, if $\cosi{}$ is asymptotically invalid, then the Wald interval is asymptotically invalid.

Similar to the classical Wald and resampling confidence sets, the miscoverage bound for $\cosi{a}$ in Theorem~\ref{thm:finite-sample-validity} is additive. This implies that whenever $\alpha$ is chosen too small (e.g., smaller in order than the rate of convergence of the sum of the distances to $\mathfrak{P}$), the miscoverage achieved is not of the same order as $\alpha$. In contrast, the $\cosi{m}$ confidence set has a multiplicative miscoverage error that yields order $\alpha$ miscoverage for a large range of $\alpha$'s. 

Theorem~\ref{thm:finite-sample-validity} implies that $B$ can diverge with the sample size as long as $\sum_{1\le j\le B}\mbox{dist}_{\mathcal{C}_j}(\cdot, \mathfrak{P})$ converges to zero. 
Theorem~\ref{thm:finite-sample-validity}, in its formulation, also relaxes the assumption of convergence in distribution of the estimators. Along different subsequences of sample sizes, the estimator can converge to different distributions as long as all these limiting distributions belong to $\mathfrak{P}.$ This fact can be seen by observing that the distance using~\eqref{def:dist-scale-invariance} can be bounded just in terms of the distribution of $W$. The following result illustrate this fact.
\begin{proposition}\label{prop:dist-charact-zero-median}
    If $W$ is a real-valued random variable and $\mathfrak{P}$ is the scale-invariance property of zero median, then setting $\mathcal{C}$ as the set of all Borel subsets of $\mathbb{R}$, we have
    \begin{equation}\label{eq:distance-zero-median}
    \mathrm{dist}_{\mathcal{C}}(W, \mathfrak{P}) ~=~ \left(\frac{1}{2} - \min\{\mathbb{P}(W \ge 0),\, \mathbb{P}(W \le 0)\}\right)_+.
    \end{equation}
\end{proposition}
See Section~\ref{appsec:proof-dist-charact-zero-median} for a proof.
In Section~\ref{appsec:proof-dist-charact-symmetry}, we provide a similar result for the scale-invariant property of central symmetry and angular symmetry.

\begin{remark}[Centering of Estimators.] 
    {\em In our discussion of the~$\cosi{}$ framework, we assumed that the estimators lie in a vector space allowing a proper interpretation of $\widehat{\theta} - \theta(P)$. In some cases, such a difference may not be meaningful. Examples include directional data or manifold data. In such cases, one might have a {\em known} function $R(\cdot, \cdot)$ such that $r_nR(\widehat{\theta}_n, \theta(P))$ converges in distribution to $L\in \mathfrak{P}$. All our methods and results follow for this setting as well by changing the confidence set in~\eqref{eq:confidence-interval-savari} to}
    \begin{equation}\label{eq:confidence-interval-savari-general}
    \begin{split}
        \cosici{n}{\alpha}{m} ~&:=~ \left\{\theta:\, \sum_{j=1}^B \phi_{\mathfrak{P},1/2}(R(\widehat{\theta}_j, \theta)) \le c^*_{B,\alpha}\right\}.
    \end{split}
    \end{equation}
    {\em Section~\ref{subsec:metric-space-mean} provides an illustration of our framework when the estimators lie in a metric space.}
\end{remark}

Theorem~\ref{thm:finite-sample-validity} also indicates that 
the $\cosi{}$ framework, in general, decouples the validity and efficiency/power of the confidence sets. Although we obtained validity guarantees for $\cosi{}$ confidence sets without the precise form of the test, it is unclear whether general results on the volume/diameter of the confidence sets can be obtained without further structure. Moreover, Theorem~\ref{thm:finite-sample-validity} already shows improved coverage guarantees for $\cosi{}$ confidence sets compared to Wald and resampling sets, but, interestingly, much stronger coverage guarantees can be obtained for specific scale-invariant properties. 

In the following, we first consider analogues of the HulC confidence interval with other scale-invariant classes on the real line. Then assuming a certain ``projection stability'' of the scale-invariant class, we find $\cosi{}$~confidence sets for parameters taking values in a general topological space.  

\section{$\cosi{}$ for real-valued functionals}\label{sec:univariate-HulC-like}
Let $\mathcal{D}(\mathbb{R})$ be the collection of all probability measures on the real line. Define 
\begin{equation}\label{eq:univariate-classes}
\begin{split}
    \mathfrak{m} &:= \{\mbox{zero median distributions in }\mathcal{D}(\mathbb{R})\},\\ 
    \mathfrak{s} &:= \{\mbox{symmetric around zero distributions in }\mathcal{D}(\mathbb{R})\},\\
    \mathfrak{u} &:= \{\mbox{unimodal distributions in }\mathcal{D}(\mathbb{R})\mbox{ with mode at zero}\},\\ 
    \mathfrak{n} &:= \{\mbox{zero mean Gaussian distributions in }\mathcal{D}(\mathbb{R})\}.
\end{split}
\end{equation}
We leave the definitions informal as their formal description can be obtained by specializing the examples in Section~\ref{sec:SAVARI}. Note that
\[
\mathfrak{m} ~\supsetneq~ \mathfrak{s} ~\supsetneq~ \mathfrak{n}\quad\mbox{and}\quad \mathfrak{u} ~\supsetneq~ \mathfrak{n}. 
\]
Hence, for instance, $\cosi{}$ confidence intervals constructed using the scale-invariance class $\mathfrak{m}$ work even when the limiting distribution is in $\mathfrak{s}$ or $\mathfrak{n}$. 

\subsection{Zero median: HulC}\label{subsec:HulC-recap}
If $W_1, \ldots, W_B$ are independent real-valued random variables with $W_j\in\mathfrak{m}$ for $1\le j\le B$, then the classical sign test yields a valid confidence interval. Formally, we have that $\sum_{j=1}^B \mathbf{1}\{W_j \le 0\}$ and $\sum_{j=1}^B \mathbf{1}\{W_j \ge 0\}$ both stochastically dominate Binom($B, 1/2$); see, e.g., (1.1) of~\cite{klenke2010stochastic}. Hence, 
\begin{align*}
&\min\left\{\mathbb{P}\left(\sum_{j=1}^B \mathbf{1}\{W_j \le 0\} \ge B - c^*_{B,\alpha}\right),\,\mathbb{P}\left(\sum_{j=1}^B \mathbf{1}\{W_j \ge 0\} \ge B - c^*_{B,\alpha}\right)\right\}\\
&\ge \mathbb{P}\left(\mbox{Binom}(B, 1/2) \ge B - c^*_{B,\alpha}\right) = \mathbb{P}(\mbox{Binom}(B,1/2) \le c^*_{B,\alpha})\\
&= 1 - \alpha.
\end{align*}
Here the last equality follows from the definition of $c^*_{B,\alpha}$ in Algorithm~\ref{alg:proposed-conf-int}. This holds for all $B \ge 1$. For $B \ge \log_2(1/\alpha)$, the sets can be written more succinctly in terms of the order statistics: $W_{(1)} \le W_{(2)} \le \cdots \le W_{(B)}$, with ties broken randomly. (Tie breaking is inconsequential for the intervals in this subsection.)
Formally, if $B \ge \log_2(1/\alpha)$, then $B - c^*_{B,\alpha} \ge 1$ almost surely, which implies that 
\begin{equation}\label{eq:under-zero-median-miscoverage}
\min\left\{\mathbb{P}\left(W_{(B - c^*_{B,\alpha})} \le 0\right),\, \mathbb{P}\left(0 \le W_{(c^*_{B,\alpha}+1)}\right)\right\} \ge 1 - \alpha.
\end{equation}
This yields one-sided confidence intervals for the zero median. It is not difficult to obtain miscoverage bounds when the zero median property is only approximately satisfied. From Proposition~\ref{prop:poisson-binomial-to-binomial}, it follows that, for $B \ge \log_2(1/\alpha)$ with $q_j = \mathbb{P}(W_j \le 0)$,
\begin{equation}\label{eq:one-sided-median-hulc}
\exp\left(-\frac{2\sum_{j=1}^B (q_j - 1/2)_+}{(1 - 2\max_j(q_j - 1/2))_+}\right)\le \frac{\mathbb{P}(W_{(B - c^*_{B,\alpha})} > 0)}{\alpha} \le \exp\left(2\sum_{j=1}^B (1/2 - q_j)_+\right).
\end{equation}
This is the two-sided version miscoverage bound~\eqref{eq:finite-sample-multiplicative} and reduces to~\eqref{eq:under-zero-median-miscoverage} when $W_j\in\mathfrak{m}$.
Because zero median is also sign-invariant, i.e., $W_j\in\mathfrak{m}$ implies $-W_j\in\mathfrak{m}$, we can create a two-sided interval which yields a better miscoverage bound. Using the stochastic dominance result above, Proposition~\ref{prop:binomial-probabilities} implies that, whenever $B \ge \log_2(2/\alpha)$,
\begin{equation}\label{eq:two-sided-median-HulC}
\exp\left(-\frac{4\sum_{j=1}^B \Delta_j^2}{(1 - 4\max_j \Delta_j^2)_+}\right)\le \frac{\mathbb{P}(0\notin[W_{(B - c^*_{B,\alpha/2})},\, W_{(c^*_{B,\alpha/2}+1)}])}{\alpha} \le \exp\left(2(B\Delta)^2\right),
\end{equation}
where
\begin{align*}
    \Delta &:= \frac{1}{B}\sum_{j=1}^B \left(\frac{1}{2} - \min\{\mathbb{P}(W_j \ge 0),\,\mathbb{P}(W_j \le 0)\}\right)_+,\\
    \Delta_j^2 &:= \left(\mathbb{P}(W_j > 0) - \frac{1}{2}\right)^2 + \mathbb{P}(W_j = 0).
\end{align*}
Note that the miscoverage error in~\eqref{eq:two-sided-median-HulC} is second-order in $\Delta$ whereas the error in~\eqref{eq:one-sided-median-hulc} is first-order. On the other hand, the requirement on $B$ shifted from $B \ge \log_2(1/\alpha)$ to $B \ge \log_2(2/\alpha) = 1 + \log_2(1/\alpha)$.

Applying these inequalities to $W_j = \widehat{\theta}_j - \theta(P)$ for independent estimators $\widehat{\theta}_j, 1\le j\le B$ yields one-sided and two-sided HulC intervals discussed in~\cite{kuchibhotla2024hulc} and~\cite{paul2024inference}:
\begin{align*}
\mathbb{P}\left(\theta(P)\notin[\widehat{\theta}_{(B - c^*_{B,\alpha})},\,\infty)\right) ~&\le~ \alpha\times\exp\left(2B\Delta_{P,1}^{\mathfrak{m}}\right),\\
\mathbb{P}\left(\theta(P)\notin\widehat{\mathrm{CI}}_{n,\alpha}^{\mathfrak{m}}\right) ~&\le~ \alpha\times\exp\left(2(B\Delta_{P,2}^{\mathfrak{m}})^2\right),
\end{align*}
where 
\begin{align*}
\widehat{\mathrm{CI}}_{n,\alpha}^{\mathfrak{m}} &:= [\widehat{\theta}_{(B - c^*_{B,\alpha/2})},\,\widehat{\theta}_{(c^*_{B,\alpha/2} + 1)}],\\
\Delta_{P,1}^{\mathfrak{m}} &= \frac1B\sum_{j=1}^B \left(\frac12 - \mathbb{P}(\widehat{\theta}_j \le \theta(P))\right)_+\\
\Delta_{P,2}^{\mathfrak{m}} &= \frac1B\sum_{j=1}^B \left(\frac{1}{2} - \min\{\mathbb{P}(\widehat{\theta}_j \le \theta(P)),\,\mathbb{P}(\widehat{\theta}_j \ge \theta(P))\}\right)_+,
\end{align*}
represent the one-sided and two-sided median biases of the estimators $\widehat{\theta}_j, 1\le j\le B$ around $\theta(P)$.
As discussed in~\cite{kuchibhotla2024hulc}, $\Delta_{P,k}^{\mathfrak{m}}, k = 1, 2$ converge to zero for a large class of problems even when asymptotic normality does not hold. The width of this confidence interval was also discussed in that paper. 

\subsection{Symmetric around zero}\label{subsec:symmetry}
If $W_1, \ldots, W_B$ are independent real-valued random variables with $W_j\in\mathfrak{s}$ for $1\le j\le B$, then the Wilcoxon signed rank test yields a valid confidence interval. This is based on the fact that conditional on $(|W_1|, \ldots, |W_B|)$, the vector of signs of $W_j$ are independent Rademacher random variables. Formally, if $W_1, \ldots, W_B$ are independent and continuously distributed with $W_i\in\mathfrak{s}$, then
\[
\sum_{i=1}^B R_i\mathbf{1}\{W_i \ge 0\} ~\overset{d}{=}~ \sum_{i=1}^B i\xi_i\quad\mbox{where}\quad \xi_1, \ldots, \xi_B\overset{\mathrm{iid}}{\sim}\mbox{Ber}(1/2).
\]
In fact, the equality of the distributions holds conditional on $|W_1|, \ldots, |W_B|$. Hence, using the quantiles of $\sum_{i=1}^B i\xi_i$, one can construct a finite-sample valid confidence interval for $0$. This can be generalized by transforming the ranks by an arbitrary monotonic function. Formally, if $h:\{0, 1, \ldots, B\}\to\mathbb{R}$ satisfies
\begin{equation}\label{eq:non-decreasing-weights}
0 = h(0) < h(1) \le \cdots \le h(B),
\end{equation}
then, for independent and continuously distributed random variables $W_1, \ldots, W_B$ with $W_j\in\mathfrak{s}$ for $1\le j\le B$, we have
\begin{equation}\label{eq:general-signed-rank-statistic}
\sum_{i=1}^B h(R_i)\mathbf{1}\{W_i \ge 0\} ~\overset{d}{=}~ \sum_{i=1}^B h(i)\xi_i\quad\mbox{where}\quad \xi_1, \ldots, \xi_B\overset{\mathrm{iid}}{\sim}\mbox{Ber}(1/2).
\end{equation}
The usefulness of the transformation is understood by considering two special cases: (1) if $h(j) = h(i)$ for all $0 < i < j$, then the resulting confidence interval is valid for all $W_1, \ldots, W_B\in\mathfrak{m}$; this is equivalent to the sign test, and (2) if $h(i) = \Phi^{-1}(1/2 + i/(2(B+1)))$ for $i\in\{0, 1, \ldots, B\}$ (i.e., van der Waerden score), then the resulting confidence interval is asymptotically equivalent to the Wald interval (as $B\to\infty$), and hence, efficient when $W_1, \ldots, W_B\in\mathfrak{n}$. Construction of transformations $h(\cdot)$ so as to attain asymptotic efficiency for different location-scale families is discussed in~\cite{fraser1957most} and~\cite{hajek1962asymptotically}.

For exact validity under potentially discrete distributions of $W_1, \ldots, W_B$, we make use of randomization. Formally, let $U_1, \ldots, U_B$ be a sequence of independent $U(-1, 1)$ random variables independent of $W_1, \ldots, W_B$. Equipping $\mathbb{R}^2$ with lexicographic ordering, let the augmented order statistics be
\[
(W_{[1]}, U_{[1]}) \prec (W_{[2]}, U_{[2]}) \prec \cdots \prec (W_{[B]}, U_{[B]}),
\]
and let the randomized sign of $W_i$ be
\[
S_i := \mathbf{1}\{(W_i, U_i) \succ (0, 0)\} - \mathbf{1}\{(W_i, U_i)\prec (0, 0)\} \in \{\pm 1\}.
\]
The fact that $\mathbb{P}(S_i = 0) = 0$ follows from the fact that $\mathbb{P}((W_i, U_i) = (t, 0)) = 0$ for all $t\in\mathbb{R}$ including $t = 0$. (Note that the augmented order statistics are all distinct with probability one, because $\mathbb{P}(U_i = U_j) = 0$ for all $i \neq j$.)
Define the (augmented) Walsh averages as 
\[
A_{ij} ~:=~ \left(\frac{W_{[i]} + W_{[j]}}{2},\, \frac{U_{[i]} + U_{[j]}}{2}\right),\quad\mbox{for}\quad 1 \le i \le j\le B,
\]
and weights as
\[
\omega_{ij} ~:=~ h(j-i+1) - h(j-i)\quad\mbox{for}\quad 1\le i \le j\le B.
\]
From~\eqref{eq:non-decreasing-weights}, it follows that $\omega_{ij} \ge 0$ for all $i \le j$. 
Also, it is easy to see that
\[
\sum_{1\le i \le j\le B} \omega_{ij} = \sum_{r=1}^B h(r) ~=:~ N_h.
\]
Following the arguments of~\cite{bauer1972constructing} (also, see Theorem 2.3.1 of~\cite{Hettamansperger}), it can be proved that
\begin{equation}\label{eq:Walsh-average-representation-of-signed-rank}
\sum_{1\le i\le j\le B} \omega_{ij}\mathbf{1}\{A_{ij} \succ (0, 0)\} = \sum_{i=1}^B h(R_i)\mathbf{1}\{(W_i, U_i) \succ (0,0)\},
\end{equation}
where $R_i$ is the (lexicographic) rank of $(|W_i|, S_iU_i)$ among $(|W_j|, S_jU_j), 1\le j\le B$; cf.~\eqref{eq:general-signed-rank-statistic}. Define
\[
G_{B,h} := \sum_{i=1}^B h(i)\xi_i,\quad\mbox{where}\quad \xi_1, \ldots, \xi_B ~\overset{\mathrm{iid}}{\sim}~ \mbox{Ber}(1/2).
\]
Let $\mathcal{S}_h = \{-1\}\cup(\mathrm{supp}(G_{B,h})\cap[0, N_h/2]),$
and
\[
P_h(c) = \mathbb{P}\left(\left|G_{B,h} - \frac{N_h}{2}\right| \ge \frac{N_h}{2} - c\right)\quad\mbox{for}\quad c\in\mathcal{S}_h.
\]
(We set $P_h(-1) = 0$.) Given $\alpha\in(0, 1)$, let $c_+ := \min\{c\in\mathcal{S}_h:\, P_h(c) \ge \alpha\}.$
Letting $c_- \le c_+$ be the predecessor of $c_+$ in $\mathcal{S}_h$ so that $P_h(c_-) < \alpha \le P_h(c_+)$, 
and independently of all data and $U_1, \ldots, U_B$, let
\begin{equation}\label{eq:quantile-definition-Wilcoxon-h}
\kappa_{B,\alpha}^h := \begin{cases}
    c_+, &\mbox{ with probability }\tau_{h,\alpha},\\
    c_-, &\mbox{ with probability }1 - \tau_{h,\alpha},
\end{cases}\quad\mbox{with}\quad \tau_{h,\alpha} := \frac{\alpha - P_h(c_-)}{P_h(c_+) - P_h(c_-)}.
\end{equation}
Define weighted Walsh average quantile as
\begin{align*}
    Q_h(\gamma) := \min\left\{z\in\{A_{lk}:\,1\le l\le k\le B\}:\,\sum_{1\le i\le j\le B}\,\omega_{ij}\mathbf{1}\{A_{ij} \succ z\} \le N_h - \gamma\right\}\quad\mbox{for}\quad \gamma\in(0, N_h],
\end{align*}
with the convention $Q_h(\gamma) = (-\infty, 0)$ if $\gamma \le 0$ and $Q_h(\gamma) = (\infty,0)$ if $\gamma > N_h$. Here the minimum is computed with respect to the lexicographic ordering.
Also, define $Q_h(\gamma+) = \lim_{\varepsilon\downarrow0}\,Q_h(\gamma + \varepsilon).$ If $\omega_{ij}$'s are non-negative integers and $\gamma$ is an integer, then $Q_h(\gamma+) = Q_h(\gamma + 1)$. Finally, define the confidence interval based on signed (transformed) rank statistic as
\begin{equation}\label{eq:general-Wilcoxon-signed-rank-test}
    \widehat{\mathrm{CI}}_{n,\alpha}^{\mathtt{SR}} := \{t\in\mathbb{R}:\, Q_h(\kappa_{B,\alpha}^h+) \prec (t, 0) \prec Q_h(N_h - \kappa_{B,\alpha}^{h})\}.
\end{equation}
It is worth noting that the external randomness from $U_1, \ldots, U_B$ is essentially used only to break ties properly and in the context of the confidence interval, the lexicographic ordering only dictates whether the confidence interval $\widehat{\mathrm{CI}}_{n,\alpha}^{\mathtt{SR}}$ is closed or either side open. It can be verified that
\begin{equation}\label{eq:sign-rank-superset}
\widehat{\mathrm{CI}}_{n,\alpha}^{\mathtt{SR}} \subseteq \left[Q_{h,1}(\kappa_{B,\alpha}^{h}+),\, Q_{h,1}(N_h - \kappa_{B,\alpha}^h)\right],
\end{equation}
where $Q_{h,1}(\gamma)$ is the first coordinate of $Q_h(\gamma)$.
Then Proposition~\ref{prop:Wilcoxon-signed-rank-test} implies that $\widehat{\mathrm{CI}}_{n,\alpha}^{\mathtt{SR}}$ satisfies
\begin{equation}\label{eq:miscoverage-gurantee-Wilcoxon-sign}
\exp\left(-\frac{B^4\Delta^2}{2(1 - 7B^4\Delta^2/12)_+}\right) \le \frac{\mathbb{P}(0 \notin \widehat{\mathrm{CI}}_{n,\alpha}^{\mathtt{SR}})}{\alpha} \le \exp\left(\frac{B^4\Delta^2}{2}\right),
\end{equation}
where
\begin{equation}\label{eq:error-metric-symmetric}
\Delta = \frac{1}{B}\sum_{j=1}^B \Delta_i,\quad\mbox{with}\quad \Delta_i := \min\left\{1, 2\sup_{u\in\mathbb{R}}|\mathbb{P}(W_i \le u) - \mathbb{P}(-W_i \le u)|\right\}.
\end{equation}
It follows that the superset in~\eqref{eq:sign-rank-superset} also satisfies the upper bound on miscoverage, but the lower bound does not hold. A lower bound miscoverage of this superset involves probabilities of ties among $W_1, \ldots, W_B$, similar to how the lower bound on miscoverage in~\eqref{eq:two-sided-median-HulC} involves $\mathbb{P}(W_j = 0)$.

Applying these inequalities to $W_j = \widehat{\theta}_j - \theta(P)$ for independent estimators $\widehat{\theta}_j, 1\le j\le B$ yields a two-sided $\cosi{}$ intervals under approximate symmetry: Define
\begin{align*}
\widehat{A}_{ij} &:= \left(\frac{\widehat{\theta}_{[i]} + \widehat{\theta}_{[j]}}{2},\,\frac{U_{[i]}+ U_{[j]}}{2}\right),\quad 1\le i \le j\le B,\\
\widehat{Q}_h(\gamma) &:= \inf\left\{z\in\{\widehat{A}_{lk}:1\le l\le k\le B\}:\,\sum_{1\le i\le j\le B}\omega_{ij}\mathbf{1}\{\widehat{A}_{ij} \succ z\} \le N_h - \gamma\right\}.
\end{align*}
Then
\begin{equation}\label{eq:COSI-under-symmetry}
\mathbb{P}\left(\theta(P)\notin\widehat{\mathrm{CI}}_{n,\alpha}^{\mathfrak{s}}\right) ~\le~ \alpha\times\exp(B^4(\Delta_P^{\mathfrak{s}})^2/2),    
\end{equation}
where
\begin{equation}\label{eq:cosi-under-symmetry-CI}
\begin{split}
\widehat{\mathrm{CI}}_{n,\alpha}^{\mathfrak{s}} &:= \{t\in\mathbb{R}:\, \widehat{Q}_h(\kappa_{B,\alpha}^h+) \prec (t, 0) \prec \widehat{Q}_h(N_h - \kappa_{B,\alpha}^h)\},\\
\Delta_P^{\mathfrak{s}} &:= \frac{1}{B}\sum_{j=1}^B \min\left\{1,\,2 \sup_{u\in\mathbb{R}}|\mathbb{P}(\widehat{\theta}_j - \theta(P) \le u) - \mathbb{P}(\theta(P) - \widehat{\theta}_j \le u)|\right\}.
\end{split}
\end{equation}
Finally, we note that using symmetry of limiting distribution to construct confidence intervals for parameters was discussed in~\cite{canay2017randomization}. However, their results only prove asymptotic validity of the confidence interval without a finite-sample result. We stress here that while a simple additive miscoverage error can be proved under approximate symmetry, for arbitrary confidence intervals that are valid under symmetry, the additive error depends on the total variation distance between the distributions of $W_i$'s and $\mathfrak{s}$. Under certain smoothness assumptions on the confidence interval, the additive error depends on the Kolmogorov-Smirnov distance between $W_i$'s and $\mathfrak{s}$; multiplicative miscoverage bounds as proved for weighted sign rank tests are non-trivial and to what extent they apply to other families of confidence intervals is unclear. 

\subsection{Mean zero normality}
If $W_1, \ldots, W_B$ are independent {\em and identically} distributed real-valued random variables with $W_j\in\mathfrak{n}$ for $1\le j\le B$, then 
\[
\frac{B^{1/2}\overline{W}_B}{\sqrt{(B-1)^{-1}\sum_{j=1}^B (W_j - \overline{W}_B)^2}} \sim t_{B-1},\quad\mbox{where}\quad \overline{W}_B = \frac{1}{B}\sum_{j=1}^B W_j.
\]
Using the symmetry of $t$-distribution, this implies
\[
\mathbb{P}\left(0 \in \left[\overline{W}_B \pm t_{B-1,\alpha/2}\sqrt{\frac{1}{B(B-1)}\sum_{j=1}^B (W_j - \overline{W}_B)^2}\right]\right) = 1 - \alpha.
\]
This is, unfortunately, false if $W_1, \ldots, W_B$ are independent but non-identically distributed with $W_j\in\mathfrak{n}$ for all $j$. \cite{ibragimov2010tstatistic} consider mean zero normality of limiting distribution to perform inference for parameters and they rely on the conservative bounds of~\cite{Bakirov2005} to provide asymptotically valid (but conservative) confidence interval for small enough $\alpha.$ In general, from Theorem 1.1 of~\cite{Bentkus_2015} and Theorem 1.1 of~\cite{Pinelis2012GaussianRademacher}, we conclude that for all $\alpha\in(0, 1)$ and $W_1, \ldots, W_B\in\mathfrak{s}$, it follows that 
\begin{equation}\label{eq:general-tail-bound-self-normalized}
\mathbb{P}\left(\frac{\sum_{j=1}^B W_j}{\sqrt{\sum_{j=1}^B W_j^2}} > t\right) \le (1 - \Phi(t))\min\left\{3.2,\, 1 + \frac{14.11}{t}\right\},\quad\mbox{for all}\quad t \ge 0
\end{equation}
Given~\eqref{eq:general-tail-bound-self-normalized}, one could set $t$ (conservatively) so that the right hand side is $\alpha/2$. Even then, for arbitrary independent random variables $W_1, \ldots, W_B$ that only approximately belong to $\mathfrak{s}$ or $\mathfrak{n}$, the miscoverage error is additive in $\alpha$ and the Kolmogorov-Smirnov distance. 
The use of Student's $t$-statistic is justified by its optimality as $B\to\infty$ when $W_1, \ldots, W_B\in\mathfrak{n}$; see Section 4 of~\cite{ibragimov2010tstatistic}. Alternatively, it is easy to see that the ratio of the length of this $t$-statistic based confidence interval converges to that of the classical Wald interval converges to one as $B\to\infty$. 

For exactness under potential heterogeneous distributions of $W_1, \ldots, W_B\in\mathfrak{n}$ and to retain asymptotic efficiency (in comparison to the Wald interval), we propose the use of signed rank test with van der Waerden scores. Formally, as in Section~\ref{subsec:symmetry}, set
\[
A_{ij} = \left(\frac{W_{[i]} + W_{[j]}}{2},\, \frac{U_{[i]} + U_{[j]}}{2}\right),\quad 1\le i\le j\le B,
\]
with weights
\[
\omega_{ij}^{\mathfrak{n}} = \Phi^{-1}\left(\frac{1}{2} + \frac{j-i+1}{2(B+1)}\right) - \Phi^{-1}\left(\frac{1}{2} + \frac{j-i}{2(B+1)}\right),\quad\mbox{for}\quad 1\le i\le j\le B.
\]
Set $N_{\mathfrak{n}} = \sum_{1\le i\le j\le B}\omega_{ij}^{\mathfrak{n}}$, and
\[
Q_{\mathfrak{n}}(\gamma) := \min\left\{z\in\{A_{lk}: 1\le l\le k\le B\}:\, \sum_{1\le i\le j\le B}\omega_{ij}^{\mathfrak{n}}\mathbf{1}\{A_{ij}\succ z\} \le N_{\mathfrak{n}} - \gamma\right\}.
\]
Let $\kappa_{B,\alpha}^{\mathfrak{n}}$ be $\kappa_{B,\alpha}^h$ defined in~\eqref{eq:quantile-definition-Wilcoxon-h} with $h(i) = \Phi^{-1}(1/2 + i/(2(B+1)))$ for $i\in\{0, 1, \ldots, B\}$.
Then, setting,
\[
\widehat{\mathrm{CI}}_{n,\alpha}^{\mathfrak{n}} := \left\{t\in\mathbb{R}:\, Q_{\mathfrak{n}}(\kappa_{B,\alpha}^{\mathfrak{n}}+) \prec (t, 0)\prec Q_{\mathfrak{n}}(N_{\mathfrak{n}} - \kappa_{B,\alpha}^{\mathfrak{n}})\right\},
\]
we conclude from~\eqref{eq:miscoverage-gurantee-Wilcoxon-sign}
\[
\mathbb{P}\left(0 \notin \widehat{\mathrm{CI}}_{n,\alpha}^{\mathfrak{n}}\right) \le \alpha\times\exp\left(B^4\Delta^2/2\right),
\]
where $\Delta$ is as in~\eqref{eq:error-metric-symmetric}. Note that $\Delta_i$ defined in~\eqref{eq:error-metric-symmetric} satisfies
\[
\Delta_i ~\le~ \inf_{\sigma \ge 0}\,\min\left\{1, 4\sup_{u\in\mathbb{R}}|\mathbb{P}(W_i \le u) - \Phi(u/\sigma)|\right\}.
\]
The resulting $\cosi{}$ confidence interval for $\theta(P)$ based on estimators $\widehat{\theta}_1, \ldots, \widehat{\theta}_B$ can be obtained similarly to~\eqref{eq:cosi-under-symmetry-CI} with coverage guarantee as in~\eqref{eq:COSI-under-symmetry}.

\subsection{Unimodality with Mode at zero}\label{sec:unimodal}
Recall that a real-valued random variable $W$ is unimodal with mode at zero if the CDF of $W$ is convex on $(-\infty, 0]$ and concave on $[0, \infty)$. \cite{abbott1962two} proposed a confidence interval for the mode of a unimodal distribution based on one observation; see, also,~\cite{BlachmanMachol1987},~\cite{Edelman1990},~\cite{Andel1991}, and~\cite{Portnoy2019} for some historical context. If $W$ is a continuously distributed unimodal random variable with mode at zero, then 
\[
\mathbb{P}\left(W - \left(\frac{2}{\gamma} - 1\right)|W - a| \le 0 \le W + \left(\frac{2}{\gamma} - 1\right)|W - a|\right) \ge 1 - \gamma\quad\mbox{for all}\quad \gamma\in(0, 1), a\in\mathbb{R}.
\]
See the main result (A.1) of~\cite{Edelman1990}. Following the proof, it can be realized that unimodality and continuity of the distribution can be replaced by a weaker condition. Define the logarithmic L{\'e}vy concentration function as
\begin{equation}\label{eq:logarithmic-Levy-concentration}
\lQ_{W}(\varepsilon) := \sup_{t \in \mathbb{R}, \eta\in\{\pm 1\}}\,\mathbb{P}(\eta W > 0,\, \ln(\eta W)\in (t, t + \varepsilon)),\quad\mbox{for}\quad \varepsilon > 0.
\end{equation}
From Theorem 2.2 of~\cite{Andel1991}, it follows that if $W$ is a continuous unimodal random variable with mode at zero, then 
\begin{equation}\label{eq:unimodality-implication}
\lQ_W(\varepsilon) \le 1 - e^{-\varepsilon},\quad\mbox{for all}\quad \varepsilon > 0.
\end{equation}
Following the proof of Theorem 2.1 of~\cite{Andel1991}, we get that for any random variable $W$, 
\begin{equation}\label{eq:generalized-Andel-inequality}
    \sup_{a\in\mathbb{R}}\,\mathbb{P}\left(0 \notin \left[W \pm \left(2/\gamma - 1\right)|W - a|\right]\right) ~=~ \lQ_W\left(-\ln(1-\gamma)\right).
\end{equation}
For completeness, we provide a proof of these two facts in Section~\ref{appsec:proof-of-generalized-Andel}. It is interesting to note that the collection of real-valued random variables $W$ satisfying~\eqref{eq:unimodality-implication} is strictly bigger than $\mathfrak{u}$ and it even contains some multimodal distributions. In fact, the random variable $Y = Re^{\sigma Z}$, where $\mathbb{P}(R = 1) = \mathbb{P}(R = -1) = 1/2$ and $Z\sim N(0, 1)$, with $\sigma \ge 1/\sqrt{2\pi}$ satisfies~\eqref{eq:unimodality-implication}, has two modes, but a zero density at zero. 

Given these inequalities, we propose a larger scale-invariant class of distributions
\[
\mathfrak{u}^{\star}
:=
\left\{
P\in\mathcal{D}(\mathbb{R}):
\,\lQ_W(\ln 2)\le 1/2
\mbox{ whenever }W\sim P
\right\}.
\]
Note that the distributions in~$\mathfrak{u}^{\star}$ only need to
satisfy~\eqref{eq:unimodality-implication} at $\varepsilon=\ln 2$.
There exist several distributions that belong to
$\mathfrak{u}^{\star}$ but do not satisfy
\eqref{eq:unimodality-implication} for all $\varepsilon>0$.
In fact, we have
$\mathfrak{m}\subsetneq\mathfrak{u}^{\star}$.
The inclusion follows because
\[
\lQ_W(\varepsilon)
\le
\max\{\mathbb{P}(W>0),\,\mathbb{P}(W<0)\}
\quad\mbox{for all }\varepsilon>0.
\]
For strictness, consider a random variable satisfying
\[
\mathbb{P}(W=0)=\frac25,
\qquad
\mathbb{P}(W=1)=\mathbb{P}(W=3)=\frac3{10}.
\]
No interval $(t,2t)$ with $t>0$ contains both $1$ and $3$, so
$\lQ_W(\ln 2)=3/10$, while zero is not a median of $W$.
Moreover, if $W\in\mathfrak{u}^{\star}$, then
$W_+,W_-\in\mathfrak{u}^{\star}$, where
$W_+=\max\{W,0\}$ and $W_-=\max\{-W,0\}$.
Every nondegenerate such part is asymmetric around zero; either part
can be degenerate even when $W$ is nondegenerate.

If $W_1,\ldots,W_B\in\mathfrak{u}^{\star}$ are independent random
variables, then, for every fixed $a\in\mathbb{R}$, it follows
from~\eqref{eq:generalized-Andel-inequality} that
\begin{equation}\label{eq:generalized-Andel-inequality-implication}
\mathbb{E}\!\left[
\mathbf{1}\{0\notin[W_j\pm3|W_j-a|]\}
\right]
\le\frac12.
\end{equation}
In fact, one can slightly improve this interval in length using the confidence interval of~\cite{BlachmanMachol1987}. Formally, we have
\[
\mathbb{E}[\mathbf{1}\{0\notin[\min\{W_j, a\} - (1+\sqrt{2})|W_j - a|,\, \max\{W_j, a\} + (1+\sqrt{2})|W_j - a|]\}] \le \frac{1}{2}.
\]
This interval has length $(3 + 2\sqrt{2})|W_j - a|$ compared to the length $6|W_j - a|$ of confidence interval in~\eqref{eq:generalized-Andel-inequality-implication}.
Hence, applying $\cosi{m}$ method, we get
\begin{equation}\label{eq:COSI-for-unimodal*}
\widehat{\mathrm{CI}}_{n,\alpha}^{\mathfrak{u}^\star}(a) := \left\{\theta:\,\sum_{j=1}^B \mathbf{1}\{\theta \notin [(W_j\wedge a,\, W_j\vee a) \pm (1+\sqrt{2})|W_j - a|]\} \le c^*_{B,\alpha}\right\}, 
\end{equation}
and from Theorem~\ref{thm:finite-sample-validity} that
\begin{equation}\label{eq:coverage-guarantee-unimodal*}
\mathbb{P}\left(0\notin \widehat{\mathrm{CI}}_{n,\alpha}^{\mathfrak{u}^\star}(a)\right) \le \alpha\times\exp\left(2\sum_{j=1}^B (\lQ_{W_j}(\ln 2) - 1/2)_+\right).
\end{equation}

The confidence interval in~\eqref{eq:COSI-for-unimodal*} is not shift
equivariant if the auxiliary choice $a$ is held fixed.
Although it is implicitly defined
in~\eqref{eq:COSI-for-unimodal*}, it can be expressed in closed form.
For $1\le j\le B$, define
\begin{equation}\label{eq:L-j-U-j-unimodal}
L_j(a) = \min\{W_j,a\}-(1+\sqrt2)|W_j-a|,
\qquad U_j(a) = \max\{W_j,a\}+(1+\sqrt2)|W_j-a|.
\end{equation}
Further, let the order statistics of these quantities be
\[
-\infty = L_{(0)}(a) < L_{(1)}(a)\le\cdots\le L_{(B)}(a)
\quad\mbox{and}\quad
U_{(1)}(a)\le\cdots\le U_{(B)}(a) \le U_{(B+1)}(a) = \infty.
\]
Then
\[
\widehat{\mathrm{CI}}_{n,\alpha}^{\mathfrak{u}^{\star}}(a) = \left[L_{(B-c^*_{B,\alpha})}(a),\,U_{(c^*_{B,\alpha}+1)}(a)\right].
\]
Alternatively, we can remove the auxiliary choice $a$ by intersecting
the intervals over all $a\in\mathbb{R}$. This requires a cutoff calibrated for the entire intersection. Define
\begin{equation}\label{eq:COSI-for-unimodal*-optimized}
\begin{aligned}
\widehat{\mathrm{CI}}_{n,\alpha}^{\mathfrak{u}^{\star}}
&:=
\bigcap_{a\in\mathbb{R}}
\left[
L_{(B-1-c^*_{B-1,\alpha/B})}(a),
\,
U_{(c^*_{B-1,\alpha/B}+2)}(a)
\right]\\
&=
\left[\max_{i\ge1}\,\{2W_{(i)}-W_{(i+1+c^*_{B-1,\alpha/B})}\},
\, \min_{i\ge1} \{2W_{(i+1+c^*_{B-1,\alpha/B})}-W_{(i)}\}\right].
\end{aligned}
\end{equation}
Here we use the convention that $W_{(j)} = -\infty$ if $j \le 0$ and $W_{(j)} = \infty$ for $j \ge B$.
When $c^*_{B-1,\alpha/B}=B-1$, define the interval to be $\mathbb{R}$. In contrast to the confidence interval in~\eqref{eq:COSI-for-unimodal*}, taking the intersection over all $a\in\mathbb{R}$ yields a shift and scale equivariant confidence interval. The resulting confidence interval in~\eqref{eq:COSI-for-unimodal*-optimized} is also similar to the one studied in~\cite{lanke1974interval} and~\cite{kuchibhotla2024hulc}. When $c^*_{B-1,\alpha/B}=B-2$, the interval~\eqref{eq:COSI-for-unimodal*-optimized}
reduces to $\left[2W_{(1)}-W_{(B)},\,2W_{(B)}-W_{(1)}\right],$ which is precisely the unimodal HulC confidence interval studied in~\cite{kuchibhotla2024hulc}. With the revised class of distributions $\mathfrak{u}^{\star}$, one can improve Theorem~4 of~\cite{kuchibhotla2024hulc}. It can be proved that
\begin{equation}\label{eq:unimodal-HulC-guarantee}
    \mathbb{P}(0\notin\widehat{\mathrm{CI}}_{n,\alpha}^{\mathfrak{u}^\star}) ~\le~ \alpha\times\exp\left(2\sum_{j=1}^{B_{\alpha}} (\lQ_{W_j}(\ln 2) - 1/2)_+\right). 
\end{equation}
See Section~\ref{appsec:proof-unimodal-HulC} for a proof. From the reasoning for
$\mathfrak{m}\subsetneq\mathfrak{u}^{\star}$, we have
\[
\sum_{j=1}^B
\bigl(\lQ_{W_j}(\ln 2)-1/2\bigr)_+
\le
\sum_{j=1}^B
\left(
\frac12-
\min\{\mathbb{P}(W_j\ge0),\,\mathbb{P}(W_j\le0)\}
\right)_+.
\]
Additionally, the left-hand side can also be controlled by the distance
of the laws of $W_1,\ldots,W_B$ to $\mathfrak{u}^{\star}$ with respect
to the version of Kolmogorov--Smirnov distance defined over the
collection of all open intervals.
Specifically, if $d_{\mathrm{int}}(P,Q) := \sup_{r<s}|P((r,s))-Q((r,s))|,$ then
\[
\bigl(\lQ_{W_j}(\ln 2)-1/2\bigr)_+
\le
\inf_{Q\in\mathfrak{u}^{\star}}
d_{\mathrm{int}}\bigl(\mathcal{L}(W_j),Q\bigr).
\]
This discrepancy is at most twice the usual Kolmogorov--Smirnov
distance and at most the total variation distance.
Regarding the width, unlike the confidence intervals under zero median or symmetry or
normality, the confidence intervals under unimodality presented above
need not shrink in length to zero as $B\to\infty$ when the distribution
of each $W_j$ remains fixed.
The confidence interval
in~\eqref{eq:generalized-Andel-inequality-implication} has a diameter of
$6|W_j-a|$, which remains on the scale of an individual observation. The same holds true for the confidence interval in~\eqref{eq:COSI-for-unimodal*-optimized}.

For better finite-sample valid confidence intervals under ordinary
unimodality, with shrinking widths under suitable density regularity,
see~\cite{paul2025finite}.
For better confidence intervals under zero median
or symmetry and unimodality, see~\cite{SatheShenoy1979}
and~\cite{Birnbaum1970}.
The constructions there are more complicated and use assumptions
beyond membership in $\mathfrak{u}^{\star}$.
For our purposes of applying these confidence intervals in the context
of $\cosi{}$, the constructions above and below suffice.

Applying these guarantees to $W_j = \widehat{\theta}_j - \theta(P)$ for independent estimators $\widehat{\theta}_j, 1\le j\le B$ yields a two-sided $\cosi{}$ interval under approximate ``unimodality.'' If a pilot/reference estimator $\vartheta$ is available, then we take $a = \vartheta - \theta(P)$ so that $|W_j - a| = |\widehat{\theta}_j - \vartheta|$ does not involve $\theta(P)$. Define 
\[
\widehat{L}_j = \min\{\widehat{\theta}_j,\vartheta\} - (1+\sqrt{2})|\widehat{\theta}_j - \vartheta|\quad\mbox{and}\quad \widehat{U}_j =
\max\{\widehat{\theta}_j,\vartheta\} + (1 + \sqrt{2})|\widehat{\theta}_j - \vartheta|.
\]
Then the $\cosi{}$ confidence interval under approximate unimodality is given by
\[
\widehat{\mathrm{CI}}_{n,\alpha}^{\mathfrak{u}^\star}(\vartheta) = \left[\widehat{L}_{(B - c^*_{B,\alpha})},\, \widehat{U}_{(c^*_{B,\alpha}+1)}\right],
\]
and
\begin{equation}\label{eq:coverage-guarantee-cosi-unimodal*}
\mathbb{P}\left(\theta(P)\notin\widehat{\mathrm{CI}}_{n,\alpha}^{\mathfrak{u}^\star}(\vartheta)\big|\vartheta\right) \le \alpha\times\exp(2B\Delta_P^{\mathfrak{u}^\star}),
\end{equation}
where
\[
\Delta_P^{\mathfrak{u}^\star} = \frac{1}{B}\sum_{j=1}^B \left(\lQ_{\widehat{\theta}_j - \theta(P)}(\log 2) - 1/2\right)_+.
\]
With no pilot estimator, we have
\[
\widehat{\mathrm{CI}}_{n,\alpha}^{\mathfrak{u}^\star} = \left[\max_{j\ge1}\, 2\widehat{\theta}_{(j)} - \widehat{\theta}_{(j + 1 + c^*_{B-1,\alpha/B})},\, \min_{j\ge1}\, 2\widehat{\theta}_{(j+1+c^*_{B-1,\alpha/B})} - \widehat{\theta}_{(j)}\right],
\]
which has the same guarantee as~\eqref{eq:coverage-guarantee-cosi-unimodal*}.
When a pilot is available, we prefer the former interval as it uses significantly inner quantiles than the latter. 

Although we have only provided upper bounds on miscoverage, from Proposition~\ref{prop:poisson-binomial-to-binomial}, one can obtain lower bounds on the miscoverage, but unfortunately, such a lower bound need not be $\alpha$ even when $W_1, \ldots, W_B\in\mathfrak{u}^\star$. This is not just an artifact, but a fundamental impossibility. There is no deterministic function $\phi:\mathbb{R}\to\{0, 1\}$ such that $\mathbb{E}[\phi(W)] = 1/2$ for all $W\in\mathfrak{u}\subseteq\mathfrak{u}^*$; this continues to hold even when $W$ is restricted further to be continuously distributed. The proof follows, by contradiction, from noting that $W \sim (1-\varepsilon)U(0, b) + \varepsilon N(0,1)$ for every $\varepsilon > 0$ and $b > 0$ is a continuous unimodal distribution with mode at $0$, which implies that $\lim_{\varepsilon\downarrow0}\mathbb{E}[\phi(W)] = 1/2$ and this is equivalent to $\int_0^b \phi(u)du = b/2$ for all $b > 0$ implying $\phi(u) = 1/2$ for all $u > 0$, a contradiction because $\phi(u)\in\{0, 1\}$ for almost every $u\in\mathbb{R}.$ A similar argument also works to show that no deterministic function $\phi:\mathbb{R}^B\to\{0, 1\}$ can achieve exact coverage over the class of continuous unimodal distributions with mode at zero.
\subsection{Summary and Simulations}\label{sec:applications-univariate}
We have provided confidence intervals for real-valued functionals of distributions when their estimators are either approximately zero median or symmetric or mean zero normal or unimodal. These confidence intervals have several advantages compared to the traditional Wald or resampling based confidence interval. Below we list of a few of them:
\begin{enumerate}
    \item $\cosi{}$ confidence intervals have a multiplicative miscoverage guarantee which implies that they have close to $\alpha$ miscoverage even as $\alpha$ converges to zero. This was previously illustrated for the case of zero median scale-invariance in~\cite{kuchibhotla2024hulc}. It is interesting to see that it extends to general classes of symmetry and mean zero normality as well; unimodality was already discussed in~\cite{kuchibhotla2024hulc}. For the cases of zero median, symmetry, and mean zero normality, the miscoverage errors are quadratic in the corresponding ``distance'' metric $\Delta$. For unimodality, it is linear in $\Delta$. 
    \item $\cosi{}$ confidence intervals are rate adaptive. For all the confidence intervals discussed above, it can be verified that the length is bounded above by a constant multiple of $\max_{j}\widehat{\theta}_j - \min_{j}\widehat{\theta}_j$; for $\mathfrak{u}^\star$, the width is a constant multiple of $\max_j |\widehat{\theta}_j - \vartheta|$. If $B = O(1)$ as $n\to\infty$, then the length of these confidence intervals converges to zero at the same rate as $\widehat{\theta}_j - \theta(P)$. Hence, even without the knowledge of rate of convergence of the estimator, the confidence intervals shrink at that rate. This is not the case with resampling based estimators where the rate of convergence is explicitly required; for empirical ($n$-out-of-$n$) bootstrap, the rate of convergence is not required but its validity holds only in narrow settings such as differentiable functionals. 
    \item $\cosi{}$ confidence intervals are computationally efficient. Resampling can be computationally expensive if the computation of the estimator is expensive. Because $\cosi{}$ uses estimators computed on smaller sample sizes, the computational cost of $\cosi{}$ intervals is often comparable if not better than the computational cost of $\widehat{\theta}_{[n]}$ (the estimator computed on $n$ observations).
\end{enumerate}
In the following, we illustrate the $\cosi{}$ confidence intervals in the context of mean and regression coefficients. 
\paragraph{Application 1: Inference for Real-valued Mean}\label{sec:mean-inference-R} 
Suppose $X_1, \ldots, X_n$ are independent random variables with mean $\mu$ and variance $\sigma^2$. Under the Lindeberg-Feller condition, it is known that
\[
\frac{n^{1/2}(\overline{X}_n - \mu)}{\widehat{\sigma}_n} \overset{d}{\to} N(0, 1),\quad\mbox{where}\quad \overline{X}_n = \frac{1}{n}\sum_{i=1}^n X_i,\; \widehat{\sigma}_n^2 = \frac{1}{n-1}\sum_{i=1}^n (X_i - \overline{X}_n)^2. 
\]
See~\cite{bentkus1996berry} and~\cite{shao2018necessary} for more general results. Hence, under general conditions, the sample mean centered at the population mean is close in distribution to $\mathfrak{m}, \mathfrak{s}, \mathfrak{n},$ and $\mathfrak{u}^\star$. Even if no moment conditions hold, if $X_i - \mu$ is symmetric and unimodal around zero, then the sample mean centered at $\mu$ belongs to (not just close to) $\mathfrak{m}, \mathfrak{s},$ and $\mathfrak{u}^\star$. We now compare the performance of each of these methods together with bootstrap, subsampling, and Wald intervals. 

\begin{figure}
    \centering
    \includegraphics[width=\linewidth,height=3.5in]{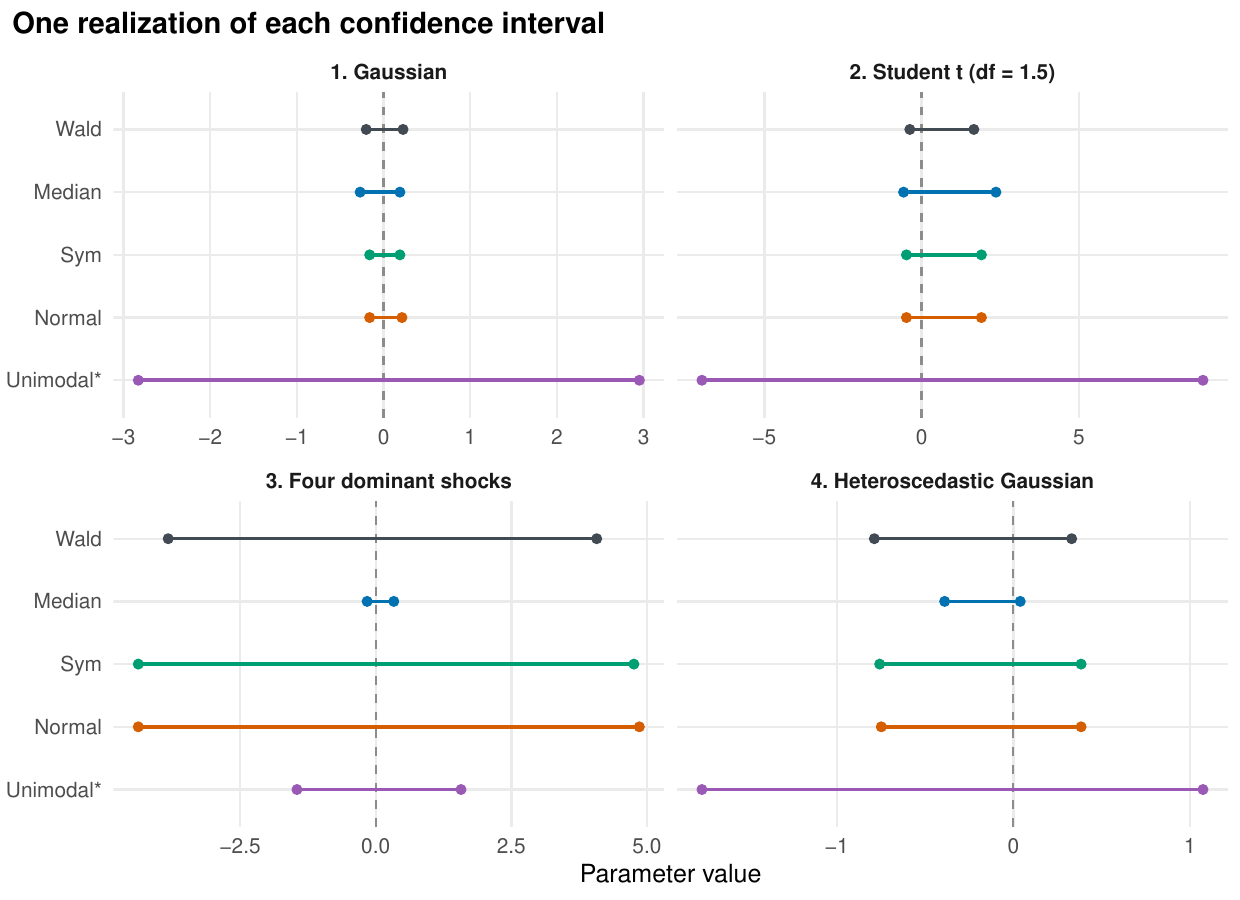}
    \caption{Illustration of $\cosi{}$ confidence intervals for the mean under four data generating processes with 100 observations. Dashed line represents the true mean of the random variables.}
    \label{fig:illustration-1}
\end{figure}

We consider four data generating schemes to illustrate the methods.
\begin{align*}
    \mathrm{DGP\text{-}1}: &\; X_1, \ldots, X_n \overset{iid}{\sim} N(0, 1),\\
    \mathrm{DGP\text{-}2}: &\; X_1, \ldots, X_n \overset{iid}{\sim} t_{1.5},\\
    \mathrm{DGP\text{-}3}: &\; X_i = Z_i + nS_i\mathbf{1}\{i \le 4\},\quad Z_1, \ldots, Z_n \overset{iid}{\sim} N(0, 1), S_1, \ldots, S_4 \overset{iid}{\sim} \mathrm{Rademacher},\\
    \mathrm{DGP\text{-}4}:&\; X_i = \sigma_iZ_i,\quad Z_1, \ldots, Z_n\overset{iid}{\sim} N(0, 1), \sigma_i = 1 + 7\mathbf{1}\{i\mbox{ mod }10 = 1\}.
\end{align*}
DGP-2 uses heavy-tailed data, DGP-3 uses data with four dominant shocks, DGP-4 uses hetereoscedastic Gaussian variables. When one replication of $n = 100$ observations, with $\alpha = 0.05,$ and $B=9$ (for all methods except for $\mathfrak{u}^\star$ for which we take $B = 10$ to compute the pilot estimator), the confidence intervals are illustrated in Figure~\ref{fig:illustration-1}.
The minimal choice of $B$ for these methods to return a bounded confidence interval is $\lceil\log_2(2/\alpha)\rceil$. But setting $B$ slightly larger can help reduce the width. To understand the effect of larger number of splits, we find the smallest $B$ such that $|c^*_{B,\alpha/2}/B - 1/2| \le \delta$\footnote{Note that $c^*_{B,\alpha/2}$ is a two-point random variable and we use the largest element of the support for this condition.} for $\delta\in(0, 1/2]$. In the following, we show the effect of $\delta$ on the coverage and width of the $\cosi{}$ confidence intervals under the four data generating processes discussed above. For each data generating process, we perform 500 Monte Carlo replications with $n=100$ observations and nominal coverage $1-\alpha=0.95$. We consider $\delta\in\{0.10,0.15,0.20,0.25,0.30,1/3\}$, corresponding to $B\in\{90,37,20,12,9,6\}$, respectively. For the $\mathfrak{u}^{\star}$ method, we use one additional block to compute the independent pilot estimator. Thus, smaller values of $\delta$ correspond to more splits and smaller blocks. We use randomized critical values for all $\cosi{}$ methods.

We compare these intervals with the Wald interval,\footnote{The code to compute $\cosi{}$ confidence sets and competitive methods can be found at \url{https://github.com/Arun-Kuchibhotla/COSI}.} the basic bootstrap interval using 500 bootstrap resamples, and subsampling with an estimated convergence rate. For subsampling, we use subsample size $b=\operatorname{round}(n^{2/3})=22$ for interval construction and 500 subsamples at each size used for rate estimation and interval construction. Within each replication, the same data are used for every method and every value of $\delta$. The Wald, bootstrap, and subsampling intervals are computed once per replication and reused across $\delta$.

\begin{figure}[h]
    \centering
    \includegraphics[width=\linewidth]{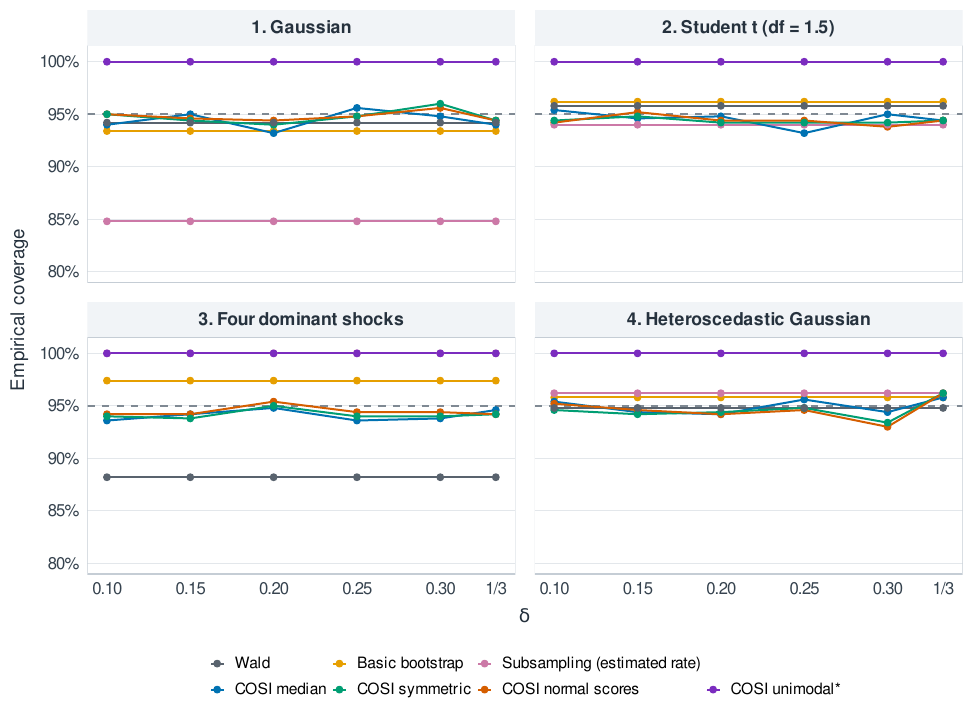}
    \caption{Empirical coverage over 500 Monte Carlo replications with $n=100$ observations under the four data generating processes. The dashed horizontal line marks the nominal coverage of $95\%$. The Wald, basic bootstrap, and rate-estimated subsampling intervals do not depend on $\delta$, so their empirical coverage curves are horizontal.}
    \label{fig:illustration-2-mean}
\end{figure}

Figure~\ref{fig:illustration-2-mean} shows that the median, symmetric, and normal-score $\cosi{}$ intervals have empirical coverages close to the nominal level across all four data generating processes and all six choices of $\delta$. Their empirical coverages range from $93.0\%$ to $96.2\%$; with 500 replications, the Monte Carlo standard error at $95\%$ coverage is approximately one percentage point. The $\mathfrak{u}^{\star}$ interval covers the target in every replication. The comparison methods behave differently across the settings: rate-estimated subsampling has only $84.8\%$ coverage under DGP-1, whereas the Wald interval has $88.2\%$ coverage under DGP-3. The basic bootstrap has empirical coverage between $93.4\%$ and $97.4\%$ across the four settings.

\begin{figure}[h]
    \centering
    \includegraphics[width=\linewidth]{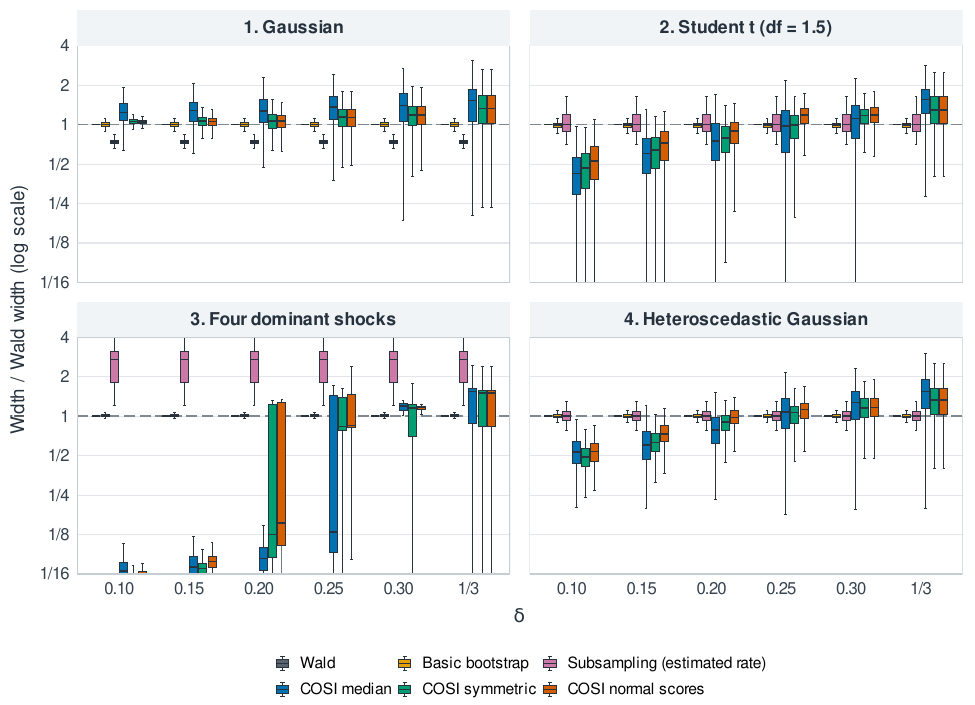}
    \caption{Interval widths relative to the Wald interval computed from the same data, over 500 Monte Carlo replications with $n=100$ observations. The vertical axis is logarithmic, and the dashed horizontal line marks a width ratio of one. Boxplot statistics are computed before restricting the displayed range to $[1/16,4]$; outlier points are omitted. The unimodal$^{\star}$ intervals are generally much wider, with median width ratios exceeding $5.8$, and are therefore omitted from the plot. The exception is DGP-3 at the tested values $\delta\le0.20$: the median width ratios are approximately $0.7$, specifically $0.735$, $0.640$, and $0.660$ for $\delta=0.10$, $0.15$, and $0.20$, respectively.}
    \label{fig:illustration-3-mean}
\end{figure}

To compare widths, we divide the width of each interval by the width of the Wald interval computed from the same data. Figure~\ref{fig:illustration-3-mean} displays boxplots of these paired width ratios.

Under DGP-1, the symmetric and normal-score $\cosi{}$ intervals have widths close to those of the Wald interval when the number of splits is large: their median width ratios lie between $1.04$ and $1.07$ for $\delta\in\{0.10,0.15,0.20\}$. The median-based interval is somewhat wider, with median ratios between $1.24$ and $1.29$ at these values of $\delta$. Although subsampling produces shorter intervals in this setting, its reduced width accompanies substantial undercoverage.

Under DGP-2, DGP-3, and DGP-4, all three displayed $\cosi{}$ methods have median width ratios below one for $\delta\in\{0.10,0.15,0.20\}$. The reductions are particularly pronounced when many splits are used. At $\delta=0.10$, their median width ratios range from $0.42$ to $0.53$ under DGP-2 and from $0.49$ to $0.54$ under DGP-4. Under DGP-3, the corresponding ratios are approximately $0.06$, while empirical coverage remains close to the nominal level. Here, the four dominant shocks affect at most four block estimates, so increasing the number of splits leaves a larger proportion of block estimates unaffected by the shocks.

The basic bootstrap intervals have median widths close to those of the Wald intervals in all four settings. Under DGP-3, rate-estimated subsampling instead produces substantially wider intervals, with a median width ratio of $2.72$ and empirical coverage of $100\%$. Overall, the smaller tested values of $\delta$ generally yield shorter $\cosi{}$ intervals without a substantial change in empirical coverage.

\paragraph{Application 2: Inference for a Linear Contrast of Projection Parameter}\label{sec:linear-constrast}
Suppose $(X_i, Y_i), 1\le i\le n$ are independent and identically distributed random vectors in $\mathbb{R}^{d+1}$ from distribution $P$. Define
\[
\theta(P) := \argmin_{\theta\in\mathbb{R}^d}\,\mathbb{E}_P[(Y - X^{\top}\theta)^2].
\]
(The assumption of identical distribution can be ignored if $\theta(P) = \argmin_{\theta\in\mathbb{R}^d}\,\mathbb{E}[(Y_i - X_i^{\top}\theta)^2]$ for all $i$.)
The natural estimator for $\theta(P)$ is the ordinary least squares:
\[
\widehat{\theta}_{[n]} := \argmin_{\theta\in\mathbb{R}^d}\,\frac{1}{n}\sum_{i=1}^n (Y_i - X_i^{\top}\theta)^2.
\]
For any vector $u\in\mathbb{R}^d$, we have
\[
u^{\top}(\widehat{\theta}_{[n]} - \theta(P)) = u^{\top}\left(\frac{1}{n}\sum_{i=1}^n X_iX_i^{\top}\right)^{-1}\left(\frac{1}{n}\sum_{i=1}^n X_i(Y_i - X_i^{\top}\theta(P))\right).
\]
Under sufficient moment conditions, with proper scaling, the left hand side converges in distribution to a mean zero Gaussian, and hence, as $n\to\infty$, $u^{\top}(\widehat{\theta}_{[n]} - \theta(P))$ is close in distribution to $\mathfrak{m}, \mathfrak{s}, \mathfrak{n}$, and $\mathfrak{u}^\star$. Even without any moment conditions, if $Y_i - X_i^{\top}\theta(P)$ conditional on $X_i$ is symmetric and unimodal around zero, then $u^{\top}(\widehat{\theta}_{[n]} - \theta(P))$ belongs to $\mathfrak{m}, \mathfrak{s},$ and $\mathfrak{u}^\star$. Note that the classical linear model with heteroscedastic Gaussian errors (i.e., $Y_i = X_i^{\top}\theta(P) + \xi_i$ with $\xi_i|X_i \sim N(0, \sigma_i^2)$) satisfy this assumption of symmetry and unimodality, and in this setting, it is known that classical Wald and bootstrap inference fail to be uniformly valid~\citep{potscher2025valid}. Interestingly, even under a misspecified model symmetry can hold in finite samples. Formally, if we observe a sequence of independent real-valued random variables $Y_1, \ldots, Y_n$ symmetric around $\mu_1, \ldots, \mu_n$, respectively, then with $\widehat{\theta}_{[n]}$ defined as above (for a sequence of non-stochastic covariates $x_1, \ldots, x_n\in\mathbb{R}^d$) and $\theta(P) = \argmin_{\theta\in\mathbb{R}^d}\,\sum_{i=1}^n \mathbb{E}[(Y_i - x_i^{\top}\theta)^2]$, we have that $u^{\top}(\widehat{\theta}_{[n]} - \theta(P))$ is symmetric around zero. 

We consider four data generating schemes for linear regression with five dependent covariates. Let
\[
G_1,\ldots,G_n \overset{iid}{\sim} N_5(0,\Sigma), \qquad \Sigma_{jk}=0.7^{|j-k|}, \qquad \beta=(1,-1,1/2,1/2,-1/2)^\top.
\]
Define
\[
g(x)=\frac12x_1^2+\frac14x_1^3+\frac12x_1x_2, \qquad \sigma(x)=\frac12+\frac12|x_1|.
\]
Let $\varepsilon_i\sim N(0,1)$, $E_i\sim\operatorname{Exp}(1)$, $T_i\sim t_3$, and $U_i\sim\operatorname{Unif}(0,1)$, independently of the covariates and independently across observations. The four data generating processes are
\begin{align*}
\mathrm{DGP\text{-}1}:&\quad X_i=G_i,\qquad Y_i=1+X_i^\top\beta+\varepsilon_i,\\
\mathrm{DGP\text{-}2}:&\quad X_i=\exp(G_i/2)-e^{1/8}\mathbf{1}_5,\qquad Y_i=1+X_i^\top\beta+\sigma(X_i)(E_i-1),\\
\mathrm{DGP\text{-}3}:&\quad X_i=G_i,\qquad Y_i=1+X_i^\top\beta+g(X_i)+\sigma(X_i)T_i/\sqrt{3},\\
\mathrm{DGP\text{-}4}:&\quad X_i=G_i,\qquad Y_i=1+X_i^\top\beta+g(X_i)+\sigma(X_i)\left(\frac13U_i^{-2/3}-1\right).
\end{align*}
The exponential transformation in DGP-2 is applied coordinatewise. DGP-1 is a correctly specified homoscedastic Gaussian linear model. DGP-2 has a linear conditional mean, dependent lognormal covariates, and asymmetric, heteroscedastic errors. DGP-3 and DGP-4 have nonlinear conditional means, so fitting a linear model entails misspecification. Their errors are heteroscedastic and heavy-tailed: DGP-3 uses symmetric errors with finite variance, whereas DGP-4 uses centered Pareto errors with tail index $3/2$, finite mean, and infinite variance.

In every setting, we fit ordinary least squares with an intercept and all five covariates. The parameter of interest is the first population regression slope,
\[
\theta(P)=e_2^{\top}\{\mathbb{E}_P[Z_iZ_i^\top]\}^{-1}\mathbb{E}_P[Z_iY_i], \qquad Z_i=(1,X_i^\top)^\top.
\]
Thus, the target is defined by the population normal equations, including under misspecification. It equals $1$ under DGP-1 and DGP-2, and $7/4$ under DGP-3 and DGP-4. For the latter two settings, Gaussian moment identities imply that the cubic term contributes $3/4$ to the first population slope; the quadratic and interaction terms do not change the slopes. In DGP-4, the normal-equation definition remains meaningful even though the squared-error risk is infinite.

We use 500 Monte Carlo replications at each sample size
\[
n\in\{100,200,300,500,750,1000\},
\]
with $\alpha=0.05$ and $\delta=0.30$. This choice of $\delta$ gave coverage comparable to $\delta=0.25$ in the mean simulations while requiring nine rather than twelve blocks, leaving more observations for each blockwise regression fit. The median, symmetric, and normal-score $\cosi{}$ intervals use $B=9$ block estimates and the same random partition. The $\mathfrak{u}^{\star}$ interval uses ten blocks in total: one supplies an independent pilot estimator and the remaining nine supply the block estimates used in the interval. At $n=100$, the former methods therefore use blocks of size 11 or 12, while the pilot-based construction uses blocks of size 10. Randomized cutoffs are used throughout.

We compare these intervals with a Wald interval using the HC0 sandwich covariance estimator, the basic pairs-bootstrap interval using 500 bootstrap resamples, and subsampling with an estimated convergence rate. For subsampling, the interval-construction subsample size is $b=\operatorname{round}(n^{2/3})$. Rate estimation uses the distinct sizes $\operatorname{round}(n^r)$ for $r\in\{0.1,0.2,\ldots,0.9\}$, retaining only sizes between 18 and $n-1$. We use 500 subsamples at each retained size and another 500 for interval construction. Within each replication at a given sample size, every method uses the same data.

\begin{figure}[h]
\centering
\includegraphics[width=\linewidth]{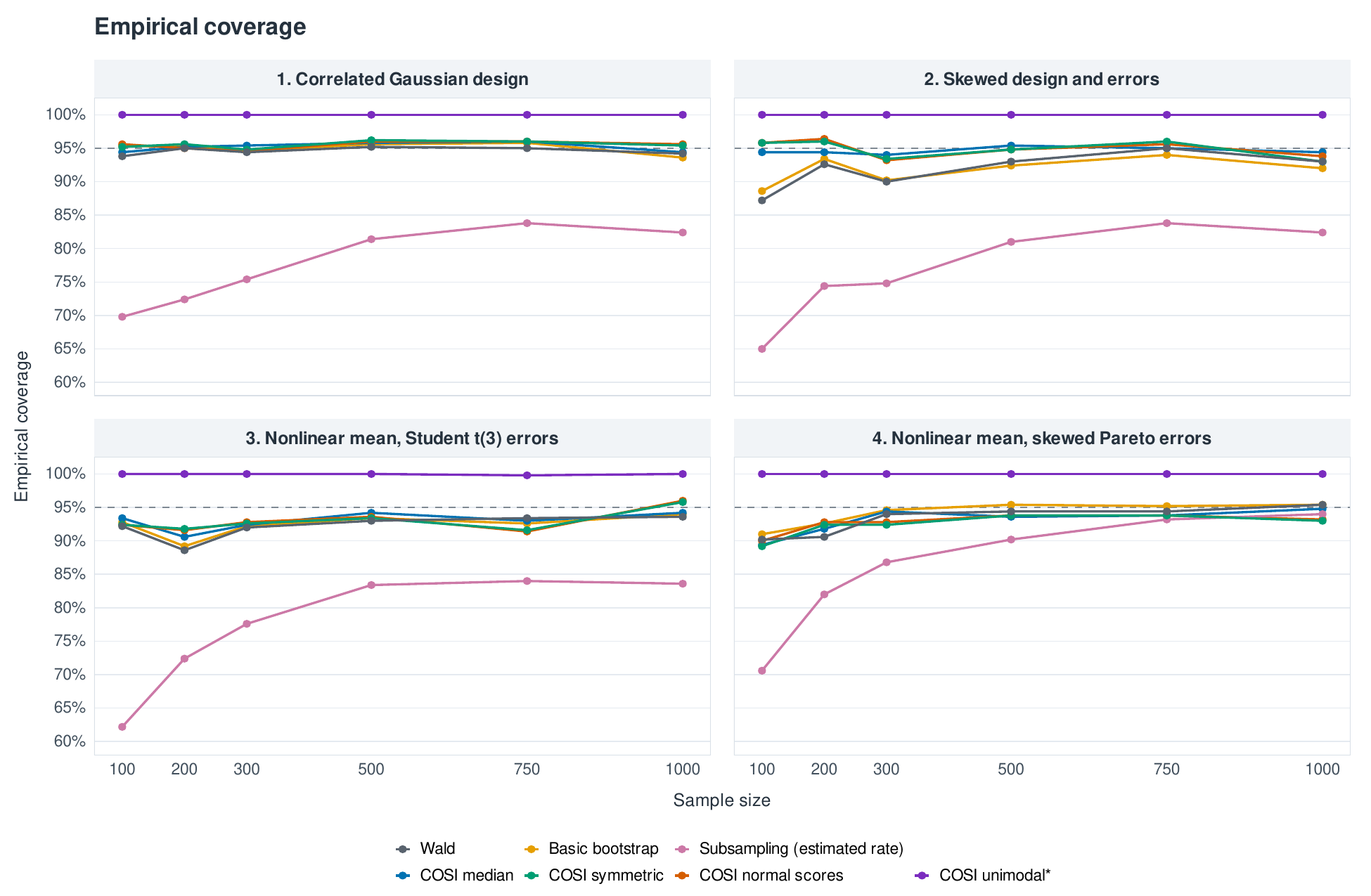}
\caption{Empirical coverage over 500 Monte Carlo replications under the four regression data generating processes, with $\alpha=0.05$ and $\delta=0.30$. The sample size varies from 100 to 1000. The dashed horizontal line marks the nominal coverage of $95\%$.}
\label{fig:coverage-regression}
\end{figure}

Figure~\ref{fig:coverage-regression} shows that the median, symmetric, and normal-score $\cosi{}$ intervals have empirical coverages close to the nominal level under DGP-1: across the tested sample sizes, their coverages range from $94.4\%$ to $96.2\%$. With 500 replications, the Monte Carlo standard error at $95\%$ coverage is approximately one percentage point.

Under DGP-2, the $\cosi{}$ intervals improve coverage over the Wald and basic bootstrap intervals, particularly at small sample sizes. At $n=100$, their empirical coverages are $94.4\%$, $95.8\%$, and $95.8\%$ for the median, symmetric, and normal-score methods, respectively, compared with $87.2\%$ for Wald and $88.6\%$ for the basic bootstrap. Across all tested sample sizes, the three $\cosi{}$ methods have empirical coverages between $93.0\%$ and $96.4\%$ in this setting.

Coverage is less uniform under the nonlinear conditional means. Under DGP-3, the three central $\cosi{}$ methods have empirical coverages between $90.6\%$ and $96.0\%$ across the tested sample sizes. At $n=1000$, their coverages are $94.2\%$, $95.8\%$, and $96.0\%$, respectively. Under DGP-4, the corresponding coverages at $n=100$ are $89.4\%$, $89.2\%$, and $90.0\%$. They increase to $94.8\%$, $93.0\%$, and $93.2\%$ at $n=1000$, where both Wald and the basic bootstrap have $95.4\%$ coverage. Thus, the central $\cosi{}$ intervals do not attain nominal coverage uniformly across these regression examples. In particular, symmetry of the observation errors, as in DGP-3, does not by itself ensure exact symmetry or median unbiasedness of the blockwise OLS estimators under misspecification.

Rate-estimated subsampling substantially undercovers at small sample sizes: its empirical coverage at $n=100$ ranges from $62.2\%$ to $70.6\%$ across the four settings. At $n=1000$, its coverages are $82.4\%$, $82.4\%$, $83.6\%$, and $94.0\%$ under DGP-1 through DGP-4, respectively. The pilot-based $\mathfrak{u}^{\star}$ interval is much more conservative, with empirical coverage between $99.8\%$ and $100\%$. Its only observed exclusion occurs under DGP-3 at $n=750$.

\begin{figure}[h]
\centering
\includegraphics[width=\linewidth]{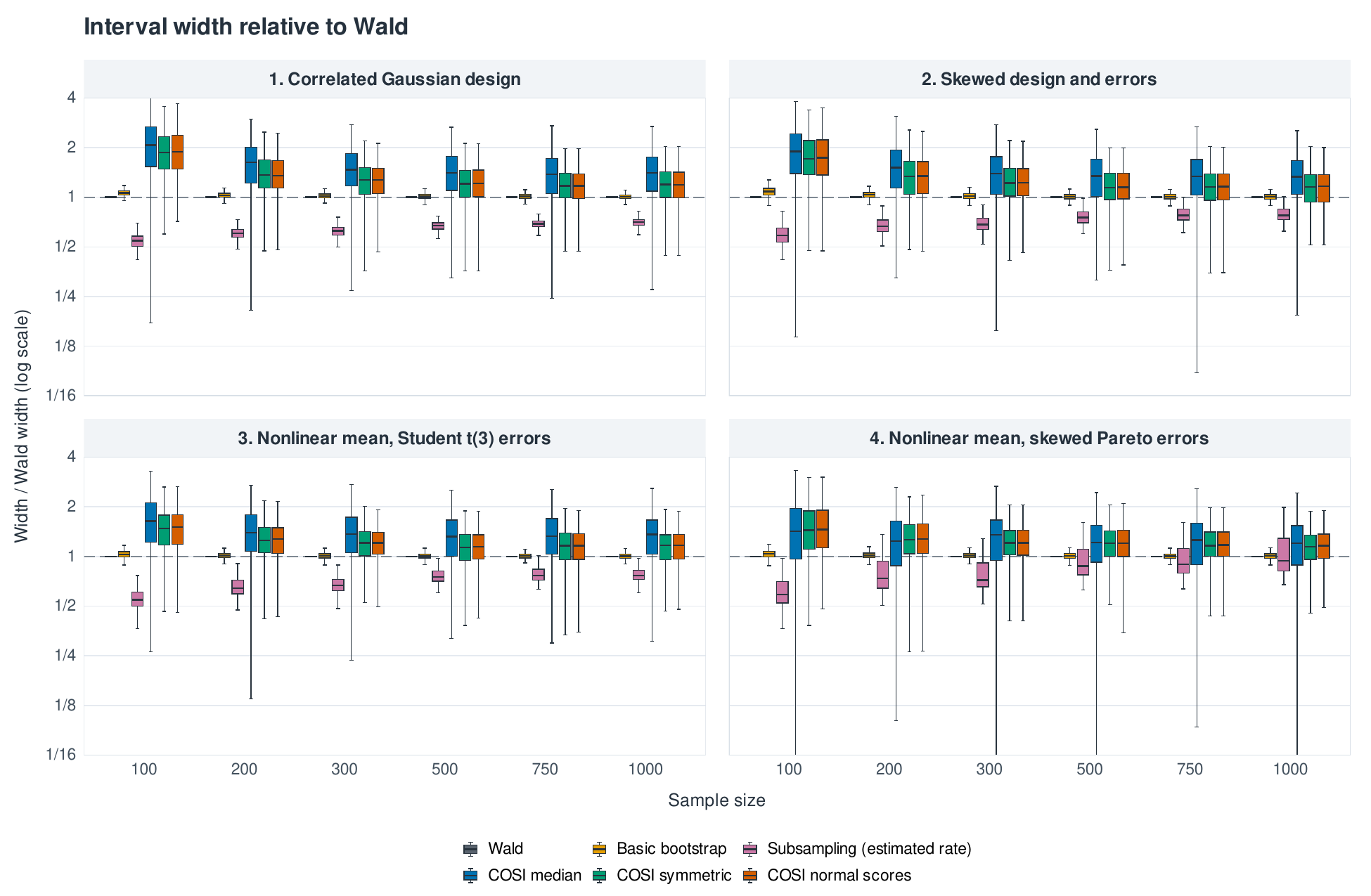}
\caption{Interval widths relative to the Wald interval computed from the same data, over 500 Monte Carlo replications, with $\alpha=0.05$ and $\delta=0.30$. The vertical axis is logarithmic, and the dashed horizontal line marks a width ratio of one. Boxplot statistics are computed before restricting the displayed range to $[1/16,4]$; outlier points are omitted. The pilot-based unimodal$^{\star}$ intervals are omitted because they are substantially wider. Their median width ratios under DGP-1 through DGP-4 are approximately $12.92$, $12.01$, $9.28$, and $8.33$ at $n=100$, and $7.55$, $7.39$, $7.41$, and $7.23$ at $n=1000$, respectively.}
\label{fig:width-regression}
\end{figure}

To compare widths, we divide each interval's width by the width of the Wald interval computed from the same data. Figure~\ref{fig:width-regression} displays boxplots of these paired width ratios.

The median width ratios of the three displayed $\cosi{}$ methods exceed one at every tested sample size in all four settings. Their relative widths generally decrease as the sample size increases. Under DGP-1, for example, the median width ratios decrease from $2.07$, $1.86$, and $1.88$ at $n=100$ to $1.40$, $1.20$, and $1.19$ at $n=1000$ for the median, symmetric, and normal-score methods, respectively. At $n=1000$, the symmetric and normal-score methods have median width ratios between $1.14$ and $1.20$ across the four settings, whereas the median-based method has ratios between $1.20$ and $1.40$.

The basic bootstrap intervals have median widths close to those of Wald, with median ratios between $1.00$ and $1.08$ across all settings and sample sizes. Subsampling produces shorter intervals: its median width ratios range from approximately $0.55$ to $0.59$ at $n=100$ and from $0.71$ to $0.94$ at $n=1000$. These reductions accompany substantial undercoverage, particularly under DGP-1, DGP-2, and DGP-3. The larger widths of the central $\cosi{}$ intervals yield a clear coverage improvement in the skewed linear setting, but do not provide a uniform coverage advantage under nonlinear misspecification.

\section{$\cosi{}$ for Topological Spaces under Projection Stability}\label{sec:hunt-and-test}
In this section, we describe a general ``hunt-and-test''~\citep{moran1973dividing,guo2025rank} scheme to develop $\cosi{}$ confidence sets when the scale-invariance class satisfies a stability property we refer to as ``projection stability.'' Before presenting the most general version of the method, let us provide a simple example to illustrate the $\cosi{}$ hunt-and-test framework for mean estimation. Suppose $X_1, \ldots, X_n$ are IID random vectors in $\mathbb{R}^d$ and we want to construct a confidence interval for the mean $\mu\in\mathbb{R}^d$. Traditionally, one would derive a CLT for the sample mean centered at $\mu$ in some norm and invert this CLT to obtain an asymptotically valid confidence set. Unfortunately, the assumptions and rate of convergence for such a CLT in $\mathbb{R}^d$ may severely restrict the applicability of the method in practice. Alternatively, it is easy to see that for any $u\in\mathbb{R}^d$, $u^{\top}(X_i - \mu), 1\le i\le n$ are mean zero univariate random variables. Hence, one could use the $t$-test to test the null hypothesis that $u^{\top}(X_i - \theta)$ has mean zero and collect all $\theta$'s for which the test is not rejected to construct a confidence set for $\theta$. Of course, this confidence set is unbounded for an arbitrary choice of $u$. This can be rectified by choosing $u$ depending on $\theta$. Formally, splitting data into two parts, computing estimator $\vartheta$ from the first split, we can test the null hypothesis that $(\vartheta - \theta)^{\top}(X_i - \theta)$ has mean zero. This yields a bounded confidence set in the Euclidean norm; this was the approach considered in~\cite{kim2024dimension,chang2024assumption}. In this example, we know that the property of zero mean continues to hold true when considering linear projections and we used the first part of the data to ``hunt'' for a good projection, while the second part of the data is used to test the property along with hunted projection. This is what is done in this section for general estimators but using scale-invariant classes instead of properties such as scale-invariance.

For greater generality, the following discussion provides an abstract construction of confidence set; one may skip to the next section for specific examples in first reading.
\begin{definition}\label{def:scale-invariance-projection-stable}
    Let $V$ be a topological vector space over the reals. Let $\mathcal{L}$ be a collection of functionals mapping $V$ to $\mathbb{R}$. A scale-invariant collection of distributions $\mathfrak{P}$ on $V$ is said to be {\em $\mathcal{L}$-projection stable} if the collection of distributions $\mathcal{L}(\mathfrak{P})$ generated by the real-valued random variables $\ell(W), W\in\mathfrak{P}$ is {\em scale-invariant} on the real line.
\end{definition}
Many of the examples of scale-invariant classes discussed in Section~\ref{sec:SAVARI} are projection stable with respect to some natural collection $\mathcal{L}$. To elaborate on $\mathcal{L}$, we repeat some of the examples here:
\begin{enumerate}
    \item Half-space symmetry: If $\mathfrak{P}$ is the collection of half-space symmetric random variable on an inner product space $(\mathbb{H}, \langle\cdot, \cdot\rangle)$, then setting $\ell_u(x) = \langle u, x\rangle$ for $u, x\in\mathbb{H}$ and $\mathcal{L} = \{\ell_u(\cdot):\, u\in\mathbb{H}\}$, we get $\mathcal{L}(\mathfrak{P}) = \mathfrak{m}$. Hence, $\mathfrak{P}$ is $\mathcal{L}$-projection stable.  
    \item Central symmetry: If $\mathfrak{P}$ is the collection of centrally symmetric random variables on an inner product space $(\mathbb{H}, \langle\cdot, \cdot\rangle)$, then setting $\mathcal{L}$ as above, we get $\mathcal{L}(\mathfrak{P})=\mathfrak{s}$.
    \item Zero rectilinear median: If $\mathfrak{P}$ is the collection of zero rectilinear median distributions, then setting $\delta_t(x) = x(t)$ for all $x\in\mathbb{R}^{\mathcal{T}}$ and $t\in\mathcal{T}$, and $\mathcal{L} = \{\delta_t(\cdot):\, t\in\mathcal{T}\}$, we get that $\mathcal{L}(\mathfrak{P}) = \mathfrak{m}.$ Note that half-space symmetry is a special case of zero rectilinear median by considering $\mathbb{H}$ as the index set.
\end{enumerate}
It is evident that in several interesting examples, $\mathcal{L}$ is a collection of {\em linear} functionals.
Given a collection of real-valued functionals $\mathcal{L}$ on the vector space $V$, we define 
\begin{equation}\label{eq:convex-norm}
\rho(x) ~:=~ \sup_{\ell\in\mathcal{L}}\,\ell(x)\quad\mbox{for all}\quad x\in V.
\end{equation}
Additionally, for every $x\in V$, let $\ell_x\in\mathcal{L}$ be an arbitrary element satisfying
\begin{equation}\label{eq:xi-maximizer-ell}
\frac{1}{2}\rho(x) \le \ell_x(x) \le \rho(x).
\end{equation}
By the definition of the supremum, such an $\ell_x\in\mathcal{L}$ must exist and we implicitly assume that there a systematic way to choose such an $\ell_x$.
If the supremum in the definition of $\rho(\cdot)$ is attained, then we choose $\ell_x(\cdot)$ to be the maximizer. Existence of the maximizer can be established, e.g., from Hahn-Banach extension theorem. Formally, if $\rho: V\to\mathbb{R}$ is a sub-linear function, i.e., $\rho(x + y) \le \rho(x) + \rho(y)$ and $\rho(tx) = r\rho(x)$ for all $t \ge 0$. Then Theorem 3.2 of~\cite{Rudin1991FunctionalAnalysis} implies that for each $x\in V$, there exists a linear functional $\ell_x: V\to\mathbb{R}$ such that $\rho(x) = \ell_x(x)$ and $|\ell_x(u)| \le \max\{\rho(u),\,\rho(-u)\}$ for all $u, x\in V$. Although $\rho(\cdot)$ and $\ell_x(\cdot)$ depend on $\mathcal{L}$, we suppress that dependence in the notation for simplicity. 

The following result provides a hunt-and-test $\cosi{}$ confidence set for general topological spaces based on tests for real-valued random variables. For this, we need the following notation. Recall that $\mathcal{L}(\mathfrak{P})$ is a scale-invariant collection of distributions supported on the real line. We assume that for any $B \ge 1$ and any $\alpha\in[0, 1]$, one can construct a valid confidence interval for zero based on independent random variables $W_1, \ldots, W_B$ in $\mathcal{L}(\mathfrak{P})$. Formally, there exists measurable maps (on an extended measurable space, if necessary\footnote{One may need to extend the measurable space to allow for external randomization, e.g., for tie-breaking.}) $(W_1, \ldots, W_B) \mapsto \widehat{L}_{\alpha}(\{W_j\})$ and $(W_1, \ldots, W_B) \mapsto \widehat{U}_{\alpha}(\{W_j\})$ such that 
\begin{enumerate}[label=\textbf{(HT\arabic*)},leftmargin=*]
    \item if $W_1, \ldots, W_B$ are arbitrary independent real-valued random variables, then\label{cond:approximate-HT} 
    \[
    \mathbb{P}\left(0\notin[\widehat{L}_{\alpha}(\{W_j\}),\, \widehat{U}_{\alpha}(\{W_j\})]\right) ~\le~ \overline{\mathsf{MC}}_{\mathcal{L}(\mathfrak{P})}(\alpha,\,\{W_j\}),\quad\mbox{for all}\quad \alpha\in[0, 1].
    \]
    Here $\overline{\mathsf{MC}}_{\mathcal{L}(\mathfrak{P})}(\alpha, \{W_j\})$ is a function of miscoverage level $\alpha\in[0, 1]$ and the distributions of $W_1, \ldots, W_B$.
\end{enumerate}
It is worth emphasizing that $\overline{\mathsf{MC}}_{\mathcal{L}(\mathfrak{P})}(\alpha, \cdot)$ are not functions of the random variables $W_1, \ldots, W_B$, but only of the distributions of $W_1, \ldots, W_B$. In a way, $\{W_j\}$ is a stand-in for the distributions of $W_j, 1\le j\le B$. 
Ideally, for any {\em fixed} $\alpha\in[0, 1]$, the miscoverage upper bound converges to $\alpha$ as the ``distance'' between the distribution of $W_j$ and $\mathcal{L}(\mathfrak{P})$ converges to zero for all $1\le j\le B$. If $\alpha$ is also allowed to tend to zero, then this convergence to $\alpha$ can be either multiplicative or additive depending on the confidence interval procedure and the scale invariance class $\mathcal{L}(\mathfrak{P})$. From the results in Section~\ref{sec:univariate-HulC-like}, we get that when $\mathcal{L}(\mathfrak{P})$ is either $\mathfrak{m}, \mathfrak{s}, \mathfrak{u},$ or $\mathfrak{n}$, there exist confidence intervals whose miscoverage upper and lower bounds are multiplicative in $\alpha$. Note that, although assumption~\ref{cond:approximate-HT} only requires an upper bound on miscoverage, the results in Section~\ref{sec:univariate-HulC-like} also provide lower bounds on miscoverage.

The following algorithm provides the construction of confidence set under $\cosi{}$ hunt-and-test framework.
\begin{algorithm2e}[h]
    \caption{$\cosi{}$ hunt-and-test~framework for confidence sets}
    \label{alg:cosi-hunt-and-test}
    \SetAlgoLined
    \SetEndCharOfAlgoLine{}
    \KwIn{Sample: $X_1,\cdots,X_n$, confidence level: $1-\alpha$, estimation procedure $\mathcal{A}(\cdot)$ that takes a data set as input and return an estimator for $\theta(P)$, an $\mathcal{L}$-projection stable class $\mathfrak{P}$ along with a test $\phi_{\mathfrak{P},\alpha}(\cdot)$, and integer $B \ge 1$.}
    \KwOut{Confidence set for $\theta(P)$ with diameter controlled with respect to $\rho(\cdot)$.}
    Randomly split $\{1, 2, \ldots, n\}$ into $B+1$ disjoint batches $S_0, S_1, \ldots, S_B$.\;
    
    Compute estimators $\vartheta = \mathcal{A}(\{X_i: i\in S_0\})$ and $\widehat{\theta}_j = \mathcal{A}(\{X_i: i\in S_j\}),$ for $1\le j\le B$.\;

    For each $\theta\in V$, find $\ell_{\vartheta - \theta}(\cdot)\in\mathcal{L}$ satisfying~\eqref{eq:xi-maximizer-ell}.\; 
    
    \Return the confidence sets 
    \begin{equation}\label{eq:confidence-interval-cosi-ht}
    \begin{split}
    \cosici{n}{\alpha}{\mathtt{ht}} ~&:=~ \left\{\theta\in V:\, 0 \in \left[\widehat{L}_{\alpha}(\{\ell_{\vartheta - \theta}(\widehat{\theta}_j - \theta)\}),\, \widehat{U}_{\alpha}(\{\ell_{\vartheta - \theta}(\widehat{\theta}_j - \theta)\})\right]\right\}.
    \end{split}
    \end{equation}
\end{algorithm2e}

The following result provides miscoverage guarantee, and diameter analysis of the confidence set from Algorithm~\ref{alg:cosi-hunt-and-test}. It may be noted here that the miscoverage guarantees are mostly tautological.
\begin{theorem}\label{thm:cosi-hunt-and-test}
    Consider the setting and notation above. Suppose $\widehat{\theta}_1, \ldots, \widehat{\theta}_B$ are random variables in a real topological space $V$ that are mutually independent conditional on $\vartheta\in V$. 
    Then, for all $\alpha\in[0, 1]$,
    \[
    \mathbb{P}_P\left(\theta(P)\notin\cosici{n}{\alpha}{\mathtt{ht}}\big|\vartheta\right) ~\le~ \overline{\mathsf{MC}}_{\mathcal{L}(\mathfrak{P})}(\alpha,\,\{\ell_{\vartheta - \theta(P)}(\widehat{\theta}_j - \theta(P))|\vartheta\})\quad\mbox{almost surely }\vartheta. 
    \]
    Here $\{\ell_{\vartheta - \theta(P)}(\widehat{\theta}_j - \theta(P))|\vartheta\}$ is a stand-in for the conditional distributions of $\ell_{\vartheta - \theta(P)}(\widehat{\theta}_j - \theta(P))$ conditional on $\vartheta$.
    If $\vartheta, \widehat{\theta}_1, \ldots, \widehat{\theta}_B$ are mutually independent, then 
    \begin{equation}\label{eq:mutually-independent-reference}
    \begin{split}
    \mathbb{P}_P\left(\theta(P)\notin\cosici{n}{\alpha}{\mathtt{ht}}\big|\vartheta\right) ~&\le~ \sup_{\ell\in\mathcal{L}}\,\overline{\mathsf{MC}}_{\mathcal{L}(\mathfrak{P})}(\alpha,\,\{\ell(\widehat{\theta}_j - \theta(P))\})\quad\mbox{almost surely }\vartheta.
    \end{split}
    \end{equation}
    Without any assumptions on the independence of $\vartheta, \widehat{\theta}_j, 1\le j\le B$, we have
    \begin{align*}
        \sup_{\theta\in\cosici{n}{\alpha}{\mathtt{ht}}}\,\rho(\vartheta - \theta) ~&\le~ 2\sup_{\ell\in\mathcal{L}}\sup_{\theta\in\cosici{n}{\alpha}{\mathtt{ht}}}\, \ell(\vartheta - \theta) - \widehat{L}_{\alpha}(\{\ell(\widehat{\theta}_j - \theta)\}_{j=1}^B).
    \end{align*}
    If, additionally, $\widehat{L}_{\alpha}(\{w_j + c\}_{j=1}^B) = \widehat{L}_{\alpha}(\{w_j\}_{j=1}^B) + c$ for all $w_1, \ldots, w_B, c\in\mathbb{R}$, and $\mathcal{L}$ is a collection of linear functionals, i.e., for all $\ell\in\mathcal{L}$, $\ell(u + v) = \ell(u) + \ell(v)$, then 
    \[
    \sup_{\theta\in\cosici{n}{\alpha}{\mathtt{ht}}}\,\rho(\vartheta - \theta) ~\le~ 2\sup_{\ell\in\mathcal{L}}\,-\widehat{L}_{\alpha}(\{\ell(\widehat{\theta}_j-\vartheta)\}_{j=1}^B)
    \]
\end{theorem}
See Section~\ref{appsec:proof-hunt-and-test} for a proof of Theorem~\ref{thm:cosi-hunt-and-test}. The constant factor 2 in the radius bound can be made arbitrarily close to 1 by changing the factor $1/2$ in~\eqref{eq:xi-maximizer-ell} to close to 1.

There are several note worthy aspects of the $\cosi{\mathtt{ht}}$ confidence sets. Firstly, nothing about the ``reference'' $\vartheta$ is needed for the coverage guarantee. But most often $\ell_{\vartheta - \theta}(\cdot)$ is identically zero if $\vartheta = \theta$, and hence, $\vartheta$ belongs to $\cosi{\mathtt{ht}}$ confidence set. Therefore, for good diameter properties of the set, $\vartheta$ should be ``close'' to $\theta(P)$ and hence, we almost always choose this to be a ``good'' estimator from $S_{0}$. It may even be preferable to have $|S_{0}|/n$ independent of $B$ for good performance. Secondly, the miscoverage bound~\eqref{eq:mutually-independent-reference} only depends on how close $\ell(\widehat{\theta}_j - \theta(P))$ is to $\mathcal{L}(\mathfrak{P})$ marginally. More importantly, we do not need any measure of distance between $\widehat{\theta}_j - \theta(P)$ to $\mathfrak{P}$ as distributions in a potentially high/infinite-dimensional space $V$; this is the advantage of the hunt-and-test framework. Finally, and perhaps surprisingly, the diameter of the confidence set with respect to $\rho(\cdot)$ depends only on $\widehat{L}_{\alpha}$. In particular, one could use $\widehat{U}_{\alpha}(\cdot) \equiv \infty$, i.e., one-sided confidence intervals satisfying~\ref{cond:approximate-HT}. Using two-sided confidence intervals can often lead to significantly stronger (second order) miscoverage guarantees as illustrated in Section~\ref{sec:univariate-HulC-like}. We also note here that, as with previous results, we do not need $X_1, \ldots, X_n$ to be identically distributed.

Before proceeding to the examples, we mention that one can have an asymmetric ``pseudo-metric'' $\rho(\cdot)$ that takes two arguments instead of one. This would mean a confidence set that has a bound on $\rho(\theta, \theta(P))$ uniformly over all elements in the confidence set. See~\eqref{eq:convex-linear-function-rep} for details. Additionally, the vector space structure on $V$ is not strictly necessary for Theorem~\ref{thm:cosi-hunt-and-test}; see Section~\ref{subsec:metric-space-mean} for an application of $\cosi{}$ confidence set for Fr{\'e}chet mean in Alexandrov spaces.

\section{Illustrations of $\cosi{}$ Framework}\label{sec:illustrations-hunt-and-test}

\subsection{Inference for Functionals in $\mathbb{R}^{\mathcal{T}}$}\label{sec:half-space-symmetry}
Recall that $w\in\mathbb{R}^{\mathcal{T}}$ is identified through the ``coordinate'' projections $\delta_t(w) = w(t)$ for all $t\in\mathcal{T}$. For any $w\in\mathbb{R}^{\mathcal{T}}$, let 
\[
\rho(w) = \sup_{t\in\mathcal{T}}\, w(t),
\]
so that the collection of functionals $\mathcal{L}$ from $\mathbb{R}^{\mathcal{T}}$ to $\mathbb{R}$ is $\mathcal{L} := \{\delta_t:\,t\in\mathcal{T}\}$.
Considering projection stable scale invariance class such that $\mathcal{L}(\mathfrak{P}) = \mathfrak{m}$, we have the following confidence set combining~\eqref{eq:two-sided-median-HulC} and Algorithm~\ref{alg:cosi-hunt-and-test}. Compute estimators $\vartheta, \widehat{\theta}_j, 1\le j\le B$ as in Algorithm~\ref{alg:cosi-hunt-and-test}. For every $\theta\in\mathbb{R}^{\mathcal{T}}$, set $t_{\theta}\in\mathcal{T}$ such that 
\[
\frac{1}{2}\rho(\vartheta - \theta) \le \delta_{t_{\theta}}(\vartheta - \theta) \le \rho(\vartheta - \theta).
\]
Define the confidence set
\begin{equation}\label{eq:cosi-median}
\cosici{n}{\alpha}{\mathfrak{m}} := \left\{\theta\in\mathbb{R}^{\mathcal{T}}:\, (\delta_{t_{\theta}}(\widehat{\theta}))_{(B-c^*_{B,\alpha/2})} \le \delta_{t_{\theta}}(\theta) \le (\delta_{t_{\theta}}(\widehat{\theta}))_{(c^*_{B,\alpha/2}+1)}\right\},
\end{equation}
where $(\delta_{t_{\theta}}(\widehat{\theta}))_{(k)}$ for $1 \le k\le B$ is the $k$-th smallest value among $\delta_{t_{\theta}}(\widehat{\theta}_j), 1\le j\le B$. If $c^*_{B,\alpha/2} = B$, then report $\mathbb{R}^{\mathcal{T}}$ as the confidence set. 

From Theorem~\ref{thm:cosi-hunt-and-test}, it follows that
\begin{equation}\label{eq:miscoverage-bound-COSI-m}
\mathbb{P}(\theta(P)\notin\cosici{n}{\alpha}{\mathfrak{m}}\big|\vartheta) ~\le~ \alpha\times\exp\left(2(B\Delta_P^{\mathfrak{m}})^2\right),
\end{equation}
where
\[
\Delta_P^{\mathfrak{m}} := \sup_{t\in\mathcal{T}}\frac{1}{B} \sum_{j=1}^B\left(\frac{1}{2} - \min\{\mathbb{P}(\delta_t(\widehat{\theta}_j - \theta(P)) \ge 0),\,\mathbb{P}(\delta_t(\widehat{\theta}_j - \theta(P)) \le 0)\}\right)_+.
\]
In words, $\cosici{n}{\alpha}{\mathfrak{m}}$ maintains asymptotic validity as long as the estimator along each ``coordinate'' is median unbiased for the target functional $\theta(P)$. Moreover, from the two-sided inequality~\eqref{eq:two-sided-median-HulC}, we get that the confidence set maintains asymptotic exact validity, if in addition to ``coordinate''-wise median unbiasedness, the estimator places asymptotically negligible mass as the target.

From the diameter analysis in Theorem~\ref{thm:cosi-hunt-and-test}, it follows that
\begin{equation}\label{eq:first-diameter-bound-cosi-m}
\begin{split}
\sup_{\theta\in\cosici{n}{\alpha}{\mathfrak{m}}}\,\rho(\vartheta - \theta) ~&\le~ 2\sup_{t\in\mathcal{T}}\, -(\delta_t(\widehat{\theta}_j - \vartheta))_{(B-c^{*}_{B,\alpha/2})}\\ 
~&=~ 2\sup_{t\in\mathcal{T}}\, (\delta_t(\vartheta - \widehat{\theta}_j))_{(c^*_{B,\alpha/2} + 1)}.
\end{split}
\end{equation}
Before analyzing the diameter bound~\eqref{eq:first-diameter-bound-cosi-m} in terms of the estimation accuracy of the estimators, we note an interesting consequence of this bound. It is easy to see that
\begin{equation}\label{eq:quantile-and-supremum-swap}
\sup_{t\in\mathcal{T}}\, (\delta_t(\vartheta - \widehat{\theta}))_{(c^*_{B,\alpha/2} + 1)} \le  \left(\sup_{t\in\mathcal{T}}\,\delta_t(\vartheta - \widehat{\theta})\right)_{(c^*_{B,\alpha/2} + 1)} = (\rho(\vartheta - \widehat{\theta}))_{(c^*_{B,\alpha/2} + 1)}.
\end{equation}
This implies that there is a simpler superset of the $\cosici{n}{\alpha}{\mathfrak{m}}$ that may be easier to compute. For example, we get
\begin{equation}\label{eq:superset-cosi-m}
\begin{split}
\cosici{n}{\alpha}{\mathfrak{m}} &\subseteq \left\{\theta
\in\mathbb{R}^{\mathcal{T}}:\, \rho(\vartheta - \theta) \le 2\sup_{t\in\mathcal{T}}\,(\delta_t(\vartheta - \widehat{\theta}))_{(c^*_{B,\alpha}+1)}\right\}\\
&\subseteq \left\{\theta\in\mathbb{R}^{\mathcal{T}}:\, \rho(\vartheta - \theta) \le 2(\rho(\vartheta - \widehat{\theta}_j))_{(c^*_{B,\alpha/2} + 1)}\right\}.
\end{split}
\end{equation}
Both supersets inherit the (second-order multiplicative) miscoverage bound~\eqref{eq:miscoverage-bound-COSI-m}, but not the lower bound on miscoverage. Interestingly, if one cares about a confidence set of this type depending on $\rho(\vartheta - \theta)$, one could get an improvement using the one-sided confidence interval for the median from Section~\ref{subsec:HulC-recap}. Additionally, it can be described in a way similar to conformal prediction. 
\begin{enumerate}
    \item Partition $\{1, 2, \ldots, n\}$ into $B+1$ sets, $S_0, S_1, \ldots, S_B$.
    \item Compute the reference estimator $\vartheta$ based on $X_i, i\in S_0$.
    \item For $1\le j\le B$, compute the estimators $\widehat{\theta}_j$ based on $X_i, i\in S_j$.
    \item Compute the scores of these $B$ estimators relative to the reference estimator $\vartheta$ as $R_j = \rho(\vartheta - \widehat{\theta}_j), 1\le j\le B$. Let the order statistics be
    \[
    R_{(1)} \le R_{(2)} \le \cdots \le R_{(B)}.    
    \]
    \item Compute the confidence sets
    \begin{equation}\label{eq:cosi-m+}
    \begin{split}
        \cosici{n}{\alpha}{\mp} &:= \left\{\theta\in\mathbb{R}^{\mathcal{T}}:\, \rho(\vartheta - \theta) \le \sup_{t\in\mathcal{T}}\, (\delta_t(\vartheta - \widehat{\theta}))_{(c^*_{B,\alpha}+1)}\right\},\\
        \cosici{n}{\alpha}{\mpp} &:= \left\{\theta\in\mathbb{R}^{\mathcal{T}}:\, \rho(\vartheta - \theta) \le R_{(c^*_{B,\alpha}+1)}\right\}.
    \end{split}
    \end{equation}
\end{enumerate}
Clearly, $\cosi{\mathfrak{m}+}$ set is a subset of $\cosi{\mpp}$; note that there is no factor of 2 which stems from the choice of $t_{\theta}$.
By a direct proof, we obtain the following result on the miscoverage of $\cosi{\mathfrak{m}+}$ and $\cosi{\mathfrak{m}++}$ confidence sets; we also include the diameter analysis of both $\cosi{\mathfrak{m}}$ and $\cosi{\mathfrak{m}+}$ confidence sets. The main conclusion for diameter is that swapping the roles of supremum and quantiles as in~\eqref{eq:quantile-and-supremum-swap} results in an inflation factor that depends on $\alpha$.
\begin{theorem}\label{thm:coverage-of-cosi-m+-diameter-analysis}
    Suppose $\vartheta, \widehat{\theta}_j, 1\le j\le B$ are mutually independent random elements in $\mathbb{R}^{\mathcal{T}}$ computed from independent data $X_1, \ldots, X_n$. Then 
    \[
    \mathbb{P}(\theta(P)\notin\cosici{n}{\alpha}{\mpp}\big|\vartheta) ~\le~ \mathbb{P}(\theta(P)\notin\cosici{n}{\alpha}{\mp}\big|\vartheta) ~\le~ \alpha\times\exp\left(2B\Delta_P^{\mathfrak{m}}\right)\quad\mbox{almost surely }\vartheta.
    \]
    Suppose there exists a continuous decreasing function $Q:(0,1)\to[0,\infty)$ such that 
    \[
    \mathbb{P}(\rho(\theta(P) - \widehat{\theta}_j) \le Q(\beta)) \ge 1 - \beta,\quad\mbox{for}\quad \beta\in(0, 1).
    \]
    Then, conditional on $\vartheta$, for any $\gamma\in(0,1)$ with probability at least $1-\gamma$,
    \begin{equation}\label{eq:diameter-of-cosi-m+}
    \begin{split}
        \sup_{\theta\in\cosici{n}{\alpha}{\mp}}\,\rho(\vartheta - \theta) ~&\le~ \sup_{\theta\in\cosici{n}{\alpha}{\mpp}}\,\rho(\vartheta - \theta)\\ 
        ~&\le~ \rho(\vartheta - \theta(P)) + Q\left(\gamma^{1/(B-c^*_{B,\alpha})}\left(\frac{B-c^*_{B,\alpha}}{eB}\right)\right).
    \end{split}
    \end{equation}
    Moreover, conditional on $\vartheta$, for any $\gamma\in(0,1)$, with probability at least $1-\gamma$,
    \begin{equation}\label{eq:diameter-of-cosi-m}
    \begin{split}
        \frac{1}{2}\sup_{\theta\in\cosici{n}{\alpha}{\mathfrak{m}}}\,\rho(\vartheta - \theta) &\le~ \sup_{\theta\in\cosici{n}{\alpha}{\mp}}\,\rho(\vartheta - \theta)\\
        &\le~
        \rho(\vartheta - \theta(P))\\ 
        &\quad+ \frac{4(e^{4\log(1/\gamma)/(B-c^*_{B,\alpha/2}-1)} - 1)}{\log(1/\gamma)/B}\mathbb{E}\left[\sup_{t\in\mathcal{T}}\frac{1}{B}\sum_{j=1}^B \varepsilon_j\delta_t(\widehat{\theta}_j - \theta(P))\right]\\
        &\quad+ \sqrt{\frac{4(e^{4\log(1/\gamma)/(B-c^*_{B,\alpha/2}-1)} - 1)}{\log(1/\gamma)/B}}\sup_{t\in\mathcal{T}}\left(\frac{1}{B}\sum_{j=1}^B \mathbb{E}[\delta_t^2(\widehat{\theta}_j - \theta(P))]\right)^{1/2},
    \end{split}
    \end{equation}
    where $\varepsilon_1, \ldots, \varepsilon_B$ are independent Rademacher random variables, independent of $X_1, \ldots, X_n$.
\end{theorem}
See Section~\ref{appsec:proof-of-cosi-m-diameter} for a proof. To better understand these bounds, consider the following comments. Firstly, for the bounds to be finite, we need $c^*_{B,\alpha} < B$ and $c^*_{B,\alpha/2} < B$, respectively. These conditions are equivalent to $\log_2(1/\alpha) < B$ and $\log_2(2/\alpha) < B$; see Theorem 1 of~\cite{paul2024inference}. Secondly, because $c^*_{B,\alpha}$ is the quantile of the Binomial($B, 1/2$) distribution, it follows from the normal approximation that $c^*_{B,\alpha} \approx B/2 + z_{\alpha}\sqrt{B}$ for $B$ large. Thirdly, from~\eqref{eq:diameter-of-cosi-m+}, for any $\alpha$, one should choose $B$ such that $(B-c^*_{B,\alpha})/B$ is away from $0$ and $1$. Finally, to compare and contrast the sizes of $\cosi{\mathfrak{m}}$ and $\cosi{\mathfrak{m}+}$ confidence sets, take $\gamma = \alpha$ and $B \ge \log_2(2/\alpha)$ such that 
\begin{equation}\label{eq:small-requirement-on-B}
\left|\frac{B - c^*_{B,\alpha/2}}{B} - \frac{1}{2}\right| \le \frac{1}{3}.
\end{equation}
(This inequality holds if, e.g., $B \approx 2.9\log_2(2/\alpha)$.)
This implies that
\[
\alpha^{1/(B-c^*_{B,\alpha})} = \exp\left(\frac{\log(1/\alpha)}{B-c^*_{B,\alpha}}\right) \le \mathfrak{C}\frac{\log(1/\alpha)}{B - c^*_{B,\alpha}}\quad\mbox{and}\quad \frac{(e^{4\log(1/\gamma)/(B-c^*_{B,\alpha/2}-1)} - 1)}{\log(1/\gamma)/B} \le \mathfrak{C},
\]
for some constant $\mathfrak{C}$. Further, using Chebyshev's inequality, we can take 
\[
Q(\beta) = \beta^{-1/2}\max_{1\le j\le B}{(\mathbb{E}[\rho^2(\theta(P) - \widehat{\theta})])^{1/2}},
\]
we get (with $\gamma = \alpha$)
\begin{equation}\label{eq:general-diameter-bound-Rademacher-complexity}
\begin{split}
    \sup_{\theta\in\cosici{n}{\alpha}{\mpp}}\,\rho(\vartheta - \theta) &\le \rho(\vartheta - \theta(P)) + \mathfrak{C}'\max_{1\le j\le B}{(\mathbb{E}[\rho^2(\theta(P) - \widehat{\theta}_j)])^{1/2}},\\
    \sup_{\theta\in\cosici{n}{\alpha}{\mp}}\,\rho(\vartheta - \theta) &\le \rho(\vartheta - \theta(P)) + \mathfrak{C}'\mathbb{E}\left[\sup_{t\in\mathcal{T}}\,\frac{1}{B}\sum_{j=1}^B \varepsilon_j\delta_t(\widehat{\theta}_j - \theta(P))\right]\\
    &\quad+ \mathfrak{C}'\sup_{t\in\mathcal{T}}\,\left(\frac{1}{B}\sum_{j=1}^B \mathbb{E}[\delta_t^2(\widehat{\theta}_j - \theta(P))]\right)^{1/2},
\end{split}
\end{equation}
for some other constant $\mathfrak{C}'$. If $\rho(\cdot)$ represents the norm in $\mathbb{R}^{\mathcal{T}}$, then the radius of $\cosi{\mathfrak{m}+}$ sets is controlled by the radius of $\rho(\theta(P) - \widehat{\theta}_j)$, while the radius of $\cosi{\mathfrak{m}}$ is split into two terms depending on Rademacher complexity and weak variance. This makes the $\cosi{\mathfrak{m}}$ set smaller asymptotically. 

To further illustrate this, suppose the estimators $\widehat{\theta}_j, 1\le j\le B$ random vectors in an inner product space $(\mathbb{H}, \langle\cdot, \cdot\rangle)$ identified via $\delta_t(\widehat{\theta}_j) = \langle t,\,\widehat{\theta}_j\rangle$ for $t\in\mathcal{T}:=\{u\in\mathbb{H}:\,\langle u, u\rangle = 1\}$, so that $\rho(w) = \langle w, w\rangle^{1/2}$ for $w\in\mathbb{H}$. Further, suppose that
\[
\sqrt{\frac{n}{B}}(\widehat{\theta}_j - \theta(P)) \overset{d}{\to} N(0, \Sigma),\quad\mbox{as}\quad n\to\infty,
\]
for some covariance operator $\Sigma$. Then, under suitable integrability conditions, with $B \asymp \log(1/\alpha)$, it can be verified that
\begin{equation}\label{eq:diameter-bounds}
\begin{split}
\sup_{\theta\in\cosici{n}{\alpha}{\mpp}}\,\rho(\vartheta - \theta) &= \rho(\vartheta - \theta(P)) + O_p\left(\sqrt{\frac{\log(1/\alpha)\mbox{tr}(\Sigma)}{n}}\right)\quad\mbox{and}\\
\sup_{\theta\in\cosici{n}{\alpha}{\mp}}\,\rho(\vartheta - \theta) &= \rho(\vartheta - \theta(P)) + O_p\left(\sqrt{\frac{\mbox{tr}(\Sigma)}{n}} + \sqrt{\frac{\log(1/\alpha)\|\Sigma\|_{\mathrm{op}}}{n}}\right).
\end{split}
\end{equation}
In particular, choosing $|S_0| \asymp n$ and $\vartheta$ as a sub-Gaussian $\alpha$-dependent estimator such that
\[
\rho(\vartheta - \theta(P)) = O_p\left(\sqrt{\frac{\mbox{tr}(\Sigma)}{n}} + \sqrt{\frac{\log(1/\alpha)\|\Sigma\|_{\mathrm{op}}}{n}}\right),
\]
we get $\cosi{\mathfrak{m}}$ is superior to $\cosi{\mathfrak{m}+}$ with a higher computational cost. It also follows that $\cosi{\mathfrak{m}}$ matches the radius of the CLT based confidence set if $\rho(\widehat{\theta} - \theta(P))$ converges in distribution to $\rho(N(0,\Sigma))$~\citep{Yurinskii,Rhee-Talagrand,Bentkus-Gotze}. 

In the context of inner product spaces, it is worth comparing the $\cosi{}$~approach with the theoretically natural plug-in approach to inference which first upper-estimates $\text{tr}(\Sigma)$ and $\|\Sigma\|_{\text{op}}$, and then constructs a confidence ball centered at a robust point estimator from, e.g.,~\cite{lugosi2019sub} with a plug-in upper-estimate of its deviation. This approach has many drawbacks: first, estimating the trace and (especially) the operator norm can be technically challenging, and can require strong conditions -- the work of \citet{abdalla2026} estimates these quantities under an $L_4-L_2$ moment condition. Their estimator requires the sample-size to be large relative to $\text{tr}(\Sigma)/\|\Sigma\|_{\text{op}}$ and requires knowledge of the $L_4-L_2$ moment constant. The other major drawback of the plug-in procedure is that it directly uses the bad numerical constants from the sub-Gaussian mean estimation literature and is consequently far from practical.
Instead, $\cosi{}$~uses independent block estimators as data-driven “witnesses”: coverage is obtained from a binomial/sign-test calibration, while the diameter is controlled indirectly through the same trace and operator-norm quantities that govern sub-Gaussian mean estimation. The method can adapt to the unknown covariance geometry without ever estimating it explicitly, and it can deliver asymptotically valid inference under weaker moment assumptions without knowing the moment constant. The tradeoff is that the resulting set is less standard than a Euclidean ball (although it could, in principle, be replaced by a ball), and its finite-sample coverage is only exact up to the Tukey median-bias term unless that bias is known or bounded. 
Interestingly, one can also construct a different confidence set based on the ideas of~\cite{lugosi2019sub}, but it requires significantly larger $B$ for valid coverage and consequently has a sub-optimal diameter; see Section~\ref{appsec:Lugosi-Mendelson-set} for details. Similar comments apply when comparing to sub-Gaussian mean estimators in $\mathbb{R}^{\mathcal{T}}$~\citep{lugosi2019near}.

\paragraph{Choice of $\rho(\cdot)$.} For functionals in $\mathbb{R}^{\mathcal{T}}$, we studied the construction of confidence sets with controlled radius in terms of $\rho(\cdot)$. The choice $\rho(w) = \sup_{t\in\mathcal{T}} w(t)$ is only one of many. Given that all the miscoverage bounds are conditional on the reference estimator $\vartheta$, it is possible to choose $\rho(\cdot)$ based on $\vartheta$. More generally, one could use $\{X_i: i\in S_0\}$ to compute the reference estimator $\vartheta$ and also a class $\mathcal{L}$ so that $\ell(\widehat{\theta}_j - \theta(P))$ is asymptotically in $\mathfrak{m}$ for all $\ell\in\mathcal{L}$. For example, in the context of inner product spaces, one could identify $(\mathbb{H}, \langle \cdot, \cdot\rangle)$ via $\delta_t(w) = \langle t, Mw\rangle$ for $t\in\{u\in\mathbb{H}:\,\langle u, u\rangle = 1\}$, where $M:\mathbb{H}\to\mathbb{H}$ is a self-adjoint operator in the sense that $\langle a, Mb\rangle = \langle Mb, a\rangle = \langle Ma, b\rangle$. In the finite-dimensional Euclidean setting, the ideal choice of $M$ is $\Sigma^{-1}$, but such a choice is not possible in infinite-dimensional spaces. Even then, one can take $M = (\Sigma + \lambda I)^{-1}$ (following the ideas of resolvent of an operator). Of course, given only the data, $\Sigma$ is unknown and cannot be used to create $M$. Using $\{X_i: i\in S_0\}$ that yields the reference estimator $\vartheta$, one can also compute an estimate $\widehat{\Sigma}$ and use it in place of $\Sigma$. All the results continue to hold with this choice. More generally, in $\mathbb{R}^{\mathcal{T}}$, one could construct $\widehat{\sigma}_t$, an ``estimator'' of ``scale'' of $\delta_t(\widehat{\theta}_j - \theta(P))$ based only on $\{X_i: i\in S_0\}$ and use
\[
\rho(w) = \sup_{t\in\mathcal{T}}\,\frac{\delta_t(w)}{\widehat{\sigma}_t}.
\]
All the results continue to hold with this choice. For a more general result, see Section~\ref{sec:rectilinear}.

\paragraph{Connection to Subsampling and Conformal Prediction.} Ignoring Step 1 in the construction of $\cosi{\mathfrak{m}+}$ confidence set, the construction closely parallels the classical subsampling procedure. For a more elaborate comparison, let us briefly recall the subsampling procedure with the root as $\rho(\widehat{\theta}_{[n]} - \theta(P))$, where $\widehat{\theta}_{[n]}$ represents the estimator computed using $n$ observations. 
\begin{enumerate}
    \item Compute $\widehat{\theta}_{[n]}$ based on the data $X_i, 1\le i\le n$. Let $r_n$ be such that $r_n\rho(\widehat{\theta}_{[n]} - \theta(P))$ converges in distribution as $n\to\infty$.
    \item Fix $1\le m \le n$, and set 
    \begin{equation}\label{eq:subsample-splits}
    \widetilde{S}_j := \left\{(j-1)m + 1, \ldots, jm\right\}\quad\mbox{for}\quad 1\le j\le \lfloor n/m\rfloor.
    \end{equation}
    (We consider disjoint subsamples instead of all possible subsamples of sizes $m$; asymptotically, this does not give rise any additional sub-optimality.)
    \item Compute estimators $\widehat{\theta}^{(j)}_{[m]}$ for $1\le j\le \lfloor n/m\rfloor$ based on $X_i, i\in \widetilde{S}_j$.
    \item Compute the subsampling roots
    $R_j := r_m\rho(\widehat{\theta}_{[m]}^{(j)} - \widehat{\theta}_{[n]}).$
    Let the ordered roots be
    \[
    R_{(1)} \le R_{(2)} \le \ldots \le R_{(\lfloor n/m\rfloor)}.
    \]
    \item Then the subsampling based confidence interval is given by
    \[
    \widehat{\mathrm{CI}}_{n,\alpha}^{\mathtt{subsamp}} := \left\{\theta\in\mathbb{R}^{\mathcal{T}}:\, r_n\rho(\widehat{\theta}_{[n]} - \theta) \le R_{(\lceil\lfloor n/m\rfloor(1-\alpha)\rceil)}\right\}.
    \]
\end{enumerate}
For the similarity to $\cosi{\mpp}$, one can view $\widehat{\theta}_{[n]}$ and $\widetilde{S}_j, 1\le j\le \lfloor n/n\rfloor$ playing the role of the reference estimator $\vartheta$ and the partition $S_j, 1\le j\le B$. Instead of just computing the $\rho(\cdot)$ as score, subsampling uses the rate normalized $\rho(\cdot)$. Finally, the set is obtained by thresholding the $\rho(\cdot)$ evaluation of difference between $\theta$ and the reference estimator at some quantile of the scores. There are some notable differences as well. Firstly, implementing subsampling as presented requires the knowledge of the rate of convergence $r_n$ and validity of subsampling (as in~\cite{Politis1999}) requires the existence of a limiting distribution. Secondly, the quantile used in subsampling is the $(1-\alpha)$-th quantile of scores, while $\cosi{}$ uses $c^*_{B,\alpha}$-th quantile (note that $c^*_{B,\alpha}/B \approx 1/2$ as $B\to\infty$). Finally, validity of subsampling requires $m\to\infty$ and $m/n \to 0$, i.e., all the estimators computed in Step 3 are sub-optimal for estimator $\theta(P)$, even if $\widehat{\theta}_{[n]}$ is optimal. In $\cosi{}$, we do not require a diverging $B$. 

Connections between bootstrap/subsampling and conformal prediction were spelled out explicitly in~\cite{paul2026probability}. For our setting, the connection can be made more formal. Fix $\alpha\in(0, 1)$. Let us recall that a prediction set $\widehat{C}_{N,\alpha}$ constructed from IID observations $Z_1, \ldots, Z_N$ from a probability measure $\mu$ is said to be valid if
\[
\mathbb{P}(Z_{N+1} \in\widehat{C}_{N,\alpha}) \ge 1 - \alpha\quad\mbox{whenever}\quad Z_{N+1}\sim \mu\quad\mbox{and}\quad Z_{N+1} \perp (Z_1, \ldots, Z_N).
\]
Here the expectation is taken with respect to $Z_1, \ldots, Z_{N+1}$. This can be equivalently written as $\mathbb{E}[\mu(\widehat{C}_{N,\alpha})] \ge 1 - \alpha$. A prediction set $\widehat{C}_{N,\alpha,\delta}$ is said to be $(\alpha, \delta)$-PAC valid if 
\[
\mathbb{P}(\mu(\widehat{C}_{N,\alpha,\delta}) \ge 1 - \alpha) \ge 1 - \delta.
\]
If $W$ is a real-valued random variable, then it is well-known that any interval $[a, b]$ satisfying $\mathbb{P}(W \in [a, b]) \ge 1/2$ contains at least one median of $W$. Moreover, if $\mathbb{P}(W \in [a, b]) > 1/2$, then $[a, b]$ contains all medians of $W$. More generally, Theorem 3.1 of~\cite{merkle2010jensen} implies that for any random variable $W\in\mathbb{R}^{\mathcal{T}}$ with measure $\mu$, if
\begin{equation}\label{eq:half-probab-constraint}
\mathbb{P}\left(W \in \bigotimes_{t\in\mathcal{T}}[a_t, b_t]\right) = \mu\left(\bigotimes_{t\in\mathcal{T}} [a_t, b_t]\right) \ge \frac{1}{2},
\end{equation}
then $[a_t, b_t]$ contains at least one median of $\delta_t(W)$ for all $t\in\mathcal{T}$, i.e., 
\[
\bigcap_{t\in\mathcal{T}}\, \mbox{Med}(\delta_t(W))\cap[a_t, b_t] \neq \emptyset.
\]
Here $\mbox{Med}(Y)$ for a real-valued random variable $Y$ represents the closed set of all medians of $Y$. Let us call sets of the form $\otimes_{t\in\mathcal{T}}[a_t, b_t]$ rectangles in $\mathbb{R}^{\mathcal{T}}$. Hence, if $C$ is a rectangle in $\mathbb{R}^{\mathcal{T}}$ with $\mu(C) \ge 1/2$, then $C$ contains at least one rectilinear median of $W\sim \mu$. If $C$ were constructed from independent data, then $\mu(C)$ is a random variable and may not satisfy $\mu(C) \ge 1/2$ almost surely. However, if $\widehat{C}$ is a {\em rectangular} $(1/2, 1/2)$-PAC valid prediction set from independent data, then by definition $\mu(\widehat{C}) \ge 1/2$ with probability at least $1/2$. Then Theorem 3.1 of~\cite{merkle2010jensen} implies that
\[
\mathbb{P}\left(\bigcap_{t\in\mathcal{T}}\mbox{Med}(\delta_t(W))\cap[a_t, b_t] \neq \emptyset\right) \ge \frac{1}{2}.
\]
This yields a formal connection between prediction sets and confidence sets.
The construction of $(1/2, 1/2)$-PAC valid prediction sets is quite easy. Formally, we have the following result:
\begin{proposition}\label{prop:prediction-to-confidence}
    Suppose $G(\cdot)$ is a measurable function mapping $\mathbb{R}^{\mathcal{T}}$ to $\mathbb{R}$. Suppose $W\in\mathbb{R}^{\mathcal{T}}$ is a random variable with probability measure $\mu$. Define
    \[
    \widehat{C} := \{w\in\mathbb{R}^{\mathcal{T}}:\, G(w) \le G(W)\}.
    \]
    Then $\mathbb{P}(\mu(\widehat{C}) > 1/2) \ge 1/2$ and consequently, if $\widehat{C}$ is a rectangle in $\mathbb{R}^{\mathcal{T}}$, then it contains at least one rectilinear median of $W$ with probability at least $1/2$. In fact, if $\mathrm{Med}_{\mu}\subseteq\mathbb{R}^{\mathcal{T}}$ is the set of all rectilinear medians of $W\in\mathbb{R}^{\mathcal{T}}$, then $\mathbb{P}(\mathrm{Med}_{\mu} \subseteq \widehat{C}) \ge 1/2$.
\end{proposition}
See Section~\ref{appsec:proof-of-prediction-to-confidence} for a proof. Proposition~\ref{prop:prediction-to-confidence} implies that, irrespective of the cardinality or complexity of $\mathcal{T}$, there is always a finite-sample valid confidence interval for the rectilinear median at confidence level $1/2.$ With $B$ independent observations $W_1, \ldots, W_B$, confidence boosting as in Theorem~\ref{thm:finite-sample-validity} implies confidence interval at arbitrary confidence $1-\alpha$. Note that $\widehat{C}$ in Proposition~\ref{prop:prediction-to-confidence} is a rectangle if the sublevel sets of $G(\cdot)$ are rectangles. It is interesting to note that $\{w: G(w) \le c\}$ is a rectangle in $\mathbb{R}^d$ for arbitrary $c\in\mathbb{R}$ if and only if 
\[
G(w) = \sup_{t\in\mathcal{T}}\, q_t(\delta_t(w)),
\]
for some extended real-valued quasi-convex functions $q_t:\mathbb{R}\to[-\infty, \infty]$. A more general result than Theorem~\ref{thm:coverage-of-cosi-m+-diameter-analysis} can be obtained for this framework as illustrated in Section~\ref{sec:rectilinear}. Any arbitrary $G(\cdot)$ of this form yields a valid confidence set but for asymptotic ``radius'' efficiency in our $\cosi{}$ construction, we use the reference estimator $\vartheta$ to anchor $G(\cdot)$. This generalization also makes clear why one can allow for asymmetric $\rho(\cdot)$ functions.

In both conformal prediction and bootstrap/subsampling, one tries to construct a set that contains the reference statistic such as $r_n\rho(\widehat{\theta}_{[n]} - \theta(P))$ with an approximate $(1-\alpha)$ probability. In $\cosi{}$, we only construct a set that contains the ``median'' of an estimator with an approximate $(1-\alpha)$ probability. This is an intuitive reason why subsampling/bootstrap use the $(1-\alpha)$-th quantile of resampled roots and $\cosi{}$ uses a much smaller quantile of the scores.

\paragraph{Necessity of Estimators satisfying $\Delta_P^{\mathfrak{m}} \to 0$.} \cite{kuchibhotla2023median} prove that, in the univariate case, the existence of asymptotically uniformly (over $P\in\mathcal{P}$) valid confidence intervals for a function $\theta(P)\in\mathbb{R}$ is equivalent to the existence of an asymptotically uniformly median unbiased estimator. An argument extending such a result to multivariate functionals $\theta(P)\in\mathbb{R}^d$ was eluded to in~\cite{jain2023rectangular}. The argument extends to general $\mathbb{R}^{\mathcal{T}}$-valued functionals almost verbatim.
\begin{theorem}\label{thm:median-unbiased-requirement}
    Suppose $\mathcal{P}$ is any collection of distributions on some measurable space. Let $\mathcal{T}$ be an index set and let $\theta:\mathcal{P}\to\mathbb{R}^{\mathcal{T}}$ is a functional of interest. Fix $\gamma\in(0, 1)$. Suppose, for each $P\in\mathcal{P}$, with IID sequence $X_1, X_2, \ldots$ from $P$, for each $m \ge1$, there exists a confidence set $\widetilde{\mathrm{CI}}_{m,\gamma}$ based on $X_1, \ldots, X_m$ such that
    \begin{equation}\label{eq:asymptotic-validity-gamma}
    \lim_{m\to\infty}\,\sup_{P\in\mathcal{P}}\,\left(\mathbb{P}_P(\theta(P)\notin\widetilde{\mathrm{CI}}_{m,\gamma}) - \gamma\right)_+ = 0,
    \end{equation}
    and
    \begin{equation}\label{eq:boundedness}
    \inf_{P\in\mathcal{P}}\,\mathbb{P}_P\left(\sup_{t\in\mathcal{T}}\,\sup_{\theta\in\widetilde{\mathrm{CI}}_{m,\gamma}}\,\delta_t(\theta) < \infty\right) = 1,
    \end{equation}
    then there exists an estimator $\widehat{\theta}_{[n]}\in[-\infty, \infty)^\mathcal{T}, n \ge 1$ such that
    \[
    \lim_{n\to\infty}\,\sup_{t\in\mathcal{T}}\,\left(\frac{1}{2} - \min\left\{\mathbb{P}_P(\delta_t(\widehat{\theta}_{[n]} - \theta(P)) \le 0),\,\mathbb{P}_P(\delta_t(\widehat{\theta}_{[n]} - \theta(P)) \ge 0)\right\}\right)_+ = 0.
    \]
\end{theorem}
See Section~\ref{appsec:proof-of-median-unbiased-require} for a proof. If $\sup_{t\in\mathcal{T}}\delta_t(\theta) = \sup_{t\in\mathcal{T}} \delta_t(-\theta)$ for all $\theta\in \mathbb{R}^{\mathcal{T}}$, then the estimator $\widehat{\theta}_{[n]}$ obtained in Theorem~\ref{thm:median-unbiased-requirement} is $\mathbb{R}^{\mathcal{T}}$-valued. Theorem~\ref{thm:median-unbiased-requirement} proves that the univariate scale-invariance class $\mathfrak{m}$ plays a fundamental role in statistical inference. 

\paragraph{Choice of $B$ and Split Sizes.} In order to obtain non-trivial ``bounded'' in $\rho(\cdot)$ confidence sets, it is necessary that $B \ge \log_2(1/\alpha)$ so that $c^*_{B,\alpha} < B$. This necessitates $n \ge \log_2(1/\alpha)$. \cite{lanke1974interval} proved that such a sample size restriction is in fact necessary to obtain a bounded confidence interval even in the univariate case; see \citet[Remark 1]{paul2024inference} for some corrections to Lanke's result. When sample size $n$ is large enough, to make the estimators as efficient as possible, one might be tempted to choose $B = \lceil\log_2(1/\alpha)\rceil$ (or $\lceil\log_2(2/\alpha)\rceil$ to ensure $c^*_{B,\alpha/2} \le B$). This, in the univariate case, reduces precisely to the HulC confidence intervals. In this case, the confidence intervals depend on extreme order statistics and can be unnecessarily large (not conservative though). Using $B$ slightly larger can improve the confidence intervals as already illustrated in~\cite{paul2024inference}. In the general context of $\cosi{}$ as well, we suggest choosing $B$ so that $c^*_{B,\alpha}/B$ is close to $1/2$. For example, the smallest integer $B$ satisfying an inequality of the type~\eqref{eq:small-requirement-on-B}. 

Regarding the split sizes, we note that although the miscoverage bounds do not depend on the properties or accuracy of the reference estimator $\vartheta$, the bounds on the radius~\eqref{eq:diameter-bounds} depend on it. Comparing the diameter bounds~\eqref{eq:diameter-bounds}, it is preferable that $\vartheta$ is computed based on $X_i, i\in S_0$ such that $\underline{e} \le |S_0|/n \le \overline{e}$ as $n$ changes for fixed ($\alpha$ independent) non-negative reals $\underline{e}, \underline{e}$. For example, one could use $|S_0| = n/2$ and $|S_1| = |S_2| = \cdots = |S_B| = n/(2B)$. 

\subsection{Inference for Mean in a Hilbert Space}\label{subsec:mean-Hilbert-space}
Suppose $X_1, \ldots, X_n$ are independent random variables in an inner product space $(\mathbb{H},\,\langle\cdot,\cdot\rangle)$ with a common mean $\mu\in\mathbb{H}$ and a common covariance operator $\Sigma$. We want to construct a confidence set for $\mu.$ We apply the $\cosi{}$ hunt-and-test framework with $\mathcal{T} = \{u\in\mathbb{H}:\, \langle u, u\rangle = 1\}$ and $\delta_u(w) = \langle u, w\rangle$ for $u\in\mathcal{T}$. 
For $\alpha\in(0, 1),$ following Algorithm~\ref{alg:cosi-hunt-and-test}, compute $\vartheta, \widehat{\theta}_j, 1\le j\le B$ based on data in $S_0, S_j, 1\le j\le B$, respectively, i.e., take
\begin{equation}\label{eq:definition-of-the-estimator-mean-HS}
\widehat{\theta}_j = \frac{1}{|S_j|}\sum_{i\in S_j} X_i,\; 1\le j\le B.
\end{equation}
We leave the choice of $\vartheta$ arbitrary for now.  
In this case, $\rho(w) = \|w\| := \langle w, w\rangle^{1/2}$ and $t_{\theta} = (\vartheta - \theta)/\|\vartheta - \theta\|$. Compute the confidence sets
\begin{align*}
    \cosici{n}{\alpha}{\mathfrak{m}} &= \left\{\theta\in\mathbb{H}:\, \langle \vartheta - \theta,\, \theta\rangle \in \left[(\langle \vartheta - \theta, \widehat{\theta}\rangle)_{(B-c^*_{B,\alpha/2})},\, (\langle \vartheta - \theta,\, \widehat{\theta}\rangle)_{(c^*_{B,\alpha/2}+1)}\right]\right\},\\
    \cosici{n}{\alpha}{\mp} &= \left\{\theta\in\mathbb{H}:\, \|\vartheta - \theta\| \le \sup_{u\in\mathcal{T}}\,(\langle u, \vartheta - \widehat{\theta}\rangle)_{c^*_{B,\alpha}+1}\right\},\\
    \cosici{n}{\alpha}{\mpp} &= \left\{\theta\in\mathbb{H}:\, \|\vartheta - \theta\| \le (\|\vartheta - \widehat{\theta}\|)_{c^*_{B,\alpha}+1}\right\}.
\end{align*}
Note that for any $u\in\mathbb{H}$,
\begin{align*}
\frac{1}{B}\sum_{j=1}^B\left(\frac{1}{2} - \mathbb{P}(\langle u, \widehat{\theta}_j - \mu\rangle \ge 0)\right)_+ \le \frac{1}{B}\sum_{j=1}^B \left|\mathbb{P}\left(\frac{1}{|S_j|}\sum_{i\in S_j}\frac{\langle u, X_i - \mu\rangle}{\langle u, \Sigma u\rangle} \ge 0\right) - \Phi(0)\right|
\end{align*}
Hence, from the Berry--Esseen bound for univariate mean zero independent random variables~\citet[Ineq. (1.5) with Corollary 2]{korolev2017bounds}, we conclude that 
\begin{align*}
&\sup_{u\in\mathbb{H}}\,\frac{1}{B}\sum_{j=1}^B\left(\frac{1}{2} - \mathbb{P}(\langle u, \widehat{\theta}_j - \mu\rangle \ge 0)\right)_+\\ 
&\le~ 1.87\sup_{u\in\mathbb{H}}\,\frac{1}{B}\sum_{j=1}^B \sum_{i\in S_j} \mathbb{E}\left[\frac{\langle \lambda, X_i-\mu\rangle^2}{|S_j|\langle \lambda, \Sigma\lambda\rangle}\min\left\{1, \frac{|\langle \lambda, X_i-\mu\rangle|}{|S_j|^{1/2}\langle \lambda, \Sigma\lambda\rangle^{1/2}}\right\}\right]\\
&\le \frac{1.87}{\min_j|S_j|^{\delta/2}}\,\sup_{\lambda\in\mathbb{H}}\,\max_{1\le i\le n}\frac{\mathbb{E}[|\langle \lambda,\, X_i - \mu\rangle|^{2+\delta}]}{(\mathbb{E}[|\langle \lambda,\, X_i - \mu\rangle|^2])^{(1+\delta/2)}}.
\end{align*}
(One can significantly strengthen these bounds by weakening the assumptions with a potential loss in the rate of convergence; see, for example, Eq. (5) and Theorem 1 of~\cite{gabdullin2018esseen}.)
In particular, if there exists $\delta \in (0, 1]$ and $L \ge 1$ such that
\begin{equation}\label{eq:moment-equivalence}
\mathbb{E}[|\langle\lambda, X_i - \mu\rangle|^{2 + \delta}] \le L(\mathbb{E}[|\langle \lambda, \Sigma\lambda\rangle|^2])^{1+\delta/2}\quad\mbox{for all}\quad \lambda\in\mathbb{H},
\end{equation}
and
\begin{equation}\label{eq:split-sizes}
\min_{1\le j\le B}|S_j| \ge \frac{n}{2B},
\end{equation}
then
\begin{equation}\label{eq:simplified-BE-bound}
\begin{split}
    \mathbb{P}\left(\mu\notin\cosici{n}{\alpha}{\mathfrak{m}}|\vartheta\right) &\le \alpha\times\exp\left(14B^{2+\delta}L^2/n^{\delta}\right),\\
    \mathbb{P}\left(\mu\notin\cosici{n}{\alpha}{\mpp}|\vartheta\right) \le~ \mathbb{P}\left(\mu \notin \cosici{n}{\alpha}{\mp}|\vartheta\right)  ~&\le~ \alpha\times\exp(5.3B^{1+\delta/2}L/n^{\delta/2}).
    \end{split}
\end{equation}
Although optimal in the worst case, inequality~\eqref{eq:simplified-BE-bound} should be considered conservative in general. For example, if $X_i - \mu$ is symmetric around zero, then the left hand side is zero but the right hand side is not. In addition to proving pointwise validity, inequality~\eqref{eq:simplified-BE-bound} proves that the $\cosi{}$ confidence sets remain uniformly valid over all distributions that satisfy~\eqref{eq:moment-equivalence}. The condition~\eqref{eq:split-sizes} is made only to simplify the bound.
It should also be stressed that these sets remain valid even if the central limit theorem for $\|\widehat{\theta}_j - \mu\|$ does not hold. The diameter analysis follows from~\eqref{eq:general-diameter-bound-Rademacher-complexity} without invoking the central limit theorem (as in~\eqref{eq:diameter-bounds}). In fact, inequality~\eqref{eq:general-diameter-bound-Rademacher-complexity} along with~\eqref{eq:split-sizes} and~\eqref{eq:definition-of-the-estimator-mean-HS} implies
\begin{equation}
    \begin{split}
    \sup_{\theta\in\cosici{n}{\alpha}{\mp}}\,\|\vartheta - \theta\| &\le \|\vartheta - \theta(P)\| + \mathfrak{C}'\frac{|S_0^c|^{1/2}}{n}\sqrt{\mathrm{tr}(\Sigma)} + \mathfrak{C}'\sqrt{\frac{B\|\Sigma\|_{\mathrm{op}}}{n}},
    \end{split}
\end{equation}
for all distributions of $X_i$ such that $\mathbb{E}[X_i] = \mu$ and $\mbox{Var}(X_i) = \Sigma$. Hence, choosing $\vartheta$ to be the robust sub-Gaussian mean estimator along with $|S_0| \asymp n$ and $B \asymp \log(1/\alpha)$ implies that the $\cosi{}$ confidence sets are sub-Gaussian in $\alpha$ and rate-optimal.

\begin{remark}[Inference for Asymptotically Linear Estimators]
We say that an estimator $\widehat{\theta}_j\in\mathbb{R}^d$ is asymptotically linear around the target of inference $\theta(P)$ at rate $|S_j|^{1/2}$ if 
\begin{equation}\label{eq:Bahadur-representation}
|S_j|^{1/2}(\widehat{\theta}_j - \theta(P)) ~=~ \frac{1}{|S_j|^{1/2}}\sum_{i\in S_j} \psi(X_i) ~+~ R_{n,j},
\end{equation}
for mean zero random vectors $\psi(X_i), i\in S_j$ and a ``small'' reminder term $R_{n,j}$. For a fixed dimension $d \ge 1$, this is referred to as an influence function expansion or Bahadur representation. Such expansions have also been considered in Hilbert spaces; see, for example,~\cite{ShangCheng2013}. While the norm in which $R_{n,j}$ converges to zero is irrelevant for fixed dimension, it makes a significant difference when the dimension is allowed to change with sample size, especially when the dimension grows much faster than the sample size~\citep{belloni2018high}. If~\eqref{eq:Bahadur-representation} holds so that for any vector $c\in\mathbb{S}^{d-1}$, $c^{\top}R_{n,j} = o_p(1)$ and $|S_j|^{-1/2}\sum_{i\in S_j}c^{\top}\psi(X_i)$ converges in distribution to a symmetric (around zero) distribution, then $\cosi{}$ confidence sets are asymptotically exactly valid. In particular, $\cosi{}$ confidence sets do not require convergence of $\|R_{n,j}\|$ for any norm.
\end{remark} 

\begin{remark}[Computation of $\cosi{\mp}$ Confidence set] 
    Note that $\cosi{\mpp}$ set is easy to compute if the inner product space norm is easy to compute, which is the case for finite-dimensional inner product spaces. The $\cosi{\mathfrak{m}}$ is difficult to compute because it involves computing the quantiles for each $\theta\in\mathbb{H}$. The confidence set $\cosi{\mp}$ sits somewhere in between for computation. It is easy to see that
    \[
    \sup_{\|u\| \le 1}\, (\langle u, \vartheta - \widehat{\theta}\rangle)_{(c^*_{B,\alpha}+1)} = \max_{S\subseteq[B], |S| = B - c^*_{B,\alpha}}\,\min_{\lambda_j \ge 0, \sum_{j\in S}\lambda_j = 1}\,\left\|\sum_{j\in S} \lambda_j(\vartheta - \widehat{\theta}_j)\right\|.
    \]
    Hence, for $\binom{B}{B - c^*_{B,\alpha}}$ is not too large, one can compute the right hand side exactly by enumeration of all subsets of $[B] := \{1, \ldots, B\}$. Recall that $\binom{B}{B-c^*_{B,\alpha}} \le \binom{B}{\lfloor B/2\rfloor} \le 2^B/\sqrt{\pi B/2}$, i.e., for $B = k\log_2(2/\alpha)$, we have at most $(2/\alpha)^k$ many subsets. For example, with $k = 2$ and $\alpha = 0.1$, we get 400 subsets to compute over. Similar enumeration based computation can be done for other spaces with other norms; one can use the dual norms in case of non-inner product spaces.
\end{remark}

\subsection{Inference for High-dimensional Mean}\label{sec:high-dimensional-mean} 
Suppose $X_1, \ldots, X_n$ are independent random vectors in $\mathbb{R}^d$ with mean $\mu$ and variance-covariance matrix $\Sigma\in\mathbb{R}^{d\times d}$. We want to construct a confidence set for $\mu$. This problem has received a lot of attention in the recent times, especially with the advances in high-dimensional central limit theorem and bootstrap. Existing results show that high-dimensional CLT and bootstrap can fail to retain validity once dimension $d$ grows faster than exponential in the sample size. Additionally, the moment conditions and variance restrictions are non-trivial to argue for in practice. Alternatively, our $\cosi{}$ confidence set retains validity irrespective of the dimension as long as certain moment equivalence holds along each coordinate. Following Algorithm~\ref{alg:cosi-hunt-and-test}, set   
\[
\widehat{\theta}_{j,k} = \frac{1}{|S_j|}\sum_{i\in S_j} X_{i,k},\quad\mbox{for}\quad 1\le j\le B,\; 1\le k\le d.
\]
Based on $X_i, i\in S_0$, let $\vartheta = (\vartheta_1, \ldots, \vartheta_k)\in\mathbb{R}^d$ be an estimator of $\mu$ and let $\widehat{\sigma} = (\widehat{\sigma}_1, \ldots, \widehat{\sigma}_k)\in\mathbb{R}^d$ be an estimator of $(\Sigma_{kk}^{1/2}: 1\le k\le d)$.
We take $\mathcal{T} = \{1, 2, \ldots, d, d+1, \ldots, 2d\}$ and 
\[
\delta_k(w) := \begin{cases}
w_k/\widehat{\sigma}_k, &\mbox{if }1\le k\le d,\\
-w_{k-d}/\widehat{\sigma}_{k-d}, &\mbox{if }d+1 \le k\le 2d,
\end{cases}
\]
so that $\rho(w) = \|w/\widehat{\sigma}\|_{\infty} = \max_{1\le k\le d}|w_k/\widehat{\sigma}_k|$. Set
\[
k_{\theta} \in \argmin_{k\in\{1, 2, \ldots, 2d\}}\, \delta_k(\vartheta - \theta)\quad\mbox{and}\quad j_{\theta} = \begin{cases}
    k_{\theta}, &\mbox{if }k_{\theta} \in \{1, 2, \ldots, d\},\\
    k_{\theta} - d, &\mbox{otherwise,} 
\end{cases}\quad\mbox{for}\quad \theta\in\mathbb{R}^d
\]
(Choose the smallest value in the argmin set if it is not a singleton.)
Define the confidence set
\begin{align*}
    \cosici{n}{\alpha}{\mathfrak{m}} &= \left\{\theta:\, \theta_{j_{\theta}} \in \left[(\widehat{\theta}_{\cdot,j_{\theta}})_{(B-c^*_{B,\alpha/2})},\, (\widehat{\theta}_{\cdot,j_{\theta}})_{(c^*_{B,\alpha/2}+1)}\right]\right\},\\
    \cosici{n}{\alpha}{\mp} &= \left\{\theta\in\mathbb{R}^d:\,\max_{1\le k\le d}\frac{|\vartheta_k - \theta_k|}{\widehat{\sigma}_k} \le \max_{1\le k\le d}\left(\frac{|\vartheta_{k} - \widehat{\theta}_{\cdot,k}|}{\widehat{\sigma}_k}\right)_{(c^*_{B,\alpha}+1)}\right\}.
\end{align*}
As in the case of Hilbert space random variable, the miscoveraage probabilities of these sets can be controlled using univariate Berry--Esseen bound. Formally, if there exists $\delta\in(0, 1]$ and $L \ge 1$ such that for $1\le j\le B$,
\begin{equation}\label{eq:moment-equivalence-HD}
    \frac{1}{|S_j|}\sum_{i\in S_j}\mathbb{E}[|X_{i,k} - \mu_k|^{2 + \delta}] \le L\left(\frac{1}{|S_j|}\sum_{i\in S_j}\mathbb{E}[|X_{i,k} - \mu_k|^2]\right)^{1 + \delta/2},\quad\mbox{for all}\quad k\in\{1, 2, \ldots, d\},
\end{equation}
and~\eqref{eq:split-sizes} hold true, then
\begin{align*}
\mathbb{P}\left(\mu\notin\cosici{n}{\alpha}{\mathfrak{m}}|\vartheta\right) &\le \alpha\times\exp(14B^{2+\delta}L^2/n^{\delta}),\\
\mathbb{P}\left(\mu\notin\cosici{n}{\alpha}{\mp}|\vartheta\right) &\le \alpha\times\exp\left(5.3B^{1+\delta/2}L/n^{\delta/2}\right).
\end{align*}
Condition~\eqref{eq:moment-equivalence-HD} is too weak a condition to imply a high-dimensional central limit theorem~\citep{kock2024remark}. Most importantly, the dimension of the data does not appear in the miscoverage bounds. The diameter analysis follows from~\eqref{eq:general-diameter-bound-Rademacher-complexity}. Formally, from~\eqref{eq:general-diameter-bound-Rademacher-complexity} and~\eqref{eq:split-sizes}, we get
\begin{align*}
    \sup_{\theta\in\cosici{n}{\alpha}{\mp}}\,\left\|\frac{\vartheta - \theta}{\widehat{\sigma}}\right\| ~&\le~ \left\|\frac{\vartheta - \mu}{\widehat{\sigma}}\right\| + \mathfrak{C}'\mathbb{E}\left[\max_{1\le k\le d}\,\left|\frac{1}{n}\sum_{i\in S_0^c} \frac{X_{i,k} - \mu_k}{\widehat{\sigma}_k}\right|\right] + \mathfrak{C}'\sqrt{\frac{B}{n}}\max_k \Sigma_{kk}^{1/2}.
\end{align*}
Under moment conditions, the second term above behaves like $\sqrt{(\log d)/n}$ and it can also be controlled using complexity terms if $d = \infty$~\citep{ding2025new}. 

It is interesting to note that the $\cosi{}$ confidence intervals are easily adaptable to the setting where number of observations for each coordinate differ. For example, if each coordinate is susceptible to missing-completely-at-random or the data is curated with each coordinate obtained from a different system, the $\cosi{}$ intervals as above can be computed with $\widehat{\theta}_{j,k}$ obtained only by averaging observations for which the $k$-th coordinate is observed. To the best of our knowledge, there is no high-dimensional CLT for the setting where sample sizes for different coordinate are different.

\paragraph{Numerical Illustration.} We consider four data generating schemes for inference on a high-dimensional mean. In each setting, we generate $n=100$ independent random vectors $X_1,\ldots,X_n\in\mathbb{R}^d$, with $d\in\{10,30,100,300,1000,3000,10000\}$, and the target is $\mu=\mathbb{E}[X_i]=0$. For $1\le i\le n$ and $1\le k\le d$, the coordinates are generated as follows:
\begin{align*}
    \mathrm{DGP\text{-}1}:&\quad X_{i,k}=\frac{Z_{i,k}+F_i}{\sqrt{2}},\\
    \mathrm{DGP\text{-}2}:&\quad X_{i,k}=\frac{T_{i,k}}{\sqrt{5}},\qquad T_{i,k}\overset{iid}{\sim}t_{2.5},\\
    \mathrm{DGP\text{-}3}:&\quad X_{i,k}=\frac{Z_{i,k}+\sqrt{nd}\,R_{i,k}A_{i,k}}{\sqrt{2}},\qquad A_{i,k}\overset{iid}{\sim}\mathrm{Bernoulli}\!\left(\frac{1}{nd}\right),\\
    \mathrm{DGP\text{-}4}:&\quad X_{i,k}=\frac{Z_{i,k}+30(H_{i,k}-0.01)}{\sqrt{9.91}},\qquad H_{i,k}\overset{iid}{\sim}\mathrm{Bernoulli}(0.01).
\end{align*}
Here, the $Z_{i,k}$ and $F_i$ are standard Gaussian random variables, the $R_{i,k}$ are Rademacher random variables, and all generating variables are mutually independent. The common factor $F_i$ induces dependence between coordinates in DGP-1. All four settings have unit marginal variances; the covariance matrix is $(I_d+\mathbf{1}\mathbf{1}^{\top})/2$ under DGP-1 and $I_d$ under DGP-2--DGP-4. DGP-2 has finite second moments but infinite third absolute and fourth moments. DGP-3 has an expected total of one shock in the entire $n\times d$ data matrix and is explicitly a triangular array whose distribution depends on $n$ and $d$. DGP-4 introduces rare positive shocks while retaining mean zero.

For the $\cosi{\mathfrak{m}}$ and $\cosi{\mp}$ confidence sets, we randomly allocate 50 observations to the pilot sample $S_0$ and divide the remaining 50 observations into $B=9$ blocks of sizes five or six, corresponding to $\alpha=0.05$ and $\delta=0.30$ (cf. Figure~\ref{fig:illustration-2-mean}). The block estimators are the coordinate-wise sample means. We consider two choices of pilot: the empirical mean with the usual sample covariance matrix, and the coordinate-wise median-of-means (MOM) estimator with an entrywise MOM covariance estimator. We choose the number of blocks for MOM independently of the dimension; see Equation~(2.7) of~\cite{MINSKER2025104724}. Specifically, we use the smallest odd integer $K$ satisfying
\[
    \mathbb{P}\!\left(\operatorname{Bin}(K,1/8)>K/2\right)\le\alpha.
\]
At $\alpha=0.05$, this gives $K=3$ at every dimension, since $\mathbb{P}(\operatorname{Bin}(3,1/8)\ge2)=11/256<0.05$. This is a scalar tuning rule; we do not interpret it as an exact simultaneous $95\%$ pilot accuracy guarantee. The location estimate is the coordinate-wise median of the three group means, with groups containing 16 or 17 observations. For covariance estimation, we form 25 disjoint paired differences $D_r=(X_{a_r}-X_{b_r})/\sqrt{2}$ from the observations in $S_0$, divide these differences into three groups containing eight or nine pairs, and take the entrywise median of the group averages of $D_rD_r^{\top}$. Only the square roots of the diagonal covariance estimates enter the confidence sets. Covariance concentration requires additional moment assumptions, whereas conditional $\cosi{}$ coverage permits any independent location pilot and positive pilot scales. Both pilot choices use the same inference blocks and the same cutoff randomizations. The $\cosi{\mathfrak{m}}$ set uses the selected-coordinate membership condition and is generally nonrectangular, whereas the $\cosi{\mp}$ set is a rectangle centered at the pilot.

We compare these four $\cosi{}$ constructions with the empirical bootstrap and Gaussian multiplier bootstrap, each using 500 resamples. Both bootstrap methods use all 100 observations and calibrate the maximum absolute coordinate of the centered, standardized sample mean; the coordinate standard deviations are computed from the original sample and held fixed over resamples. We also include a Gaussian confidence box centered at the full-sample mean and calibrated using the true covariance matrix. Specifically, if $q_{d,\alpha}$ is the $(1-\alpha)$ quantile of $\|G\|_{\infty}$ for $G\sim N(0,\Sigma)$, this box is $\prod_{k=1}^d[\overline{X}_k\pm q_{d,\alpha}/\sqrt{n}]$. This is an oracle reference for assessing Gaussian approximation rather than a feasible method with unknown covariance. Every method uses the same data within each of the 500 Monte Carlo replications.

\begin{figure}[h]
    \centering
    \includegraphics[width=\linewidth]{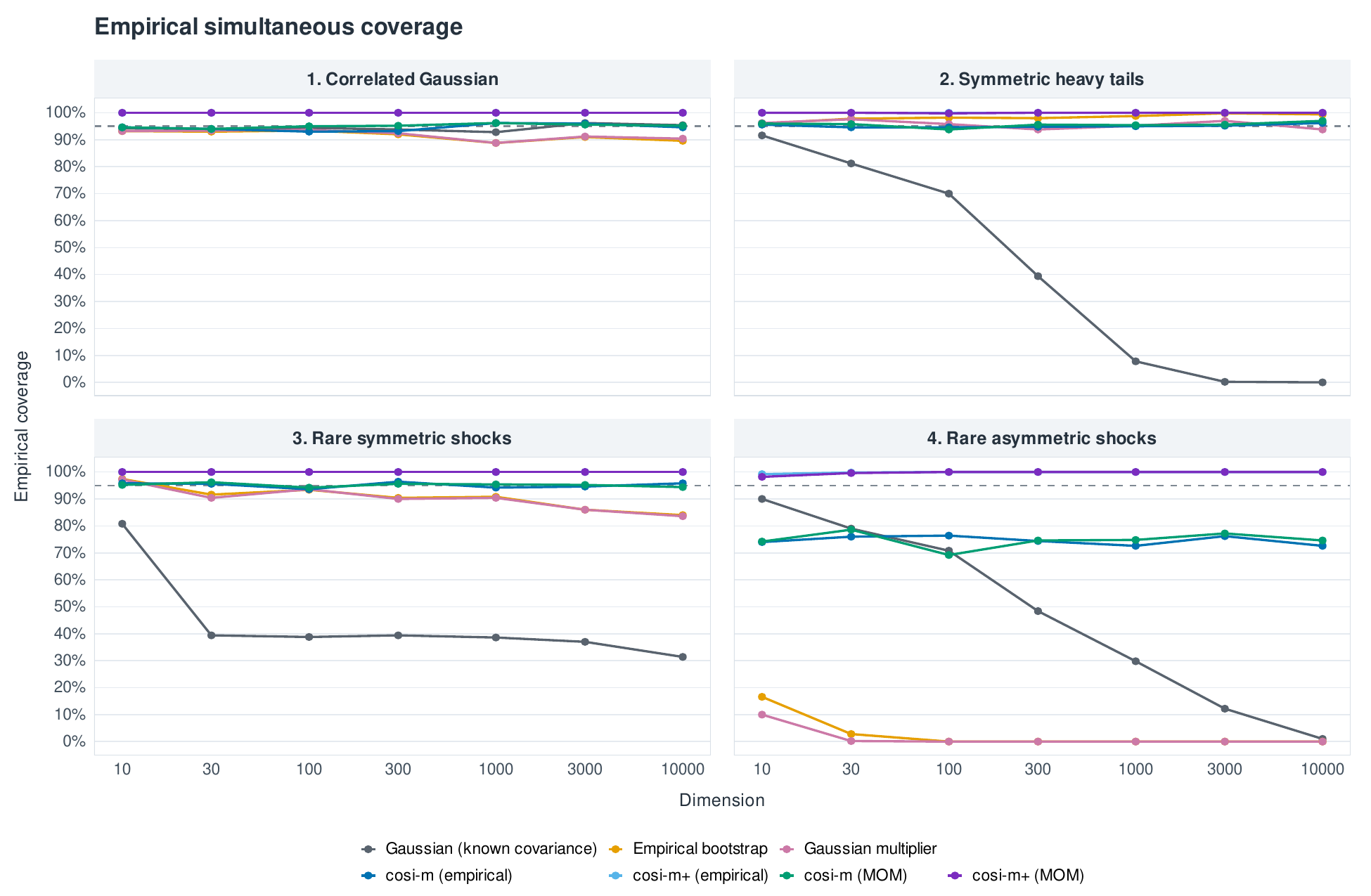}
    \caption{Empirical simultaneous coverage over 500 Monte Carlo replications with $n=100$, $\alpha=0.05$, and $\delta=0.30$, as the dimension increases under the four data generating processes. The dashed horizontal line marks the nominal coverage of $95\%$. The labels empirical and MOM distinguish the two pilot estimators used for the $\cosi{\mathfrak{m}}$ and $\cosi{\mp}$ confidence sets. The Gaussian reference uses the true covariance matrix.}
    \label{fig:coverage-highdim-mean}
\end{figure}

Figure~\ref{fig:coverage-highdim-mean} shows that $\cosi{\mathfrak{m}}$ maintains coverage near the nominal level under DGP-1--DGP-3. Across these settings and all seven settings of dimensions, its empirical coverage ranges from $93.0\%$ to $96.4\%$ with the empirical pilot and from $93.8\%$ to $97.0\%$ with the MOM pilot. At $95\%$ coverage, the Monte Carlo standard error with 500 replications is approximately one percentage point. The $\cosi{\mp}$ set with the empirical pilot covers the target in every replication under these three settings, while its MOM-pilot coverage ranges from $99.8\%$ to $100.0\%$. Their symmetry ensures that each coordinate of every inference-block mean has median zero, regardless of the dimension or the accuracy of the independent pilot.

Under DGP-1, the Gaussian reference has $95.4\%$ coverage at $d=10000$, compared with $89.6\%$ for the empirical bootstrap and $90.4\%$ for the Gaussian multiplier bootstrap. The corresponding coverages of $\cosi{\mathfrak{m}}$ are $94.6\%$ and $95.2\%$ with empirical and MOM pilots, respectively. Since the sample mean is exactly Gaussian in this setting, the bootstrap undercoverage reflects estimation and calibration at this sample size rather than failure of Gaussianity.

Under DGP-2, the Gaussian reference deteriorates sharply as the dimension increases: its coverage decreases from $91.6\%$ at $d=10$ to zero at $d=10000$. At the largest dimension, the empirical and Gaussian multiplier bootstraps have coverages of $99.4\%$ and $93.8\%$, respectively, while $\cosi{\mathfrak{m}}$ has coverages of $96.2\%$ and $97.0\%$ with the two pilots. Thus, failure of Gaussian approximation for the maximum of the unstudentized sample mean need not imply failure of the bootstrap procedures, which use estimated coordinate scales.

DGP-3 provides a different mechanism for failure of Gaussian approximation. A single shock contributes an additive term $\pm\sqrt{d/2}$ to one coordinate of $\sqrt{n}\,\overline{X}$, whereas the corresponding Gaussian maximum grows at the order of $\sqrt{\log d}$. At $d=10000$, the Gaussian reference has $31.4\%$ coverage, and the empirical and Gaussian multiplier bootstraps have coverages of $84.0\%$ and $83.6\%$ respectively. In contrast, the two versions of $\cosi{\mathfrak{m}}$ have coverages of $95.8\%$ and $94.4\%$ respectively. The marginal symmetry sufficient for these constructions is preserved despite the rare shocks.

DGP-4 illustrates the limitation of the median assumption. Although the observations have mean zero, the means of blocks containing only five or six observations have negative medians. At $d=10000$, $\cosi{\mathfrak{m}}$ has coverage of $72.6\%$ with the empirical pilot and $74.6\%$ with the MOM pilot; changing the pilot does not remove the median bias of the inference-block estimators. The Gaussian reference has $1.0\%$ coverage, and both bootstrap methods have zero coverage. The $\cosi{\mp}$ sets remain empirically conservative: across all dimensions, their coverages range from $99.2\%$ to $100.0\%$ with the empirical pilot and from $98.2\%$ to $100.0\%$ with the MOM pilot. This observed conservatism does not establish a coverage guarantee when the median condition is violated.

\begin{figure}[h]
    \centering
    \includegraphics[width=\linewidth]{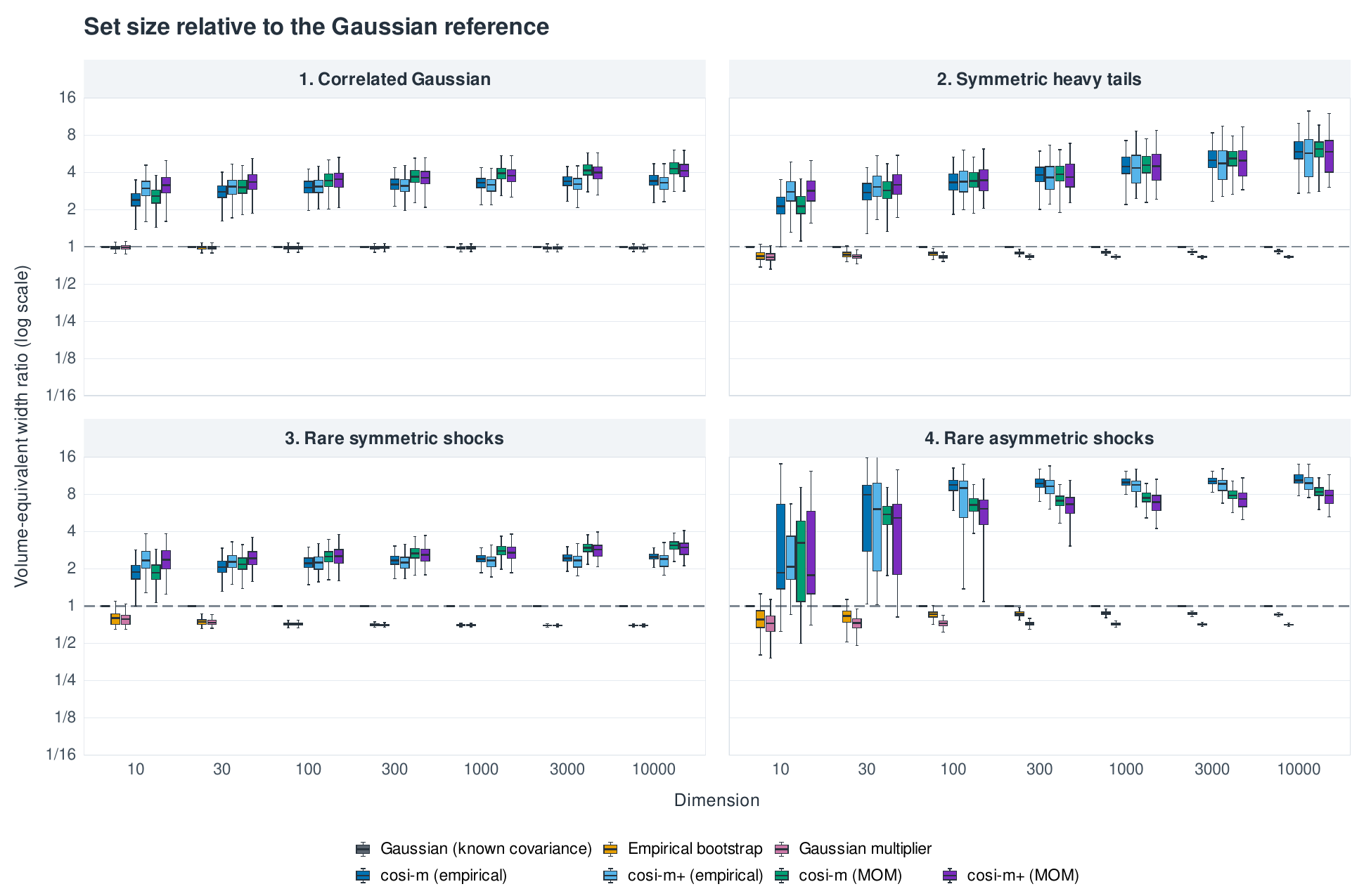}
    \caption{Boxplots of $\operatorname{Vol}(C)^{1/d}$ relative to the Gaussian reference, over 500 Monte Carlo replications with $n=100$. The vertical axis is logarithmic, and the dashed horizontal line marks a ratio of one. Boxplot statistics are computed before restricting the displayed range to $[1/16,16]$; outlier points are omitted. The Gaussian reference uses the true covariance matrix and need not attain nominal coverage outside DGP-1.}
    \label{fig:width-highdim-mean}
\end{figure}

To compare sizes across dimensions and between rectangular and nonrectangular sets, we report $\operatorname{Vol}(C)^{1/d}$, the side length of a cube having the same volume as $C$. For each replication, we divide this quantity by that of the Gaussian reference, namely $2q_{d,\alpha}/\sqrt{n}$. Note that for $\cosi{\mathfrak{m}}$, the volume can be computed by partitioning it into regions corresponding to each selected coordinate and sign and summing their volumes.

Figure~\ref{fig:width-highdim-mean} shows that the $\cosi{}$ confidence sets are substantially larger than the Gaussian reference in these small-sample, high-dimensional settings. At $d=10000$, the median ratios for $\cosi{\mathfrak{m}}$ and $\cosi{\mp}$ with the empirical pilot are, respectively, $3.42$ and $3.30$ under DGP-1, $5.87$ and $5.73$ under DGP-2, and $2.50$ and $2.41$ under DGP-3. With the MOM pilot, the corresponding ratios are $4.29$ and $4.13$, $6.16$ and $5.86$, and $3.09$ and $2.98$. The MOM pilot therefore does not uniformly reduce set size. Fixing the group count removes the explicit dependence of its tuning on $d$, while the uniform pilot error and the confidence-set sizes can still depend on dimension. Under DGP-4, the empirical-pilot ratios are $10.40$ and $9.88$, while the MOM-pilot ratios are $8.36$ and $7.86$. These size comparisons must be read together with coverage: smaller bootstrap sets can under-cover, and even large $\cosi{\mathfrak{m}}$ volume does not compensate for violation of the median assumption. The two $\cosi{}$ sets have different shapes and are not generally nested, so their volume ordering need not match their coverage ordering.

\paragraph{Implication for the Maximum/Minimum of the Mean Vector:} In the context of mean inference, it may be of interest to construct confidence intervals for the minimum or the maximum of the mean vector, i.e., the target of inference is $\max_{1\le k\le d}\,\mathbb{E}[X_{i,k}]$ or $\min_{1\le k\le d}\,\mathbb{E}[X_{i,k}]$. Such a problem arises commonly in models specified by moment inequalities~\citep{andrews2010inference,chernozhukov2013intersection,canay2023user}, where one is interested in inference for
\[
\theta^*(P) := \argmin_{\theta\in\mathbb{R}^d}\,\mathbb{E}_P[\ell(Z; \theta)]\mbox{ subject to }\max_{1\le j\le d}\,\mathbb{E}_P[f_j(Z; \theta)] \le 0.
\]
To construct a confidence set for $\theta^*(P)$, one first identifies a high-probability superset and subset of the constraint set. This can be achieved by construct a confidence upper bound and lower bound for the maximum of expectations arising in the constraint set. There are numerous other applications in the literature~\citep{huang2023efficient,pmlr-v291-muller25a,guigues2017non,pmlr-v267-tong25a,lam2018bounding,javeed2026statistical}. 

Returning to inference for maximum of means, given our $\cosi{\mp}$ confidence set, we get that
\[
\left[\max_{1\le k\le d}\,\{\vartheta_k - \widehat{\sigma}_k\widehat{R}_{\alpha}\},\, \max_{1\le k\le d}\,\{\vartheta_k + \widehat{\sigma}_k\widehat{R}_{\alpha}\}\right],
\]
where $\widehat{R}_{\alpha} = \max_{1\le k\le d}(|\vartheta_k - \widehat{\theta}_{\cdot, k}|/\widehat{\sigma}_k)_{(c^*_{B,\alpha}+1)}$ is an asymptotically valid $(1-\alpha)$ confidence set for $\max_{1\le k\le d}\mu_k$. Note that this confidence set can be significantly smaller than $[\max_{1\le k\le d}\vartheta_k \pm \max_{1\le k\le d}\widehat{\sigma}_k\widehat{R}_{\alpha}]$, obtained from the Lipschitz property of the max-operator. One can construct similar confidence sets for other quantiles of the mean vector. 

\paragraph{Implication for Random Projection Mean Testing:} Under Gaussian data, the optimal test for the mean vector is the Hotelling $T^2$-test which can only be implemented if the dimension of the data is smaller than the sample size. The maximum based tests are optimal against sparse alternatives. If one is interested in denser alternatives, it is beneficial to consider other $\ell_p$-norms~\cite{KockPreinerstorfer2026} or some aggregation of Hotelling $T^2$-tests applied on subvectors of the data or random projections of the data~\citep{srivastava2016raptt,lopes2011more}. While the $\ell_p$-norm based tests are calibrated using high-dimensional bootstrap (and hence, asymptotically exact), the random projection based tests are combined using $p$-value aggregation methods that can be conservative. One can relax the assumptions needed for $\ell_p$-norm based methods and provide asymptotically exact calibration of random projection tests while adapting to the dependence structure. 

For $\ell_p$-norm based tests, note that we have
\[
\|\vartheta - \theta\|_{\ell_p} = \sup_{u\in\mathbb{R}^d:\,\|u\|_{\ell_q} \le 1}\, u^{\top}(\vartheta - \theta)\quad\mbox{for}\quad \frac{1}{q} + \frac{1}{p} = 1.
\]
Hence, an application of Theorem~\ref{thm:coverage-of-cosi-m+-diameter-analysis}, yields an asymptotically valid $\ell_p$-ball confidence set for the mean vector. For the random projection methods, we can create an augmented space from $\mathbb{R}^d$ and apply Theorem~\ref{thm:coverage-of-cosi-m+-diameter-analysis}. Formally, fix $\mathcal{K} \subseteq \{1, 2, \ldots, d\}$ and $m \ge 1$ and let $P_{k,1}, \ldots, P_{k,m}$ be independent random projections such that $P_{k,j}\in\mathbb{R}^{k\times d}$ for $1\le j\le m$ and the entires of $P_{k,j}$ are IID standard Gaussian random variables. Partition $\{1, 2, \ldots, n\}$ into $S_0, S_1, \ldots, S_B$ are in Algorithm~\ref{alg:cosi-hunt-and-test}, and let $\widehat{\Sigma}$ be an estimator of $\Sigma$ based on $X_i, i\in S_0$. One could now apply Theorem~\ref{thm:cosi-hunt-and-test} using
\begin{equation}\label{eq:scale-independent-maximum}
\rho(u) = \max_{k\in\mathcal{K}}\,\max_{1\le j\le m}\, u^{\top}P_{k,j}^{\top}(P_{k,j}\widehat{\Sigma}P_{k,j}^{\top})^{-1}P_{k,j}u = \max_{k\in\mathcal{K}}\,\max_{1\le j\le m}\, \sup_{v\in\mathbb{R}^k}\,\frac{v^{\top}P_{k,j}u}{(v^{\top}P_{k,j}\widehat{\Sigma}P_{k,j}^{\top}v)^{1/2}}.
\end{equation}
(Here it is assumed that $P_{k,j}\widehat{\Sigma}P_{k,j}^{\top}$ is invertible with probability 1.) Because $\rho(u)$ is the supremum over a collection of linear functions of $u$, Theorem~\ref{thm:coverage-of-cosi-m+-diameter-analysis} yields asymptotically exact validity guarantee. In the construction above~\eqref{eq:scale-independent-maximum}, all the dimensions $k\in\mathcal{K}$ are treated the same and it is conceivable that the maximum is dominated by larger values of $k\in\mathcal{K}$. One way to avoid this is by applying a debiasing at each $k$:
\[
\rho(u) = \max_{k\in\mathcal{K}}\,\max_{1\le j\le m}\, u^{\top}P_{k,j}^{\top}(P_{k,j}\widehat{\Sigma}P_{k,j}^{\top})^{-1}P_{k,j}u - cD_k,
\]
for some constant $c \ge 1$ and $D_k$ is, for example, the median of the Hotelling $T^2$ distribution at dimension $k$; because $D_k$ does not depend on $k$, Theorem~\ref{thm:coverage-of-cosi-m+-diameter-analysis} holds. 

\paragraph{Implication for Probabilities of Sets.} Suppose $X_1, \ldots, X_n$ are independent and identically distributed random variables in some measurable space and let $\mathcal{A}$ be a collection of measurable sets in that space. A common problem in empirical process theory is to infer about $\{\mathbb{P}(X \in A):\, A\in\mathcal{A}\}$. If the collection of sets $\mathcal{A}$ is VC or satisfies Donsker conditions, then one could apply either concentration inequalities or Gaussian approximation results to construct simultaneous confidence intervals for $\mathbb{P}(X\in A), A\in\mathcal{A}$. Owing to severe conservativeness of concentration inequalities and to Donskerity assumptions for Gaussian approximation, one could alternatively apply the $\cosi{}$ framework. A direct application of the construction above is not suitable because condition~\eqref{eq:moment-equivalence} is not satisfies for $X_{i,A} = \mathbf{1}\{X_i \in A\}$ when $\mathbb{P}(X_i \in A)$ is too small compared to the sample size. In fact, this example can be seen as a infinite-dimensional extension of the binomial proportion example discussed in~\cite{kuchibhotla2024hulc}. 

We construct a finite-sample valid confidence set, irrespective of the complexity of $\mathcal{A}$, using Proposition~\ref{prop:prediction-to-confidence}. As in Algorithm~\ref{alg:cosi-hunt-and-test}, partition $\{1, 2, \ldots, n\}$ as $S_0\cup S_1\cup\cdots\sup S_B$. For simplicity, let $|S_0| = n/2, |S_j| = n/(2B)$ for $1\le j\le B$. Let $\phi:[0,1]^2\to\mathbb{R}_+$ be a non-negative function that is convex in the first argument so that
\begin{equation}\label{eq:convex-linear-function-rep}
\phi(x, y) = \sup_{\ell\in\mathcal{L}_y}\,\ell(x),\quad\mbox{for all}\quad x, y\in[0, 1],
\end{equation}
for a collection $\mathcal{L}_y$ of linear functions $\ell:[0, 1]\to\mathbb{R}$. 
For each $A\in\mathcal{A}$, let
\[
p_j(A) := \frac{1}{|S_j|}\sum_{i\in S_j} \mathbf{1}\{X_i\in A\}\quad\mbox{for}\quad 0 \le j\le B,
\]
and
\[
C_j(A) := \left\{y\in[0, 1]:\,\phi(y,\,p_0(A)) \le \phi(p_j(A),\, p_0(A))\right\}\quad\mbox{for}\quad 1\le j\le B.
\]
From Proposition~\ref{prop:prediction-to-confidence}, we get that $C_j(A)$ contains all the medians of $p_j(A)$ for all $A\in\mathcal{A}$ with a probability of at least $1/2$. From the results of~\cite{hamza1995smallest}, we get that the median and mean of $p_j(A)$ differ at most by $\log 2/|S_j|$ and hence, from Proposition~\ref{prop:prediction-to-confidence}, it follows that the inflation of $C_j(A)$ by $2B\log 2/n$ contains $\mathbb{P}(X_i\in A) = \mathbb{E}[p_j(A)]$ for all $A\in\mathcal{A}$ with probability at least $1/2$. 

Consider
\[
\cosicit{n}{\alpha}{\mp} := \left\{\omega\in[0,1]^{\mathcal{A}}:\,\sup_{A\in\mathcal{A}}\phi(\omega_A,\,p_0(A)) \le \sup_{A\in\mathcal{A}}(\phi(p_{\cdot}(A),\,p_0(A)))_{(c^*_{B,\alpha}+1)}\right\},
\]
where the quantile on the right is computed from $\phi(p_j(A), p_0(A)), 1\le j\le B$. By the convexity of $x\mapsto\phi(x, y)$, we get that
\[
\cosicit{n}{\alpha}{\mp} = \bigotimes_{A\in\mathcal{A}}\left[\overline{\ell}_{\alpha}(A),\, \overline{u}_{\alpha}(A)\right],
\]
for some $\overline{\ell}_{\alpha}(\cdot), \overline{u}_{\alpha}(\cdot)\in[0, 1]$. Finally, define
\[
\cosici{n}{\alpha}{\mp} := \bigotimes_{A\in\mathcal{A}} \left[\overline{\ell}_{\alpha}(A) - \frac{2B\log 2}{n},\, \overline{u}_{\alpha}(A) + \frac{2B\log 2}{n}\right].
\]
It follows from (the logic of) Theorem~\ref{thm:cosi-rm} that
\[
\mathbb{P}\left((\mathbb{P}(X\in A), A\in\mathcal{A}) \notin \cosici{n}{\alpha}{\mp}\right) \le \alpha\quad\mbox{for all}\quad n\ge \log_2(2/\alpha), \alpha\in(0, 1).
\]
Note that~\eqref{eq:convex-linear-function-rep} allows for many interesting choices of $\phi(\cdot)$. For example, one could use $\phi(x, y) = |x - y|$~\citep[KS distance]{massart2000constants} $\phi(x, y) = |x - y|/\sqrt{y(1-y)}$~\citep{eicker1979asymptotic}, $\phi(x, y) = (\sqrt{x} - \sqrt{y})^2 + (\sqrt{1-x} - \sqrt{1-y})^2$ (Hellinger distance), $\phi(x, y) = x\ln(x/y) + (1-x)\ln((1-x)/(1-y))$~\citep[KL divergence]{dumbgen2023new}; it also allows for debiased KL divergences introduced in~\cite{dumbgen2023new} or arbitrary $\phi$-divergences. 

\subsection{Inference for Gram Matrix}\label{sec:gram-matrix}
Suppose $X_1, \ldots, X_n$ are independent random vectors in $\mathbb{R}^d$. Let $\Sigma = \mathbb{E}[X_iX_i^{\top}]$ (a common Gram matrix). There are multiple index sets feasible in this case, corresponding to different norm balls for confidence sets:
\begin{enumerate}
    \item Setting $\mathcal{T} = \{A\in\mathbb{R}^{d\times d}:\, A\mbox{ is symmetric with }\mbox{tr}(A^2) \le 1\}$ and 
    \[
    \delta_A(M) = \mbox{tr}(AM)\quad\mbox{for}\quad A\in\mathcal{T}\mbox{ and }M\mbox{ symmetric},
    \]
    yields the Frobenius norm (or Hilbert-Schmidt norm)
    \[
    \rho(M) = \|M\|_{\mathrm{F}} = \sqrt{\mbox{tr}(M^2)}.
    \]
    \item Setting $\mathcal{T} = \{(\eta, v)\in\{\pm 1\}\times\mathbb{R}^d:\, \|v\|_2 \le 1\}$ and
    \[
    \delta_t(M) = \eta v^{\top}Mv,\quad\mbox{for}\quad t = (\eta, v)\in\mathcal{T}\mbox{ and }M\mbox{ symmetric}, 
    \]
    yields the operator norm
    \[
    \rho(M) = \|M\|_{\mathrm{op}} = \sup_{\|v\|_2 \le 1}\,|v^{\top}Mv|.
    \]
    \item Setting $\mathcal{T} = \{(\eta, i, j):\, \eta\in\{\pm 1\}, 1\le i \le j\le d\}$ and
    \[
    \delta_t(M) = \eta M_{ij},\quad\mbox{for}\quad t = (\eta, i, j)\mbox{ and }M\mbox{ symmetric},
    \]
    yields the maximum elementwise norm
    \[
    \rho(M) = \max_{1\le i\le j \le d}|M_{ij}|.
    \]
\end{enumerate}
These examples can be easily extended to include other matrix norms. 
For any of these choices of $\mathcal{T}$, we can define $Y_1, \ldots, Y_n \in \mathbb{R}^{\mathcal{T}}$ through their evaluation maps
\[
\delta_t(Y_i) = \delta_t(X_iX_i^{\top})\quad\mbox{for all}\quad t\in\mathcal{T}.
\]
This can be considered a general version of vectorizing. Note that $\delta_t(X_iX_i^{\top})$ is linear in $X_iX_i^{\top}$ and hence, the random variables $Y_i$'s satisfy
\[
\mathbb{E}[\delta_t(Y_i)] = \delta_t(\Sigma)\quad\mbox{for all}\quad t\in\mathcal{T}.
\]
Following Algorithm~\ref{alg:cosi-hunt-and-test}, set
\[
\widehat{\theta}_{j,t} := \frac{1}{|S_j|}\sum_{i\in S_j}\, \delta_t\left(X_iX_i^{\top}\right) = \delta_t\left(\frac{1}{|S_j|}\sum_{i\in S_j}X_iX_i^{\top}\right)\quad\mbox{for all}\quad t\in\mathcal{T}.
\]
This can be alternatively described as computing the empirical Gram matrix on each of the partitions $S_1, \ldots, S_B$ as $\widehat{\Sigma}_1, \ldots, \widehat{\Sigma}_B$ and then finding the evaluations:
\[
\widehat{\theta}_{j,t} = \delta_t(\widehat{\Sigma}_j)\quad\mbox{where}\quad \widehat{\Sigma}_j := \frac{1}{|S_j|}\sum_{i\in S_j} X_iX_i^{\top}.
\]
Let $\widehat{\Sigma}_0\in\mathbb{R}^{d\times d}$ be an arbitrary estimator of $\Sigma$ based on $X_i, i\in S_0$. Define the $\cosi{\mp}$ confidence set
\[
\cosici{n}{\alpha}{\mp} = \left\{M\in\mathbb{R}^{d\times d}\mbox{ symmetric}:\sup_{t\in\mathcal{T}}\,\delta_t(\widehat{\Sigma}_0 - M) \le \sup_{t\in\mathcal{T}}\,\left(\delta_t(\widehat{\Sigma}_0 - \widehat{\Sigma}_{\cdot})_{(c^*_{B,\alpha}+1)}\right)\right\}.
\]
Similar to the setting of high-dimensional mean, if there exists $\delta \in (0, 1]$ and $L \ge 1$ such that
\begin{equation}\label{eq:moment-equivalence-covariance}
\frac{1}{|S_j|}\sum_{i\in S_j}\mathbb{E}[|\delta_t(X_iX_i^{\top} - \Sigma)|^{2 + \delta}] \le L\left(\frac{1}{|S_j|}\sum_{i\in S_j}\mathbb{E}[|\delta_t(X_iX_i^{\top} - \Sigma)|^2]\right)^{1 + \delta/2},\quad\mbox{for all}\quad t\in\mathcal{T},
\end{equation}
and~\eqref{eq:split-sizes} hold true, then
\[
\mathbb{P}(\Sigma \notin \cosici{n}{\alpha}{\mathfrak{m}+}|\widehat{\Sigma}_0) \le \alpha\times\exp(5.3B^{1 + \delta/2}L/n^{\delta/2}).
\]
This confidence set depending on the choice of $\mathcal{T}$ can yield either Frobenius ball or operator norm ball or elementwise rectangle confidence regions. Similar to the high-dimensional mean example, there is no dimension constraint in the coverage guarantee, and in particular, dimension can be arbitrarily high and even infinite as long as~\eqref{eq:moment-equivalence-covariance} holds. It is interesting to contrast this with increasing dimension CLTs for Frobenius or operator norm of $\widehat{\Sigma}_{[n]} - \Sigma$ where $\widehat{\Sigma}_{[n]}$ is the empirical Gram matrix computed on $n$ observations. \cite{lopes2023bootstrapping}, and~\cite{lopes2025improved} have derived dimension independent CLTs under significantly restrictive conditions such as decaying eigenvalues of $\Sigma$ and that $X_i$'s are obtained as linear combinations of independent random variables. We do not require any such conditions as long as~\eqref{eq:moment-equivalence-covariance} holds. Note that all of the discussion above carries over seamlessly to infinite-dimensional spaces.
\paragraph{Implication for Eigenvalues.} If $\mathcal{T}$ is chosen so as to obtain operator norm balls as confidence sets of $\Sigma$, then one can obtain inference for all eigenvalues simultaneous without additional cost, thanks to Weyl's inequality:
\[
\max_{1\le j\le d}|\lambda_j(M) - \lambda_j(\widehat{\Sigma}_0)| \le \|M - \widehat{\Sigma}_0\|_{\mathrm{op}},
\]
where $\lambda_1(M) \ge \lambda_2(M) \ge \cdots \ge \lambda_d(M)$ are the ordered eigenvalues of the symmetric matrix $M$. Therefore, 
\[
\left\{v\in\mathbb{R}^d:\, v_1 \ge v_2 \ge \cdots \ge v_d\mbox{ and }\max_{1\le j\le d}|v_j - \lambda_j(\widehat{\Sigma}_0)| \le \sup_{u\in\mathbb{S}^{d-1}}\,|u^{\top}(\widehat{\Sigma}_0 - \widehat{\Sigma}_{\cdot})u|_{(c^*_{B,\alpha} + 1)}\right\},
\]
is an asymptotically valid confidence set for the ordered eigenvalues if $\cosi{\mp}$ is an asymptotically valid confidence set for $\Sigma$; see Section 3.2 of~\cite{lopes2023bootstrapping} for a related discussion. Note that this also yields asymptotically valid confidence intervals for the operator norm of $\Sigma$. We refer to this article for further applications of operator norm ball confidence sets for the Gram matrix.
\paragraph{Extension for Tensor norms.} All of the discussion above also extends to tensors. If, instead, of the Gram matrix one is interested in estimating the $q$-th order tensor of $X_i$'s, i.e., the target of inference is
\[
\Sigma_q := \mathbb{E}[X\otimes X\otimes\cdots\otimes X] = \mathbb{E}[X^{\otimes q}].
\]
Let $X_1, \ldots, X_n$ be independent random vectors in $\mathbb{R}^d$ with a common $\Sigma_q$. Let $\mathrm{Sym}_{q}^d(\mathbb{R}^d)$ be the space of symmetric tensors of order $q$ over $\mathbb{R}^d$ endowed with its operator norm:
\[
\|M\|_{\mathrm{op}} := \sup_{u\in\mathbb{S}^{d-1}}\,|\langle M,\, u^{\otimes q}\rangle|
\]
As before, define
\[
\widehat{\Sigma}_{q,j} := \frac{1}{|S_j|}\sum_{i\in S_j} X_i^{\otimes q}\quad\mbox{and}\quad \delta_t(\widehat{\Sigma}_{q,j}) = \eta\frac{1}{|S_j|}\sum_{i\in S_i} \langle X_i, u\rangle^q\quad\mbox{for}\quad t = (\eta, u)\in\{\pm1\}\times\mathbb{S}^{d-1}.
\]
Then the $\cosi{\mp}$ confidence set as defined above yields a confidence set with a dimension-free asymptotic validity under a condition analogues to~\eqref{eq:moment-equivalence-covariance}. See~\cite{OliveiraRicoThompson2026} for references and a related robust estimation scheme. Also, see~\cite{mikulincer2022clt} for some high-dimensional CLTs for higher-order tensors. 
\subsection{Inference for High-dimensional Mean under Adversarial Cellwise Contamination}\label{sec:mean-cellwise-contamination}
Suppose $X_1', X_2', \ldots, X_n'$ are independent random vectors in $\mathbb{R}^d$ with mean $\mu$ and variance-covariance matrix $\Sigma\in\mathbb{R}^{d\times d}$. We want to construct a confidence set for $\mu\in\mathbb{R}^d$. Unlike the traditional high-dimensional mean inference setting, we do not observe $X_1', \ldots, X_n'$ but $X_1, \ldots, X_n$ which were manipulated by an adversary. \cite{kock2025high} have considered the question of robust estimation and robust high-dimensional CLTs under this setting for winsorized means. This work improves on its predecessors~\cite{liu2024robust} (where the focus is on reducing moment assumption in high-dimensional CLTs) and~\cite{resende2024robust} (where the focus is on adversarial contamination but the estimator is based on trimmed means.) The only restriction on the adversary is that he/she can change a known number of observations, i.e., 
\begin{equation}\label{eq:Kock-Prienerstofer-adversarial}
    \sum_{i=1}^n \mathbf{1}\{X_i \neq X_i'\} \le n\eta_n,\quad\mbox{for all}\quad n\ge 1,\quad\mbox{for a known sequence }\{\eta_n\}_{n\ge1}.
\end{equation}
Under this setting,~\cite{kock2025high} have established a high-dimensional CLT for the winsorized mean estimator in $\mathbb{R}^d$. We describe it below as it is also needed for our construction. For any $m\ge 1$ and vector $z = (z_1, \ldots, z_m)\in\mathbb{R}^m$, let the winsorized mean at rank $r \le \lfloor m/2\rfloor$ be defined as
\[
W_r(z) := \frac{1}{m}\left[rz_{(r)} + \sum_{\ell = r+1}^{m-r} z_{(\ell)} + rz_{(m-r+1)}\right],
\]
where $z_{(1)} \le z_{(2)} \le \cdots \le z_{(m)}$ are the order statistics of $z$. 
Let $X_i = (X_{i,1}, X_{i,2}, \ldots, X_{i,d})^{\top}$ represent the $i$-th observation. For any $\varepsilon_n \in (0, 1/2)$, the high-dimensional winsorized mean estimator of~\cite{kock2025high} then takes the form
\[
\widehat{\theta}_{[n]} = (\widehat{\theta}_{[n],1}, \ldots, \widehat{\theta}_{[n],d})^{\top},
\]
with
\[
\widehat{\theta}_{[n],k} = W_{\lceil \varepsilon_n n\rceil}((X_{i,k})_{1\le i\le n}),\quad\mbox{and}\quad \varepsilon_n = \lambda_1\eta_n + \lambda_2\frac{\log(dn)}{n},
\]
for some $\lambda_1\in(1, \infty)$ and $\lambda_2\in(0,\infty)$ that are data/distribution independent; the authors suggest using $\lambda_1 = 1.01$ and $\lambda_2 = 0.1$. Without significant calculations, it is evident that this construction can be significantly sub-optimal when the coordinates of $X_i'$ (uncorrupted data) are highly correlated. For instance, if all coordinates of $X_i'$ are precisely the same, then one should true $d$ as 1 and hence the second component of $\varepsilon_n$ imposes significant winsorization. Also, to ensure $\varepsilon_n < 1/2$, it is necessary that $\log(dn)/n \ll 1$, which imposes an explicit dimension restriction even without the consideration of moments of $X_i'$. From~\cite{kock2025high}, it follows that the rate of convergence of the high-dimensional CLT for winsorized mean is significantly slower and imposes strong requirements on the dimension. In what follows, we show that the $\cosi{}$ framework can yield a confidence set under much weaker conditions on the data and the contamination scheme. 

We consider a much more general adversarial contamination setting called cellwise adversarial contamination~\citep{chen2013robust,agostinelli2015robust,hubert2026cellwise}:
\begin{equation}\label{eq:cellwise-adversarial}
    \max_{1\le k\le d}\,\sum_{i=1}^n \mathbf{1}\{X_{i,k} \neq X_{i,k}'\} \le n\eta_n\quad\mbox{for all}\quad n\ge 1,\quad\mbox{for a known sequence }\{\eta_n\}_{n\ge1}.
\end{equation}
This allows for the case where each coordinate of data is obtained from a different source and $d$ different adversaries can change the coordinates arbitrarily. For example, if an adversary changes only one coordinate in each observation, then~\eqref{eq:cellwise-adversarial} holds with $\eta_n = 1/n$ while~\eqref{eq:Kock-Prienerstofer-adversarial} can only hold with $\eta_n = 1$. In this general setting, we are not aware of any central limit theorem results, nevertheless we can apply $\cosi{}$ framework and obtain an asymptotically valid confidence hyperrectangle for $\mu$ irrespective of the dimension $d$. Another advantage is that winsorization used for each coordinate now becomes independent of dimension $d$.

As in Algorithm~\ref{alg:cosi-hunt-and-test}, partition $\{1, 2, \ldots, n\}$ as
\[
\{1, 2, \ldots, n\} = S_0 \cup S_1 \cup \cdots \cup S_B.
\]
Unlike the settings considered above, the data $X_i, i\in S_j$ is not independent of $X_i, i\in S_{j'}$ for $j \neq j'$ because the adversary has access to the entire data for contamination and could potentially introduce dependence. Hence, one cannot obtain $B+1$ independent estimators of the mean. 

For any vector $z = (z_1, \ldots, z_m)\in\mathbb{R}^m$ and $q < r \le m/2$, define
\begin{equation}\label{eq:bracketing-winsorized-means}
\begin{split}
{W}_{r,q}^{-}(z) &= \frac{1}{m}\left[rz_{(r-q)} + \sum_{\ell = r-q+1}^{m-r-q} z_{(\ell)} + rz_{(m-r-q+1)}\right],\\
{W}_{r,q}^{+}(z) &= \frac{1}{m}\left[rz_{(r + q)} + \sum_{\ell = r+q+1}^{m-r+q} z_{(\ell)} + rz_{(m-r+q+1)}\right].
\end{split}
\end{equation}
Fix $\lambda \ge 1$ and define
\begin{equation}\label{eq:winsorization-rank}
r := \lceil\lambda\eta_nn + 1\rceil,\quad\mbox{and}\quad r_0 = \lceil \lambda\eta_nn + h\rceil,
\end{equation}
with $\eta_n$ as in~\eqref{eq:cellwise-adversarial}. For $1\le j\le B$ and $1\le k\le d$, set
\begin{align*}
    \widehat{\theta}_{j,k}^- := {W}_{r,\eta_nn}^{-}((X_{i,k})_{i\in S_j}),\quad\mbox{and}\quad \widehat{\theta}_{j,k}^+ := {W}_{r,\eta_nn}^{+}((X_{i,k})_{i\in S_j}).
\end{align*}
Also, for $1\le k\le $d, set
\[
\vartheta_k^- := W_{r_0, \eta_nn}^{-}((X_{i,k})_{i\in S_0})\quad\mbox{and}\quad\vartheta_k^{+} := W_{r_0,\eta_nn}^+((X_{i,k})_{i\in S_0}) 
\]
Finally, define
\[
\cosici{n}{\alpha}{\mathtt{m}\mbox{-}\mathtt{rob}+} := \bigotimes_{k=1}^d \left[\vartheta_{k}^- - R_{\alpha},\, \vartheta_{k}^+ + R_{\alpha}\right],
\]
where
\[
R_{\alpha} := \max_{1\le \ell\le d}\left(\max\left\{|\vartheta_{\ell}^+ - \widehat{\theta}_{\cdot,\ell}^-|,\,|\vartheta_{\ell}^- - \widehat{\theta}_{\cdot,\ell}^+|\right\}\right)_{(c^*_{B,\alpha}+1)}.
\]
Even though the estimators $\widehat{\theta}_j^{\pm}$ are not independent over $1\le j\le B$, we have the following result regarding coverage and diameter. 
\begin{theorem}\label{thm:cover-diameter-cellwise-contamination}
    Suppose $X_1', \ldots, X_n'$ are independent random vectors in $\mathbb{R}^d$ with mean $\mu$ and variance-covariance matrix $\Sigma\in\mathbb{R}^{d\times d}$. Further, suppose for some $\delta > 0$ and $L \ge 1$,
    \begin{equation}\label{eq:moment-equivalence-cellwise}
    \frac{1}{|S_j|}\sum_{i\in S_j}\mathbb{E}[|X_{i,k}' - \mu_k|^{2 + \delta}] \le L\left(\frac{1}{|S_j|}\sum_{i\in S_j} \mathbb{E}[|X_{i,k}' - \mu_k|^2]\right)^{1 + \delta/2}\quad\mbox{for all}\quad 0\le j\le B, 1\le k\le d. 
    \end{equation}
    Then, under~\eqref{eq:split-sizes}, for any $\lambda \ge 1$ and $h \ge 1$, 
    \[
    \mathbb{P}\left(\mu\notin\cosici{n}{\alpha}{\mathtt{m}\mbox{-}\mathtt{rob}+}\right) ~\le~ \alpha\times\exp\left(\frac{C_{\delta}LB^{\delta/(4+2\delta)}(\lambda\eta_nn + 2 + \log(n/B))^{(1+\delta)/(2 + \delta)}}{n^{\delta/(4+2\delta)}}\right),
    \]
    for a constant $C_{\delta}$ depending only on $\delta > 0$.
    Additionally, suppose $\lambda > 2$, $\eta_nn \ge 1$, $|S_0| \ge n/4$. For $\gamma\in(0, 1)$ with $h = 1 + \log(3/\gamma)$, we have that with probability at least $1 - \gamma$,
    \begin{align*}
    \frac{1}{C}\sup_{\theta\in\cosici{n}{\alpha}{\mathtt{m}\mbox{-}\mathtt{rob}+}}\,\|\theta-\mu\|_\infty &\le
    \frac{B(e^{4\log(3/\gamma)/(B-c^*_{B,\alpha}-1)}-1)}
    {\log(3/\gamma)}
    \left[\mathfrak{E}_d+
    \mathfrak{v}(B\eta_n)^{(1+\delta)/(2+\delta)}\right]\\
    &\quad+\frac{B\mathfrak{v}}{\sqrt n}
    \sqrt{\frac{e^{4\log(3/\gamma)/(B-c^*_{B,\alpha}-1)}-1}
    {\log(3/\gamma)}}.
    \end{align*}
    where $C$ is a constant depending only on $\delta, L, \lambda$, and
    \[
    \mathfrak{E}_d := \mathbb{E}\left[\max_{1 \le k\le d}\,\left|\frac{1}{n}\sum_{i=1}^n (X_{i,k}' - \mu_k)\right|\right],\quad\mbox{and}\quad \mathfrak{v}^2 := \max_{1\le k\le d}\max_{0 \le j\le B}\,\frac{1}{|S_j|}\sum_{i\in S_j} \mathbb{E}[Y_{i,k}^2].
    \]
\end{theorem}
See Section~\ref{appsec:proof-of-cover-diameter-cellwise-contamination} for a proof. Note that the assumption $\eta_nn \ge 1$ can be made without loss of generality because even if there is no contamination ($\eta_n = 0$), inequality~\eqref{eq:cellwise-adversarial} holds with $\eta_n = 1/n$. Observe also that the asymptotic coverage guarantee requires $\eta_n$ to converge to zero so that
\begin{equation}\label{eq:condition-on-contamination}
\eta_n = o\left(n^{-(2+\delta)/(2+2\delta)}\right)\quad\mbox{as}\quad n\to\infty,
\end{equation}
assuming $B$ is fixed as $n\to\infty$. It is easy to verify that if $d$ is fixed as $n\to\infty$, condition~\eqref{eq:condition-on-contamination} is precisely the same requirement in Theorem 2.1 of~\cite{kock2025high}, and~\eqref{eq:condition-on-contamination} can be significantly weaker if $d \to \infty$ with $n\to\infty$. 

As a special case of the diameter bound, taking $\gamma = \alpha$ and choosing $B$ so as to satisfy~\eqref{eq:small-requirement-on-B} simplifies the radius bound as
\[
C\left[\mathfrak{E}_d + \mathfrak{v}\sqrt{\frac{\log(1/\alpha)}{n}}
    +\mathfrak{v}(\log(1/\alpha)\eta_n)^{(1+\delta)/(2+\delta)}\right],
\]
which closely matches the rate of convergence of robust mean estimator under adversarial (row-wise) contamination in~\cite{oliveira2025trimmed}, with a slight sub-optimality in the term depending on contamination level $\eta_n$. Note that this radius bound is derived without requiring~\eqref{eq:condition-on-contamination}. If~\eqref{eq:condition-on-contamination} holds, then for fixed $\alpha\in(0,1)$, the third term in the radius bound is dominated by the second term. Whether inference is possible without $\eta_n\to0$ (in the absence of knowledge of $L$ and $\delta$) is unclear at the moment and we believe it to be impossible. 

\subsection{Inference for Fr{\'e}chet Mean in ``nice'' Metric Spaces}\label{subsec:metric-space-mean}
Algorithm~\ref{alg:cosi-hunt-and-test} is designed for estimators that lie in a topological vector space. Nevertheless, the ideology can be ported to general metric spaces with some structure. In this section, we illustrate this idea by constructing a confidence set for Fr{\'e}chet means in metric spaces. We describe two approaches one that does not require any structure on the metric space but yields larger confidence sets, and the second that requires local Hilbert space structure as in~\cite{le2022fast} and yields ``optimal'' radius. 

Let $(\mathcal{X}, d)$ be a separable metric space and $X$ be a random variable with probability measure $P$ taking values in the measurable space $(\mathcal{X}, \mathcal{B}(\mathcal{X}))$. Assuming $\mathbb{E}[d^2(X, x)] < \infty$ for some $x\in\mathcal{X}$, a Fr{\'e}chet mean is defined as a minimizer:
\begin{equation}\label{eq:Frechet-mean}
\mathcal{M}(P) = \argmin_{x\in\mathcal{X}}\,\mathbb{E}_P[d^2(X, x)].
\end{equation}
We assume that at least one minimizer exists and for $\theta^*\in\mathcal{M}(P)$ and $\theta\in\mathcal{X}$, we have
\begin{equation}\label{eq:zero-order-criterion}
\mathbb{E}[d^2(X, \theta^*) - d^2(X, \theta)] \le 0.
\end{equation}
Hence, one could construct a confidence set for $\mathcal{M}(P)$ by collecting all $\theta^*$ such that the test for~\eqref{eq:zero-order-criterion} is not rejected against a well-designed $\theta$. This strategy is well-developed in the literature for parametric models and general $M$-estimation problems; see~\cite{wasserman2020universal} and~\cite{takatsu2025bridging,takatsu2025honest} for details and literature review. As in Algorithm~\ref{alg:cosi-hunt-and-test}, let $S_0, S_1, \ldots, S_B$ denote the partition of $\{1, 2, \ldots, n\}$ and let $\vartheta\in\mathcal{X}$ be an estimator of $\mu$ based on $X_i, i\in S_0$. Consider
\begin{equation}\label{eq:risk-inversion-set}
\widetilde{\mathrm{CI}}_{n,\alpha}^{\mathtt{RI}} := \left\{\theta\in\mathcal{X}:\, \sum_{j=1}^{B} \mathbf{1}\left\{\sum_{i\in S_j}\, [d^2(X_i, \theta) - d^2(X_i, \vartheta)] \le 0\right\} \ge B - c^*_{B,\alpha}\right\}.
\end{equation}
Theorem 3 of~\cite{takatsu2025honest} provides conditions such that
\begin{equation}\label{eq:median-valid}
\inf_{\theta^*\in\mathcal{M}(P)}\,\mathbb{P}\left(\sum_{i\in S_j}[d^2(X_i, \theta^*) - d^2(X_i, \vartheta)] \le 0\big|\vartheta\right) \ge \frac{1}{2} - o_p(1)\quad\mbox{as}\quad |S_j| \to \infty.
\end{equation}
Hence, from Proposition~\ref{prop:binomial-probabilities}, it follows that $\widetilde{\mathrm{CI}}_{n,\alpha}^{\mathtt{RI}}$ has an asymptotic coverage of at least $1 - \alpha$. The confidence set in~\eqref{eq:risk-inversion-set} can be seen as a slight generalization of the confidence set described in Section 4.1 of~\cite{takatsu2025honest}. Diameter analysis also follows from Section 5 of that paper, which we do not pursue here. The confidence set~\eqref{eq:risk-inversion-set} is understood to be based on the zeroth order criterion~\eqref{eq:zero-order-criterion} for Fr{\'e}chet mean. The advantage of this construction is that uniqueness of Fr{\'e}chet mean is not required. The disadvantage is that the validity guarantee~\eqref{eq:median-valid} can be significantly conservative because $\mathbb{E}[d^2(X_i, \theta^*) - d^2(X_i, \vartheta)|\vartheta]$ can be significantly negative; this, in turn, implies that~\eqref{eq:risk-inversion-set} can be significantly conservative as already illustrated in Section 5 of~\cite{takatsu2025bridging}.

Under certain structure on the metric space, there is a first order criterion for Fr{\'e}chet mean as a solution to an equation. This yields sharper control on the coverage guarantee and also a better diameter bounds. For this, we assume that the argmin set $\mathcal{M}(P)$ is a singleton and denote the Fr{\'e}chet mean by $\mu$. We need the following background to define the first order criterion and the subsequent confidence set.
\paragraph{Geodesics, tangent cones, and logarithm maps.}
We now introduce the first-order geometric objects used in the second
construction. The discussion follows~\citet[Parts II--III]{alexander2024alexandrov},~\citet[Section 2]{le2022fast}, and~\citet[Section 4.1]{bartl2025robust}.

For the remainder of this subsection, suppose that $(\mathcal X,d)$ is a complete separable geodesic space whose curvature is bounded below in the Alexandrov sense: $\operatorname{curv}(\mathcal X)\ge \underline{K}$ for some $\underline{K}\in\mathbb R$. Informally, this means that geodesic triangles in $\mathcal X$ are no thinner than comparison triangles in the two-dimensional space of constant curvature $\underline{K}$; see~\citet[Chapter 7]{alexander2024alexandrov} and~\citet[Section 2.1]{le2022fast} for the precise comparison inequality. A constant-speed minimizing geodesic is a curve $\gamma\colon[0,\varepsilon]\to\mathcal X$, with $\varepsilon>0$, such that
\[
d(\gamma(s),\gamma(t))=r|s-t| \quad\mbox{for all}\quad s,t\in[0,\varepsilon],
\]
for some speed $r\ge0$. Geodesicity means that every pair of points can be joined by such a curve.

For $b\in\mathcal X$, let $\mathcal G_b$ contain all such geodesics starting at $b$, allowing arbitrary speeds and positive domain
lengths. Under the curvature assumption, 
\[
d_b^{\mathrm{tan}}(\gamma,\eta) := \lim_{t\downarrow0}\frac{d(\gamma(t),\eta(t))}{t}
\]
exists and defines a pseudometric on $\mathcal G_b$. Identifying $\gamma$ and $\eta$ when $d_b^{\mathrm{tan}}(\gamma,\eta)=0$, and
taking the metric completion, gives the tangent cone $T_b\mathcal X$. Write $[\gamma]$ for the class of $\gamma$ and
$d_{T_b\mathcal X}$ for the induced metric. The class of constant geodesics is its tip $o_b$, and we write
$\|u\|_b:=d_{T_b\mathcal X}(u,o_b)$. Thus, tangent vectors record initial directions and speeds;
see~\citet[Section~6E]{alexander2024alexandrov}.

Fix a selected minimizing geodesic $\gamma_{b,x}\colon[0,1]\to\mathcal X$ from $b$ to $x$ for each
$(b,x)$. Define 
\[
\log_b(x):=[\gamma_{b,x}], \quad \log_b(b)=o_b, \quad \|\log_b(x)\|_b=d(b,x).
\]
We assume that $(b,x)\mapsto\gamma_{b,x}$ is Borel measurable for the uniform topology on $C([0,1],\mathcal X)$. The scalar quantities
below are then jointly measurable, as follows from their expressions through distances along selected paths. Fixed-base
measurability is discussed in \citet[Remark~6]{le2022fast}. The selector is fixed throughout; logarithms and the resulting
confidence sets may depend on it when minimizing geodesics are not unique.

For $u,v\in T_b\mathcal X$, define the bracket
\begin{equation}\label{eq:tangent-cone-bracket}
\langle u,v\rangle_b := \frac{1}{2}\left(\|u\|_b^2+\|v\|_b^2-d_{T_b\mathcal X}^2(u,v)\right).
\end{equation}
It satisfies $|\langle u,v\rangle_b|\le\|u\|_b\|v\|_b$, but need not be a bilinear inner product because the tangent cone
need not be a vector space; see~\citet[Section~4.1.2]{bartl2025robust}.
In a Hilbert space, $\log_b(x)=x-b$ and the bracket is the usual inner product (independent of $b$).

For $y\ne b$, the hugging function is
\begin{equation}\label{eq:hugging-function}
\kappa_b^y(x) := 1- \frac{d_{T_b\mathcal X}^2(\log_b(x),\log_b(y))-d^2(x,y)}{d^2(b,y)}.
\end{equation}
It measures the discrepancy between distances in $\mathcal{X}$ and distances between the selected logarithms; in a Hilbert
space, it equals one. Direct substitution gives 
\begin{equation}\label{eq:hugging-distance-identity}
d^2(x,y)-d^2(x,b) = \kappa_b^y(x)d^2(b,y) - 2\langle\log_b(x),\log_b(y)\rangle_b.
\end{equation}
We set $\kappa_b^b(x)=1$ by convention.

If $\mathbb{E}[d^2(X_i,\mu)]<\infty$ and $\mu$ is a Fr{\'e}chet mean of each $X_i$, the first-order optimality condition is
\begin{equation}\label{eq:mean-zero-Frechet-mean}
\mathbb E[\langle\log_\mu(X_i),\log_\mu(y)\rangle_\mu]=0\quad\mbox{for all}\quad y\in\mathcal{X},\ 1\le i\le n.
\end{equation}
Moreover, the logarithms of all observations lie almost surely in a common Hilbert subspace $L_\mu\mathcal X\subseteq
T_\mu\mathcal X$, with inner product given by the restricted bracket. These facts follow from
\citet[Theorem~2.4]{le2022fast} and \citet[Proposition~4.12]{bartl2025robust}. The first-order identity yields centered scores for coverage; the Hilbert structure is used in the radius proof.

With this background, we are now ready to describe the $\cosi{}$ confidence set for Fr{\'e}chet mean. Given independent observations $X_1, \ldots, X_n$, following Algorithm~\ref{alg:cosi-hunt-and-test}, compute $\vartheta$ an estimator of $\mu$ based on $X_i, i\in S_0$. Consider 
\begin{equation}\label{eq:cosi-m-inner-product-metric-space}
\cosicit{n}{\alpha}{\mp} := \left\{\theta\in\mathcal{X}:\, \sum_{j=1}^B \mathbf{1}\left\{\sum_{i\in S_j} \langle \log_{\theta}(X_i), \log_{\theta}(\vartheta)\rangle_{\theta} \le 0\right\} \ge B - c^*_{B,\alpha}\right\},
\end{equation}
with the inner product as defined in~\eqref{eq:tangent-cone-bracket}. Note that, after fixing a jointly measurable geodesic selector, the confidence set is well defined even if minimizing geodesics are not unique; however, it may depend on the chosen selector. From~\eqref{eq:mean-zero-Frechet-mean}, it follows from Lindeberg-Feller CLT that 
\[
\frac{1}{|S_j|^{1/2}\sigma_{j}}\sum_{i\in S_j}\langle \log_{\mu}(X_i), \log_{\mu}(\vartheta)\rangle_{\mu} ~\overset{d}{\to}~ N(0, 1),
\]
for some $\sigma_j \ge 0$ as $|S_j| \to \infty$; see Theorem~\ref{thm:metric-space-mean} for a formal assumption. Hence, $\cosicit{n}{\alpha}{\mp}$ has an asymptotic coverage of exactly $1-\alpha$ as $n\to\infty$ (e.g., under~\eqref{eq:split-sizes}). However, a control of the radius of this confidence set requires a non-negativity assumption on $(\theta, x, y)\mapsto \kappa_{\theta}^y(x)$ over $\mathcal{X}^3$ which can be restrictive. But it is interesting to note that one can combine~\eqref{eq:risk-inversion-set} and~\eqref{eq:cosi-m-inner-product-metric-space} for sharper coverage and radius guarantee. Consider
\begin{equation}\label{eq:cosi-metric-space-final}
\cosici{n}{\alpha}{\mp} ~:=~ \left\{\theta\in\mathcal{X}:\,\sum_{j=1}^B \mathbf{1}\{\max\{S_j(\theta),\, R_j(\theta)\} \le 0\} \ge B - c^*_{B,\alpha}\right\},
\end{equation}
where, for $1\le j\le B$,
\[
S_j(\theta) := \sum_{i\in S_j} \langle\log_{\theta}(X_i),\,\log_{\theta}(\vartheta)\rangle_{\theta}\quad\mbox{and}\quad R_j(\theta) := \frac{1}{2}\sum_{i\in S_j} [d^2(X_i, \theta) - d^2(X_i, \vartheta)].
\]
Clearly,
\[
\cosici{n}{\alpha}{\mp} ~\subseteq~ \cosicit{n}{\alpha}{\mathtt{RI}} \mathbin{\scalebox{2}{$\cap$}} \cosicit{n}{\alpha}{\mp}\quad\mbox{for all}\quad n \ge 1, \alpha\in [0, 1].
\]

The following result provides miscoverage and radius bounds for this confidence set. The following assumption on the hugging function plays a role in the diameter analysis.
\begin{equation}\label{eq:lower-bound-hugging-function}
\inf_{y\in\mathcal{X},\,y\neq \mu}\,\kappa_{\mu}^y(X_i) \ge \kappa_{\min} > 0,\quad\mbox{for all}\quad i\notin S_0\quad\mbox{almost surely}.
\end{equation}
If~\eqref{eq:lower-bound-hugging-function} holds, then the Fr{\'e}chet mean is unique; see Theorem 3.3 of~\cite{le2022fast}.
\begin{theorem}\label{thm:metric-space-mean}
    Suppose $X_1, \ldots, X_n$ are independent random variables in the metric space $(\mathcal{X}, d)$ with a common Fr{\'e}chet mean $\mu$ such that~\eqref{eq:mean-zero-Frechet-mean} holds. If there exists $L \ge 1$ and $\delta\in(0, 1]$ such that for $1\le j\le B$ and all $b\in\mathcal{X}$,
    \begin{equation}\label{eq:Berry--Esseen-Metric-space}
    \frac{1}{|S_j|}\sum_{i\in S_j}\mathbb{E}[|\langle \log_{\mu}(X_i), \log_{\mu}(b)\rangle_{\mu}|^{2 + \delta}] \le L\left(\frac{1}{|S_j|}\sum_{i\in S_j}\mathbb{E}[|\langle \log_{\mu}(X_i), \log_{\mu}(b)\rangle_{\mu}|^2]\right)^{1 + \delta/2},
    \end{equation}
    and~\eqref{eq:split-sizes} hold,
    then
    \[
    \mathbb{P}(\mu\notin\cosicit{n}{\alpha}{\mp}) ~\le~ \alpha\times\exp\left(5.3B^{1 + \delta/2}L/n^{\delta/2}\right).
    \]
    In addition, if $\kappa_{\mu}^{\vartheta}(X_i) \ge 0$ almost surely (which is implied by~\eqref{eq:lower-bound-hugging-function}) holds, then
    \[
    \mathbb{P}(\mu\notin\cosici{n}{\alpha}{\mp}) = \mathbb{P}(\mu\notin\cosicit{n}{\alpha}{\mp}) \le \alpha\times\exp\left(5.3B^{1 + \delta/2}L/n^{\delta/2}\right).
    \]
    Set $e_{\kappa, n} := (2/n)\sum_{i\notin S_0}\,\mathbb{E}[\kappa_{\mu}^{\vartheta}(X_i)|\vartheta]$, and
    \begin{align*}
    V_{2,n} &:= \frac{1}{n}\sum_{i=1}^n \mathbb{E}[d^2(X_i, \mu)] < \infty,\quad\mbox{and}\quad \sigma_{w,n}^2 := \sup_{\theta\in\mathcal{X}\setminus\{\mu\}}\,\frac{1}{n}\sum_{i=1}^n \frac{\mathbb{E}[|\langle \log_{\mu}(X_i),\,\log_{\mu}(\theta)\rangle_{\mu}|^2]}{d^2(\mu, \theta)}.
    \end{align*}
    If~\eqref{eq:lower-bound-hugging-function} holds, then for any $\gamma\in(0, 1]$, conditional on $\vartheta$, with probability at least $1 - \gamma$,
    \begin{align*}
    \sup_{\theta\in\cosici{n}{\alpha}{\mp}}\,d(\theta,\mu) 
    &\le~ \frac{16B(e^{4\log(1/\gamma)/(B-c^*_{B,\alpha} - 1)} - 1)}{\log(1/\gamma)}\sqrt{\frac{V_{2,n}}{n\kappa_{\min}^2}}\\ 
    &\quad+ \sqrt{\frac{B(e^{4\log(1/\gamma)/(B-c^*_{B,\alpha} - 1)} - 1)}{\log(1/\gamma)}}\left(20\sqrt{\frac{B\sigma_{w,n}^2}{n\kappa_{\min}^2}} + 2d(\mu,\vartheta)(e_{\kappa,n}/\kappa_{\min}+1)^{1/2}\right).
    \end{align*}
    The same radius bound also holds for $\widetilde{\mathrm{CI}}_{n,\alpha}^{\mathtt{RI}}$.
    If, in addition, $\kappa_{\theta}^y(x) \ge 0$ for all $\theta, y, x\in\mathcal{X}$ with $y\neq \theta$ holds, then the same radius bound holds for $\cosicit{n}{\alpha}{\mp}$.
\end{theorem}
See Section~\ref{appsec:proof-of-thm-metric-space-mean} for a proof. 
As mentioned, condition~\eqref{eq:lower-bound-hugging-function} holds for Hilbert spaces with $\kappa_{\min} = 1$. The same also holds true for Hadamard spaces (also called global NPC spaces or complete CAT(0) spaces), but they do not imply the first order criterion~\eqref{eq:mean-zero-Frechet-mean}. However, for the coverage guarantee of Theorem~\ref{thm:metric-space-mean}, we only need $\mathbb{E}[\langle\log_{\mu}(X_i),\,\log_{\mu}(y)\rangle_{\mu}] \le 0$ for all $y\in\mathcal{X}, 1\le i\le n$ and this holds for all CAT(0) spaces. From CAT(0) variance inequality~\citep{gietl2026generalized}, one can obtain bounds on the radius of the risk inversion confidence set~\eqref{eq:risk-inversion-set} which implies radius bound for the $\cosi{\mp}$ confidence set~\eqref{eq:cosi-metric-space-final}. This implies confidence sets for spaces discussed in~\citet[Section 1.2]{schotz2025transformed}. 

Condition~\eqref{eq:lower-bound-hugging-function} can be somewhat relaxed for both coverage and radius analysis. Firstly, note from the definition of Fr{\'e}chet mean and~\eqref{eq:tangent-cone-bracket}, it follows that
\begin{equation}\label{eq:equation-for-e-kappa-n}
\mathbb{E}[\kappa_{\mu}^{y}(X_i)d^2(\mu,y)] = \mathbb{E}[d^2(X_i, y) - d^2(X_i, \mu)] \ge 0\quad\mbox{for all}\quad y\in\mathcal{X},
\end{equation}
and hence, for all $y \neq \mu$, $\mathbb{E}[\kappa_{\mu}^y(X_i)] \ge 0$. In particular, $\mathbb{E}[\kappa_{\mu}^{\vartheta}(X_i)|\vartheta] \ge 0$ for all $i\notin S_0$. For the coverage guarantee of $\cosici{n}{\alpha}{\mp}$, the proof implies
\[
\mathbb{P}(\mu\notin\cosici{n}{\alpha}{\mp}|\vartheta) \le \alpha\times\exp\left(5.3B^{1+\delta/2}L/n^{\delta/2}\right) + \mathbb{P}\left(\min_{1\le j\le B}\,\sum_{i\in S_j}\,\kappa_{\mu}^{\vartheta}(X_i) < 0\big|\vartheta\right).
\]
Hence, if the law of large numbers holds for $\kappa_{\mu}^{\vartheta}(X_i), i\in S_j, 1\le j\le B$ and $\mathbb{E}[\kappa_{\mu}^{\vartheta}(X_i)] > 0$, then the last term in the miscoverage bound converges to zero. For the radius analysis, from the proof, it follows that the same bound as stated holds without~\eqref{eq:lower-bound-hugging-function} but with probability 
\[
1 - \gamma - \mathbb{P}\left(\min_{1\le j\le B}\inf_{\theta\in\mathcal{X}}\,\frac{1}{|S_j|}\sum_{i\in S_j}\kappa_{\mu}^{\theta}(X_i) < \kappa_{\min}\right).
\]
Section 3.3 and Theorem 3.5 of~\cite{le2022fast} proposes a similar idea to allow for negative hugging functions; they assume $\kappa_{\mu}^{\theta}(x) \ge \underline{\kappa}$ for some $\underline{\kappa}\in\mathbb{R}$ and that $\mathbb{E}[\inf_{\theta\in\mathcal{X}}\kappa_{\mu}^{\theta}(X_i)] > 0$.
Finally, note that the radius bound in Theorem~\ref{thm:metric-space-mean} depends on $e_{\kappa,n}$. From~\eqref{eq:equation-for-e-kappa-n}, we have
\begin{equation}\label{eq:inequality-for-e-kappa}
e_{\kappa,n} = \frac{2}{n}\sum_{i\notin S_0}\,\frac{\mathbb{E}[d^2(X_i, \vartheta) - d^2(X_i, \mu)|\vartheta]}{d^2(\mu, \vartheta)} ~\le~ \frac{2(n-|S_0|)}{n} + \frac{4}{n}\sum_{i\notin S_0}\frac{\mathbb{E}[d(X_i, \mu)]}{d(\mu, \vartheta)},
\end{equation}
where the last inequality follows from triangle inequality:
\[
d^2(X_i, \vartheta) - d^2(X_i, \mu) \le (d(X_i, \mu) + d(\vartheta, \mu))^2 - d^2(X_i, \mu) \le 2d(X_i, \mu)d(\mu, \vartheta) + d^2(\mu, \vartheta).
\]
The inequality~\eqref{eq:inequality-for-e-kappa} can be severely sub-optimal, in general. For example, in Hilbert spaces, $d^2(X_i, \vartheta) = \|X_i - \vartheta\|^2$ and hence, $\mathbb{E}[d^2(X_i, y) - d^2(X_i, \mu)] = d^2(\mu, y)$ for all $y\in\mathcal{X}$, which implies $e_{\kappa, n} = 1$ irrespective of the pilot estimator $\vartheta$. Indeed, for Alexandrov spaces with positive curvature, $\kappa_{\mu}^{y}(x) \le 1$ for all $x, y$ and hence, $e_{\kappa,n} \le 1$; see Remark 3.2 of~\cite{le2022fast}. Furthermore, combined with Lemma 5 of~\cite{zhang2016first}, we get that for all Alexandrov spaces with $\mbox{curv}(\mathcal{X}) \ge \underline{K}$ and all $x, \mu, y\in\mathcal{X}$,
\[
\kappa_{\mu}^y(x) \le \begin{cases}
    \sqrt{-\underline{K}}d(\mu, x)\coth(\sqrt{-\underline{K}}d(\mu, x)), &\mbox{if }\underline{K} < 0,\mbox{ and }\mu \neq y\\
    1, &\mbox{if }\underline{K} \ge 0,\mbox{ or }\mu = y.
\end{cases}
\]
More generally, if $\mathbb{E}[d^2(X_i, y) - d^2(X_i, \mu)] \le Cd^2(\mu, y)$ holds for all $y\in\mathcal{X}$ (``close'' to $\mu$), then $e_{\kappa, n} \le 2$ whenever $\vartheta$ is close to $\mu$.

If $\vartheta$ is chosen to the robust subGaussian mean estimator from~\cite{bartl2025robust} and $e_{\kappa,n} = O_p(1)$, then Theorem~\ref{thm:metric-space-mean} implies a confidence set with optimal subGaussian radius. 
While we do not study the problem of inference under contamination as in~\cite{bartl2025robust}, one can follow the approach of Section~\ref{sec:mean-cellwise-contamination} to obtain valid confidence set for the Fr{\'e}chet mean under adversarial contamination of the data. 

The proposed framework covers several applications including separable Hilbert spaces, complete simply connected Riemannian manifolds with negatively bounded sectional curvature~\citep[Theorem 4.1]{bartl2025robust}, and small closed geodesic balls in Riemannian manifolds with bounded curvature~\citep{bhattacharya2005large,afsari2013convergence}. The same holds for Hilbert spaces equipped with non-inner-product norms, e.g., the space of positive definite matrices equipped with log-Euclidean metric which has applications in medical imaging~\citep{arsigny2006logeuc,pennec2006riemannian}. Finally, we note that while the results above do not directly apply, one could follow the same approach to construct confidence sets for the mean in a uniformly convex Banach spaces as discussed in Section 5 of~\cite{bartl2025robust}; this yields, for example, confidence sets for $\ell_p$ means in $\mathbb{R}^d$. In the following, we provide one explicit example for the Wasserstein barycenter in $\mathbb{R}^d$, in the context of template deformation model studied in~\cite{boissard2015distribution} and~\cite{zemel2019frechet}.

\paragraph{Template deformation model.} Suppose $\mu^\star$ is a probability measure on $\mathbb{R}^d$. In a template deformation model, one observes $n$ independent copies of $\mu = T_{\#}\mu^\star$ for a random map $T:\mathbb{R}^d\to\mathbb{R}^d$ with distribution, say, $Q$; this means that $\mu$ is the distribution of $T(X)$ when $X\sim\mu^\star$ and we observe $\mu_i = (T_i)_{\#}\mu^\star$ for $T_1, \ldots, T_n$ IID from $Q$. The goal is to estimate $\mu^\star$. Under some regularity conditions, $\mu^\star$ is the Wasserstein barycenter of $\mu$; see, e.g., Theorem 5 of~\cite{zemel2019frechet}. For instance, if $T(x) = x + \xi\nabla\varphi(x)$ for a Rademacher random variable $\xi$ and a random function $\varphi:\mathbb{R}^d\to\mathbb{R}$ supported on the space of twice differentiable functions with $\|\nabla^2\varphi(x)\|_{\mathrm{op}} \le \varepsilon < 1/2$ for all $x$, then from Theorem 5 of~\cite{zemel2019frechet}, it follows that $\mu^\star$ is the Wasserstein barycenter and further from Theorems 4.2, 4.3 of~\cite{le2022fast}, it follows that
\[
1 - 2\varepsilon \le \kappa_{\mu^\star}^{\vartheta}(\nu) \le 1,
\]
which implies $\kappa_{\min} = 1 - 2\varepsilon > 0$ and $e_{\kappa,n} \le 2$. Additionally, it can be verified that $\log_{\mu^*}((T_i)_{\#}\mu^*) = T_i - \mathrm{Id} = \xi\nabla\varphi_i$~\citep[Section 6.2]{bartl2025robust}. This implies that $\langle\log_{\mu^\star}(\mu_i),\,\log_{\mu^\star}(b)\rangle_{\mu^*}, 1\le i\le n$ are independent symmetric around zero random variables, and hence, the confidence set in Theorem~\ref{thm:metric-space-mean} are finite-sample exactly valid. One can, of course, relax the assumption of symmetry of $\xi$ to get an asymptotically exactly valid confidence set. Indeed, if $T(x) = x + \sum_{\ell=1}^m \xi_{\ell}\nabla\varphi_{\ell}(x)$ for independent $[-1, 1]$-valued mean zero random variables $\xi_1, \ldots, \xi_m$ and {\em deterministic functions} $\varphi_{\ell}, 1\le \ell \le m$, then assuming $\mathbb{E}[|\xi_{\ell}|^3] \le C(\mathbb{E}[|\xi_{\ell}|^2])^{3/2}$ for all $\ell$, assumption~\eqref{eq:Berry--Esseen-Metric-space} is also verified.

\section{Dependent Data}\label{sec:cosi-dependent}

In this section, we extend the method
to handle dependent data. The basic idea is to split the dependent data into approximately independent blocks and apply the \SAV-confidence set on these blocks. Formally, the algorithm is as stated in Algorithm~\ref{alg:proposed-conf-int-dependent}. The setting is as follows: we have observations $X_i \sim P_i$ with the joint distribution of $(X_1, \ldots, X_n)$ denoted by $P^n$. Let $\theta(P^n) = \theta(P_i)$ for all $1\le i\le n$, i.e., there is a target parameter of interest that is the same for all distributions $P_i$.
\begin{algorithm2e}[h]
    \caption{$\cosi{}$~framework for confidence sets (with dependent data)}
    \label{alg:proposed-conf-int-dependent}
    \SetAlgoLined
    \SetEndCharOfAlgoLine{}
    \KwIn{Sample: $(X_1,\cdots,X_n)\sim P^n$, confidence Level: $1-\alpha$, estimation procedure $\mathcal{A}(\cdot)$ that takes a data set as input and return an estimator for $\theta(P^n)$, a scale-invariant property $\mathfrak{P}$ along with a test $\phi_{\mathfrak{P},\alpha}(\cdot)$, and integers $k_n,\, B \ge 1$.}
    \KwOut{Confidence set for $\theta(P^n)$.}
    Split $\{1, 2, \ldots, n\}$ into $B$ disjoint batches with gap of size $k_n$
    \[
    S_j = \left\{m\in\mathbb{Z}:\, (j-1)\lfloor n/B\rfloor +1 \le m \le j\lfloor n/B\rfloor - k_n\right\},\quad 1\le j\le B-1.
    \]
    Finally, $S_B = \{(B-1)\lfloor n/B\rfloor + 1, \ldots, n\}$.\;
    
    Compute estimators $\widehat{\theta}_j = \mathcal{A}(\{X_i: i\in S_j\}),$ for $1\le j\le B$\;
    
    \Return the confidence set 
    \begin{equation}\label{eq:confidence-interval-savari-dependent}
    \begin{split}
    \cosici{n}{\alpha}{dep\mbox{-}a} ~&:=~ \left\{\theta:\, \phi_{\mathfrak{P},\alpha}(\{\widehat{\theta}_j - \theta\}_{j=1}^B) = 0\right\},\\
    \cosici{n}{\alpha}{dep\mbox{-}m} ~&:=~ \left\{\theta:\, \sum_{j=1}^B \phi_{\mathfrak{P},1/2}(\widehat{\theta}_j - \theta) \le c_{B,\alpha}^*\right\},
    \end{split}
    \end{equation}
    with $c^*_{B,\alpha}$ is as in Algorithm~\ref{alg:proposed-conf-int}.
\end{algorithm2e}
Algorithm~\ref{alg:proposed-conf-int-dependent} specifically uses the structure of the indices $\{1, 2, \ldots, n\}$ to create approximately independent blocks by creating gaps of size $k_n$. Similar blocking can be done when the indices are nodes of a dependency graph. 

To prove valid coverage bounds for the dependent data \SAV-confidence set, we use the classical notions of mixing processes. Before describing the mixing coefficients, let us briefly mention other notions of ``weak'' dependence and why they are not so suitable for a general study of $\cosi{}$ framework. While general notions of functional dependence~\citep{wu2005nonlinear}, near epoch-dependence (NED)~\citep{jenish2012spatial}, and mixingales~\citep{mcleish1975maximal,andrews1988laws} are more general than mixing processes, these general notions are tied to the scale of the random process. Formally, these notions of dependence can be described as approximating the given dependent sequence by a more manageable dependent sequence with tractable structure. Hence, if the original data is assumed to be functionally dependent (or NED or mixingale), then one cannot claim that the estimators $\widehat{\theta}_1, \ldots, \widehat{\theta}_B$ are ``weakly'' dependent in a suitable sense without assuming some smoothness on the estimation algorithm $\mathcal{A}(\cdot)$. In fact, if one restricts the smoothness of the algorithm $\mathcal{A}(\cdot)$, then one could obtain guarantees for $\widehat{\mathrm{CI}}_{n,\alpha}^{\texttt{COSI}}$ from Algorithm~\ref{alg:proposed-conf-int} (without gaps in the blocks). See, for example,~\cite{lunde2019sample}. The advantage of mixing conditions is that they are invariant to arbitrary transformations of random variables and hence, a mixing condition on the data implies that the estimators $\widehat{\theta}_j, 1\le j\le B$ are also mixing. This will also be clear from the following brief description of the mixing conditions we use. Such invariance to transformations is especially useful when dealing with estimators in a high/infinite-dimensional spaces.

For any $1\le a \le b\le n$, let $\mathcal{F}_{a}^b$ represent the $\sigma$-field generated by $\{X_{a}, \ldots, X_b\}$. Define the mixing coefficients as
\begin{align*}
    \alpha_X(k) &:= \sup\{|\mathbb{P}(A\cap B) - \mathbb{P}(A)\mathbb{P}(B)|:\, A\in \mathcal{F}_1^{m},\, B\in\mathcal{F}_{m+k}^{n},\,m\ge1\},\\
    \beta_X(k) &:= \mathbb{E}\left[\sup\left\{|\mathbb{P}(B|\mathcal{F}_{1}^m) - \mathbb{P}(B)|:\, B\in\mathcal{F}_{m+k}^n,\, m\ge1\right\}\right]
\end{align*}
A sequence $X_1, \ldots, X_n$ is said to be $\alpha$-mixing or strongly mixing if $\alpha_X(k)\to0$ as $k\to\infty$. Similarly, it is called $\beta$-mixing or absolutely regular if $\beta_X(k)\to0$ as $k\to\infty$. 
Defining, alternatively,
\begin{align*}
    \alpha'_X(k) &:= \sup\left\{\mathbb{E}[|\mathbb{P}(B|\mathcal{F}_{1}^m) - \mathbb{P}(B)|]:\, B\in\mathcal{F}_{m+k}^n,\, m\ge1\right\},
\end{align*}
it is relatively easy to see that $\alpha'_X(k) \le \beta_X(k)$, and additionally, $\alpha_X(k) \le \alpha'_X(k) \le 4\alpha_X(k)$. See, for example, Section 3.1 of~\cite{Dehling2002} for more details. Similar to how total variation distance reduces when the same transformation is applied to random variables, the mixing coefficients defined above decrease when $X_1, \ldots, X_n$ is replaced with $g(X_1), \ldots, g(X_n)$ for an arbitrary transformation $g(\cdot)$. Section 3.2 of~\cite{Dehling2002} provides several examples of sequences that are $\beta$-mixing, including $m$-dependent sequences, irreducible and aperiodic strictly stationary Markov chains on countable state spaces, ARMA, ARCH/GARCH, and iterative processes. 

The following result is an extension of Theorem~\ref{thm:finite-sample-validity} to the case of $\beta$-mixing sequences. 
\begin{theorem}[$\cosi{}$ under $\beta$-mixing]\label{thm:beta-dependent-COSI-validity}
    For any $\alpha\in[0,1]$, and any collection of distributions $\mathcal{P}^n$, let $\widehat{\theta}_1, \ldots, \widehat{\theta}_{B}$ be the estimators computed in Step 2 of Algorithm~\ref{alg:proposed-conf-int-dependent}. Let $\widetilde{\theta}_1, \ldots, \widetilde{\theta}_B$ be mutually independent random variables such that $\widetilde{\theta}_j$ is identically distributed as $\widehat{\theta}_j$ for each $1\le j\le B$. Finally, set 
    \[
    \begin{split}
    \cosicit{n}{\alpha}{dep\mbox{-}a} ~&:=~ \left\{\theta:\, \phi_{\mathfrak{P},\alpha}(\{\widetilde{\theta}_j - \theta\}_{j=1}^B) = 0\right\},\\
    \cosicit{n}{\alpha}{dep\mbox{-}m} ~&:=~ \left\{\theta:\, \sum_{j=1}^B \phi_{\mathfrak{P},1/2}(\widetilde{\theta}_j - \theta) \le c_{B,\alpha}^*\right\}.
    \end{split}
    \]
    Then, for $\mathtt{*}\in\{\mathtt{a},\,\mathtt{m}\}$ and $P^n\in\mathcal{P}^n$,
    \[
    \mathbb{P}_{P^n}\left(\theta(P^n)\notin\cosicit{n}{\alpha}{dep\mbox{-}*}\right) ~\le~ \mathbb{P}_{P^n}\left(\theta(P^n) \notin \cosicit{n}{\alpha}{dep\mbox{-}*}\right) ~+~ (B-1)\beta_X(k_n).
    \]
\end{theorem}
See Section~\ref{appsec:proof-of-beta-dependent-COSI-validity} for a proof. One could write a similar statement for the $\cosi{}$ hunt-and-test framework.
With no further structure on the test $\phi_{\mathfrak{P},\alpha}(\cdot)$, Theorem~\ref{thm:beta-dependent-COSI-validity} cannot be improved by replacing $\beta(k_n)$ with anything smaller. In particular, for $B = 2$, there exists a test such that the miscoverage error of the $\cosi{}$-confidence set for dependent data and for independent data differ exactly by $\beta(k_n)$. This follows from the coupling based characterization of $\beta$-mixing coefficient~\citep[Berbee's Lemma, Lemma 5.1]{rio2017asymptotic}. The miscoverage error from Theorem~\ref{thm:beta-dependent-COSI-validity} is given in an implicit form. One could use Theorem~\ref{thm:finite-sample-validity} to control the miscoverage probability on the right hand side of Theorem~\ref{thm:beta-dependent-COSI-validity}.

With some structure on the test, guarantees can be obtained when the data is strongly mixing (instead of absolutely regular). To illustrate this, we present the coverage guarantee of HulC for univariate estimators under alpha-mixing. Recall from~\cite{kuchibhotla2024hulc} and~\cite{paul2024inference} that the HulC interval is obtained by taking the order statistics of the $B$ estimators. Formally, the HulC interval (for $B \ge \log_2(2/\alpha)$) under dependence is
\[
\widehat{\mathrm{CI}}_{n,\alpha}^{\mathtt{HulC}\mbox{-}\mathtt{dep}} = \left[\widehat{\theta}_{(\lfloor B/2\rfloor - c^*_{B,\alpha/2})},\, \widehat{\theta}_{(\lceil B/2\rceil + c^*_{B,\alpha/2} + 1)}\right],
\]
where $\widehat{\theta}_{(1)} \le \widehat{\theta}_{(2)} \le \ldots \le \widehat{\theta}_{(B)}$ are order statistics of estimators computed in Algorithm~\ref{alg:proposed-conf-int-dependent}, and $c^*_{B,\alpha}$ is as in Algorithm~\ref{alg:proposed-conf-int}. 

\begin{theorem}[HulC under Strong Mixing]\label{thm:HulC-strong-mixing}
    For any $\alpha\in[0, 1]$, $B\ge\log_2(2/\alpha)$, and any collection of distributions $\mathcal{P}^n$, we have
    \[
    \sup_{P^n\in\mathcal{P}^n}\,\mathbb{P}_{P^n}\left(\theta(P^n)\notin\widehat{\mathrm{CI}}_{n,\alpha}^{\mathtt{HulC}\mbox{-}\mathtt{dep}}\right) ~\le~ \alpha e^{2(B\Delta)^2} ~+~ 3(B-1)\sup_{P^n\in\mathcal{P}^n}\alpha_X(k_n),
    \]
    where
    \[
    \Delta := \sup_{P^n\in\mathcal{P}^n}\frac{1}{B}\sum_{j=1}^B \left(\frac{1}{2} - \min\left\{\mathbb{P}_{P^n}(\widehat{\theta}_j - \theta(P)\ge 0),\,\mathbb{P}_{P^n}(\widehat{\theta}_j - \theta(P) \le 0)\right\}\right)_+.
    \]
\end{theorem}
See Section~\ref{appsec:proof-of-Theorem-HulC-strong-mixing} for a proof. Considering Theorems~\ref{thm:beta-dependent-COSI-validity} and~\ref{thm:HulC-strong-mixing}, it is interesting to observe that the implementation of $\cosi{}$ framework requires no knowledge of the dependence parameters, i.e., the analyst does not need to know how fast $\beta(k_n)$ or $\alpha(k_n)$ decays to zero as $n\to\infty$. We also do not need any summability assumption on the mixing coefficients. This is in contrast to several existing methods in the literature including block bootstrap or heteroscedasticity based variance estimation methods~\citep{andrews1991heteroskedasticity} that require either of these conditions.  

\paragraph{Application 2 (revisited): Inference for a Linear Contrast of Projection Parameter}\label{sec:linear-constrast-dependence}
Suppose $(X_i, Y_i), 1\le i\le n$ are identically distributed random vectors in $\mathbb{R}^{d+1}$ from distribution $P$ that strong mixing. Define
\[
\theta(P) := \argmin_{\theta\in\mathbb{R}^d}\,\mathbb{E}_P[(Y - X^{\top}\theta)^2].
\]
(The assumption of identical distribution can be ignored if $\theta(P) = \argmin_{\theta\in\mathbb{R}^d}\,\mathbb{E}[(Y_i - X_i^{\top}\theta)^2]$ for all $i$.)
The natural estimator for $\theta(P)$ is the ordinary least squares:
\[
\widehat{\theta}_{[n]} := \argmin_{\theta\in\mathbb{R}^d}\,\frac{1}{n}\sum_{i=1}^n (Y_i - X_i^{\top}\theta)^2.
\]
As in Algorithm~\ref{alg:proposed-conf-int-dependent}, obtain $S_1, \ldots, S_B$ from $\{1, 2, \ldots, n\}$ and compute
\[
\widehat{\theta}_j = \left(\sum_{i\in S_j} X_iX_i^{\top}\right)^{-1}\sum_{i\in S_j} X_iY_i.
\]
As before, this implies that for any vector $u\in\mathbb{R}^d$, we have
\[
u^{\top}(\widehat{\theta}_{j} - \theta(P)) = u^{\top}\left(\sum_{i\in S_j} X_iX_i^{\top}\right)^{-1}\left(\sum_{i\in S_j} X_i(Y_i - X_i^{\top}\theta(P))\right).
\]
Under sufficient moment conditions, with proper scaling, the left hand side converges in distribution to a mean zero Gaussian. Moreover, $u^{\top}(\widehat{\theta}_j - \theta(P)), 1\le j\le B$ are strong-mixing. Hence, HulC confidence intervals are asymptotically exactly valid if the CLT within each batch holds. Note that the classical linear model with heteroscedastic and correlated Gaussian errors (i.e., $Y_i = X_i^{\top}\theta(P) + \xi_i$ with independent vectors $X_1, \ldots, X_n$ and $(\xi_1, \ldots, \xi_n)|X_1, \ldots, X_n \sim N(0, \Sigma)$) satisfy the assumption of symmetry and unimodality within each batch even if the CLT may not hold. Strong mixing and even absolute regularity of the estimators holds if $\Sigma$ is close to a banded matrix~\citep[Theorem 5.4]{bickel2012approximating} (e.g., $\Sigma_{ij} = \sigma_i\sigma_j\rho^{|i-j|}$ with $|\rho| < 1$). It is known that classical Wald and bootstrap inference fail to be uniformly valid~\citep{potscher2025valid}. In the example of $\Sigma_{ij} = \rho^{|i-j|}$, if $\rho = 1$, then the data cannot be split into approximately independent clusters. But with a simple detrending trick, one can recover asymptotically exact inference. Formally, suppose we observe $Y_i = x_i^{\top}\beta_0 + \xi_i$ with non-stochastic covariate vectors $x_1, \ldots, x_n$ and $(\xi_1, \ldots, \xi_n) \sim N(0, \Sigma)$ with $\Sigma_{ij} = \rho^{|i-j|}$, then for $2 \le i\le n$
\[
Y_i' = \beta_0^{\top}x_i' + \xi_i'\quad\mbox{with}\quad Y_i' = Y_i - Y_{i-1},\quad x_i' = x_i - x_{i-1},\quad \xi_i' = \xi_i - \xi_{i-1}.
\]
This gives $n-1$ observations and one can apply HulC (for dependent data) on this dataset to recover asymptotic exact coverage. Concretely, obtain $S_1, \ldots, S_B$ from $\{1, 2, \ldots, n-1\}$ as in Algorithm~\ref{alg:proposed-conf-int-dependent}. Compute
\[
\widehat{\theta}_j = \left(\sum_{i\in S_j} (x_i')(x_i')^{\top}\right)^{-1}\sum_{i\in S_j} x_i'Y_i',\quad 1\le j\le B.
\]
Then $(u^{\top}(\widehat{\theta}_j - \beta_0): 1\le j\le B)$ is a jointly Gaussian random vector with a $\beta$-mixing coefficient of order $|\rho|^{k_n}$. Hence, Theorem~\ref{thm:HulC-strong-mixing} applies. The same trick applies of $\xi_i$'s are generated from an ARIMA process.

\paragraph*{Numerical Illustation.} We consider four data generating schemes for inference on a linear regression projection under serial dependence. Each observation has five covariates, and every ordinary least squares fit includes an intercept. Let $G_i\in\mathbb{R}^5$ and $H_i\in\mathbb{R}$ be independent stationary Gaussian autoregressive processes satisfying
\[
    G_i=0.6G_{i-1}+0.8V_i,\qquad H_i=0.6H_{i-1}+0.8\eta_i,
\]
where $V_i\overset{iid}{\sim}N_5(0,\Gamma)$, $\Gamma_{jk}=0.7^{|j-k|}$, and $\eta_i\overset{iid}{\sim}N(0,1)$. The initial observations have their stationary laws, $G_1\sim N_5(0,\Gamma)$ and $H_1\sim N(0,1)$. Set $U_i=\Phi(H_i)$, $b=(1,-1,0.5,0.5,-0.5)^{\top}$, and define $m_0(x)=1+x^{\top}b$, $q(x)=0.5x_1^2+0.25x_1^3+0.5x_1x_2$, and $s(x)=0.5+0.5|x_1|$. We generate $Y_i=m(X_i)+\sigma(X_i)\varepsilon_i$ using
\begin{align*}
    \mathrm{DGP\text{-}1}:&\quad X_i=G_i,\quad m=m_0,\quad \sigma=1,\quad \varepsilon_i=H_i,\\
    \mathrm{DGP\text{-}2}:&\quad X_i=\exp(G_i/2)-e^{1/8}\mathbf{1},\quad m=m_0,\quad \sigma=s,\quad \varepsilon_i=-\log(1-U_i)-1,\\
    \mathrm{DGP\text{-}3}:&\quad X_i=G_i,\quad m=m_0+q,\quad \sigma=s,\quad \varepsilon_i=F_{t_3}^{-1}(U_i)/\sqrt{3},\\
    \mathrm{DGP\text{-}4}:&\quad X_i=G_i,\quad m=m_0+q,\quad \sigma=s,\quad \varepsilon_i=e_a(U_i),\qquad a=2.5,
\end{align*}
where exponentiation is coordinate-wise, $F_{t_3}$ is the distribution function of a Student $t_3$ random variable, and
\[
    e_a(u)=\sqrt{a(a-2)}\left\{\frac{a-1}{a}(1-u)^{-1/a}-1\right\}.
\]
All error margins have mean zero and variance one. DGP-1 uses a correctly specified linear conditional mean with Gaussian covariates and errors. DGP-2 uses skewed covariates and exponential errors, together with heteroscedasticity. DGP-3--DGP-4 combine nonlinear misspecification with symmetric or skewed heavy-tailed errors. We use Pareto shape $2.5$, rather than $1.5$, so that the squared-error population objective remains finite. All four processes are stationary and geometrically mixing. The target is the first slope in the population least-squares projection of $Y_i$ on $(1,X_i^{\top})^{\top}$. This target is $1$ under DGP-1--DGP-2 and $1.75$ under DGP-3--DGP-4; the cubic term contributes $3/4$ to the first slope. The population intercept in the nonlinear settings is $1.85$.

We use $\alpha=0.05$, fix $\delta=0.30$, and consider $n\in\{100,200,300,500,750,1000\}$. As in the preceding regression experiment (Figure~\ref{fig:coverage-regression}), this choice gives $B=9$ inference estimates; the unimodal$^{\star}$ construction uses ten blocks, reserving the first for its pilot. We use Algorithm~\ref{alg:proposed-conf-int-dependent} with $k_n=\max\{1,\lfloor\log(n)/2\rfloor\},$ which gives gaps of $2,2,2,3,3,3$ observations at the six respective sample sizes. The median, symmetric, and normal-score constructions use the same block estimates. The unimodal$^{\star}$ method has a separate balanced partition because it uses an additional pilot block. At $n=100$, the retained sample sizes are 84 and 82, respectively, and the smallest blocks contain nine and eight observations. With fixed $B$, the gaps diverge while the proportion of omitted observations tends to zero. This supplies asymptotic separation under the mixing assumptions, but does not give an exact finite-sample $95\%$ coverage guarantee at these particular gap lengths. In the Gaussian setting, block estimation errors are Gaussian conditional on the complete design and have centered symmetric and unimodal marginal distributions unconditionally.

We compare the four $\cosi{}$ methods with a Wald interval using a heteroscedasticity and autocorrelation consistent (HAC) covariance estimator, a basic moving-block pairs bootstrap, and subsampling with an estimated convergence rate. For Wald inference, we use the Bartlett-kernel and  Newey--West estimator~\citep{newey1987simple,zeileis2004econometric}. Write $Z_i=(1,X_i^{\top})^{\top}\in\mathbb{R}^{q}$, with $q=6$, $Q_n=\sum_{i=1}^nZ_iZ_i^{\top}$, $\widehat r_i=Y_i-Z_i^{\top}\widehat\theta_{[n]}$, and $\widehat\psi_i=Z_i\widehat r_i$. Define the unnormalized sample score autocovariances by $\widehat\Gamma_h=\sum_{i=h+1}^n\widehat\psi_i\widehat\psi_{i-h}^{\top}$. We estimate the covariance matrix of the full-sample ordinary least squares estimator by
\[
\widehat V_{\mathrm{HAC}}
=\frac{n}{n-q}\,Q_n^{-1}
\left[\widehat\Gamma_0+\sum_{h=1}^{L_n}
\left(1-\frac{h}{L_n+1}\right)
\bigl(\widehat\Gamma_h+\widehat\Gamma_h^{\top}\bigr)\right]Q_n^{-1}.
\]
The kernel is $K(u)=(1-|u|)_+$, the maximum included lag is $L_n=\left\lfloor4(n/100)^{2/9}\right\rfloor$, and the kernel bandwidth is $L_n+1$. Thus, the lag weights are $K(h/(L_n+1))$, and the six tested sample sizes use maximum lags $4,4,5,5,6,6$. This is a fixed rule depending only on sample size, also documented as the default lag rule for the \texttt{statsmodels} function \texttt{cov\_hac}~\citep{statsmodelsHAC}; it is not an automatically estimated bandwidth. We use no prewhitening and include the finite-sample covariance adjustment $n/(n-q)$. These choices correspond to calling \texttt{sandwich::NeweyWest} with an explicit lag $L_n$, \texttt{prewhite=FALSE}, and \texttt{adjust=TRUE}~\citep{sandwichNeweyWest}. The first-slope Wald interval is
\[
\left[\widehat\theta_{[n],2}\pm z_{1-\alpha/2}\sqrt{\widehat V_{\mathrm{HAC},22}}\right],
\]
where $z_{1-\alpha/2}$ is the standard Gaussian quantile.

The bootstrap uses the moving-block pairs procedure of~\cite{kunsch1989jackknife}, with 500 resamples formed by sampling with replacement from overlapping blocks of $(X_i,Y_i)$ pairs of length $\operatorname{round}(n^{1/3})$, truncating the last block to obtain exactly $n$ observations. Subsampling follows~\cite[Chapter~3, pp.~65--100]{politis1999subsampling} and uses 500 randomly selected consecutive windows at each size, with $b=\operatorname{round}(n^{2/3})$ for interval construction and feasible sizes $\operatorname{round}(n^{\beta})$, $\beta\in\{0.9,0.8,\ldots,0.1\}$, for rate estimation. Windows must contain at least 18 observations. The convergence rate is estimated from the dependence of interquantile widths on subsample size, following~\cite{bertail1999subsampling}. We retain the 100-point inverse-CDF grid and interquantile regression used in the preceding rate-estimated subsampling implementation. All three benchmark methods use the full sample. Every method uses the same data within each of the 500 Monte Carlo replications, and each setting/sample-size cell has a fixed random seed.

\begin{figure}[h]
    \centering
    \includegraphics[width=\linewidth]{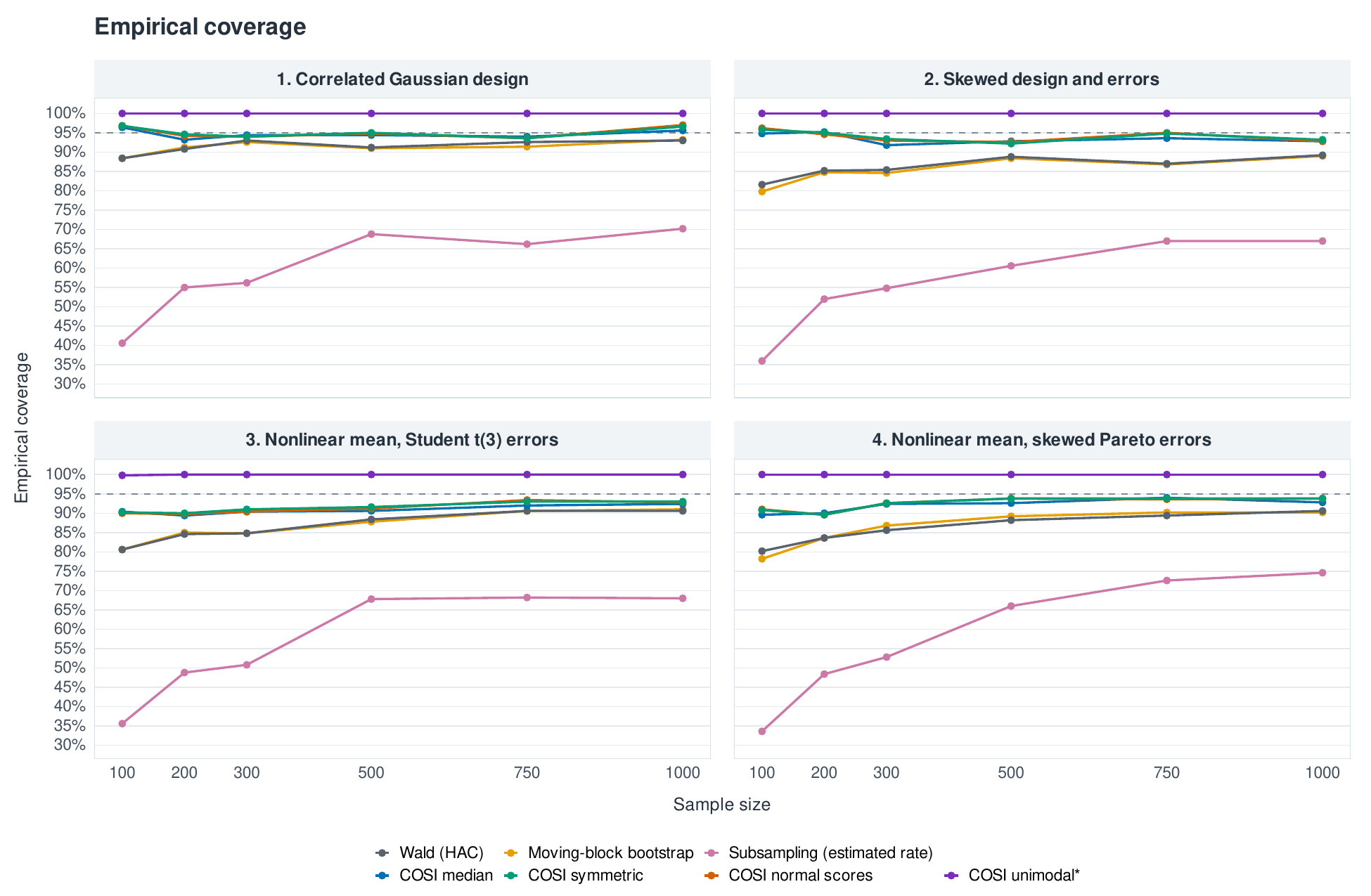}
    \caption{Empirical coverage over 500 Monte Carlo replications for the first population projection slope under serial dependence. The horizontal dashed line marks $95\%$ coverage. COSI uses $\delta=0.30$ and ordered blocks separated by $g_n=\max\{1,\lfloor\log(n)/2\rfloor\}$ omitted observations. The Wald interval uses the Bartlett-kernel HAC covariance with maximum lag $L_n=\lfloor4(n/100)^{2/9}\rfloor$, no prewhitening, and finite-sample adjustment $n/(n-6)$.}
    \label{fig:coverage-dependent-regression}
\end{figure}

Figure~\ref{fig:coverage-dependent-regression} shows that the tested HAC Wald and moving-block bootstrap intervals under-cover in these finite samples. At $n=1000$, HAC Wald coverage under DGP-1--DGP-4 is $93.0\%$, $89.2\%$, $90.6\%$, and $90.6\%$, respectively. Moving-block bootstrap coverage ranges from $89.0\%$ to $93.2\%$ at this sample size.

Under DGP-1, the median, symmetric, and normal-score $\cosi{}$ intervals have coverages ranging from $93.2\%$ to $97.0\%$ across the tested sample sizes. At $n=1000$, their coverages are $95.6\%$, $96.6\%$, and $97.0\%$. The non-Gaussian settings are more challenging. At $n=1000$, these three coverages are $92.8\%$, $93.2\%$, and $92.8\%$ under DGP-2; $92.4\%$, $93.0\%$, and $92.8\%$ under DGP-3; and $92.8\%$, $93.8\%$, and $93.8\%$ under DGP-4. Thus, these experiments do not support a claim of uniformly nominal coverage over the tested non-Gaussian settings. The asymptotic argument does not remove finite-block approximation errors, median bias, or residual dependence between separated blocks. The unimodal$^{\star}$ interval is conservative, with empirical coverage between $99.8\%$ and $100\%$ across all settings and sample sizes. The Monte Carlo standard error at $95\%$ coverage is approximately one percentage point.

Rate-estimated subsampling has especially low coverage. At $n=100$, its coverage ranges from $33.6\%$ to $40.6\%$, increasing to $67.0\%$--$74.6\%$ at $n=1000$. Its estimated rates are substantially above the root-$n$ exponent $1/2$: the median estimates range from $1.24$ to $1.34$ at $n=100$, and from $0.66$ to $0.72$ at $n=1000$. Since the interval scales the unnormalized subsample deviations by $(b/n)^{\widehat\gamma}$, larger rate estimates produce shorter intervals. 

To compare interval lengths, we divide each interval width by the Wald (HAC) width computed from the same data. The paired ratio is calculated within every replication before computing the reported medians and boxplot summaries. The Wald (HAC) ratio is therefore exactly one in every replication.

\begin{figure}[h]
    \centering
    \includegraphics[width=\linewidth]{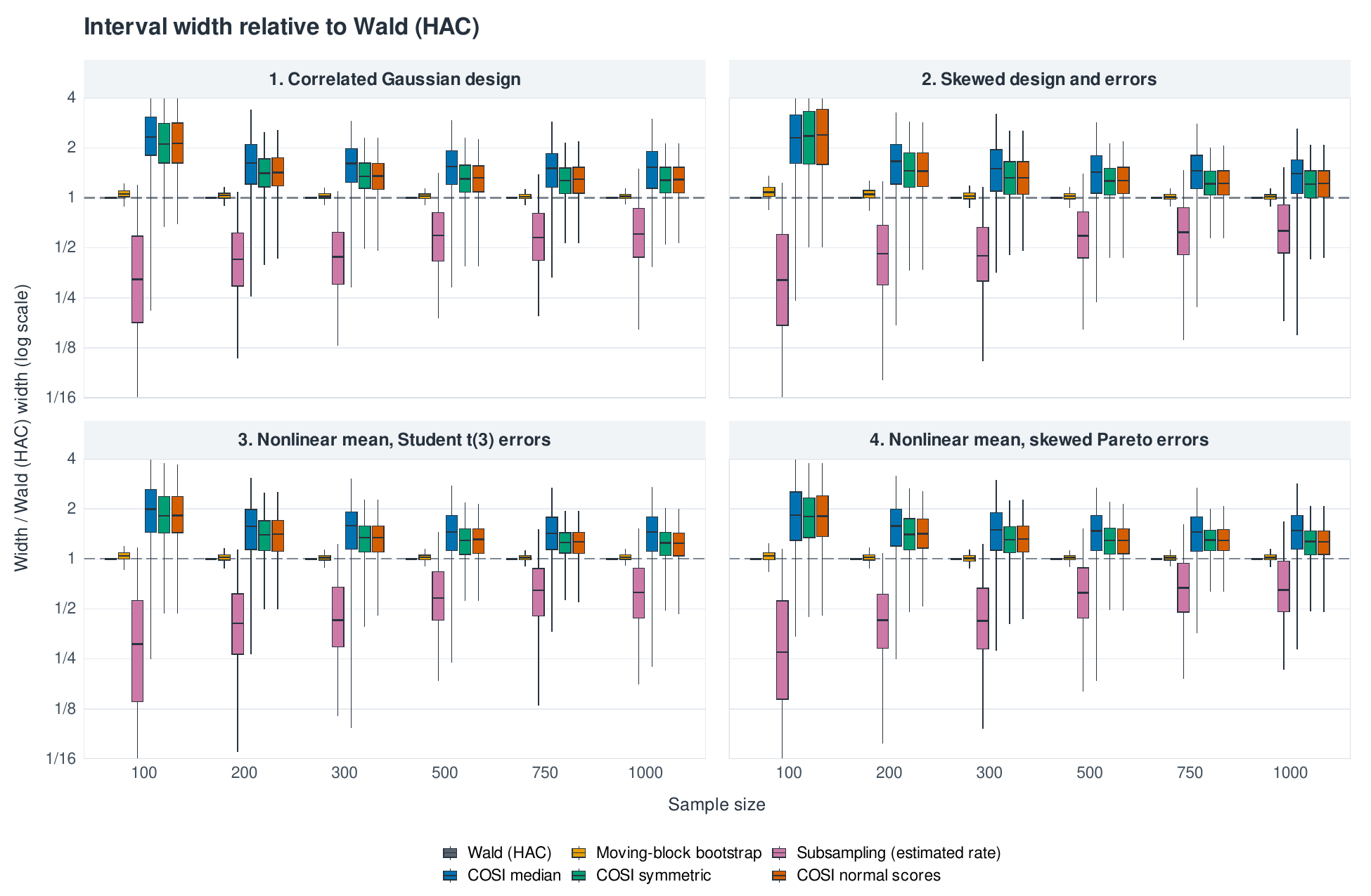}
    \caption{Interval widths relative to Wald (HAC) over 500 Monte Carlo replications. Boxplot statistics are computed before restricting the logarithmic display to $[1/16,4]$; outlier points are omitted. The dashed line marks a width ratio of one. The unimodal$^{\star}$ method is omitted because its median width ratios range from $7.85$ to $14.89$ across the tested settings and sample sizes. These medians are saved separately for use in the caption.}
    \label{fig:width-dependent-regression}
\end{figure}

Figure~\ref{fig:width-dependent-regression} shows that the $\cosi{}$ intervals are particularly wide at $n=100$, where each regression block has few residual degrees of freedom. At $n=1000$, the median-based width ratios under DGP-1--DGP-4 are $1.53$, $1.40$, $1.46$, and $1.49$. The corresponding ratios for the symmetric construction are $1.27$, $1.21$, $1.25$, and $1.28$, while the normal-score ratios are $1.29$, $1.22$, $1.25$, and $1.27$. Moving-block bootstrap intervals have median width ratios of approximately $1.01$--$1.02$ at this sample size. Estimated-rate subsampling has smaller median ratios, approximately $0.61$--$0.65$, together with the substantial undercoverage reported above. The unimodal$^{\star}$ median ratios at $n=1000$ are $8.42$, $7.86$, $8.28$, and $8.00$.

\section{Conclusions}\label{sec:conclusions}
In this paper, we proposed a general framework for constructing confidence sets for parameters in real line, Euclidean spaces, general topological spaces, and some metric spaces, without relying on distributional approximation results such as the CLT. In most cases, our confidence sets are asymptotically exact and achieve minimax rate optimal diameters. We provide computationally easier versions of our confidence sets that retain asymptotic validity under weaker conditions. We also prove that our conditions are the weakest possible for the existence of asymptotically valid confidence sets. We also make an interesting connection between conformal prediction and our confidence sets. 

We illustrate the framework by constructing confidence sets for high-dimensional mean, Hilbert or Banach space means, and Gram matrices without any restrictions on the dimension or complexity of the underlying space. In addition, we construct valid confidence sets under adversarial cellwise contamination and also for Fr{\'e}chet means in Alexandrov spaces.  

The methods proposed in this paper can be extended, generalized, or used in several other settings. Firstly, we proposed all the methods through sample splitting. While this reduces the sample size used for each batch, the resulting loss in sample size exactly matches the width scaling of the confidence set. The main role of sample splitting is to obtain (approximately) independent estimators of the target of inference. This could also be achieved using data fission/thinning ideas, but this requires significantly more structure on the data generating process; we refer the readers to~\cite{leiner2025data} and~\cite{neufeld2024data} for details. Note that data fission/thinning retains the sample size $n$ for each batch but increases the variability of the estimator obtained from each batch. Understanding the trade-off between sample splitting and data thinning could be an interesting future direction. We also note here that our confidence sets are already rate-optimal and hence, any improvement must stem from improved constants. Secondly, in the context of dependent data, we restricted ourselves to $\beta$-mixing data on a linear time scale (indices are ordered). The method can be easily extended to spatial data or network data but finding a good splitting for arbitrary spatial structures or network seems difficult if not impossible. In the same vein, it is possible to extend the framework to perform inference for martingale data assuming the availability of certain conditional CLTs. It would be interesting to see how far the methods can be extended to other types of dependence. As mentioned in Section~\ref{sec:cosi-dependent}, if the underlying tests are not suitably restricted, then the assumption of $\beta$-mixing is indispensable. But extensions beyond $\beta$-mixing would greatly extend the scope of $\cosi{}$ framework. 

Finally, we concede that the proposed confidence sets while rate-optimal can be significantly wider than the classical Wald/resampling based confidence sets. Of course, this does not mean that the Wald/resampling based inference is valid. The main motivation for the $\cosi{}$ framework is to propose confidence sets with validity under much weaker conditions. Nevertheless, if the practitioner strongly believes that the assumptions required for uniform validity of Wald/resampling based methods is justified, then we suggest he/she to use them instead of the $\cosi{}$ framework. When those strong regularity conditions are not satisfied, it maybe preferable to use an alternative method such as $\cosi{}$. Moreover, in examples where constants in the diameter of the confidence set are of less importance, such as preliminary confidence sets for nuisance components, then $\cosi{}$ might be a cheaper confidence set to compute and retain validity.

\section*{Acknowledgments} AKK thanks the Sloan Foundation that supported some part of this research. AKK also thanks the participants of EcoSta 2026 including Miles Lopes and Anders Kock (for discussions related to Sections~\ref{sec:gram-matrix} and~\ref{sec:mean-cellwise-contamination}) and of IMS Annual Meeting 2026 including Daniel Bartl and Gabor Lugosi (for discussions related to Section~\ref{subsec:metric-space-mean}). AKK gratefully acknowledges the discussions with Benedikt M. P{\"o}tscher that led to the dependent data linear regression example. 

\section*{On the use of AI}
The methods of this paper were developed over the past four years and the first author has given talks on the COSI framework since 2022 at least. The validity of confidence sets in Hilbert spaces and Banach spaces were also developed during this period. The use of AI (in the past few months) is mostly restricted to doing routine calculations in the proofs for simplification purposes. There are three major aspects that AI (in particular, ChatGPT 5.6 Sol) has greatly enhanced or helped the authors. Below are more details on these. 
\begin{enumerate}
    \item The fact that the Wilcoxon signed rank test can have a multiplicative coverage in terms of the KS distance under deviations from symmetry was proved by AI. Note that for any test or confidence interval that is valid under symmetry, it is easy to get an additive miscoverage error bound under general distribution where the error is bounded by the total variation distance to the collection of symmetric distributions. In some of the presentations by the first author, the following multiplicative error bound was presented: let $h:[0,\infty)\to\mathbb{R}$ be a non-negative function. Let $\widehat{\theta}_j, 1\le j\le B$ as in Algorithm~\ref{alg:proposed-conf-int}. Consider
    \[
    \widehat{\mathrm{CI}}_{n,\alpha} := \left\{t\in\mathbb{R}:\, \sum_{j=1}^B \mbox{sign}(\widehat{\theta}_j - t)h(|\widehat{\theta}_j - t|) \in \left[\widehat{q}_{\alpha/2}(t),\widehat{q}_{1-\alpha/2}(t)\right]\right\},
    \]
    where $\widehat{q}_{\gamma}(t)$ is a suitably computed $\gamma$-th quantile of $\sum_{j=1}^B e_jh(|\widehat{\theta}_j - t|)$ over $(e_1, \ldots, e_B) \in \{\pm1\}^B$. Clearly, this confidence interval has a coverage of at least $1 - \alpha$ for $\theta(P)$ if $\widehat{\theta}_j - \theta(P)$ are symmetric around zero. It is not difficult to prove that 
    \[
    \mathbb{P}\left(\theta(P)\notin\widehat{\mathrm{CI}}_{n,\alpha}\right) \le \alpha\mathbb{E}\left[\exp(2B\Delta)\right],
    \]
    where 
    \[
    \Delta = \max_{1\le j\le B}\,\left(\frac{1}{2} - \min\{\mathbb{P}(\widehat{\theta}_j \ge \theta(P)||\widehat{\theta}_j - \theta(P)|),\,\mathbb{P}(\widehat{\theta}_j \le \theta(P)||\widehat{\theta}_j - \theta(P)|)\}\right)_+.
    \]
    Note that for symmetric random variables, the conditional probabilities above are at least $1/2$. While this is a reasonable result, conditional probabilities such as those above are hard to argue for based on asymptotic distributions unlike the KS distances used in Section~\ref{subsec:symmetry}. After some prompts, ChatGPT 5.6 Sol figured out that the Wilcoxon signed rank test yields a confidence interval that has a multiplicative miscoverage bound depending only on the KS distance to symmetric random variables. The extension to general scores like the van der Waerden scores is subsequently figured out by the authors by noticing the the proof generalizes. The original proof given by AI was rather cryptic and the current proof was rewritten by the authors. 
    \item The authors have been aware of confidence intervals for the mode of a unimodal distribution based on one-observation for a while. This result discussed in Section~\ref{sec:unimodal} was the motivation/inspiration for~\cite{paul2025finite}. These results required continuity of the underlying distribution which again is difficult to argue for from convergence in distribution. Initially, Section~\ref{sec:unimodal} was planned to be about the scale-invariant collection of continuous unimodal distributions with mode at zero. After some interactions with AI, we figured out the more general formulation in the work~\cite{Andel1991}. This is how the logarithmic L{\'e}vy concentration function was developed. Subsequent results in that section were developed by the authors. 
    \item Finally, Section~\ref{subsec:metric-space-mean} greatly benefitted from interactions with ChatGPT 5.6 Sol. While the construction of $\cosicit{n}{\alpha}{\mp}$ was developed by the authors using the definitions from~\cite{le2022fast} and~\cite{bartl2025robust}, spelling out the details for which conditions are reasonable for the study of radius and the necessary measurability/existence issues of tangent cones, local Hilbert space like structure were greatly supported by the AI. The authors have first written Theorem~\ref{thm:metric-space-mean} for $\cosicit{n}{\alpha}{\mp}$ and the assumption that $\kappa_{\theta}^y(x) \ge 0$ for all $\theta, x, y$ seemed reasonable. With increased interactions with the AI, we realized that this is far more restrictive and the geometry at the barycenter greatly benefits the assumption $\kappa_{\mu}^y(x) \ge \kappa_{\min}$ and cannot be expected at arbitrary $\theta$. This is when AI suggested to consider the current definition of $\cosici{n}{\alpha}{\mp}$. This was a rather mysterious combination of zeroth order and first order criteria of Fr{\'e}chet mean that works under weaker conditions.
\end{enumerate}

\bibliography{references}
\bibliographystyle{apalike}
\newpage
\addtocontents{toc}{\protect\setlength{\cftsecnumwidth}{3em}}
\setcounter{section}{0}
\setcounter{equation}{0}
\setcounter{figure}{0}
\renewcommand{\thesection}{S.\arabic{section}}
\renewcommand{\theequation}{E.\arabic{equation}}
\renewcommand{\thefigure}{A.\arabic{figure}}
\renewcommand{\theHsection}{S.\arabic{section}}
\renewcommand{\theHequation}{E.\arabic{equation}}
\renewcommand{\theHfigure}{A.\arabic{figure}}
  \begin{center}
  \Large {\bf Supplement to ``$\SAV$: A Framework for Robustly Valid Inference''}
  \end{center}
       
\begin{abstract}
This supplement contains the proofs of all the main results in the paper and some supporting lemmas. 
\end{abstract}

\section{Confidence Sets from Tournaments}\label{appsec:Lugosi-Mendelson-set}
With estimators $\widehat{\theta}_1, \ldots, \widehat{\theta}_B, \vartheta$ as in the construction of $\cosi{HS}$ set, consider 
\begin{equation}\label{eq:Lugosi-Mendelson-set}
\widehat{\mathrm{CI}}_{n,\alpha}^{\mathtt{LM}} ~:=~ \left\{\theta\in\mathbb{H}:\, \sum_{j=1}^B \mathbf{1}\left\{\langle\theta - \vartheta,\,\widehat{\theta}_j - \theta\rangle \ge -\frac{\|\theta - \vartheta\|^2}{2}\right\} > \frac{B}{2}\right\}. 
\end{equation}
This is precisely $T_{\vartheta}^c$ with $T_{\vartheta}$ defined as in Eq. (2.1) of~\cite{lugosi2019sub}. Theorem 2 of~\cite{lugosi2019sub} implies a coverage guarantee for this set (conditional on $\vartheta$), in the context of mean estimation, assuming $\|\vartheta - \theta(P)\|$ is sufficiently large. In the following, we provide a more refined coverage guarantee.

The confidence set $\widehat{\mathrm{CI}}_{n,\alpha}^{\mathtt{LM}}$ can be readily compared to the $\cosi{HS}$ confidence set, with two differences: (1) Inside the indicator, $0$ is replaced with $-\|\theta - \vartheta\|^2/2$ and hence, the expected value of each of the indicator cannot be controlled in terms of the Tukey median bias; and (2) $c^*_{B,\alpha}$ is replaced with $B/2$ and hence, the only way ${\mathtt{LM}}$ set achieves the nominal coverage is through a proper choice of $B$. (Note that, for $\cosi{HS}$, choice of $B$ is not so critical and it can be large.)
\begin{theorem}[Coverage of $\mathtt{LM}$ confidence set]\label{thm:coverage-of-LM}
    Consider the setting of Algorithm~\ref{alg:cosi-hunt-and-test}. Define
    \[
    p_j(\vartheta) := \mathbb{P}_P\left(\langle \theta(P) - \vartheta,\, \widehat{\theta}_j - \theta(P)\rangle \ge -\|\theta(P) - \vartheta\|^2/2\,\mid\,\vartheta\right)\quad\mbox{and}\quad \overline{P}(\vartheta) := \frac{1}{B}\sum_{j=1}^B p_j(\vartheta).
    \] 
    Then
    \begin{equation}\label{eq:miscoverage-of-LM}
        \mathbb{P}_P(\theta(P)\notin\widehat{\mathrm{CI}}_{n,\alpha}^{\mathtt{LM}}) \le \mathbb{P}_P\left(\overline{P}(\vartheta) < 1/2\right) + \mathbb{P}(\mathrm{Binom}(B, \max\{1/2, \overline{P}(\vartheta)\}) \le B/2).
    \end{equation}
\end{theorem}
See Section~\ref{appsec:proof-of-coverage-of-LM} for a proof.
Theorem~\ref{thm:coverage-of-LM} shows that the miscoverage probability is controlled in terms of the tails of the univariate projections of individual estimators, which cannot be controlled from the Tukey median bias. Under additional assumption on asymptotic normality of these univariate projections, one can control the miscoverage. Although the analysis can be extended to other limiting distributions, the asymptotic Gaussian case is already illustrative. For any two real-valued random variables $W_1, W_2$, let the Kolmogorov-Smirnov distance be defined as
\[
\mathrm{KS}(W_1, W_2) := \sup_{t\in\mathbb{R}}|\mathbb{P}(W_1 \le t) - \mathbb{P}(W_2 \le t)|.
\]
Also, let $Z$ stand for the standard Gaussian random variable, i.e., $Z\sim N(0, 1)$, and recall $z_{\gamma} = 1 - \Phi^{-1}(\gamma)$. 
\begin{corollary}\label{cor:coverage-of-LM-asym-normality}
    Consider the setting of Theorem~\ref{thm:coverage-of-LM}. Suppose, for simplicity, that $\widehat{\theta}_j, 1\le j\le B$ are computed based on the same number of observations $m$ and $\vartheta$ is computed based on $km$ observations for some $k\ge 0$; without loss of generality, assume $km$ is an integer. Fix any covariance operator $\Sigma$ and any sequence $\{r_m\}_{m\ge1}$ and set
    \[
    \Delta = \frac{1}{B}\sum_{j=1}^B \mathrm{KS}\left(\frac{r_m}{\|\lambda\|_{\Sigma}}\langle \lambda, \widehat{\theta}_j - \theta(P)\rangle,\, Z\right),\quad\mbox{and}\quad \Delta' = \sup_{\lambda\in\mathbb{H}}\mathrm{KS}\left(\frac{\langle\lambda,\,\theta(P) - \vartheta\rangle}{\|\lambda\|_{\Sigma}},\, Z\right).
    \]
    Then for any $\delta \ge 1/2$,
    \begin{align*}
    \mathbb{P}\left(\overline{P}(\vartheta) \le \delta\right) ~&\le~ \mathbb{P}_P\left(\frac{\|\theta - \vartheta\|^2}{\|\theta - \vartheta\|_{\Sigma}} \le \frac{2z_{\delta + \Delta}}{r_m}\right)
    ~\le~ \sup_{\lambda\in\mathbb{H}}\,\mathbb{P}_P\left(\frac{|\langle\lambda,\,\theta - \vartheta\rangle|}{\|\lambda\|_{\Sigma}} \le \frac{2z_{\delta + \Delta}}{r_m}\right). 
    \end{align*}
    Moreover,
    \[
    \mathbb{P}_P(\theta(P)\notin\widehat{\mathrm{CI}}_{n,\alpha}^{\mathtt{LM}}) ~\le~ \mathbb{P}\left(\mathrm{Binom}\left(B, \Phi\left(\frac{r_m|Z|}{2r_{km}}\right)\right) \le \frac{B}{2}\right) + (2\sqrt{B} + 2r_{km}/r_m)\Delta + 4\Delta'.
    \]
\end{corollary}
See Section~\ref{appsec:proof-coverage-LM-asymp-normality} for a proof. Corollary~\ref{cor:coverage-of-LM-asym-normality} proves that $\mathbb{P}(\mbox{Binom}(B, \Phi(r_m|Z|/(2r_{km}))) \le B/2)$ is the asymptotic miscoverage of $\mathtt{LM}$ confidence set, assuming $\Delta + \Delta'\to 0$. While Corollary~\ref{cor:coverage-of-LM-asym-normality} only yields an upper bound on miscoverage, the proof shows that for the univariate case this is a two-sided bound. Even in a Hilbert space, for the worst case covariance operator, this yields a two-sided bound on miscoverage. Unlike the $\cosi{HS}$ confidence set, controlling the miscoverage error requires some knowledge of the rate of convergence and limiting distribution. If $k = 1$, i.e., $r_{km} = r_m$, the asymptotic miscoverage becomes
\[
\mathbb{P}\left(\mbox{Binom}\left(B,\,\Phi\left(\frac{|Z|}{2}\right)\right) \le \frac{B}{2}\right).
\]
Hence, one can choose $B$ (without the knowledge of the rate of convergence) such that this probability is at most $\alpha$ to ensure asymptotic nominal coverage of $1-\alpha$. Unfortunately, such a $B$ asymptotes as $1/\alpha^2$ as $\alpha\to0$; this can be seen from the normal approximation to the binomial or beta distribution. This is an exponential degradation in the number of splits required when compared to $\cosi{}$ sets. With knowledge of the rate of convergence in the sense that $r_m/r_{km} \to k^{-\gamma}$ as $m\to\infty$ (for some $\gamma \ge 0$), the asymptotic miscoverage for arbitrary $k$ becomes 
\[
\mathbb{P}\left(\mbox{Binom}\left(B,\,\Phi\left(\frac{k^{\gamma}|Z|}{2}\right)\right) \le \frac{B}{2}\right).
\]
Assuming a known $\gamma$, one can choose $(k, B)$ so as to minimize $(k + B)$ and that the asymptotic miscoverage is $\alpha$. (Note that with a total of $n$ observations, the number of observations used for $\widehat{\theta}_j$ is $n/(k+B)$, and the number of observations used for $\vartheta$ is $kn/(k+B)$.) Such an optimization still yields $B \asymp \alpha^{-2/(2\gamma + 1)}$, which is almost exponentially worse than that required for $\cosi{}$ sets. Unless $B\asymp \log(1/\alpha)$, the $\mathtt{LM}$ confidence set cannot achieve the optimal sub-Gaussian diameter as for the tournament estimator in~\cite{lugosi2019sub}. 
\section{Auxiliary Results on Supremum of Stochastic Processes}
The following result provides a deviation inequality for supremum of quantiles. It can be seen as a generalization of Eq. (2.7) of~\cite{minsker2025uniform} (and more generally, results on the supremum of median of means). Some elements of the proof are similar to that of Theorem 1.5 of~\cite{koltchinskii2015bounding}.
\begin{theorem}[Supremum of Quantiles.]\label{thm:supremum-of-quantiles}
    Suppose $\mathcal{F}$ is a collection of functions from some measurable space $\mathcal{W}$ to $\mathbb{R}$. Let $W_1, \ldots, W_B$ be independent random variables taking values in $\mathcal{W}$, and let $\varepsilon_1, \ldots, \varepsilon_B$ be independent Rademacher random variables independent of $W_1, \ldots, W_B$. For each $f\in\mathcal{F}$ and $c\in\{1, 2, \ldots, B\}$, define $(f(W))_{(c)}$ as the $c$-th smallest element of $\{f(W_1), \ldots, f(W_B)\}$. Let 
    \[
    F(w) := \sup_{f\in\mathcal{F}}\,(f(w))_+,\quad\mbox{for}\quad w\in\mathcal{W},
    \]
    be the envelope of $\mathcal{F}$. Then setting 
    \[
    \mathfrak{R}(\mathcal{F}) := \mathbb{E}\left[\sup_{f\in\mathcal{F}}\frac{1}{B}\sum_{j=1}^B \varepsilon_j f(W_j)\right],\quad \sigma^2 := \sup_{f\in\mathcal{F}}\frac{1}{B}\sum_{j=1}^B \mathbb{E}[f^2(W_j)]\quad\mbox{and}\quad V = \left(\frac{1}{B}\sum_{j=1}^B \mathbb{E}[F^q(W_j)]\right)^{1/q}, 
    \]
    we have with probability at least $1 - \delta$,
    \begin{align*}
        \sup_{f\in\mathcal{F}}(f(W))_{(c)} \le \min\{T_1,\,T_2,\,T_3\},
    \end{align*}
    where
    \begin{align*}
        T_1 &:= \frac{20\mathfrak{R}(\mathcal{F})}{1-c/B} + \sqrt{\frac{10\sigma^2}{3(1 - c/B)}} + VB^{1/q}\left(\frac{e^{7\log(1/\delta)/(B-c)} - 1}{\log(1/\delta)}\right)^{1/q},\\
        T_2 &:= \frac{4B(e^{4\log(1/\delta)/(B-c)} - 1)}{\log(1/\delta)}\mathfrak{R}(\mathcal{F}) + 2\sigma\sqrt{\frac{B(e^{4\log(1/\delta)/(B-c)} - 1)}{\log(1/\delta)}},\\
        T_3 &:= V(B/\delta)^{1/q}.
    \end{align*}
\end{theorem}
\paragraph{Some Comments on Theorem~\ref{thm:supremum-of-quantiles}.} We provide three bounds that share different characteristics. The first two bounds $T_1, T_2$ are finite only if $c < B$, while the final bound is finite for all $c \le B$. Unfortunately, $T_3$ does not depend on $c$ and is sub-optimal, in general. For the following comparisons, we assume $1 - c/B$ is bounded away from zero. If $\log(1/\delta) \le \mathfrak{C}B$ for some constant $\mathfrak{C}$, then 
\[
e^{4\log(1/\delta)/(B-c)} - 1 \asymp \frac{\mathfrak{C}'\log(1/\delta)}{B-c},
\]
for some other constant $\mathfrak{C}'$. In this case, (further assuming $1-c/B \gg 0$)
\[
T_1 = O\left(\mathfrak{R}(\mathcal{F}) + V + \sigma\right)\quad\mbox{and}\quad T_2 = O(\mathfrak{R}(\mathcal{F}) + \sigma).
\]
Here $T_1$ could be worse than $T_2$ if $V$ is of larger order than $\mathfrak{R}(\mathcal{F})$. 
If $\log(1/\delta)/B \gg 1$, then 
\[
T_1 = O\left(\mathfrak{R}(\mathcal{F}) + \sigma + {V}\left(\frac{e^{7\log(1/\delta)/(B-c)}}{\log(1/\delta)/(B-c)}\right)^{1/q}\right),
\]
and
\[
T_2 = O\left(\mathfrak{R}(\mathcal{F})\frac{e^{4\log(1/\delta)/(B-c)}}{\log(1/\delta)/(B-c)} + \sigma\sqrt{\frac{e^{4\log(1/\delta)/(B-c)}}{\log(1/\delta)/(B-c)}}\right). 
\]
Here $T_2$ could be worse than $T_1$ if $V$ is of the same order as or lower order than $\mathfrak{R}(\mathcal{F})$ (for example, if $\mathcal{F}$ is a bounded non-Donsker class).
\begin{proof}[Proof of Theorem~\ref{thm:supremum-of-quantiles}]
    Note that
    \begin{equation}\label{eq:quantile-to-CDF-process}
    \begin{split}
    &\sup_{f\in\mathcal{F}}\, (f(W))_{(c)} \le K \\
    \Leftrightarrow\quad &\inf_{f\in\mathcal{F}}\, \frac{1}{B}\sum_{j=1}^B \mathbf{1}\{f(W_j) \le K\} \ge \frac{c}{B}\\
    \Leftrightarrow\quad &\sup_{f\in\mathcal{F}}\,\frac{1}{B}\sum_{j=1}^B \mathbf{1}\{f(W_j) > K\} \le 1 - \frac{c}{B}.
    \end{split}
    \end{equation}
    Hence, it suffices to find a $K$ to satisfy the last inequality. We now prove bounds for the supremum of the empirical survival functions. For all these bounds, we make use of a smooth sandwich of indicator function; formally, for any $K_1 \le K$, we choose a Lipschitz function $\phi(\cdot)$ such that
    \[
    \mathbf{1}\{x > K\} \le \phi(x) \le \mathbf{1}\{x > K_1\}\quad\mbox{and}\quad \sup_{x\neq y}\,\frac{|\phi(x) - \phi(y)|}{|x - y|} \le \frac{1}{K-K_1}. 
    \]
    Such a function exists; for example, one can take $\phi(x) =\mathbf{1}\{x>K\} +(x-K_1)\mathbf{1}\{K_1<x\le K\}/(K-K_1)$.
    \paragraph{First bound.} For the first bound on $K$, we apply Corollary 2 of~\cite{pinelis1986remarks}. For this, we first observe that
    \begin{equation}\label{eq:expectation-bound}
    \begin{split}
    &\mathbb{E}\left[\sup_{f\in\mathcal{F}}\,\frac{1}{B}\sum_{j=1}^B \mathbf{1}\{f(W_j) > K\}\right]\\ 
    &\le \mathbb{E}\left[\sup_{f\in\mathcal{F}}\,\frac{1}{B}\sum_{j=1}^B \left[\mathbf{1}\{f(W_j) > K\} - \mathbb{P}(f(W_j) > K_1)\right]\right] + \sup_{f\in\mathcal{F}}\,\frac{1}{B}\sum_{j=1}^B \mathbb{P}(f(W_j) > K_1)\\
    &\overset{(a)}{\le} \mathbb{E}\left[\sup_{f\in\mathcal{F}}\,\frac{1}{B}\sum_{j=1}^B \{\phi(f(W_j)) - \mathbb{E}[\phi(f(W_j))]\}\right] + \sup_{f\in\mathcal{F}}\frac{1}{B}\sum_{j=1}^B \frac{\mathbb{E}[f^2(W_j)]}{K_1^2}\\
    &\overset{(b)}{\le} 2\mathbb{E}\left[\sup_{f\in\mathcal{F}}\,\frac{1}{B}\sum_{j=1}^B \varepsilon_j\phi(f(W_j))\right] + \frac{\sigma^2}{K_1^2}\\
    &\overset{(c)}{\le} \frac{2}{K-K_1}\mathbb{E}\left[\sup_{f\in\mathcal{F}}\,\frac{1}{B}\sum_{j=1}^B \varepsilon_jf(W_j)\right] + \frac{\sigma^2}{K_1^2}\\
    &= \frac{2\mathfrak{R}(\mathcal{F})}{K-K_1} + \frac{\sigma^2}{K_1^2}.
    \end{split}
    \end{equation}
    Here inequality~(a) follows from Chebyshev's inequality, inequality~(b) follows from symmetrization principle~\citep[Theorem 3.1.21]{GineNickl2015}, and inequality~(c) follows from the contraction principle~\citep[Lemma 8]{mohri2014learning}.
    
    On the other hand, from Corollary 2 of~\cite{pinelis1986remarks}, we get
    \[
    \mathbb{P}\left(\sup_{f\in\mathcal{F}}\,\frac{1}{B}\sum_{j=1}^B \mathbf{1}\{f(W_j) > K\} - \mathbb{E}\left[\sup_{f\in\mathcal{F}}\frac{1}{B}\sum_{j=1}^B\mathbf{1}\{f(W_j) > K\}\right] \ge x\right) ~\le~ \exp\left(-B\Sigma^2h(x/\Sigma^2)\right),
    \]
    where $\Sigma^2 = B^{-1}\sum_{j=1}^B \mathbb{P}(F(W_j) > K)$ and $h(u) = (1 + u)\ln(1 + u) - u$. Hence, inverting this inequality, we conclude that with probability at least $1 - \delta$,
    \[
    \sup_{f\in\mathcal{F}}\,\frac{1}{B}\sum_{j=1}^B \mathbf{1}\{f(W_j) > K\} - \mathbb{E}\left[\sup_{f\in\mathcal{F}}\frac{1}{B}\sum_{j=1}^B\mathbf{1}\{f(W_j) > K\}\right] < \Sigma^2h^{-1}\left(\frac{\log(1/\delta)}{B\Sigma^2}\right).
    \]
    Combining all the inequalities, it suffices to choose $K_1, K$ such that
    \[
    \frac{2\mathfrak{R}(\mathcal{F})}{K - K_1} + \frac{\sigma^2}{K_1^2} + \Sigma^2h^{-1}\left(\frac{\log(1/\delta)}{B\Sigma^2}\right) \le 1 - \frac{c}{B}.
    \]
    It is not difficult to verify that
    \begin{equation}\label{eq:inverse-Bennett-bound}
    h^{-1}(t) \le \sqrt{2t} + \frac{2t}{\log(1 + t)}\quad\mbox{for all}\quad t\ge 0.
    \end{equation}
    See, for example, Lemma 6.3 of~\cite{wellner2017bennett}. Hence, it suffices to find $K_1, K$ such that
    \begin{equation}\label{eq:penultimate-sufficient-condition-bound-I}
    \frac{2\mathfrak{R}(\mathcal{F})}{K - K_1} + \frac{\sigma^2}{K_1^2} + \sqrt{\frac{2\Sigma^2\log(1/\delta)}{B}} + \frac{2\log(1/\delta)}{B\log(1 + \log(1/\delta)/(B\Sigma^2))} \le 1 - \frac{c}{B}.
    \end{equation}
    Taking $K = K_1 + 20\mathfrak{R}(\mathcal{F})/(1 - c/B)$,~\eqref{eq:penultimate-sufficient-condition-bound-I} becomes
    \[
    \frac{\sigma^2}{K_1^2} + \sqrt{\frac{2\Sigma^2\log(1/\delta)}{B}} + \frac{2\log(1/\delta)}{B\log(1 + \log(1/\delta)/(B\Sigma^2))} \le \frac{9}{10}\left(1 - \frac{c}{B}\right).
    \]
    It suffices to choose $K_1$ such that
    \begin{align*}
        \frac{\sigma^2}{K_1^2} &\le \frac{3}{10}\left(1 - \frac{c}{B}\right),\\
        \sqrt{\frac{2\Sigma^2\log(1/\delta)}{B}} &\le \frac{3}{10}\left(1 - \frac{c}{B}\right),\\
        \frac{2\log(1/\delta)}{B\log(1 + \log(1/\delta)/(B\Sigma^2))} &\le \frac{3}{10}\left(1 - \frac{c}{B}\right).
    \end{align*}
    By Markov's inequality, we have
    \[
    \Sigma^2 \le \frac{1}{B}\sum_{j=1}^B \frac{\mathbb{E}[F^q(W_j)]}{K^q} \le \frac{V^q}{K^q}.
    \]
    Substituting this inequality, the condition becomes
    \begin{align*}
        K &\ge \max\left\{\frac{20\mathfrak{R}(\mathcal{F})}{1-c/B} + \sqrt{\frac{10\sigma^2}{3(1 - c/B)}},\, V\left(\frac{200\log(1/\delta)}{9(B-c)^2}\right)^{1/q},\,VB^{1/q}\left(\frac{e^{20\log(1/\delta)/(3(B-c))} - 1}{\log(1/\delta)}\right)^{1/q}\right\}\\
        &= \max\left\{\frac{20\mathfrak{R}(\mathcal{F})}{1-c/B} + \sqrt{\frac{10\sigma^2}{3(1 - c/B)}},\,VB^{1/q}\left(\frac{e^{20\log(1/\delta)/(3(B-c))} - 1}{\log(1/\delta)}\right)^{1/q}\right\}.
    \end{align*}
    The equality follows from the fact that $e^{x} - 1 \ge x^2/2$ for all $x \ge 0$.
    Conservatively, we can take
    \begin{align*}
    K = \frac{20\mathfrak{R}(\mathcal{F})}{1-c/B} + \sqrt{\frac{10\sigma^2}{3(1 - c/B)}} + VB^{1/q}\left(\frac{e^{7\log(1/\delta)/(B-c)} - 1}{\log(1/\delta)}\right)^{1/q}.
    \end{align*}
    This completes the first bound.
    \paragraph{Second Bound.} Here instead of applying Corollary 2 of~\cite{pinelis1986remarks}, we apply Theorem 10 of~\cite{massart2000constants}. It follows that
    \[
    \mathbb{P}\left(\sup_{f\in\mathcal{F}}\frac{1}{B}\sum_{j=1}^B \mathbf{1}\{f(W_j) > K\} - E \ge x\right) \le \exp(-BEh(x/E)),
    \]
    where
    \[
    E := \mathbb{E}\left[\sup_{f\in\mathcal{F}}\frac{1}{B}\sum_{j=1}^B \mathbf{1}\{f(W_j) > K\}\right].
    \]
    Hence, inverting the inequality, we conclude that with probability at least $1 - \delta$,
    \[
    \sup_{f\in\mathcal{F}}\,\frac{1}{B}\sum_{j=1}^B \mathbf{1}\{f(W_j) > K\} \le E + Eh^{-1}\left(\frac{\log(1/\delta)}{BE}\right).
    \]
    As before, using~\eqref{eq:inverse-Bennett-bound}, it suffices to choose $K$ such that
    \[
    E + E\left(\sqrt{\frac{2\log(1/\delta)}{BE}} + \frac{2\log(1/\delta)}{BE\log(1 + \log(1/\delta)/(BE))}\right) \le 1 - \frac{c}{B},
    \]
    or equivalently, we need to choose $K$ such that
    \[
    E + \sqrt{\frac{2E\log(1/\delta)}{B}} + \frac{2\log(1/\delta)}{B\log(1 + \log(1/\delta)/(BE))} \le 1 - \frac{c}{B}.
    \]
    We shall choose $K$ such that
    \begin{align*}
        E + \sqrt{\frac{2E\log(1/\delta)}{B}} &\le \frac{1}{2}\left(1 - \frac{c}{B}\right),\\
        \frac{2\log(1/\delta)}{B\log(1 + \log(1/\delta)/(BE))} &\le \frac{1}{2}\left(1 - \frac{c}{B}\right).
    \end{align*}
    Inverting these inequalities, we get the sufficient condition as 
    \begin{equation}\label{eq:expectation-sufficient-condition-bound-2}
    \begin{split}
        E ~&\le~ \min\left\{\frac{(1-c/B)^2}{4(1-c/B + 2\log(1/\delta)/B)},\,\frac{\log(1/\delta)}{B(\exp(4\log(1/\delta)/(B-c)) - 1)}\right\},\\
        ~&=~ \frac{\log(1/\delta)}{B(\exp(4\log(1/\delta)/(B-c)) - 1)}.
    \end{split}
    \end{equation}
    Here the equality follows from the fact that $1/(1 + y) \ge 2y/(e^{2y} - 1)$ for all $y \ge 0$ and that the first term in the minimum in~\eqref{eq:expectation-sufficient-condition-bound-2} can be written as $(1 - c/B)/4(1 + y)$ and the second term can be written as $(1-c/B)y/(2e^{2y} - 2)$ for $y = 2\log(1/\delta)/(B-c)$.
    From~\eqref{eq:expectation-bound} with $K_1 = K/2$, we obtain
    \[
    E \le \frac{4\mathfrak{R}(\mathcal{F})}{K} + \frac{4\sigma^2}{K^2}.
    \]
    Substituting this bound in~\eqref{eq:expectation-sufficient-condition-bound-2}, we conclude a sufficient choice of $K$ as
    \[
    K \ge \frac{4B(e^{4\log(1/\delta)/(B-c)} - 1)}{\log(1/\delta)}\mathfrak{R}(\mathcal{F}) + 2\sigma\sqrt{\frac{B(e^{4\log(1/\delta)/(B-c)} - 1)}{\log(1/\delta)}}.
    \]
    This completes the second bound.

    \paragraph{Third bound.} Both the bounds above become infinity if $c = B$, but it is not difficult to obtain a finite bound on the quantile. For any $c \le B$, we have
    \[
    \sup_{f\in\mathcal{F}}\,(f(W))_{(c)} \le (F(W))_{(c)} \le \max_{1\le j\le B} F(W_j).
    \]
    By Chebyshev's inequality, 
    \[
    \mathbb{P}\left(\max_{1\le j\le B}F(W_j) \ge t\right) \le \frac{\sum_{j=1}^B \mathbb{E}[F^q(W_j)]}{t^q} = \frac{BV^q}{t^q}.
    \]
    Hence, with probability at least $1 - \delta$, $\sup_{f\in\mathcal{F}}\,(f(W))_{(c)} \le V(B/\delta)^{1/q}$. This completes the third bound. 
    
    This completes the proof. 
\end{proof}

\section{Auxiliary Results on Tail Inequalities}
The following result compares probabilities under Poisson-Binomial distribution to that under the binomial distribution.
\begin{proposition}\label{prop:poisson-binomial-to-binomial}
    Suppose $Q_1, \ldots, Q_B$ are independent Bernoulli random variables with $\mathbb{P}(Q_j = 1) = q_j \in [0, 1]$. Let $c^*_{B,\alpha}$ be as defined in Algorithm~\ref{alg:proposed-conf-int}. Then 
    \[
    \alpha\exp\left(-\frac{2\sum_{j=1}^B (q_j - 1/2)_+}{(1 - 2\max_{j} (q_j - 1/2)_+)_+}\right) ~\le~ \mathbb{P}\left(\sum_{j=1}^B Q_j < B - c^*_{B,\alpha}\right) ~\le~ \alpha\exp\left(2\sum_{j=1}^B (1/2 - q_j)_+\right). 
    \]
\end{proposition}
\begin{proof}
    It is easy to see that
    \[
    \mathbb{P}(\mbox{Bernoulli}(q_j) < t) \le \max\{2(1-q_j), 1\}\mathbb{P}(\mbox{Bernoulli}(1/2) < t) \quad\mbox{for all}\quad t\in\mathbb{R}.
    \]
    In fact, if $t > 1$ or $t \le 0$, then both probabilities are equal to one or zero, and if $t \in (0, 1]$, then the probability on the left is $(1-q_j)$ while the one on the right is $1/2$. 

    Because $Q_1, \ldots, Q_B$ are independent, by a conditioning argument, we get
    \begin{align*}
        \mathbb{P}\left(\sum_{j=1}^B Q_j < B - c^*_{B,\alpha}\right) &\le \mathbb{P}\left(\sum_{j=1}^B \mathrm{Bernoulli}(1/2) < B - c^*_{B,\alpha}\right)\prod_{j=1}^B \max\{2(1-q_j), 1\}\\
        &= \alpha\prod_{j=1}^B \max\{1,\, 2(1-q_j)\}\\
        &= \alpha\prod_{j=1}^B (1 + 2(1/2 - q_j)_+)\\
        &\le \alpha\exp\left(2\sum_{j=1}^B (1/2 - q_j)_+\right).
    \end{align*}
    To prove the lower bound, note that
    \[
    \mathbb{P}(\mbox{Bernoulli}(1/2) < t) \le \max\left\{\frac{1}{2(1-q_j)},\, 1\right\}\mathbb{P}(\mbox{Bernoulli}(q_j) < t) \quad\mbox{for all}\quad t\in\mathbb{R}.
    \]
    Hence, we conclude
    \[
    \mathbb{P}(\mbox{Binom}(B, 1/2) < B - c^*_{B,\alpha}) \le \mathbb{P}\left(\sum_{j=1}^B Q_j < B - c^*_{B,\alpha}\right)\prod_{j=1}^B \max\left\{1,\, \frac{1}{2(1-q_j)}\right\}.
    \]
    Noting that
    \[
    \max\left\{1,\, \frac{1}{2(1-q_j)}\right\} = \frac{1}{\min\{1,\, 2(1-q_j)\}} = \frac{1}{1 - 2(q_j-1/2)_+},
    \]
    we get
    \begin{align*}
    \mathbb{P}\left(\sum_{j=1}^B Q_j < B - c^*_{B,\alpha}\right) &\ge \alpha\prod_{j=1}^B (1 - 2(q_j - 1/2)_+)\\
    &\ge \alpha\exp\left(-\frac{2\sum_{j=1}^B (q_j - 1/2)_+}{(1 - 2\max_{j} (q_j - 1/2)_+)_+}\right).
    \end{align*}
    Here the last inequality follows from the fact that $\log(1-x) \ge -x/(1-\rho)$ whenever $x\in[0, \rho]$.
    This completes the proof.
\end{proof}
The following result controls the miscoverage of confidence interval for median under a bound on median bias around zero. A similar, but slightly weaker, result is given in the proof of Theorem 7 of~\cite{paul2024inference}; the proof techniques are also different.
\begin{proposition}\label{prop:binomial-probabilities}
    Suppose $V_1, \ldots, V_B$ are independent real-valued random variables. 
    Let $c^*_{B,\alpha}$ be as defined in Algorithm~\ref{alg:proposed-conf-int}.
    Then, for any $\alpha\in(0, 1)$,
    \[
    \mathbb{P}\left(\min\left\{\sum_{i=1}^B \mathbf{1}\{V_i \le 0\},\,\sum_{i=1}^B \mathbf{1}\{V_i \ge 0\}\right\} < B - c^*_{B,\alpha/2}\right) \le \alpha\exp(2(B\Delta)^2),
    \]
    where
    \[
    \Delta = \frac{1}{B}\sum_{i=1}^B \left(\frac{1}{2} - \min\{\mathbb{P}(V_i \le 0),\,\mathbb{P}(V_i \ge 0)\}\right)_+.
    \]
    Moreover,
    \[
    \mathbb{P}\left(\min\left\{\sum_{i=1}^B \mathbf{1}\{V_i \le 0\},\,\sum_{i=1}^B \mathbf{1}\{V_i \ge 0\}\right\} < B - c^*_{B,\alpha/2}\right) \ge \alpha\exp\left(-\frac{4\sum_{i=1}^B \Delta_i^2}{(1 - 4\max_i \Delta_i^2)_+}\right),
    \]
    where
    \[
    \Delta_i^2 := \left(\mathbb{P}(V_i > 0) - \frac{1}{2}\right)^2 + \mathbb{P}(V_i = 0).
    \]
\end{proposition}
\begin{proof}
    Let
    \[
    p_i = \mathbb{P}(V_i > 0),\quad q_i = \mathbb{P}(V_i < 0),\quad\mbox{and}\quad r_i = \mathbb{P}(V_i = 0).
    \]
    We have $p_i + q_i + r_i = 1$. This implies that $\min\{p_i, q_i\} \le 1/2$. Set
    \[
    \Delta_i := \left(\frac{1}{2} - \min\{\mathbb{P}(V_i \le 0),\,\mathbb{P}(V_i \ge 0)\}\right)_+.
    \]
    \begin{enumerate}
        \item If $\max\{p_i, q_i\} \le 1/2$, then define
        \[
        \lambda_i = \frac{1}{2}\left(\frac{(p_i - q_i) - 2\Delta_i}{r_i} + 1\right) \in (0, 1),
        \]
        so that $p_i + \lambda_i r_i = 1/2 - \Delta_i$ and $q_i + (1-\lambda_i)r_i = 1/2 + \Delta_i$.
        \item If $q_i \le 1/2 < p_i$, then define
        \[
        \lambda_i = \frac{1}{2}\left(\frac{2\Delta_i - (p_i - q_i)}{r_i} + 1\right) \in (0, 1),
        \]
        so that $p_i + \lambda_i r_i = 1/2 + \Delta_i$ and $q_i + (1 - \lambda_i)r_i = 1/2 - \Delta_i$.
        \item If both $p_i \le 1/2$ and $q_i \le 1/2$ hold true, then choose $\lambda_i$ as in item 1 above.
    \end{enumerate}
    Let $U_i \sim \mbox{Uniform}(0, 1)$ and define
    \[
    \widetilde{V}_i = \begin{cases}
        V_i, &\mbox{if either}\quad V_i < 0\mbox{ or }V_i > 0,\\
        +1, &\mbox{if }V_i = 0, U_i \le \lambda_i,\\
        -1, &\mbox{if }V_i = 0, U_i > \lambda_i.
    \end{cases}
    \]
    It follows that 
    \[
    \mathbf{1}\{V_i \le 0\} \ge \mathbf{1}\{\widetilde{V}_i \le 0\}\quad\mbox{and}\quad \mathbf{1}\{V_i \ge 0\} \ge \mathbf{1}\{\widetilde{V}_i \ge 0\},
    \]
    and
    \begin{equation}\label{eq:probabilities-Delta_i}
    \mathbb{P}(\widetilde{V}_i = 0) = 0\quad\mbox{and}\quad \mathbb{P}(\widetilde{V}_i \le 0) = q_i + (1-\lambda_i)r_i \in \left\{\frac{1}{2}\pm\Delta_i\right\}.
    \end{equation}
    This implies that
    \begin{align*}
    \min\left\{\sum_{i=1}^B \mathbf{1}\{V_i \le 0\},\, \sum_{i=1}^B \mathbf{1}\{V_i \ge 0\}\right\} &\ge \min\left\{\sum_{i=1}^B \mathbf{1}\{\widetilde{V}_i \le 0\},\,\sum_{i=1}^B \mathbf{1}\{\widetilde{V}_i \ge 0\}\right\}\\
    &= \frac{B}{2} - \left|\sum_{i=1}^B \mathbf{1}\{\widetilde{V}_i \le 0\} - \frac{B}{2}\right|.
    \end{align*}
    Define $S_i = \mathbf{1}\{\widetilde{V}_i > 0\}$ and
    \begin{equation}\label{eq:event-A-binomial-prob}
    A = \left\{\mathbf{x} = (x_1, \ldots, x_B)\in\{0,1\}^B:\, \min\left\{\sum_{i=1}^B x_i,\, \sum_{i=1}^B (1-x_i)\right\} < B - c^*_{B,\alpha/2}\right\}.
    \end{equation}
    Clearly, $\mathbf{x}\in A$ if and only if $1-\mathbf{x}\in A$. Hence, we have
    \begin{align*}
        &\mathbb{P}\left(\min\left\{\sum_{i=1}^B \mathbf{1}\{V_i \le 0\},\,\sum_{i=1}^B \mathbf{1}\{V_i \ge 0\}\right\} < B - c^*_{B,\alpha/2}\right)\\ 
        &\le \mathbb{P}\left(\min\left\{\sum_{i=1}^B \mathbf{1}\{\widetilde{V}_i \le 0\},\,\sum_{i=1}^B \mathbf{1}\{\widetilde{V}_i \ge 0\}\right\} < B - c^*_{B,\alpha/2}\right)\\
        &= \mathbb{P}\left(\min\left\{\sum_{i=1}^B \mathbf{1}\{\widetilde{V}_i < 0\},\,\sum_{i=1}^B \mathbf{1}\{\widetilde{V}_i > 0\}\right\} < B - c^*_{B,\alpha/2}\right)\\
        &= \mathbb{E}\left[\mathbf{1}\{(S_1, \ldots, S_B) \in A\}\right]\\
        &= \sum_{\mathbf{x}\in \{0,1\}^B} \mathbf{1}\{\mathbf{x}\in A\}\prod_{i=1}^B \mathbb{P}(S_i = x_i)\\
        &= \sum_{\mathbf{x}\in \{0,1\}^B} \mathbf{1}\{\mathbf{x}\in A\}\left[\frac{\prod_{i=1}^B \mathbb{P}(S_i = x_i) + \prod_{i=1}^B \mathbb{P}(S_i = 1-x_i)}{2}\right].
    \end{align*}
    Setting
    \[
    \tilde{p}_i = \mathbb{P}(S_i = 1) = \mathbb{P}(\widetilde{V}_i > 0),
    \]
    we get from~\eqref{eq:probabilities-Delta_i} that $\Delta_i = |\tilde{p}_i - 1/2|$. This implies
    \begin{align*}
    \prod_{i=1}^B \mathbb{P}(S_i = x_i) &= \prod_{i=1}^B \left(\frac{1}{2} + (2x_i - 1)(\tilde{p}_i - 1/2)\right) \le (1/2)^B\exp\left(\sum_{i=1}^B (2x_i - 1)(2\tilde{p}_i - 1)\right)\\
    \prod_{i=1}^B \mathbb{P}(S_i =1 - x_i) &= \prod_{i=1}^B \left(\frac{1}{2} - (2x_i - 1)(\tilde{p}_i - 1/2)\right) \le (1/2)^B \exp\left(-\sum_{i=1}^B (2x_i - 1)(2\tilde{p}_i - 1)\right).
    \end{align*}
    Therefore, 
    \begin{align*}
    &\mathbb{P}\left(\min\left\{\sum_{i=1}^B \mathbf{1}\{V_i \le 0\},\,\sum_{i=1}^B \mathbf{1}\{V_i \ge 0\}\right\} < B - c^*_{B,\alpha/2}\right)\\ 
    &\le \sum_{\mathbf{x}\in\{0,1\}^B}\,\mathbf{1}\{\mathbf{x}\in A\} \left(\frac{1}{2}\right)^B \sup_{\mathbf{x}\in A}\,\frac{\exp(\sum_{i=1}^B (2x_i - 1)(2\tilde{p}_i - 1)) + \exp(-\sum_{i=1}^B (2x_i - 1)(2\tilde{p}_i - 1))}{2}\\
    &= \alpha\times \sup_{\mathbf{x}\in A}\cosh\left(\sum_{i=1}^B (2x_i - 1)(2\tilde{p}_i - 1)\right)\\
    &\le \alpha\times\cosh\left(\sum_{i=1}^B |2\tilde{p}_i - 1|\right)\\
    &= \alpha\times\cosh\left(2\sum_{i=1}^B \Delta_i\right) = \alpha\cosh(2B\Delta).
    \end{align*}
    This completes the proof of the upper bound, using the inequality $\cosh(x) \le \exp(x^2/2)$.

    To prove the lower bound, consider the event
    \[
    \mathcal{O} = \{V_i \neq 0\quad\mbox{for all}\quad 1\le j\le B\}.
    \]
    From the independence of $V_i$'s, $\mathbb{P}(\mathcal{O}) = \prod_{j=1}^B \mathbb{P}(V_i \neq 0).$
    Then
    \begin{align*}
        &\mathbb{P}\left(\min\left\{\sum_{i=1}^B \mathbf{1}\{V_i \le 0\},\,\sum_{i=1}^B \mathbf{1}\{V_i \ge 0\}\right\} < B - c^*_{B,\alpha/2}\right)\\
        &\ge \mathbb{P}\left(\min\left\{\sum_{i=1}^B \mathbf{1}\{V_i \le 0\},\,\sum_{i=1}^B \mathbf{1}\{V_i \ge 0\}\right\} < B - c^*_{B,\alpha/2}\big| \mathcal{O}\right)\prod_{j=1}^B \mathbb{P}(V_j \neq 0).
    \end{align*}
    Letting $S_i = \mathbf{1}\{V_i \ge 0\}$ and recalling the definition of $A$ from~\eqref{eq:event-A-binomial-prob}, we get that for any $\mathbf{x}\in\{0, 1\}^B$, 
    \begin{align*}
        &\mathbb{P}\left(\min\left\{\sum_{i=1}^B \mathbf{1}\{V_i \le 0\},\,\sum_{i=1}^B \mathbf{1}\{V_i \ge 0\}\right\} < B - c^*_{B,\alpha/2}\right)\\
        &\ge \mathbb{E}\left[\mathbf{1}\{(S_1, \ldots, S_B)\in A\}\big| \mathcal{O}\right]\prod_{j=1}^B \mathbb{P}(V_j \neq 0)\\
        &= \sum_{\mathbf{x}\in\{0,1\}^B} \mathbf{1}\{x\in A\}\mathbb{P}\left(\bigcap_{j=1}^B \{S_j = x_j\}\big|\mathcal{O}\right)\prod_{j=1}^B \mathbb{P}(V_j \neq 0)\\
        &= \sum_{\mathbf{x}\in\{0,1\}^B} \mathbf{1}\{x\in A\}\prod_{j=1}^B \mathbb{P}\left(S_j = x_j\big|\mathcal{O}\right)\prod_{j=1}^B \mathbb{P}(V_j \neq 0).
    \end{align*}
    As before, using the ``symmetry'' of $A$, we get
    \begin{align*}
        &\mathbb{P}\left(\min\left\{\sum_{i=1}^B \mathbf{1}\{V_i \le 0\},\,\sum_{i=1}^B \mathbf{1}\{V_i \ge 0\}\right\} < B - c^*_{B,\alpha/2}\right)\\
        &\ge \sum_{\mathbf{x}\in\{0,1\}^B}\,\mathbf{1}\{\mathbf{x}\in A\}\left[\frac{\prod_{i=1}^B \mathbb{P}(S_i = x_i|V_i \neq 0) + \prod_{i=1}^B \mathbb{P}(S_i = 1 - x_i|V_i \neq 0)}{2}\right]\prod_{j=1}^B \mathbb{P}(V_j \neq 0)\\
        &\ge \sum_{\mathbf{x}\in\{0,1\}^B}\,\mathbf{1}\{\mathbf{x}\in A\}\sqrt{\prod_{i=1}^B \mathbb{P}(V_i > 0|V_i \neq 0)\mathbb{P}(V_i < 0|V_i \neq 0)}\prod_{j=1}^B \mathbb{P}(V_j \neq 0)\\
        &= \sum_{\mathbf{x}\in\{0,1\}^B}\,\mathbf{1}\{\mathbf{x}\in A\}\sqrt{\prod_{i=1}^B \mathbb{P}(V_i > 0)\mathbb{P}(V_i < 0)}\\
        &= \sum_{\mathbf{x}\in\{0,1\}^B}\,\mathbf{1}\{\mathbf{x}\in A\}\frac{1}{2^B}\sqrt{\prod_{i=1}^B (2\mathbb{P}(V_i > 0))(2\mathbb{P}(V_i < 0))}\\
        &= \alpha\sqrt{\prod_{i=1}^B (2\mathbb{P}(V_i > 0))(2\mathbb{P}(V_i < 0))}.
    \end{align*}
    Here the second inequality follows from the AM-GM inequality. Noting that
    \begin{align*}
    4\mathbb{P}(V_i > 0)\mathbb{P}(V_i < 0) &= 1 - 4(\mathbb{P}(V_i > 0) - 1/2)^2 - 4\mathbb{P}(V_i > 0)\mathbb{P}(V_i = 0)\\ 
    &\ge 1 - 4(\mathbb{P}(V_i > 0) - 1/2)^2 - 4\mathbb{P}(V_i = 0)\\
    &= 1 - 4\Delta_i^2
    \end{align*}
    we get using the inequality $\log(1-x) \ge -x/(1-\rho)$ for all $x\in[0, \rho]$ that
    \begin{align*}
    &\mathbb{P}\left(\min\left\{\sum_{i=1}^B \mathbf{1}\{V_i \le 0\},\,\sum_{i=1}^B \mathbf{1}\{V_i \ge 0\}\right\} < B - c^*_{B,\alpha/2}\right)\\ 
    &\ge \alpha\exp\left(-\frac{4\sum_{i=1}^B \Delta_i^2}{(1 - 4\max_{i} \Delta_i^2)_+}\right).
    \end{align*}
    This completes the proof.
\end{proof}

\begin{proposition}\label{prop:Wilcoxon-signed-rank-test}
    Suppose $W_1, \ldots, W_B$ are independent real-valued random variables. For any $h$ satisfying~\eqref{eq:non-decreasing-weights}, let $\widehat{\mathrm{CI}}_{n,\alpha}^{\mathtt{SR}}$ be as defined in~\eqref{eq:general-Wilcoxon-signed-rank-test}. Then
    \[
    \alpha\exp\left(-\frac{{(B-1)^2}\sum_{i < j} \Delta_i\Delta_j}{(1 - 7(B-1)^2\sum_{i < j}\Delta_i\Delta_j/6)_+}\right)~\le~ \mathbb{P}(0\notin\widehat{\mathrm{CI}}_{n,\alpha}^{\mathtt{WSR}}) ~\le~ \alpha\exp\left({(B-1)^2}\sum_{i < j} \Delta_i\Delta_j\right),
    \]
    where
    \[
    \Delta_i := \min\left\{1, 2\sup_{u\in\mathbb{R}}|\mathbb{P}(W_i \le u) - \mathbb{P}(-W_i \le u)|\right\}.
    \]
\end{proposition}
\begin{proof}
Recall
\[
S_i = \mathbf{1}\{(W_i, U_i)\succ(0,0)\} - \mathbf{1}\{(W_i, U_i)\prec(0,0)\}.
\]
Set
\[
Z_i:=S_i(W_i, U_i),
\quad\mbox{and}\quad
M_i:=|W_i|,
\quad\mbox{for}\quad 1\le i\le B.
\]
Thus, $Z_i=(M_i,S_iU_i)$ is the augmented magnitude of $(W_i, U_i)$ and
lies in the positive augmented half-line
$\mathcal Z_+:=\{z\in\mathbb R^2:z\succ(0,0)\}.$

We first establish the conditional distribution of $S_i$ given $Z_i$. We will prove that $S_i$ and $S_iU_i$ are independent conditional on $M_i$ so that the conditional distribution of $S_i$ given $Z_i$ is same as that of $S_i$ given $M_i$. In particular, this would imply that $\mathbb{E}[S_i|Z_i] = \mathbb{E}[S_i|M_i]$ and hence,
\[
\mathbb{P}(S_i = s_i|Z_i) = \frac{1 + s_i\mathbb{E}[S_i|Z_i]}{2} = \frac{1+ s_i\beta_i(M_i)}{2}\quad\mbox{where}\quad \beta_i(M_i) = \mathbb{E}[S_i|M_i].
\]
Therefore,
\begin{equation}\label{eq:conditional-sign-given-Z-general-h-final}
\mathbb P(S_i=s_i\mid Z_i)
=\frac{1+s_i\beta_i(M_i)}2,
\qquad s_i\in\{\pm1\}.
\end{equation}
Since the pairs $(S_i,Z_i)$ are independent across $i$, it follows that
\begin{equation}\label{eq:conditional-sign-vector-given-Z-general-h-final}
\mathbb P(S=s\mid Z)
=\prod_{i=1}^B\frac{1+s_i\beta_i(M_i)}2,
\qquad s\in\{\pm1\}^B.
\end{equation}
Now, to prove the conditional independence, note that if $M_i = 0$, then $S_i = \mbox{sign}(U_i)$ and $S_iU_i = |U_i|$ which are independent because $U_i$ is symmetric around zero. On the other hand, if $M_i > 0$, then for $s_i\in\{\pm 1\}$ and any Borel measurable set $I\subseteq\mathbb{R}$,
\begin{align*}
\mathbb{P}(S_i = s_i, S_iU_i \in I|M_i) &= \mathbb{P}(\mbox{sign}(W_i) = s_i, s_iU_i\in I|M_i)\\ 
&= \mathbb{P}(\mbox{sign}(W_i) = s_i|M_i)\mathbb{P}(s_iU_i\in I|M_i)\\ 
&= \left(\frac{1 + s_i\beta_i(M_i)}{2}\right)\mathbb{P}(U_i\in I),
\end{align*}
because $s_iU_i \overset{d}{=} U_i$. Hence, $S_i$ and $S_iU_i$ are independent conditional on $M_i$. This proves~\eqref{eq:conditional-sign-given-Z-general-h-final}.

Let us also note that $Z_1,\ldots,Z_B$ are almost surely distinct. Indeed,
conditional on $M_i=m>0$, the second coordinate $S_iU_i$ has a continuous
distribution on $(-1,1)$. Conditional on $M_i=0$, it has the continuous
distribution of $|U_i|$ on $(0,1)$. Therefore, for any $m\in[0,\infty)^B$, the second coordinates of $Z_i$'s are independent and continuously
distributed. Hence, with probability one, the lexicographic ranks
$R_1,\ldots,R_B$ of $Z_1,\ldots,Z_B$ satisfy
\[
\{R_1,\ldots,R_B\}=\{1,2,\ldots,B\}.
\]

Let
\[
T_h:=\sum_{j=1}^B h(R_j)\mathbf 1\{S_j=1\}.
\]
By Lemma~\ref{lem:weighted-Walsh-average-transformed-rank-representation},
\begin{equation}\label{eq:Th-weighted-Walsh-general-h-final}
T_h
=
\sum_{1\le i\le j\le B}\omega_{ij}
\mathbf 1\{A_{ij}\succ(0,0)\}.
\end{equation}
The auxiliary variables are continuously distributed, so, almost surely, no
augmented Walsh average lies on the horizontal line
$\{(t,0):t\in\mathbb R\}$. Moreover, distinct unordered index pairs yield
distinct augmented Walsh averages almost surely: equality of two distinct
augmented Walsh averages would require a nontrivial linear equality among the
independent continuous variables $U_1,\ldots,U_B$. Therefore, the weighted
quantiles are unambiguous.

From the definition of $Q_h$ and $\kappa_{B,\alpha}^{h}$,
\[
Q_h(\kappa_{B,\alpha}^h+)\preceq(0,0)
\quad\Longleftrightarrow\quad
N_h-T_h>\kappa_{B,\alpha}^h,
\]
and
\[
(0,0)\preceq Q_h(N_h-\kappa_{B,\alpha}^h)
\quad\Longleftrightarrow\quad
N_h-T_h<N_h-\kappa_{B,\alpha}^h.
\]
It follows that
\begin{equation}\label{eq:general-h-miscoverage-event-final}
\{0\notin\widehat{\mathrm{CI}}_{n,\alpha}^{\mathtt{SR}}\}
=
\{T_h\le\kappa_{B,\alpha}^h\}
\cup
\{T_h\ge N_h-\kappa_{B,\alpha}^h\}.
\end{equation}
This proves the connection between the confidence interval defined through the
weighted Walsh averages and the transformed signed-rank statistic.

If $W_1,\ldots,W_B$ are independent and symmetric around zero, then
$\beta_i(M_i)=0$ almost surely for every $i$. By
\eqref{eq:conditional-sign-vector-given-Z-general-h-final},
$(S_1,\ldots,S_B)$ conditional on $Z_1,\ldots,Z_B$ is distributed uniformly
on $\{\pm1\}^B$ almost surely. This implies that, under symmetry,
\[
T_h\overset{d}{=}G_{B,h}:=\sum_{j=1}^Bh(j)\xi_j,
\]
and hence
\begin{equation}\label{eq:confidence-under-symmetry-general-h-final}
\mathbb P\left(
G_{B,h}\le\kappa_{B,\alpha}^h
\ \mbox{or}\ 
G_{B,h}\ge N_h-\kappa_{B,\alpha}^h
\right)=\alpha.
\end{equation}

To prove the coverage guarantee without symmetry, introduce the following
notation. For any deterministic vector
$z=(z_1,\ldots,z_B)\in\mathcal Z_+^B$ with distinct coordinates, let
$R_i(z)$ be the lexicographic rank of $z_i$ among $z_1,\ldots,z_B$. Also, for
any vector $s=(s_1,\ldots,s_B)\in\{\pm1\}^B$, define
\begin{align}
\phi_z(s)
:=\mathbb E_{\kappa_{B,\alpha}^h}\Bigg[
\mathbf 1\Bigg\{
\left|\frac{N_h}{2}
-\sum_{j=1}^B h(R_j(z))\mathbf 1\{s_j=1\}\right|
\ge\frac{N_h}{2}-\kappa_{B,\alpha}^h
\Bigg\}\Bigg],
\label{eq:phi-z-general-h-final}
\end{align}
where the expectation is computed only with respect to
$\kappa_{B,\alpha}^h$. Clearly, by
\eqref{eq:general-h-miscoverage-event-final},
\begin{equation}\label{eq:miscoverage-as-phi-Z-general-h-final}
\mathbb P(0\notin\widehat{\mathrm{CI}}_{n,\alpha}^{\mathtt{SR}})
=\mathbb E[\phi_Z(S)].
\end{equation}
Also, because $\{R_j(z):1\le j\le B\}=\{1,\ldots,B\}$, the equality
\eqref{eq:confidence-under-symmetry-general-h-final} is equivalent to
\begin{equation}\label{eq:fair-sign-calibration-general-h-final}
\frac1{2^B}\sum_{s\in\{\pm1\}^B}\phi_z(s)=\alpha
\end{equation}
for every such $z$.

It follows from
\eqref{eq:conditional-sign-vector-given-Z-general-h-final} that
\begin{align*}
&\mathbb P(0\notin\widehat{\mathrm{CI}}_{n,\alpha}^{\mathtt{SR}})\\
&\quad=
\mathbb E_Z\left[
\sum_{s\in\{\pm1\}^B}\phi_Z(s)
\prod_{j=1}^B\frac{1+s_j\beta_j(M_j)}2
\right]\\
&\quad=
\mathbb E_Z\left[
\frac1{2^B}\sum_{s\in\{\pm1\}^B}\phi_Z(s)
\left(1+
\sum_{\varnothing\ne A\subseteq\{1,\ldots,B\}}
\prod_{j\in A}s_j\beta_j(M_j)\right)
\right].
\end{align*}
For $A\subseteq\{1,\ldots,B\}$, define
\begin{equation}\label{eq:aA-z-general-h-final}
a_A(z):=
\frac1{2^B}\sum_{s\in\{\pm1\}^B}
\phi_z(s)\prod_{j\in A}s_j.
\end{equation}
Then
\begin{equation}\label{eq:first-representation-general-h-final}
\mathbb P(0\notin\widehat{\mathrm{CI}}_{n,\alpha}^{\mathtt{SR}})
=\alpha+
\mathbb E_Z\left[
\sum_{\varnothing\ne A\subseteq\{1,\ldots,B\}}
a_A(Z)\prod_{j\in A}\beta_j(M_j)
\right].
\end{equation}

Because $\{R_j(z):1\le j\le B\}=\{1,\ldots,B\}$, it follows that
\[
\left|\frac{N_h}{2}
-\sum_{j=1}^B h(R_j(z))\mathbf 1\{s_j=1\}\right|
=
\left|\frac{N_h}{2}
-\sum_{j=1}^B h(R_j(z))\mathbf 1\{s_j=-1\}\right|.
\]
This implies that
\begin{equation}\label{eq:phi-even-general-h-final}
\phi_z(s)=\phi_z(-s)
\quad\mbox{for all }s\in\{\pm1\}^B.
\end{equation}
This further implies that $a_A(z)=0$ whenever $|A|$ is odd. Hence,
\eqref{eq:first-representation-general-h-final} simplifies to
\begin{equation}\label{eq:penultimate-general-h-final}
\mathbb P(0\notin\widehat{\mathrm{CI}}_{n,\alpha}^{\mathtt{SR}})
=\alpha+
\sum_{\substack{A\subseteq\{1,\ldots,B\}\\
|A|\ge2,\ |A|\ \mathrm{even}}}
\mathbb E\left[
a_A(Z)\prod_{j\in A}\beta_j(M_j)
\right].
\end{equation}
Observe that
\begin{equation}\label{eq:bound-aA-general-h-final}
|a_A(z)|
\le\frac1{2^B}\sum_{s\in\{\pm1\}^B}\phi_z(s)
=\alpha
\end{equation}
for every $A\subseteq\{1,\ldots,B\}$.

It now suffices to control
\[
\mathbb E\left[
a_A(Z)\prod_{j\in A}\beta_j(M_j)
\right]
\]
for $A\subseteq\{1,\ldots,B\}$ with $|A|=k\ge2$. We will control
\begin{equation}\label{eq:fA-general-h-final}
f(A):=
\mathbb E\left[
a_A(Z)\prod_{j\in A}\beta_j(M_j)
\ \middle|\ Z_{A^c}\right].
\end{equation}
Let
\[
Z_{(1):A^c}\prec Z_{(2):A^c}\prec\cdots
\prec Z_{(B-k):A^c}
\]
be the lexicographic order statistics of $Z_j$, $j\in A^c$.
These $(B-k)$ order statistics divide $\mathcal Z_+$ into $B-k+1$ disjoint
cells. The probability that $Z_j$, $j\in A$, equals one of the complementary
boundary points is zero. When $A^c=\varnothing$, use the evident convention
that $C_1=\mathcal Z_+$. Otherwise, define
\begin{align*}
C_1&:=\{z\in\mathcal Z_+:z\prec Z_{(1):A^c}\},\\
C_q&:=\{z\in\mathcal Z_+:
Z_{(q-1):A^c}\prec z\prec Z_{(q):A^c}\},
\quad2\le q\le B-k,\\
C_{B-k+1}&:=\{z\in\mathcal Z_+:z\succ Z_{(B-k):A^c}\}.
\end{align*}
We claim that
\begin{equation}\label{eq:cell-constancy-general-h-final}
a_A(Z)
=\sum_{q\in\{1,\ldots,B-k+1\}^{A}}
\mathfrak C_q(Z_{A^c})
\prod_{j\in A}\mathbf 1\{Z_j\in C_{q_j}\}
\end{equation}
for some constants $\mathfrak C_q(Z_{A^c})$ that depend only on their
displayed arguments. We will prove the result assuming
\eqref{eq:cell-constancy-general-h-final}.

Given that the cells are disjoint, we get from
\eqref{eq:bound-aA-general-h-final} that
\[
\sup_q|\mathfrak C_q(Z_{A^c})|\le\alpha.
\]
Therefore, from the independence of $Z_j$, $j\in A$, conditional on
$Z_{A^c}$, we conclude
\begin{equation}\label{eq:fA-factor-general-h-final}
f(A)
=\sum_{q\in\{1,\ldots,B-k+1\}^{A}}
\mathfrak C_q(Z_{A^c})
\prod_{j\in A}\Theta_j(C_{q_j}),
\end{equation}
where, for every deterministic lexicographic interval
$C\subseteq\mathcal Z_+$,
\[
\Theta_j(C):=
\mathbb E\left[\beta_j(M_j)\mathbf 1\{Z_j\in C\}\right].
\]
Conditional on $Z_{A^c}$, every $C_{q_j}$ is fixed, so this notation is
justified by independence. By the law of iterated expectations and
\eqref{eq:conditional-sign-given-Z-general-h-final},
\begin{align}
\Theta_j(C)
&=\mathbb E\left[S_j\mathbf 1\{Z_j\in C\}\right]\notag\\
&=\mathbb P\bigl((W_j,U_j)\in C\bigr)
-\mathbb P\bigl((-W_j,-U_j)\in C\bigr).
\label{eq:cell-expectation-general-h-final}
\end{align}
To see the second equality, note that
$Z_j=S_j(W_j,U_j)$ lies in the positive augmented half-line. Hence,
pointwise,
\[
S_j\mathbf 1\{Z_j\in C\}
=\mathbf 1\{(W_j,U_j)\in C\}
-\mathbf 1\{(-W_j,-U_j)\in C\}.
\]
For $v\in(-1,1)$, let
\[
I_v(C):=\{w\in\mathbb R:(w,v)\in C\}.
\]
Because $C$ is a lexicographic interval, $I_v(C)$ is an interval in
$\mathbb R$, possibly empty, unbounded, or a singleton. Since $U_j$ is
independent of $W_j$ and symmetric around zero,
\begin{align*}
&\mathbb P\bigl((W_j,U_j)\in C\bigr)
-\mathbb P\bigl((-W_j,-U_j)\in C\bigr)\\
&\qquad=\frac12\int_{-1}^1
\left[
\mathbb P\{W_j\in I_v(C)\}
-\mathbb P\{-W_j\in I_v(C)\}
\right]dv.
\end{align*}
Noting that
\[
\sup_{u\in\mathbb R}
|\mathbb P(W_j\le u)-\mathbb P(-W_j\le u)|
=
\sup_{u\in\mathbb R}
|\mathbb P(W_j<u)-\mathbb P(-W_j<u)|,
\]
and that the reflection discrepancy of any interval is at most twice the
reflection discrepancy of a half-line, we conclude that
\begin{equation}\label{eq:cell-bound-general-h-final}
|\Theta_j(C_{q_j})|\le\Delta_j,
\qquad j\in A.
\end{equation}
Therefore,
\begin{align}
|f(A)|
&\le\sum_{q\in\{1,\ldots,B-k+1\}^{A}}
|\mathfrak C_q(Z_{A^c})|
\prod_{j\in A}\Delta_j\notag\\
&\le\alpha(B-|A|+1)^{|A|}
\prod_{j\in A}\Delta_j.
\label{eq:fA-bound-general-h-final}
\end{align}
Substituting this in \eqref{eq:penultimate-general-h-final}, we obtain
\begin{equation}\label{eq:relative-polynomial-general-h-final}
\left|
\frac{\mathbb P(0\notin
\widehat{\mathrm{CI}}_{n,\alpha}^{\mathtt{SR}})}{\alpha}-1
\right|
\le \sum_{\substack{A\subseteq\{1,\ldots,B\}\\
|A|\ge2,\ |A|\ \mathrm{even}}}
(B-|A|+1)^{|A|}\prod_{j\in A}\Delta_j.
\end{equation}
To bound the right hand side, define
\[
e_r(\Delta)
=\sum_{\substack{A\subseteq\{1,\ldots,B\}\\|A|=r}}
\prod_{j\in A}\Delta_j.
\]
Using the inequality
\[
e_{2\ell}(\Delta)
\le\frac{2^\ell e_2^\ell(\Delta)}{(2\ell)!},
\]
we conclude
\begin{align*}
1+\sum_{\substack{A\subseteq\{1,\ldots,B\}\\
|A|\ge2,\ |A|\ \mathrm{even}}}
(B-|A|+1)^{|A|}\prod_{j\in A}\Delta_j
&\le1+\sum_{\ell\ge1}(B-1)^{2\ell}
\frac{2^\ell e_2^\ell(\Delta)}{(2\ell)!}\\
&=\cosh\left((B-1)\sqrt{2e_2(\Delta)}\right).
\end{align*}
Finally, the inequality $\cosh(x)\le\exp(x^2/2)$ yields
\begin{equation}\label{eq:Lambda-general-h-final}
1+\sum_{\substack{A\subseteq\{1,\ldots,B\}\\
|A|\ge2,\ |A|\ \mathrm{even}}}
(B-|A|+1)^{|A|}\prod_{j\in A}\Delta_j
\le\exp\left((B-1)^2
\sum_{1\le i<j\le B}\Delta_i\Delta_j\right).
\end{equation}
This implies the second inequality in the result. To prove the first inequality, note that~\eqref{eq:relative-polynomial-general-h-final} implies
\[
\frac{\mathbb P(0\notin
\widehat{\mathrm{CI}}_{n,\alpha}^{\mathtt{SR}})}{\alpha}-1 \ge \left(2 - \cosh\left((B-1)\sqrt{2e_2(\Delta)}\right)\right)_+.
\]
It is easy to verify that for all $x \ge 0$,
\[
(2 - \cosh(\sqrt{2x}))_+ \ge \exp\left(-\frac{x}{(1 - 7x/6)_+}\right),
\]
which implies the first inequality of the result. 
This completes the proof, assuming
\eqref{eq:cell-constancy-general-h-final}. We now prove
\eqref{eq:cell-constancy-general-h-final}. Fix any assignment
$q=(q_j)_{j\in A}\in\{1,\ldots,B-k+1\}^{A}$. The idea is that
$Z_j\in C_{q_j}$ for $j\in A$ implies fixed ranks for $Z_i$, $i\in A^c$,
while the ranks for $Z_j$, $j\in A$, are not fixed but are fixed up to
certain permutations. Define
\[
n_\ell(q):=|\{j\in A:q_j=\ell\}|,
\qquad
N_\ell(q):=\sum_{r=1}^{\ell}n_r(q),
\qquad N_0(q):=0.
\]
It follows that the rank of $Z_{(i):A^c}$, the $i$th complementary order
statistic, among $Z_1,\ldots,Z_B$ is $i+N_i(q)$. Hence, for a fixed $q$, the
ranks of the augmented magnitudes in $A^c$ are fixed. There are $n_\ell(q)$
observations from $A$ in cell $C_\ell$, and hence the set of ranks occupied by
the observations in $A$ is
\begin{equation}\label{eq:Rq-general-h-final}
\mathcal R(q)
=\bigcup_{\ell=1}^{B-k+1}
\{\ell+N_{\ell-1}(q),\ldots,\ell+N_\ell(q)-1\},
\end{equation}
where an empty set is omitted. Note that $|\mathcal R(q)|=|A|$. Hence, the
ranks of the $A$ observations are not fixed, but the set of ranks they can
take is fixed by $q$ alone. For every realization in this product cell, there
exists a unique bijection $\pi:A\to\mathcal R(q)$ such that
\[
R_j=\pi(j),\qquad j\in A.
\]
It now suffices to show that $a_A(Z)$ depends only on $q$ and $Z_{A^c}$ and
not on the bijection $\pi$, i.e., the ordering of the observations from $A$
within each cell does not matter for $a_A(Z)$. Recall that
\[
a_A(Z)
=\frac1{2^B}\sum_{s\in\{\pm1\}^B}
g_h\left(
\sum_{j=1}^B h(R_j(Z))\mathbf 1\{s_j=1\}
\right)\prod_{j\in A}s_j,
\]
where
\[
g_h(T)
=\mathbb E_{\kappa_{B,\alpha}^h}
\left[\mathbf 1\left\{
\left|\frac{N_h}{2}-T\right|
\ge\frac{N_h}{2}-\kappa_{B,\alpha}^h
\right\}\right].
\]
Hence, for fixed $s_{A^c}$, it suffices to show that, for any $\mathfrak K$,
\[
\sum_{s_A\in\{\pm1\}^k}
g_h\left(\mathfrak K+
\sum_{j\in A}h(R_j(Z))\mathbf 1\{s_j=1\}\right)
\prod_{j\in A}s_j
\]
depends only on $q$ and not on $\pi$. Observe that
\begin{align*}
&\sum_{s_A\in\{\pm1\}^k}
g_h\left(\mathfrak K+
\sum_{j\in A}h(R_j(Z))\mathbf 1\{s_j=1\}\right)
\prod_{j\in A}s_j\\
&\qquad=
\sum_{J\subseteq A}(-1)^{k-|J|}
g_h\left(\mathfrak K+
\sum_{j\in J}h(R_j(Z))\right)\\
&\qquad=
\sum_{H\subseteq\mathcal R(q)}(-1)^{k-|H|}
g_h\left(\mathfrak K+
\sum_{r\in H}h(r)\right).
\end{align*}
The final expression depends only on $q$ and the fixed complementary ranks,
and hence \eqref{eq:cell-constancy-general-h-final} follows.
\end{proof}

\begin{lemma}\label{lem:weighted-Walsh-average-transformed-rank-representation}
Consider the setting and notation of Section~\ref{subsec:symmetry}. 
Then, almost surely,
\begin{equation}\label{eq:weighted-Walsh-rank-sum-identity-final}
\sum_{1\le i\le j\le B}
\omega_{ij}\mathbf 1\{A_{ij}\succ(0,0)\}
=
\sum_{i=1}^B h(R_i)\mathbf 1\{S_i=1\}.
\end{equation}
\end{lemma}

\begin{proof}
Because $U_1,\ldots,U_B$ are independent and continuously distributed,
with probability one no augmented Walsh average $A_{ij}$ equals $(0,0)$.
Indeed, such an equality would require $U_{[i]}+U_{[j]}=0$, and there are
only finitely many index pairs. We work on this probability-one event.

Recall that
\[
(W_{[1]}, U_{[1]}) \prec (W_{[2]}, U_{[2]}) \prec \cdots \prec (W_{[B]}, U_{[B]}).
\]
Set
\[
S_{[j]} = \mathbf 1\{(W_{[j]}, U_{[j]})\succ(0,0)\}
     -\mathbf 1\{(W_{[j]}, U_{[j]})\prec(0,0)\},
\]
and let $R_{[j]}$ be the rank of the augmented magnitude $S_{[j]}(W_{[j]}, U_{[j]}) = (|W_{[j]}|, S_{[j]}U_{[j]})$ among
$S_{[i]}(W_{[i]}, U_{[i]}), 1\le i\le B$. Note that $(S_{[1]}, \ldots, S_{[B]})$ and $(R_{[1]}, \ldots, R_{[B]})$ is just a re-labeling of indices for vectors $(S_1, \ldots, S_B)$ and $(R_1, \ldots, R_B)$, respectively. Given that either side of~\eqref{eq:weighted-Walsh-rank-sum-identity-final} is permutation invariant, we get
\begin{align*}
\sum_{j=1}^B h(R_j)\mathbf 1\{S_j=1\} = \sum_{j=1}^B h(R_{[j]})\mathbf{1}\{S_{[j]} = 1\}.
\end{align*}
Hence, it suffices to prove that
\begin{equation}\label{eq:weighted-Walsh-rank-sum-ordered-final}
\sum_{1\le i\le j\le B} \omega_{ij}\mathbf{1}\{A_{ij} \succ (0,0)\} = \sum_{j=1}^B h(R_{[j]})\mathbf{1}\{S_{[j]} = 1\}.
\end{equation}
Fix $j\in\{1, 2, \ldots, B\}$. If $(W_{[j]}, U_{[j]})\prec(0,0)$, then $(W_{[i]}, U_{[i]})\preceq (W_{[j]}, U_{[j]})\prec(0,0)$ for every
$i\le j$, and hence $A_{ij}\prec(0,0)$ for every $i\le j$. Thus the
contribution of the $j$th column on the left-hand side of
\eqref{eq:weighted-Walsh-rank-sum-ordered-final} is zero, which agrees with
the contribution on the right-hand side.

Suppose now that $(W_{[j]}, U_{[j]})\succ(0,0)$. For $i<j$, translation invariance of the
lexicographic order gives
\[
A_{ij}\succ(0,0)
\quad\Longleftrightarrow\quad
(W_{[i]}, U_{[i]})\succ-(W_{[j]}, U_{[j]}).
\]
Because $(W_{[i]}, U_{[i]})\prec (W_{[j]}, U_{[j]})$ and $(W_{[j]}, U_{[j]})\succ(0,0)$, this is equivalent to
\[
(|W_{[i]}|, {S}_{[i]}U_{[i]}) \prec (W_{[j]}, U_{[j]}) = (|W_{[j]}|, {S}_{[j]}U_{[j]}).
\] 
Also,
$A_{jj}=(W_{[j]}, U_{[j]})\succ(0,0)$. If $k>j$, then
$(W_{[k]}, U_{[k]})\succ (W_{[j]}, U_{[j]})\succ(0,0)$, and hence,
$(|W_{[k]}|, S_{[k]}U_{[k]})=(W_{[k]}, U_{[k]})\succ (W_{[j]}, U_{[j]})$. It follows that
\begin{equation}\label{eq:number-positive-Walsh-column-final}
\#\left\{i\le j:A_{ij}\succ(0,0)\right\}=R_{[j]}.
\end{equation}
Moreover, $i\mapsto A_{ij}$ is increasing in lexicographic order.
Therefore, the set of $i$ satisfying $A_{ij}\succ (0, 0)$ is
$\{j-R_{[j]}+1,\ldots,j\}.$
Consequently,
\begin{align*}
&\sum_{i=1}^j
\{h(j-i+1)-h(j-i)\}
\mathbf 1\left\{A_{ij}\succ(0,0)\right\}\\
&\qquad=
\sum_{i=j-R_{[j]}+1}^j\{h(j-i+1)-h(j-i)\}
=
\sum_{q=1}^{R_{[j]}}\{h(q)-h(q-1)\}
=h(R_{[j]}).
\end{align*}
Summing over $j=1,\ldots,B$ proves
\eqref{eq:weighted-Walsh-rank-sum-ordered-final} and hence
\eqref{eq:weighted-Walsh-rank-sum-identity-final}. 
\end{proof}
\section{Proof of~\eqref{eq:generalized-Andel-inequality} and~\eqref{eq:unimodality-implication}}\label{appsec:proof-of-generalized-Andel}
We first prove~\eqref{eq:generalized-Andel-inequality}.
Set $t_{\gamma}(a) = |a|(2/\gamma - 1)/(2/\gamma - 2).$ Note that
\begin{align*}
\mathbb{P}(0 \notin [W \pm (2/\gamma - 1)|W - a|]) &= \mathbb{P}(|W| > (2/\gamma - 1)|W - a|)\\
&= \begin{cases}
    \mathbb{P}((1-\gamma)t_{\gamma}(a) < W < t_{\gamma}(a)), &\mbox{if }a \ge 0,\\
    \mathbb{P}((1-\gamma)t_{\gamma}(a) < -W < -t_{\gamma}(a)), &\mbox{if }a < 0.
\end{cases}
\end{align*}
This implies that
\[
\mathbb{P}((1-\gamma)t_{\gamma}(a) < W < t_{\gamma}(a)) = \mathbb{P}(W > 0, \ln(1-\gamma) + \ln(t_{\gamma}(a)) < \ln(W) < \ln t_{\gamma}(a)).
\]
Therefore, the supremum over all $a\in[0,\infty)$ is given by
\[
\sup_{t\in\mathbb{R}}\,\mathbb{P}(W > 0, \ln(W) \in (t, t - \ln(1-\gamma))).
\]
Similarly, the supremum over all $a\in(-\infty, 0)$ is given by
\[
\sup_{t\in\mathbb{R}}\,\mathbb{P}(W < 0,\, \ln(-W) \in (t, t - \ln(1-\gamma))).
\]
Hence, taking the maximum over these two cases completes the proof of~\eqref{eq:generalized-Andel-inequality}.

To prove~\eqref{eq:unimodality-implication} when $W$ is continuous and unimodal, note that
\[
\mathbb{P}(W > 0,\, \ln(W) \in (t, t+ \varepsilon)) = \mathbb{P}(W \le e^{\varepsilon}e^{t}) - \mathbb{P}(W \le e^{t}).
\]
Here we do not distinguish strict and non-strict inequalities because $W$ is assumed continuous. Unimodality of $W$ at zero implies that the distribution function is concave on $[0,\infty)$, and hence, 
\[
\frac{F(e^te^{\varepsilon}) - F(e^t)}{e^t(e^{\varepsilon}  - 1)} \le \frac{F(e^{t}e^{\varepsilon}) - F(0)}{e^te^{\varepsilon}} \le \frac{1}{e^te^{\varepsilon}}.
\]
Therefore, 
\[
\mathbb{P}(W > 0,\, \ln(W) \in (t, t+ \varepsilon)) \le \frac{e^{\varepsilon} - 1}{e^\varepsilon} = 1 - e^{-\varepsilon}.
\]
Replacing $W$ with $-W$ implies the result for continuous unimodal distributions.
\section{Proof of Theorem~\ref{thm:finite-sample-validity}}\label{appsec:proof-of-finite-sample-validity}
    
    By definition, $C_j(w_1, \ldots, w_{j-1}, w_{j+1}, \ldots, w_B) \in \mathcal{C}_j$ for $1 \le j\le B$. 
    From definition~\eqref{def:dist-scale-invariance}, it follows that for any $\varepsilon > 0$, there exists a random variable $L_{\varepsilon}$ such that
    \[
    \mathbb{P}(W \in C) \le \mathbb{P}(L_{\varepsilon}\in C) + (1 + \varepsilon)\mathrm{dist}_{\mathcal{C}}(W, \mathfrak{P}).
    \]
    Note that the distribution of the random variable $L_{\varepsilon}$ is independent of $C$ but can depend on $\mathcal{C}$. Setting $W_j = \widehat{\theta}_j - \theta(P)$ for $1 \le j\le B$, we denote by $L_{j,\varepsilon}$ a random variable satisfying
    \begin{equation}\label{eq:C_j-convergence}
    \mathbb{P}(W_j \in C) \le \mathbb{P}(L_{j,\varepsilon}\in C) + (1 + \varepsilon)\mbox{dist}_{\mathcal{C}_j}(\widehat{\theta}_j - \theta(P), \mathfrak{P})\quad\mbox{for all}\quad C\in\mathcal{C}_j.
    \end{equation}
    Observe now that
    \begin{align*}
        &\mathbb{P}_P\left(\theta(P)\notin\widehat{\mathrm{CI}}_{n,\alpha}^{\mathtt{COSI}}\right)\\ &= \mathbb{P}_P\left(\phi_{\mathfrak{P},\alpha}(W_1, \ldots, W_B) = 1\right)\\
        &= \mathbb{E}_P\left[\mathbb{P}_P\left(W_1 \in C_1(W_2, \ldots, W_B)\big|W_2, \ldots, W_B\right)\right]\\
        &\le \mathbb{E}_P\left[\mathbb{P}_P\left(L_{1,\varepsilon} \in C_1(W_2, \ldots, W_B)\big|W_2, \ldots, W_B\right)\right] + (1 + \varepsilon)\mbox{dist}_{\mathcal{C}_1}(\widehat{\theta}_1 - \theta(P), \mathfrak{P})\\
        &= \mathbb{P}_P\left(\phi_{\mathfrak{P},\alpha}(L_{1,\varepsilon}, W_2, \ldots, W_B) = 1\right) + (1 + \varepsilon)\mbox{dist}_{\mathcal{C}_1}(\widehat{\theta}_1 - \theta(P), \mathfrak{P})\\
        &= \mathbb{E}_P\left[\mathbb{P}_P(W_2 \in C_1(L_{1,\varepsilon}, W_3, \ldots, W_B)|L_{1,\varepsilon}, W_3, \ldots, W_B)\right] + (1 + \varepsilon)\mbox{dist}_{\mathcal{C}_1}(\widehat{\theta}_1 - \theta(P), \mathfrak{P})\\
        &\le \mathbb{E}_P\left[\mathbb{P}_P(L_{2,\varepsilon} \in C_1(L_{1,\varepsilon}, W_3, \ldots, W_B)|L_{1,\varepsilon}, W_3, \ldots, W_B)\right] + (1 + \varepsilon)\sum_{j=1}^2 \mbox{dist}_{\mathcal{C}_j}(\widehat{\theta}_j - \theta(P), \mathfrak{P})\\
        &= \mathbb{P}_P\left(\phi_{\mathfrak{P},\alpha}(L_{1,\varepsilon}, L_{2,\varepsilon}, W_3, \ldots, W_B) = 1\right) + (1 + \varepsilon)\sum_{j=1}^2 \mbox{dist}_{\mathcal{C}_j}(\widehat{\theta}_j - \theta(P), \mathfrak{P})\\
        &\le \mathbb{P}_P\left(\phi_{\mathfrak{P},\alpha}(\{L_{j,\varepsilon}\}_{j=1}^B) = 1\right) + (1+\varepsilon)\sum_{j=1}^B \mbox{dist}_{\mathcal{C}_j}(\widehat{\theta}_j - \theta(P), \mathfrak{P})\\
        &\le \alpha + (1 + \varepsilon)\sum_{j=1}^B \mbox{dist}_{\mathcal{C}_j}(\widehat{\theta}_j - \theta(P), \mathfrak{P}).
    \end{align*}
    Because $\varepsilon > 0$ is arbitrary, letting it tend to zero yields the result. Finally,~\eqref{eq:finite-sample-multiplicative} follows from Proposition~\ref{prop:poisson-binomial-to-binomial}. 
\section{Proof of Proposition~\ref{prop:dist-charact-zero-median}}\label{appsec:proof-dist-charact-zero-median}
    Observe that for $\mathcal{C}$, the collection of all Borel subsets of $\mathbb{R}$, the distance is the same as the total variation distance. 
    First, note that if the right hand side of~\eqref{eq:distance-zero-median} is zero, then $W$ itself has zero median and hence, the left hand side of~\eqref{eq:distance-zero-median} is also zero. To deal with general random variables $W$, assume that $\mathbb{P}(W \ge 0) \ge \mathbb{P}(W \le 0)$ so that the right hand side is $\delta = 1/2 - \mathbb{P}(W \le 0) = \mathbb{P}(W > 0) - 1/2$. Define a new random variable $W'$ based on $W$ as follows: Set $U\sim\mbox{Uniform}[0, 1]$ and set
    \[
    W' = \begin{cases}
    W, &\mbox{if }W \le 0,\\
    W, &\mbox{if }W > 0, U \le q,\\
    -W, &\mbox{if }W> 0, U > q.
    \end{cases}
    \]
    It is easy to see that
    \begin{align*}
    \mathbb{P}(W' \ge 0) &= \mathbb{P}(W = 0) + \mathbb{P}(W > 0)q,\\
    \mathbb{P}(W' \le 0) &= \mathbb{P}(W \le 0) + \mathbb{P}(W > 0)(1-q),\\
    \mathbb{P}(W \neq W') &= \mathbb{P}(W > 0)(1-q).
    \end{align*}
    Taking $q$ such that $1 - q = \delta/\mathbb{P}(W > 0) = \delta/(1/2 + \delta)$, we get
    \begin{align*}
        \mathbb{P}(W' \ge 0) &= \mathbb{P}(W = 0) + \frac{1/2}{1/2 + \delta}\mathbb{P}(W > 0)\\ 
        &= \mathbb{P}(W = 0) + 1/2 \ge 1/2,\\
        \mathbb{P}(W' \le 0) &= \mathbb{P}(W \le 0) + \delta = 1/2,\\
        \mathbb{P}(W \neq W') &= \delta.
    \end{align*}
    Hence, $W'$ is a zero median random variable and the total variation distance between $W$ and $W'$ is equal to $\delta$. This proves that the distance is at most the right hand side of~\eqref{eq:distance-zero-median}. To prove the other direction, note that for any random variable $L\in\mathfrak{P}$, we have
    \begin{align*}
    \mathbb{P}(W \le 0) &\ge \mathbb{P}(L \le 0) - \mathrm{TV}(W, L) \ge 1/2 - \mathrm{TV}(W, L),\\
    \mathbb{P}(W \ge 0) &\ge \mathbb{P}(L \ge 0) - \mathrm{TV}(W, L) \ge 1/2 - \mathrm{TV}(W, L).
    \end{align*}
    This implies
    \[
    \mathrm{TV}(W, L) \ge \left(\frac{1}{2} - \min\{\mathbb{P}(W \ge 0),\,\mathbb{P}(W \le 0)\}\right)_+.
    \]
    Hence, the result follows.
    \section{Proof of Theorem~\ref{thm:cosi-hunt-and-test}}\label{appsec:proof-hunt-and-test}
The first miscoverage bound follows from an application of~\ref{cond:approximate-HT} conditional on $\vartheta$. The second miscoverage bound follows from the fact that under mutual independence the conditional distributions of $\widehat{\theta}_j - \theta(P)$ given $\vartheta$ do not depend on $\vartheta$.

For the diameter analysis, note that $\theta\in\cosici{n}{\alpha}{\mathtt{ht}}$ implies
\[
\widehat{L}_{\alpha}(\{\ell_{\vartheta - \theta}(\widehat{\theta}_j - \theta)\}_{j=1}^B) \le 0.
\]
Hence,
\begin{align*}
\ell_{\vartheta - \theta}(\vartheta - \theta) &\le \ell_{\vartheta - \theta}(\vartheta - \theta) - \widehat{L}_{\alpha}(\{\ell_{\vartheta - \theta}(\widehat{\theta}_j - \theta)\}_{j=1}^B)\\
&\le \sup_{\ell\in\mathcal{L}}\, \ell(\vartheta - \theta) - \widehat{L}_{\alpha}(\{\ell(\widehat{\theta}_j - \theta)\}_{j=1}^B).
\end{align*}
From~\eqref{eq:xi-maximizer-ell}, $\rho(\vartheta - \theta) \le 2\ell_{\vartheta - \theta}(\vartheta - \theta)$, and hence, the first diameter bound follows by taking the supremum over all $\theta\in\cosici{n}{\alpha}{\mathtt{ht}}$.

If the additional assumption of shift-invariance holds, then
\[
\sup_{\theta\in\cosici{n}{\alpha}{\mathtt{ht}}}\, \rho(\vartheta - \theta) ~\le~ 2\sup_{\theta\in\cosici{n}{\alpha}{\mathtt{ht}}}\,\sup_{\ell\in\mathcal{L}}\, -\widehat{L}_{\alpha}\left(\{\ell(\widehat{\theta}_j - \theta) - \ell(\vartheta - \theta)\}_{j=1}^B\right).
\]
Finally, if every $\ell\in\mathcal{L}$ is linear, then
\begin{align*}
\sup_{\theta\in\cosici{n}{\alpha}{\mathtt{ht}}}\, \rho(\vartheta - \theta) ~&\le~ 2\sup_{\theta\in\cosici{n}{\alpha}{\mathtt{ht}}}\,\sup_{\ell\in\mathcal{L}}\, -\widehat{L}_{\alpha}\left(\{\ell(\widehat{\theta}_j - \vartheta)\}_{j=1}^B\right)\\
~&= 2\sup_{\ell\in\mathcal{L}}\, -\widehat{L}_{\alpha}\left(\{\ell(\widehat{\theta}_j - \vartheta)\}_{j=1}^B\right).
\end{align*}
This completes the proof.

\section{Proof of Theorem~\ref{thm:coverage-of-cosi-m+-diameter-analysis}}\label{appsec:proof-of-cosi-m-diameter}
    Let $t_{m}\in\mathcal{T}, m \ge 1$ be a sequence such that
    \begin{equation}\label{eq:almost-maximizer}
    \delta_{t_{m}}(\vartheta - \theta(P)) \le \delta_{t_{m+1}}(\vartheta - \theta(P))\quad\mbox{and}\quad \delta_{t_{m}}(\vartheta - \theta(P)) \le \rho(\vartheta - \theta(P)) \le \delta_{t_m}(\vartheta - \theta(P)) + \frac{1}{m}.
    \end{equation}
    As we condition on $\vartheta$, $t_m$ is a fixed quantity. For $1\le j\le B$, define
    \begin{align*}
    \mathcal{A}_{m} &:= \left\{\delta_{t_m}(\vartheta - \theta(P)) \le \sup_{t\in\mathcal{T}}(\delta_t(\vartheta - \widehat{\theta}))_{(c^*_{B,\alpha}+1)}\right\},\\
    \mathcal{B}_{m} &:= \left\{\delta_{t_m}(\vartheta - \theta(P)) \le (\delta_{t_m}(\vartheta - \widehat{\theta}))_{(c^*_{B,\alpha}+1)}\right\}.
    \end{align*}
    From~\eqref{eq:almost-maximizer} and~\eqref{eq:quantile-and-supremum-swap},
    \[
    \mathcal{A}_{m+1} \subseteq \mathcal{A}_m,\quad \bigcap_{m=1}^{\infty} \mathcal{A}_m = \cosici{n}{\alpha}{\mp},\quad\mbox{and}\quad \mathcal{A}_{m} \supseteq \mathcal{B}_m\quad\mbox{for all}\quad m \ge 1.
    \]
    Therefore, 
    \begin{align*}
    \mathbb{P}(\theta(P)\in\cosici{n}{\alpha}{\mp}\big|\vartheta) &= \lim_{m\to\infty}\,\mathbb{P}(\mathcal{A}_{m}\big|\vartheta) \ge \lim_{m\to\infty}\,\mathbb{P}(\mathcal{B}_{m}\big|\vartheta).
    \end{align*}
    This implies
    \[
    \mathbb{P}(\theta(P)\notin \cosici{n}{\alpha}{\mp}\big|\vartheta) \le \lim_{m\to\infty}\,\mathbb{P}(\mathcal{B}_m^c\big|\vartheta).
    \]
    Observe that
    \begin{align*}
    \mathbb{P}(\mathcal{B}_m^c\big|\vartheta) &= \mathbb{P}\left(\delta_{t_m}(\vartheta - \theta(P)) > (\delta_{t_m}(\vartheta - \widehat{\theta}))_{(c^*_{B,\alpha}+1)}\big|\vartheta\right)\\
    &\overset{(i)}{=} \mathbb{P}\left(\delta_{t_m}(-\theta(P)) > (\delta_{t_m}(-\widehat{\theta}))_{(c^*_{B,\alpha}+1)}\big|\vartheta\right)\\
    &\overset{(ii)}{=} \mathbb{P}\left((\delta_{t_m}(\widehat{\theta} - \theta(P)))_{(B-c^*_{B,\alpha})} > 0\big|\vartheta\right)\\
    &\overset{(iii)}{\le} \alpha\exp\left(2B\Delta_P^{\mathfrak{m}}\right)
    \end{align*}
    where $(i)$ and $(ii)$ follow from the linearity of $\delta_t(\cdot)$, while $(iii)$ follows from~\eqref{eq:one-sided-median-hulc} (and the independence of $\vartheta$ and $(\widehat{\theta}_j: 1\le j\le B)$. This completes the proof of miscoverage for $\cosi{\mathfrak{m}+}$ confidence set.

    To prove the diameter bound for $\cosi{\mpp}$ confidence set, we note that
    \[
    \sup_{\theta\in\cosici{n}{\alpha}{\mpp}}\,\rho(\vartheta - \theta) \le \rho(\vartheta - \theta(P)) + (\rho(\theta(P) - \widehat{\theta}))_{(c^*_{B,\alpha} + 1)}.
    \]
    Thus, it suffices to control the tail of the second term.
    For any $\gamma \in (0, 1)$, we find an $\eta\in(0, 1)$ such that 
    \[
    \mathbb{P}((\rho(\theta(P) - \widehat{\theta}))_{(c^*_{B,\alpha}+1)} > Q(\eta)) \le \gamma,
    \]
    which proves the result.
    Observe that
    \begin{align*}
        \mathbb{P}((\rho(\theta(P) - \widehat{\theta}))_{(c^*_{B,\alpha}+1)} > Q(\eta)) &= \mathbb{P}\left(\sum_{j=1}^B \mathbf{1}\{\rho(\theta(P) - \widehat{\theta}_j) > Q(\eta)\} \ge B - c^*_{B,\alpha}\right).
    \end{align*}
    From the hypothesis, we know
    \[
    \mathbb{E}\left[\mathbf{1}\{\rho(\theta(P) - \widehat{\theta}_j) > Q(\eta)\}\right] \le \eta\quad\Rightarrow\quad \mathbf{1}\{\rho(\theta(P) - \widehat{\theta}_j) > Q(\eta)\} \preceq \mbox{Bernoulli}(\eta).
    \]
    This implies that
    \begin{align*}
    \mathbb{P}_P((\rho(\theta(P) - \widehat{\theta}))_{(c^*_{B,\alpha}+1)} > Q(\eta)) &\le \mathbb{P}(\mbox{Binom}(B, \eta) \ge B - c^*_{B,\alpha})\\
    &\le \binom{B}{B - c^*_{B,\alpha}}\eta^{B - c^*_{B,\alpha}}\\
    &\le \left(\frac{eB}{B - c^*_{B,\alpha}}\right)^{B - c^*_{B,\alpha}}\eta^{B-c^*_{B,\alpha}}.
    \end{align*}
    Therefore, taking
    \[
    \eta = \gamma^{1/(B-c^*_{B,\alpha})}\left(\frac{B - c^*_{B,\alpha}}{eB}\right),
    \]
    suffices to ensure that the probability is less than $\gamma$,
    which completes the proof of~\eqref{eq:diameter-of-cosi-m+}. Inequality~\eqref{eq:diameter-of-cosi-m} follows readily from using~\eqref{eq:first-diameter-bound-cosi-m} and the $T_2$ bound in Theorem~\ref{thm:supremum-of-quantiles}.

\subsection{A General Result}\label{sec:rectilinear}
Recall that a random element $W\in\mathbb{R}^{\mathcal{T}}$ is said to have a zero rectilinear median if 
\begin{equation}\label{eq:rectilinear-median-defi}
\min\left\{\mathbb{P}\left(W(t) \ge 0\right),\, \mathbb{P}(W(t) \le 0)\right\} \ge \frac{1}{2}\quad\mbox{for all}\quad t\in\mathcal{T}.
\end{equation}
For $t\in\mathcal{T}$, let $q_t:\mathbb{R}\to\mathbb{R}$ be a non-negative quasiconvex function with $q_t(0) = 0$. This implies that $q_t(\cdot)$ is non-decreasing on $[0,\infty)$ and non-increasing on $(-\infty, 0]$. (Some examples include $q_t(u) = |u|, (u)_+, (u)_-$.) Define a ``pseudo-distance'' on $\mathbb{R}^{\mathcal{T}}$ as 
\[
\|a - b\|_{q,\mathcal{T}} = \sup_{t\in\mathcal{T}}\,q_t(a(t) - b(t)),\quad\mbox{for any}\quad a, b\in\mathbb{R}^{\mathcal{T}}.
\]
Unless $q_t(\cdot)$ is a norm on $\mathbb{R}$, $\|\cdot\|$ is not a norm. In particular, $\|a\|_{q,\mathcal{T}}$ may not equal $\|-a\|_{q,\mathcal{T}}.$ 
The following is a simple result that implies a test for zero rectilinear median bias.
\begin{lemma}\label{lem:rectilinear-median-bias}
    For any random variable $W\in\mathbb{R}^{\mathcal{T}}$ with a zero rectilinear median, 
\[
\mathbb{P}(\|-a\|_{q,\mathcal{T}} \le \|W - a\|_{q,\mathcal{T}}) \ge \frac{1}{2}\quad\mbox{for any}\quad a\in\mathbb{R}^{\mathcal{T}}.
\]
More generally, for any random variable $W\in\mathbb{R}^{\mathcal{T}}$, we have
\begin{equation}\label{eq:RMB-1/2-prob}
\mathbb{P}(\|-a\|_{q,\mathcal{T}} \le \|W - a\|_{q,\mathcal{T}}) \ge \frac{1}{2} - \mathrm{RMB}(W),\quad\mbox{for all}\quad a\in\mathbb{R}^{\mathcal{T}},
\end{equation}
where $\mathrm{RMB}(W)$ represents the rectilinear median bias of $W$ defined as
\[
\mathrm{RMB}(W) := \sup_{t\in\mathcal{T}}\left(\frac{1}{2} - \min\{\mathbb{P}(W(t) \ge 0),\,\mathbb{P}(W(t) \le 0)\}\right)_+.
\]
\end{lemma}
First part of Lemma~\ref{lem:rectilinear-median-bias} can also be derived from Theorem 3.1 of~\cite{merkle2010jensen} and Example 2.3.2 of~\cite{small1987measures}. Theorem 3.1 of~\cite{merkle2010jensen} gives a more formal connection between prediction sets and our $\cosi{}$ confidence sets. Inequality~\eqref{eq:RMB-1/2-prob} is new to our knowledge.
\begin{proof}[Proof of Lemma~\ref{lem:rectilinear-median-bias}]
    We shall prove~\eqref{eq:RMB-1/2-prob} which proves the first result. Fix any $a\in\mathbb{R}^{\mathcal{T}}$. For any $m \ge 1$, let $t_m\in\mathcal{T}$ be such that
    \[
    q_{t_m}(-a(t_m)) \le \|-a\|_{q,\mathcal{T}} \le q_{t_m}(-a(t_m)) + \frac{1}{m}.
    \]
    Such a $t_m$ must exist by the definition of $\|-a\|_{q,\mathcal{T}}$. Note that non-stochasticity of $a$ implies the same for $t_m$ for any $m\ge1$. Define
    \[
    \mathcal{A}_m = \{q_{t_m}(-a(t_m)) \le \|W - a\|_{q,\mathcal{T}}\}\quad\mbox{and}\quad \mathcal{B}_m = \{q_{t_m}(-a(t_m)) \le q_{t_m}(W(t_m) - a(t_m))\}.
    \]
    As $m\uparrow\infty$, $\mathcal{A}_m$ decreases, i.e., $\mathcal{A}_{m+1}\subseteq\mathcal{A}_{m}$, and additionally, we have
    \[
    \bigcap_{m=1}^{\infty} \mathcal{A}_m = \{\|-a\|_{q,\mathcal{T}} \le \|W - a\|_{q,\mathcal{T}}\}.
    \]
    Also, note that $\mathcal{A}_m \supseteq \mathcal{B}_m$. Therefore, 
    \[
    \mathbb{P}(\|-a\|_{q,\mathcal{T}} \le \|W - a\|_{q,\mathcal{T}}) ~=~ \lim_{m\to\infty}\,\mathbb{P}(\mathcal{A}_m) ~\ge~ \lim_{m\to\infty}\,\mathbb{P}(\mathcal{B}_m).
    \]
    If $a(t_m) \ge 0$, on the event $W(t_m) \le 0$, $W(t_m) - a(t_m) \le -a(t_m) \le 0$. Non-increasing property of $q_{t_m}(\cdot)$ on $(-\infty, 0]$ implies $q_{t_m}(-a(t_m)) \le q_{t_m}(W(t_m) - a(t_m))$. Similarly, if $a(t_m) \le 0$, then on the event $W(t_m) \ge 0$, $W(t_m) - a(t_m) \ge -a(t_m) \ge 0$. Non-decreasing property of $q_{t_m}(\cdot)$ on $[0,\infty)$ implies $\mathcal{B}_m$ occurs. Therefore, we get
    \begin{equation}\label{eq:B-m-C_m}
        \mathbf{1}\{q_{t_m}(-a(t_m)) \le q_{t_m}(W(t_m) - a(t_m))\} \ge \mathbf{1}\{\mbox{sign}(a(t_m))W(t_m) \le 0\},
    \end{equation}
    and hence,
    \begin{align*}
    \mathbb{P}(\mathcal{A}_m) \ge \mathbb{P}(\mathcal{B}_m) &=\mathbb{P}(q_{t_m}(-a(t_m)) \le q_{t_m}(W(t_m) - a(t_m)))\\ &\ge \mathbb{P}(\mbox{sign}(a(t_m))W(t_m) \le 0)\\
    &\ge \frac{1}{2} - \left(\frac{1}{2} - \min\{\mathbb{P}(W(t_m) \le 0),\,\mathbb{P}(W(t_m) \ge 0)\}\right)_+\\
    &\ge \frac{1}{2} - \mathrm{RMB}(W).
    \end{align*}
    Hence, the result follows by taking the limit as $m\uparrow\infty$.
\end{proof}
Using Lemma~\ref{lem:rectilinear-median-bias}, with $\mathfrak{P}$ representing the zero rectilinear median bias class, we have the test
\[
\phi_{\mathfrak{P},1/2}(W) = \left\{\|-a\|_{q,\mathcal{T}} > \|W - a\|_{q,\mathcal{T}}\right\},
\]
for any arbitrary $a\in\mathbb{R}^{\mathcal{T}}$. Hence, one can define an instantiation of the $\cosi{m}$ confidence set, by taking $a = \vartheta - \theta$ and $W_j = \widehat{\theta}_j - \theta$, as
\begin{equation}\label{eq:COSI-RM-conf-set}
\begin{split}
    \cosici{n}{\alpha}{\rmpp} &:= \left\{\theta\in\mathbb{R}^{\mathcal{T}}:\,\sum_{i=1}^B \mathbf{1}\left\{\|\theta - \vartheta\|_{q,\mathcal{T}} \le \|\widehat{\theta}_j - \vartheta\|_{q,\mathcal{T}}\right\} \ge B - c^*_{B,\alpha}\right\}\\
    &= \left\{\theta\in\mathbb{R}^{\mathcal{T}}:\, \|\theta - \vartheta\|_{q,\mathcal{T}} \le \|\widehat{\theta} - \vartheta\|_{q,\mathcal{T},(c^*_{B,\alpha}+1)}\right\},
\end{split}
\end{equation}
where $\|\widehat{\theta} - \vartheta\|_{q,\mathcal{T},(c)}$ represents the $c$-th smallest value among $\|\widehat{\theta}_j - \vartheta\|_{q,\mathcal{T}}, 1\le j\le B$. 
Note that every choice of $q(\cdot)$ yields a confidence set and it can different for different choices. For example,
\begin{enumerate}
    \item If $q(u) = (u)_+$, then
    \begin{align*}
    \cosici{n}{\alpha}{\rmpp} &:= \bigotimes_{t\in\mathcal{T}}\left(-\infty,\, \vartheta(t) + \|\widehat{\theta} - \vartheta\|_{q,\mathcal{T},(c^*_{B,\alpha}+1)}\right].
    \end{align*}
    \item If $q(u) = (u)_-$, then
    \begin{align*}
    \cosici{n}{\alpha}{\rmpp} &:= \bigotimes_{t\in\mathcal{T}}\left[\vartheta(t) - \|\widehat{\theta} - \vartheta\|_{q,\mathcal{T},(c^*_{B,\alpha}+1)},\,\infty\right).
    \end{align*}
    \item If $q(u) = |u|$, then 
    \begin{align*}
    \cosici{n}{\alpha}{\rmpp} &:= \bigotimes_{t\in\mathcal{T}}\left[\vartheta(t) - \|\widehat{\theta} - \vartheta\|_{q,\mathcal{T},(c^*_{B,\alpha}+1)},\,\vartheta(t) + \|\widehat{\theta} - \vartheta\|_{q,\mathcal{T},(c^*_{B,\alpha}+1)}\right].
    \end{align*}
\end{enumerate}
Hence, the choice of $q(\cdot)$ allows for directional confidence sets. An interesting choice of $q_t(\cdot)$ that allows for ``adaptivity'' to marginal variances is $q_t(u) = {|u|}/{\widehat{\sigma}_t}$ for some estimator $\widehat{\sigma}_t$ of the standard deviation of $\widehat{\theta}_j(t) - \theta(P)(t)$. Unfortunately, the proof breaks down if the data used to compute $\widehat{\theta}_j$ is used again to compute $\widehat{\sigma}_t$ and so, we recommend estimate $\widehat{\sigma}_t$ based on the same data used to compute $\vartheta$. Additionally, even if $\vartheta$ or $\widehat{\theta}_j$ are the same average, it is not recommended to take $\widehat{\sigma}_t$ to be the sample standard deviation because otherwise $\|\widehat{\theta}_j - \vartheta\|_{q,\mathcal{T}}$ becomes infinite almost sure in infinite-dimensional examples; see, for example,~\cite{eicker1979asymptotic} and~\citet[Eq. (1.7)]{datta2018optimalinferencemultidimensionalmultiscale}. A simple (problem-independent) solution would be to replace ``usual standard deviation estimator'' $\widehat{\sigma}_t$ with $\sqrt{\widehat{\sigma}_t^2 + \lambda^2}$ for some $\lambda > 0$, or with some structure on the problem, one could explore multi-scale ideas as in~\cite{datta2018optimalinferencemultidimensionalmultiscale}.

The confidence set~\eqref{eq:COSI-RM-conf-set} can be improved as
\begin{equation}
    \cosici{n}{\alpha}{\rmp} := \left\{\theta\in\mathbb{R}^{\mathcal{T}}:\, \|\theta - \vartheta\|_{q,\mathcal{T}} \le \sup_{t\in\mathcal{T}}(q_t(\widehat{\theta}_{\cdot}(t) - \vartheta(t)))_{(c^*_{B,\alpha}+1)}\right\}.
\end{equation}
The following result proves a miscoverage bound analogues to Theorem~\ref{thm:coverage-of-cosi-m+-diameter-analysis}. 
\begin{theorem}\label{thm:cosi-rm}
    For any $\mu\in\mathbb{R}^\mathcal{T}$, we have
    \[
    \mathbb{P}(\mu\notin\cosici{n}{\alpha}{\rmp}) \le \alpha\exp(2B\Delta_{\mu}^{\mathfrak{m}}),
    \]
    where
    \[
    \Delta_{\mu}^{\mathfrak{m}} := \sup_{t\in\mathcal{T}}\,\frac{1}{B}\sum_{j=1}^B \left(\frac{1}{2} - \min\left\{\mathbb{P}(\widehat{\theta}_j(t) \ge \mu(t)),\,\mathbb{P}(\widehat{\theta}_j(t) \le \mu(t))\right\}\right)_+.
    \]
\end{theorem}
\begin{proof}[Proof of Theorem~\ref{thm:cosi-rm}]
    Throughout the proof, we consider $\vartheta$ as non-stochastic (or alternatively, one can consider all probabilities below as conditional on $\vartheta$). As in the proof of Lemma~\ref{lem:rectilinear-median-bias}, for any $m\ge1$, let $t_m\in\mathcal{T}$ be such that
    \[
    q_{t_m}(\mu(t_m) - \vartheta(t_m)) \le \|\mu - \vartheta\|_{q,\mathcal{T}} \le q_{t_m}(\mu(t_m) - \vartheta(t_m)) + \frac{1}{m}. 
    \]
    Define
    \begin{align*}
    \mathcal{A}_{m} &= \left\{q_{t_m}(\mu(t_m) - \vartheta(t_m)) \le \sup_{t\in\mathcal{T}}(q_t(\widehat{\theta}_{\cdot}(t) - \vartheta(t)))_{(c^*_{B,\alpha}+1)}\right\},\\
    \mathcal{B}_m &= \left\{q_{t_m}(\mu(t_m) - \vartheta(t_m)) \le (q_{t_m}(\widehat{\theta}_{\cdot}(t_m) - \vartheta(t_m)))_{(c^*_{B,\alpha}+1)}\right\}.
    \end{align*}
    Note that $\mathcal{B}_m$ is equivalent to
    \[
    \mathcal{B}_m = \left\{q_{t_m}(\mu(t_m)) \le (q_{t_m}(\widehat{\theta}_{\cdot}(t_m)))_{(c^*_{B,\alpha}+1)}\right\},
    \]
    because quantiles are translation equivariant. As in proof of Lemma~\ref{lem:rectilinear-median-bias}, $\mathcal{A}_m, m\ge1$ are a decreasing sequence of sets and
    \[
    \bigcap_{m=1}^{\infty}\mathcal{A}_m := \{\mu\in\cosici{n}{\alpha}{\rmp}\}.
    \]
    Moreover, $\mathcal{A}_m \supseteq \mathcal{B}_m$. Therefore, by the monotone convergence theorem,
    \[
    \mathbb{P}(\mu\in\cosici{n}{\alpha}{\rmp}) = \lim_{m\to\infty}\mathbb{P}(\mathcal{A}_m) \ge \lim_{m\to\infty}\mathbb{P}(\mathcal{B}_m).
    \]
    Hence, it suffices to upper bound $\mathbb{P}(\mathcal{B}_m^c).$ From the definition of $\mathcal{B}_m$, we get
    \[
    \mathcal{B}_m = \left\{\sum_{j=1}^B \mathbf{1}\{q_{t_m}(\mu(t_m) - \vartheta(t_m)) \le q_{t_m}(\widehat{\theta}_j(t_m) - \vartheta(t_m))\} \ge B - c^*_{B,\alpha}\right\},
    \]
    which by~\eqref{eq:B-m-C_m} yields
    \[
    \mathcal{B}_m \supseteq \left\{\sum_{j=1}^B \mathbf{1}\{\mbox{sign}(\mu(t_m) - \vartheta(t_m))(\widehat{\theta}_j(t_m) - \vartheta(t_m)) \le 0\}\right\}.
    \]
    The probability of the complement of the set on the right follows from Proposition~\ref{prop:poisson-binomial-to-binomial} and completes the proof of the result.
\end{proof}
\section{Proof of Proposition~\ref{prop:prediction-to-confidence}}\label{appsec:proof-of-prediction-to-confidence}
    Let $W'$ be an independent copy of $W$. Clearly, $G(W')$ and $G(W)$ are IID real-valued random variables. Let $F(x) = \mathbb{P}(G(W) \le x)$ be the cumulative distribution function of $G(W)$. 
    Note that
    \[
    \mu(\widehat{C}) = \mathbb{P}(W\in\widehat{C}|\widehat{C}) = \mathbb{P}(G(W') \le G(W)|W) = F(G(W)).
    \]
    If $G(W)$ is continuously distributed, then $F(G(W))\sim \mbox{Uniform}[0, 1]$ and hence, $\mathbb{P}(\mu(\widehat{C}) > 1/2) = 1/2$. It is well-known that without continuous distribution, $F(G(W))$ is sub-uniform, i.e., $\mathbb{P}(F(G(W)) \le u) \le u$ for all $u \in [0, 1]$. For completeness, we give a short proof here. Set
    \[
    A_u = \{x:\, F(x) \le u\}\quad\mbox{and}\quad a = \sup_{b\in A_u}\,b.
    \]
    If $a\in A_u$, then $A_u = (-\infty, a]$ and hence
    \[
    \mathbb{P}(F(G(W)) \le u) = \mathbb{P}(G(W) \le a) = F(a) \le u,
    \]
    with the last inequality stemming from $a\in A_u.$ If $a\notin A_u$, then $A_u = (-\infty, a)$ and hence,
    \[
    \mathbb{P}(F(G(W)) \le u) = \mathbb{P}(G(W) < a) = \lim_{x\uparrow a}\,F(x) \le u,
    \]
    because $x < a$ implies $x\in A_u$.

    Therefore, 
    \[
    \mathbb{P}(\mu(\widehat{C}) > 1/2) = 1 - \mathbb{P}(\mu(\widehat{C}) \le 1/2) = 1 - \mathbb{P}(F(G(W)) \le 1/2) \ge \frac{1}{2}.
    \]
    Hence, if $\widehat{C}$ is a rectangle, then Theorem 3.1 of~\cite{merkle2010jensen} implies, on the event $\{\mu(\widehat{C}) > 1/2\}$, that $\mbox{Med}_{\mu}\subseteq\widehat{C}$. This completes the proof.
\section{Proof of Theorem~\ref{thm:median-unbiased-requirement}}\label{appsec:proof-of-median-unbiased-require}
The logic of the proof is precisely that of Theorem 1 of~\cite{kuchibhotla2023median}. Take integer sequence $m_n, n\ge 1$ such that $m_n \to \infty$ and $m_n/n \to 0$. Consider the partition of $\{1, 2, \ldots, n\}$ into $\widetilde{S}_j, 1\le j\le \lfloor n/m_n\rfloor$ as in~\eqref{eq:subsample-splits}. Compute the version of $\widetilde{\mathrm{CI}}_{m_n,\gamma}$ on each of $X_i, i\in\widetilde{S}_j$ for $1\le j\le \lfloor n/m_n\rfloor$. This yields $\lfloor n/m_n\rfloor$ confidence sets 
\[
\widetilde{\mathrm{CI}}_{m_n, \gamma}^{(j)},\quad 1\le j\le \lfloor n/m_n\rfloor.
\]
Set
\[
\widetilde{\mathrm{CI}}_{n} = \bigcup_{j=1}^{\lfloor n/m_n\rfloor}\, \widetilde{\mathrm{CI}}_{m_n,\gamma}^{(j)}.
\]
Because of independence of these confidence sets, we get
\[
\mathbb{P}_P\left(\theta(P)\notin\widetilde{\mathrm{CI}}_{n}\right) \le \prod_{j=1}^{\lfloor n/m_n\rfloor}\,\mathbb{P}_P(\theta(P)\notin\widetilde{\mathrm{CI}}_{m_n,\gamma}^{(j)})
\]
From the validity guarantee~\eqref{eq:asymptotic-validity-gamma}, we get
\[
\lim_{n\to\infty}\,\sup_{P\in\mathcal{P}}\,\mathbb{P}_P\left(\theta(P)\notin\widetilde{\mathrm{CI}}_n\right) = 0.
\]
For $t\in\mathcal{T}$, define
\[
L_n(t) := \inf_{\theta\in\widetilde{\mathrm{CI}}_n}\, \delta_t(\theta),\quad\mbox{and}\quad U_n(t) = \sup_{\theta\in\widetilde{\mathrm{CI}}_n}\,\delta_t(\theta).
\]
From the non-triviality assumption~\eqref{eq:boundedness} and that $\widetilde{\mathrm{CI}}_n$ is a countable union of confidence sets, we get
\[
\mathbb{P}\left(\sup_{t\in\mathcal{T}}\,U_n(t) < \infty\right) = 1.
\]
The same holds for $L_n(t)$ as well because $L_n(t) \le U_n(t)$. Without further assumptions, one cannot guarantee that $L_n(t) > -\infty$ with probability 1. But if $\sup_{t\in\mathcal{T}}\,\delta_t(\theta) = \sup_{t\in\mathcal{T}}\,\delta_t(-\theta)$, then $L_n(t)$ is almost surely finite. 

Let $R\sim \mbox{Ber}(1/2)$ be a random variable independent of $X_1, X_2, \ldots$. Define the estimator $\widehat{\theta}_{[n]}$ via
\[
\delta_t(\widehat{\theta}_{[n]}) := \begin{cases}
    L_n(t), &\mbox{if }R = 0,\\
    U_n(t), &\mbox{if }R = 1,
\end{cases}\quad\mbox{for all}\quad t\in\mathcal{T}.
\]
We claim that $\widehat{\theta}_{[n]}$ is asymptotically median unbiased. To see this, observe that
\begin{align*}
\mathbb{P}_P\left(\delta_t(\widehat{\theta}_{[n]} - \theta(P)) \le 0\right) &= \mathbb{P}_P\left(L_n(t) \le \delta_t(\theta(P))\right)\frac{1}{2} + \mathbb{P}_P\left(U_n(t) \le \delta_t(\theta(P))\right)\frac{1}{2}\\
&\ge \mathbb{P}_P\left(\theta(P)\in\widetilde{\mathrm{CI}}_n\right)\frac{1}{2} + 0,
\end{align*}
and
\begin{align*}
    \mathbb{P}_P\left(\delta_t(\widehat{\theta}_{[n]} - \theta(P)) \ge 0\right) &= \mathbb{P}_P\left(L_n(t) \ge \delta_t(\theta(P))\right)\frac{1}{2} + \mathbb{P}_P\left(U_n(t) \ge \delta_t(\theta(P))\right)\frac{1}{2}\\
&\ge 0 + \mathbb{P}_P\left(\theta(P)\in\widetilde{\mathrm{CI}}_n\right)\frac{1}{2},
\end{align*}
Therefore, 
\begin{align*}
&\sup_{P\in\mathcal{P}}\,\sup_{t\in\mathcal{T}}\,\left(\frac{1}{2} - \min\left\{\mathbb{P}_P(\delta_t(\widehat{\theta}_{[n]} - \theta(P)) \ge 0),\,\mathbb{P}_P(\delta_t(\widehat{\theta}_{[n]} - \theta(P))\ge0)\right\}\right)_+\\ 
&\le \sup_{P\in\mathcal{P}}\,\mathbb{P}(\theta(P)\notin\widetilde{\mathrm{CI}}_n) \to 0,
\end{align*}
as $n\to\infty$. This proves the result.

\section{Proof of~\eqref{eq:unimodal-HulC-guarantee}}\label{appsec:proof-unimodal-HulC}
Write
\[
\delta_j := \bigl(\lQ_{W_j}(\ln 2)-1/2\bigr)_+, \qquad c:=1+c^*_{B-1,\alpha/B}.
\]
We use the finite index range $1\le i\le B-c$ in the definition of the interval, with the interval defined to be $\mathbb{R}$ when $c=B$.

Note that
\begin{align*}
\{0\notin\widehat{\mathrm{CI}}_{n,\alpha}^{\mathfrak{u}^\star}\} &= \bigcup_{i=1}^{B-c}\{2W_{(i)}>W_{(i+c)}\} \;\cup\; \bigcup_{i=1}^{B-c}\{2W_{(i+c)}<W_{(i)}\}.
\end{align*}
We bound the probability of the first union, and the second follows
by considering the $-W_j$'s.

Observe that $2W_{(i)}>W_{(i+c)}$ implies $W_{(i)}>0$.
Moreover, the $c+1$ observations in this consecutive cluster
all lie in $[W_{(i)},2W_{(i)})$. Therefore,
\begin{align*}
\bigcup_{i=1}^{B-c}\{2W_{(i)}>W_{(i+c)}\} &\subseteq
\bigcup_{k=1}^B \left\{W_k>0,\ \mbox{and}\ \sum_{j\ne k}\mathbf{1}\{W_k\le W_j<2W_k\}
>c^*_{B-1,\alpha/B}\right\}.
\end{align*}
Consequently,
\begin{align*}
&\mathbb{P}\left(\bigcup_{i=1}^{B-c}\{2W_{(i)}>W_{(i+c)}\}\right)\\ &\qquad\le\sum_{k=1}^B
\mathbb{E}\left[ \mathbf{1}\{W_k>0\} \mathbb{P}\left( \sum_{j\ne k}\mathbf{1}\{W_k\le W_j<2W_k\} >c^*_{B-1,\alpha/B}\,\middle|\,W_k\right)\right].
\end{align*}

Conditional on $W_k=x>0$, the complementary indicators
$\mathbf{1}\{W_j\notin[x,2x)\}$, $j\ne k$, are independent
Bernoulli random variables. Their success probabilities satisfy
\[
\left(
\frac12-\mathbb{P}(W_j\notin[x,2x))
\right)_+
=
\left(
\mathbb{P}(x\le W_j<2x)-\frac12
\right)_+
\le \delta_j,
\]
because
\[
\mathbb{P}(x\le W_j<2x)
=
\lim_{\varepsilon\downarrow0}
\mathbb{P}(xe^{-\varepsilon}<W_j<2xe^{-\varepsilon})
\le \lQ_{W_j}(\ln 2).
\]
Here the equality follows by bounded convergence. Furthermore,
\begin{align*}
&\left\{
\sum_{j\ne k}\mathbf{1}\{x\le W_j<2x\}
>c^*_{B-1,\alpha/B}
\right\}\\
&\qquad=
\left\{
\sum_{j\ne k}\mathbf{1}\{W_j\notin[x,2x)\}
<(B-1)-c^*_{B-1,\alpha/B}
\right\}.
\end{align*}
Applying the upper bound in
Proposition~\ref{prop:poisson-binomial-to-binomial}, with
$B-1$ Bernoulli variables and significance level $\alpha/B$,
therefore gives, on $\{W_k>0\}$,
\[
\mathbb{P}\left(
\sum_{j\ne k}\mathbf{1}\{W_k\le W_j<2W_k\}
>c^*_{B-1,\alpha/B}
\,\middle|\,W_k
\right)
\le
\frac{\alpha}{B}
\exp\left(2\sum_{j\ne k}\delta_j\right).
\]
The proposition averages over the cutoff randomization, which
remains independent after conditioning on $W_k$. Thus,
\begin{align*}
&\mathbb{P}\left(
\bigcup_{i=1}^{B-c}\{2W_{(i)}>W_{(i+c)}\}
\right) \le
\frac{\alpha}{B}
\sum_{k=1}^B
\mathbb{P}(W_k>0)
\exp\left(2\sum_{j\ne k}\delta_j\right).
\end{align*}
Applying the same argument to the $-W_j$'s yields
\begin{align*}
\mathbb{P}\left(
0\notin\widehat{\mathrm{CI}}_{n,\alpha}^{\mathfrak{u}^\star}
\right)
&\le
\frac{\alpha}{B}
\sum_{k=1}^B
[\mathbb{P}(W_k>0)+\mathbb{P}(W_k<0)]
\exp\left(2\sum_{j\ne k}\delta_j\right)\\
&\le
\frac{\alpha}{B}
\sum_{k=1}^B
\exp\left(2\sum_{j\ne k}\delta_j\right)\\
&\le
\alpha\times
\exp\left(
2\sum_{j=1}^B
(\lQ_{W_j}(\ln 2)-1/2)_+
\right).
\end{align*}
When $B\le B_\alpha$, enlarging the nonnegative sum in the
exponent gives~\eqref{eq:unimodal-HulC-guarantee}.

\section{Proof of Theorem~\ref{thm:cover-diameter-cellwise-contamination}}\label{appsec:proof-of-cover-diameter-cellwise-contamination}
    First, we need to the following lemma for coverage guarantee. The result essentially follows from the arguments in Theorem 2.1 of~\cite{kock2025high} and instead of applying the high-dimensional Berry--Esseen bound, we apply the univariate Berry--Esseen bound to get the final result.
\begin{lemma}[Median bias of the winsorized mean]\label{lem:winsorized-mean-median-bias}
    Let $Y_1,\ldots,Y_m$ be independent random variables with $\mathbb{E}[Y_i]=\mu$ for all $1\le i\le m$. Suppose that, for some $\sigma^2, \delta>0$, $L\ge1$,
    \[
    \frac{1}{m}\sum_{i=1}^m\operatorname{Var}(Y_i)=\sigma^2\qquad\mbox{and}\qquad \frac{1}{m}\sum_{i=1}^m\mathbb{E}[|Y_i-\mu|^{2+\delta}]\le L\sigma^{2+\delta}.
    \]
    For an integer $1\le r<m/2$, define
    \[
    W_r((Y_i)_{i=1}^m):=\frac{1}{m}\left[rY_{(r)}+\sum_{i=r+1}^{m-r}Y_{(i)}+rY_{(m-r+1)}\right],
    \]
    and let
    \[
    \Delta_{m,r}:=\left(\frac12-\min\left\{\mathbb{P}\left(W_r((Y_i)_{i=1}^m)\le\mu\right),\,\mathbb{P}\left(W_r((Y_i)_{i=1}^m)\ge\mu\right)\right\}\right)_+.
    \]
    Then there exists a finite constant $C_\delta$, depending only on $\delta$, such that
    \[
    \Delta_{m,r}\le C_\delta L\frac{(r+\log m)^{(1+\delta)/(2+\delta)}}{m^{\delta/(4+2\delta)}}.
    \]
\end{lemma}
\begin{proof}
    If $r = 1$, then the winsorized mean is the sample mean and the result follows from the usual Berry--Esseen bound. Indeed, H{\"o}lder's inequality gives
    \[
    \frac{1}{m}\sum_{i=1}^m\mathbb{E}[|Y_i-\mu|^{2+(\delta\wedge1)}]
    \le\left(\frac{1}{m}\sum_{i=1}^m\mathbb{E}[|Y_i-\mu|^{2+\delta}]\right)^{(2+(\delta\wedge1))/(2+\delta)}
    \le L^{(2+(\delta\wedge1))/(2+\delta)}\sigma^{2+(\delta\wedge1)},
    \]
    and hence, by~\citet[Ineq. (5) and Corollary 2]{korolev2017bounds}, we get
    \[
    \Delta_{m,1}\le\frac{1.87L^{(2+(\delta\wedge1))/(2+\delta)}}{m^{(\delta\wedge1)/2}}
    \le1.87L\frac{(\log m)^{(1+\delta)/(2+\delta)}}{m^{\delta/(4+2\delta)}},
    \]
    since $m\ge3$ and $\delta/(4+2\delta)\le(\delta\wedge1)/2$.
    Assume $r \ge 2$. Define
    \[
    u:=\frac{2(r-1+\log m)}{m}.
    \]
    We first consider the case in which
    \[
    u\le\frac{1}{2}\qquad\mbox{and}\qquad 2L^{1/(2+\delta)}u^{\delta/(4+2\delta)}\le\frac{1}{2}.
    \]
    Let $F_i$ denote the distribution function of $Y_i$, and define
    \[
    \overline{F}_m(x):=\frac{1}{m}\sum_{i=1}^mF_i(x),\qquad a:=\overline{F}_m^{-1}(u),\qquad b:=\overline{F}_m^{-1}(1-u).
    \]
    If $Z$ has distribution function $\overline{F}_m$, then
    \[
    \mathbb{E}[Z]=\mu,\qquad \operatorname{Var}(Z)=\sigma^2,\qquad \mathbb{E}[|Z-\mu|^{2+\delta}]\le L\sigma^{2+\delta}.
    \]
    Thus, the quantile inequality in Lemma~C.1 of~\cite{kock2026winsorized} gives
    \[
    |a-\mu|\vee|b-\mu|\le L^{1/(2+\delta)}\sigma u^{-1/(2+\delta)}.
    \]
    The definition of the quantiles also gives
    \[
    \frac{1}{m}\sum_{i=1}^m\mathbb{P}(Y_i<a)\le u,\qquad \frac{1}{m}\sum_{i=1}^m\mathbb{P}(Y_i\le a)\ge u,
    \]
    and
    \[
    \frac{1}{m}\sum_{i=1}^m\mathbb{P}(Y_i>b)\le u,\qquad \frac{1}{m}\sum_{i=1}^m\mathbb{P}(Y_i\ge b)\ge u.
    \]

    Let
    \[
    N_-:=\sum_{i=1}^m\mathbf{1}\{Y_i\le a\}\qquad\mbox{and}\qquad N_+:=\sum_{i=1}^m\mathbf{1}\{Y_i\ge b\}.
    \]
    Both are sums of independent Bernoulli random variables with expectations at least $mu$. The exponential-moment argument underlying Lemma~B.1 of~\cite{kock2026winsorized} gives
    \[
    \mathbb{P}(N_\pm\le r-1)\le\exp\left\{-\left(mu-(r-1)-(r-1)\log\left(\frac{mu}{r-1}\right)\right)\right\}.
    \]
    Since $mu=2(r-1+\log m)$,
    \begin{align*}
    mu-(r-1)-(r-1)\log\left(\frac{mu}{r-1}\right)
    &=(r-1)+2\log m-(r-1)\log\left(2+\frac{2\log m}{r-1}\right)\\
    &\ge(1-\log2)(r-1)+\log m\\
    &\ge\log m.
    \end{align*}
    Consequently,
    \[
    \mathbb{P}(Y_{(r)}>a)\le\frac{1}{m}\qquad\mbox{and}\qquad \mathbb{P}(Y_{(m-r+1)}<b)\le\frac{1}{m}.
    \]
    Define
    \[
    V_i^-:=\max\{Y_i,a\}\qquad\mbox{and}\qquad V_i^+:=\min\{Y_i,b\}.
    \]
    Then, we have
    \begin{equation}\label{eq:two-sided-inequality-winsorized-mean}
    \begin{split}
    W_r((Y_i)_{i=1}^m)\le\frac{1}{m}\sum_{i=1}^mV_i^-,\quad&\mbox{on}\quad \{Y_{(r)}\le a\},\\
    W_r((Y_i)_{i=1}^m)\ge\frac{1}{m}\sum_{i=1}^mV_i^+,\quad&\mbox{on}\quad \{Y_{(m-r+1)}\ge b\}.
    \end{split}
    \end{equation}
    Hence, we get
    \begin{align*}
    \mathbb{P}(W_r((Y_i)_{i=1}^m) \le \mu) &\ge \mathbb{P}\left(\frac{1}{m}\sum_{i=1}^m V_i^- \le \mu\right) - \frac{1}{m}\\
    \mathbb{P}(W_r((Y_i)_{i=1}^m) \ge \mu) &\ge \mathbb{P}\left(\frac{1}{m}\sum_{i=1}^m V_i^+ \ge \mu\right) - \frac{1}{m}.
    \end{align*}
    Note that $V_i^-, 1\le i\le m$ are independent random variables and similarly, $V_i^{+}, 1\le i\le m$ are also independent random variables. Therefore, \citet[Ineq. (1.5) with Corollary 2]{korolev2017bounds} implies
    \[
    \sup_{t\in\mathbb{R}}\,\left|\mathbb{P}\left(\frac{\sum_{i=1}^m (V_i^{\pm} - \mathbb{E}[V_i^{\pm}])}{\sqrt{\sum_{i=1}^m \mathrm{Var}(V_i^{\pm})}} \le t\right) - \Phi(t)\right| \le 1.87\frac{\sum_{i=1}^m \mathbb{E}[|V_i^{\pm} - \mathbb{E}[V_i^{\pm}]|^{2 + (\delta\wedge1)}]}{(\sum_{i=1}^m \mathrm{Var}(V_i^{\pm}))^{1 + (\delta\wedge1)/2}}.
    \]
    Here we used $\min\{1,x\}\le x^{\delta\wedge1}$ for $x\ge0$. The denominators are positive by the variance bound proved below. By continuity of $\Phi$, the same bound holds with $\le$ replaced by $<$.
    Therefore,
    \begin{align*}
    \Delta_{m,r} \le \frac{1}{m} + \max_{\pm}\left\{\left|\frac{1}{2} - \Phi\left(\frac{|\sum_{i=1}^m \mathbb{E}[Y_i - V_i^{\pm}]|}{(\sum_{i=1}^m \mbox{Var}(V_i^{\pm}))^{1/2}}\right)\right| + 1.87\frac{\sum_{i=1}^m \mathbb{E}[|V_i^{\pm} - \mathbb{E}[V_i^{\pm}]|^{2 + (\delta\wedge1)}]}{(\sum_{i=1}^m \mathrm{Var}(V_i^{\pm}))^{1 + (\delta\wedge1)/2}}\right\}.
    \end{align*}
    Since
    $V_i^-=\max\{Y_i,a\}$, we have
    \[
    \frac{1}{m}\sum_{i=1}^m\mathbb{E}[V_i^-]-\mu=\frac{1}{m}\sum_{i=1}^m\mathbb{E}[(a-Y_i)_+]\ge0.
    \]
    On the event $\{Y_i<a\}$,
    $a-Y_i\le|Y_i-\mu|+|a-\mu|.$
    Therefore,
    \begin{align*}
    \frac{1}{m}\sum_{i=1}^m\mathbb{E}[(a-Y_i)_+]
    &\le\frac{1}{m}\sum_{i=1}^m\mathbb{E}[|Y_i-\mu|\mathbf{1}\{Y_i<a\}]+|a-\mu|\frac{1}{m}\sum_{i=1}^m\mathbb{P}(Y_i<a).
    \end{align*}
    Applying H{\"o}lder's inequality, we obtain
    \begin{align*}
    \frac{1}{m}\sum_{i=1}^m\mathbb{E}[|Y_i-\mu|\mathbf{1}\{Y_i<a\}]
    &\le\left(\frac{1}{m}\sum_{i=1}^m\mathbb{E}[|Y_i-\mu|^{2+\delta}]\right)^{1/(2+\delta)}
    \left(\frac{1}{m}\sum_{i=1}^m\mathbb{P}(Y_i<a)\right)^{(1+\delta)/(2+\delta)}\\
    &\le L^{1/(2+\delta)}\sigma u^{(1+\delta)/(2+\delta)}.
    \end{align*}
    The quantile bound gives
    \begin{align*}
    |a-\mu|\frac{1}{m}\sum_{i=1}^m\mathbb{P}(Y_i<a)
    &\le L^{1/(2+\delta)}\sigma u^{-1/(2+\delta)}u\\
    &=L^{1/(2+\delta)}\sigma u^{(1+\delta)/(2+\delta)}.
    \end{align*}
    Therefore,
    \[
    0\le\frac{1}{m}\sum_{i=1}^m\mathbb{E}[V_i^-]-\mu = \frac{1}{m}\sum_{i=1}^m \mathbb{E}[(a - Y_i)_+]\le2L^{1/(2+\delta)}\sigma u^{(1+\delta)/(2+\delta)},
    \]
    and by a similar argument, we conclude
    \[
    0\le\mu-\frac{1}{m}\sum_{i=1}^m\mathbb{E}[V_i^+]\le2L^{1/(2+\delta)}\sigma u^{(1+\delta)/(2+\delta)}.
    \]
    The corresponding second-moment calculations give
    \[
    \frac{1}{m}\sum_{i=1}^m\mathbb{E}[(V_i^\pm-Y_i)^2]\le4L^{2/(2+\delta)}\sigma^2u^{\delta/(2+\delta)}.
    \]
    Indeed,
    \begin{align*}
    &\frac{1}{m}\sum_{i=1}^m\mathbb{E}[(V_i^--Y_i)^2]\\
    &=\frac{1}{m}\sum_{i=1}^m\mathbb{E}[(a-Y_i)_+^2]\\
    &\le2\frac{1}{m}\sum_{i=1}^m\mathbb{E}[|Y_i-\mu|^2\mathbf{1}\{Y_i<a\}]
    +2|a-\mu|^2\frac{1}{m}\sum_{i=1}^m\mathbb{P}(Y_i<a)\\
    &\le2\left(\frac{1}{m}\sum_{i=1}^m\mathbb{E}[|Y_i-\mu|^{2+\delta}]\right)^{2/(2+\delta)}
    \left(\frac{1}{m}\sum_{i=1}^m\mathbb{P}(Y_i<a)\right)^{\delta/(2+\delta)}
    +2L^{2/(2+\delta)}\sigma^2u^{1-2/(2+\delta)}\\
    &\le4L^{2/(2+\delta)}\sigma^2u^{\delta/(2+\delta)},
    \end{align*}
    and similarly for $V_i^+$.
    Therefore, writing $v_\pm^2:=\sum_{i=1}^m\operatorname{Var}(V_i^\pm)$, the reverse triangle inequality gives
    \begin{align*}
    v_\pm
    &\ge\left(\sum_{i=1}^m\operatorname{Var}(Y_i)\right)^{1/2}-\left(\sum_{i=1}^m\mathbb{E}[(V_i^\pm-Y_i)^2]\right)^{1/2}\\
    &\ge\sigma\sqrt{m}\left(1-2L^{1/(2+\delta)}u^{\delta/(4+2\delta)}\right) \ge\frac{\sigma\sqrt{m}}2.
    \end{align*}
    We also have, by Jensen's inequality, for an independent copy $\tilde{V}_i^{\pm}$ of $V_i^{\pm}$,
    \[
    \frac{1}{m}\sum_{i=1}^m\mathbb{E}[|V_i^\pm-\mathbb{E}[V_i^{\pm}]|^{2+\delta}] \le \frac{1}{m}\sum_{i=1}^m \mathbb{E}[|V_i^{\pm} - \tilde{V}_i^{\pm}|^{2 + \delta}] \le \frac{1}{m}\sum_{i=1}^m \mathbb{E}[|Y_i - \tilde{Y}_i|^{2 + \delta}] \le 2^{1 + \delta}L\sigma^{2+\delta}.
    \]
    Here the second inequality follows from the fact that $y\mapsto\max\{y, a\}$ and $y\mapsto \min\{y, b\}$ are Lipschitz functions and $\tilde{Y}_i$ is an independent copy of $Y_i.$ The last inequality follows from
    \[
    |x-y|^{2+\delta}\le2^\delta(x-y)(x|x|^\delta-y|y|^\delta),\qquad x,y\in\mathbb{R},
    \]
    applied to $x=Y_i-\mu$ and $y=\tilde{Y}_i-\mu$, using independence and their zero means. The displayed scalar inequality follows from convexity of $t\mapsto t^{1+\delta}$ on $[0,\infty)$.
    Consequently, H{\"o}lder's inequality gives
    \begin{align*}
    \frac{1}{m}\sum_{i=1}^m\mathbb{E}[|V_i^\pm-\mathbb{E}[V_i^\pm]|^{2+(\delta\wedge1)}]
    &\le\left(\frac{1}{m}\sum_{i=1}^m\mathbb{E}[|V_i^\pm-\mathbb{E}[V_i^\pm]|^{2+\delta}]\right)^{(2+(\delta\wedge1))/(2+\delta)}\\
    &\le(2^{1+\delta}L)^{(2+(\delta\wedge1))/(2+\delta)}\sigma^{2+(\delta\wedge1)}\\
    &\le2^{1+\delta}L\sigma^{2+(\delta\wedge1)}.
    \end{align*}
    Combining all these inequalities with $|\Phi(x)-1/2|\le |x|/\sqrt{2\pi}$ for $x\in\mathbb{R}$, we get
    \begin{align*}
    \Delta_{m,r}
    &\le\frac{1}{m}+\frac{1}{\sqrt{2\pi}}\max_{\pm}\frac{|\sum_{i=1}^m\mathbb{E}[Y_i-V_i^\pm]|}{v_\pm}
    +1.87\max_{\pm}\frac{\sum_{i=1}^m\mathbb{E}[|V_i^\pm-\mathbb{E}[V_i^\pm]|^{2+(\delta\wedge1)}]}{v_\pm^{2+(\delta\wedge1)}}\\
    &\le\frac{1}{m}+\frac{4L^{1/(2+\delta)}}{\sqrt{2\pi}}\sqrt{m}\,u^{(1+\delta)/(2+\delta)}
    +\frac{2^{5+\delta}L}{m^{(\delta\wedge1)/2}}.
    \end{align*}
    Substituting the definition of $u$ gives
    \[
    \Delta_{m,r}\le\frac{1}{m}+\frac{4\cdot2^{(1+\delta)/(2+\delta)}L^{1/(2+\delta)}}{\sqrt{2\pi}}\frac{(r-1+\log m)^{(1+\delta)/(2+\delta)}}{m^{\delta/(4+2\delta)}}+\frac{2^{5+\delta}L}{m^{(\delta\wedge1)/2}}.
    \]
    Since $L\ge1$, $r-1+\log m\ge1$, and $\delta/(4+2\delta)\le(\delta\wedge1)/2\le1$, this implies
    \[
    \Delta_{m,r}\le C_\delta L\frac{(r-1+\log m)^{(1+\delta)/(2+\delta)}}{m^{\delta/(4+2\delta)}}.
    \]

    Finally, if either condition imposed after the definition of $u$ fails, the expression on the right-hand side of the preceding display is bounded below by a positive numerical constant. Indeed, if $u>1/2$, then $r-1+\log m>m/4$, and
    \[
    L\frac{(r-1+\log m)^{(1+\delta)/(2+\delta)}}{m^{\delta/(4+2\delta)}}
    >4^{-(1+\delta)/(2+\delta)}L\sqrt{m}
    \ge\frac14.
    \]
    If $2L^{1/(2+\delta)}u^{\delta/(4+2\delta)}>1/2$, then
    \begin{align*}
    L\frac{(r-1+\log m)^{(1+\delta)/(2+\delta)}}{m^{\delta/(4+2\delta)}}
    &=2^{-\delta/(4+2\delta)}L^{(1+\delta)/(2+\delta)}(r-1+\log m)^{1/2}
    \left(L^{1/(2+\delta)}u^{\delta/(4+2\delta)}\right)\\
    &>\frac{2^{-\delta/(4+2\delta)}}4 \ge\frac{1}{4\sqrt2}.
    \end{align*}
    Thus, after choosing $C_\delta$ sufficiently large, it is at least $1/2$. Since $\Delta_{m,r}\le1/2$ always, the same bound continues to hold in these cases. Combining the cases $r=1$ and $r\ge2$ proves the result.
\end{proof}
The following lemma gathers a few deterministic inequalities for winsorized means and winsorized brackets that are useful for the radius bounds. We present them as separate results as they may be of independent interest. Here we note that~\cite{oliveira2025trimmed} provide bounds for the behavior of a vector of trimmed means and our bound on the radius of $\cosi{}$ set requires an analogues result for winsorized mean as one of the components. 
\begin{lemma}\label{lem:some-ineq-for-winsorized-means}
    Fix any vector $z = (z_1, \ldots, z_m)\in\mathbb{R}^m$ and any $r \in (2\eta_nn, m/2]\cap\mathbb{Z}$. Recall the definitions $W_{r,2\eta_nn}^\pm(z)$ from~\eqref{eq:bracketing-winsorized-means}. Set
    \[
    \tau_{M}(u) = \max\{-M, \min\{u, M\}\}.
    \]
    The following statements hold:
    \begin{align*}
        W^{\pm}_{r,2\eta_nn}(z + u\mathbf{1}) &= W^{\pm}_{r,2\eta_nn}(z) + u, \quad W^{+}_{r,2\eta_nn}(z) = -W^{-}_{r,2\eta_nn}(-z),\\
        |W_{r,2\eta_nn}^{\pm}(z) - \overline{z}| &\le \frac{1}{m}\left[\frac{r + 2\eta_nn - 1}{(r - 2\eta_nn)^{1/p}} + (r + 2\eta_nn - 1)^{1-1/p}\right]\inf_{u\in\mathbb{R}}\|z - u\mathbf{1}\|_p,\quad\mbox{for}\quad p\ge 1\\
        \left|W_{r,2\eta_nn}^{\pm}(z) - \frac{1}{m}\sum_{i=1}^m \tau_M(z_i)\right| &\le \frac{2(r + 2\eta_nn - 1)M}{m} + \frac{2r}{m(r-2\eta_nn)}\sum_{i=1}^m (|z_i| - M)_+,
    \end{align*}
    where in the last inequality, the last term can be omitted if $\sum_{i=1}^m \mathbf{1}\{|z_i| > M\} \le r-2\eta_nn - 1$.
\end{lemma}
\begin{proof}
    The first statement follows from translation equivariance of order statistics. 
    For the second, recall that 
    \[
    z_{(i)} = -(-z)_{(m-i+1)}\quad\mbox{for}\quad 1\le i\le m,
    \]
    and hence,
    \[
    W^{-}_{r,2\eta_nn}(-z) = \frac{1}{m}\left[-z_{(m-r+2\eta_nn+1)} + \sum_{\ell = r-q+1}^{m-r-q} (-z_{(m-\ell+1)}) - rz_{(r+2\eta_nn)}\right] = -W^+_{r,2\eta_nn}(z).
    \]

    Observe that
    \begin{align*}
    W^{-}_{r,2\eta_n n}(z)-\overline{z}
    &= \frac{1}{m}\sum_{\ell<r-2\eta_n n}(z_{(r-2\eta_n n)}-z_{(\ell)})\\
    &\quad-\frac{1}{m}\sum_{\ell>m-r-2\eta_n n+1}(z_{(\ell)}-z_{(m-r-2\eta_n n+1)})-\frac{2\eta_n n}{m}(z_{(m-r-2\eta_n n+1)}-z_{(r-2\eta_n n)}).
    \end{align*}
    The first sum and the sum of the two subtracted quantities are nonnegative. Therefore, the absolute value of their difference is bounded by their maximum.

    To control the first sum, note that
    \[
    z_{(r-2\eta_n n)}\le\frac{1}{m-r+2\eta_n n+1}\sum_{\ell\ge r-2\eta_n n}z_{(\ell)}.
    \]
    Consequently, for every $u\in\mathbb{R}$, H{\"o}lder's inequality gives
    \begin{align*}
    \sum_{\ell<r-2\eta_n n}(z_{(r-2\eta_n n)}-z_{(\ell)})
    &\le\frac{r-2\eta_n n-1}{m-r+2\eta_n n+1}\sum_{\ell\ge r-2\eta_n n}(z_{(\ell)}-u)-\sum_{\ell<r-2\eta_n n}(z_{(\ell)}-u)\\
    &\le\left[r-2\eta_n n-1+\frac{(r-2\eta_n n-1)^{p/(p-1)}}{(m-r+2\eta_n n+1)^{1/(p-1)}}\right]^{1-1/p}\|z-u\mathbf{1}\|_p.
    \end{align*}
    Here the coefficients in the linear combination sum to zero, which allows arbitrary centering by $u$. Since $r-2\eta_n n-1\le r+2\eta_n n-1$ and $m-r+2\eta_n n+1\ge r-2\eta_n n$, taking the infimum over $u$ yields
    \[
    \sum_{\ell<r-2\eta_n n}(z_{(r-2\eta_n n)}-z_{(\ell)})\le\left[r+2\eta_n n-1+\frac{(r+2\eta_n n-1)^{p/(p-1)}}{(r-2\eta_n n)^{1/(p-1)}}\right]^{1-1/p}\inf_{u\in\mathbb{R}}\|z-u\mathbf{1}\|_p.
    \]

    Next, we control the upper-tail sum and the remaining correction together. There are $r+2\eta_n n-1$ indices strictly greater than $m-r-2\eta_n n+1$, and $2\eta_n n\le r+2\eta_n n-1$. Hence,
    \begin{align*}
    &\sum_{\ell>m-r-2\eta_n n+1}(z_{(\ell)}-z_{(m-r-2\eta_n n+1)})+2\eta_n n(z_{(m-r-2\eta_n n+1)}-z_{(r-2\eta_n n)})\\
    &\qquad\le\sum_{\ell>m-r-2\eta_n n+1}(z_{(\ell)}-z_{(r-2\eta_n n)}).
    \end{align*}
    Moreover,
    \[
    z_{(r-2\eta_n n)}\ge\frac{1}{r-2\eta_n n}\sum_{\ell\le r-2\eta_n n}z_{(\ell)}.
    \]
    Thus, for every $u\in\mathbb{R}$, another application of H{\"o}lder's inequality gives
    \begin{align*}
    \sum_{\ell>m-r-2\eta_n n+1}(z_{(\ell)}-z_{(r-2\eta_n n)})
    &\le\sum_{\ell>m-r-2\eta_n n+1}(z_{(\ell)}-u)-\frac{r+2\eta_n n-1}{r-2\eta_n n}\sum_{\ell\le r-2\eta_n n}(z_{(\ell)}-u)\\
    &\le\left[r+2\eta_n n-1+\frac{(r+2\eta_n n-1)^{p/(p-1)}}{(r-2\eta_n n)^{1/(p-1)}}\right]^{1-1/p}\|z-u\mathbf{1}\|_p.
    \end{align*}
    The two index sets are disjoint, and their coefficients again sum to zero. Taking the infimum over $u$ and combining the preceding bounds, we conclude that
    \[
    |W^{-}_{r,2\eta_n n}(z)-\overline{z}|\le\frac{1}{m}\left[r+2\eta_n n-1+\frac{(r+2\eta_n n-1)^{p/(p-1)}}{(r-2\eta_n n)^{1/(p-1)}}\right]^{1-1/p}\inf_{u\in\mathbb{R}}\|z-u\mathbf{1}\|_p.
    \]
    Finally, applying the same argument to $-z$ and using $W^{+}_{r,2\eta_n n}(z)=-W^{-}_{r,2\eta_n n}(-z)$ gives
    \begin{align*}
    |W^{\pm}_{r,2\eta_n n}(z)-\overline{z}|
    &\le\frac{1}{m}\left[r+2\eta_n n-1+\frac{(r+2\eta_n n-1)^{p/(p-1)}}{(r-2\eta_n n)^{1/(p-1)}}\right]^{1-1/p}\inf_{u\in\mathbb{R}}\|z-u\mathbf{1}\|_p\\
    &\le\frac{1}{m}\left[(r+2\eta_n n-1)^{1-1/p}+\frac{r+2\eta_n n-1}{(r-2\eta_n n)^{1/p}}\right]\inf_{u\in\mathbb{R}}\|z-u\mathbf{1}\|_p.
    \end{align*}

    To prove the final statement, note that
    \[
    W^-_{r,2\eta_nn}(z) = \frac{1}{m}\sum_{\ell=1}^m \omega_{\ell}z_{(\ell)}\quad\mbox{with}\quad \omega_{\ell} := \begin{cases}
        0, &\mbox{if }\ell < r - 2\eta_nn,\mbox{ or }\ell > m -r -2\eta_nn + 1\\
        r, &\mbox{if }\ell = r - 2\eta_nn\mbox{ or }\ell = m - r - 2\eta_nn + 1,\\
        1, &\mbox{if }r - 2\eta_nn < \ell < m - r - 2\eta_nn + 1.
    \end{cases}
    \]
    Let $\tau_M(z) := (\tau_M(z_1), \ldots, \tau_M(z_m))$. Because $u\mapsto\tau_M(u)$ is non-decreasing, the $\tau_M((z)_{(\ell)}) = (\tau_M(z))_{(\ell)}$.
    This implies
    \[
    W^{-}_{r,2\eta_nn}(z) - W^{-}_{r,2\eta_nn}(\tau_M(z)) = \frac{1}{m}\sum_{\ell = 1}^m \omega_{\ell}(z_{(\ell)} - \tau_M(z_{(\ell)})) 
    \]
    Clearly, $u - \tau_{M}(u) = (u - M)_+ - (-M - u)_+$, which implies
    \begin{align*}
    |W^-_{r,2\eta_nn}(z) - W^{-}_{r,2\eta_nn}(\tau_M(z))| &\le \max\left\{\frac{1}{m}\sum_{\ell = 1}^m \omega_{\ell}(z_{(\ell)} - M)_+,\, \frac{1}{m}\sum_{\ell = 1}^m \omega_{\ell}(-M - z_{(\ell)})_+\right\}.
    \end{align*}
    If $\sum_{\ell = 1}^m \mathbf{1}\{|z_{\ell}| > M\} \le r - 2\eta_n - 1$, then all the summands on the right hand side are zero. 
    From the definition of the weights, for every $1\le\ell\le m$, we have
    \begin{align*}
    \sum_{i=1}^{\ell}\omega_i &\le \ell\max\left\{\frac{r}{r-2\eta_n n},\,\frac{m}{m-r-2\eta_n n+1}\right\},\\
    \sum_{i=\ell}^{m}\omega_i &\le (m-\ell+1)\frac{m}{m-r+2\eta_n n+1}.
    \end{align*}
    Indeed, the first cumulative sum equals zero for $\ell<r-2\eta_n n$, equals $\ell+2\eta_n n$ for $r-2\eta_n n\le\ell<m-r-2\eta_n n+1$, and equals $m$ thereafter. The second cumulative sum equals $m$ for $\ell\le r-2\eta_n n$, equals $m-\ell+1-2\eta_n n$ for $r-2\eta_n n<\ell\le m-r-2\eta_n n+1$, and equals zero thereafter.

    Because $\ell\mapsto(-M-z_{(\ell)})_+$ is non-increasing, summation by parts gives
    \begin{align*}
    &\sum_{\ell=1}^{m}\omega_\ell(-M-z_{(\ell)})_+\\
    &=m(-M-z_{(m)})_++\sum_{\ell=1}^{m-1}\left(\sum_{i=1}^{\ell}\omega_i\right)\left[(-M-z_{(\ell)})_+-(-M-z_{(\ell+1)})_+\right]\\
    &\le\max\left\{\frac{r}{r-2\eta_n n},\,\frac{m}{m-r-2\eta_n n+1}\right\}\left[m(-M-z_{(m)})_++\sum_{\ell=1}^{m-1}\ell\left[(-M-z_{(\ell)})_+-(-M-z_{(\ell+1)})_+\right]\right]\\
    &=\max\left\{\frac{r}{r-2\eta_n n},\,\frac{m}{m-r-2\eta_n n+1}\right\}\sum_{\ell=1}^{m}(-M-z_\ell)_+.
    \end{align*}
    Similarly, because $\ell\mapsto(z_{(\ell)}-M)_+$ is non-decreasing, the bound on the cumulative sums from the other end yields
    \begin{align*}
    &\sum_{\ell=1}^{m}\omega_\ell(z_{(\ell)}-M)_+\\
    &=m(z_{(1)}-M)_++\sum_{\ell=2}^{m}\left(\sum_{i=\ell}^{m}\omega_i\right)\left[(z_{(\ell)}-M)_+-(z_{(\ell-1)}-M)_+\right]\\
    &\le\frac{m}{m-r+2\eta_n n+1}\left[m(z_{(1)}-M)_++\sum_{\ell=2}^{m}(m-\ell+1)\left[(z_{(\ell)}-M)_+-(z_{(\ell-1)}-M)_+\right]\right]\\
    &=\frac{m}{m-r+2\eta_n n+1}\sum_{\ell=1}^{m}(z_\ell-M)_+.
    \end{align*}
    Since $m-r+2\eta_n n+1\ge m-r-2\eta_n n+1$, combining these inequalities gives
    \[
    |W^-_{r,2\eta_n n}(z)-W^-_{r,2\eta_n n}(\tau_M(z))|\le\frac{1}{m}\max\left\{\frac{r}{r-2\eta_n n},\,\frac{m}{m-r-2\eta_n n+1}\right\}\sum_{\ell=1}^{m}(|z_\ell|-M)_+.
    \]
    Applying the same argument to $-z$, and using $W^+_{r,2\eta_n n}(z)=-W^-_{r,2\eta_n n}(-z)$ and $\tau_M(-z)=-\tau_M(z)$, we get the result for $W^+_{r,2\eta_nn}(z)$. The final result follows from an application of the third statement (with $p = \infty, u = 0$) in bounding $|W^{\pm}_{r,2\eta_nn}(\tau_M(z)) - \overline{\tau_M(z)}|$.
\end{proof}

The following lemma is needed for our proof of Theorem~\ref{thm:cover-diameter-cellwise-contamination}.
\subsection{Proof of Theorem~\ref{thm:cover-diameter-cellwise-contamination}}
    Define
    \begin{align*}
    \vartheta^* &:= \left(W_{r_0}((X_{i,k}')_{i\in S_j})\right)_{1\le k\le d},\\
    \widetilde{\theta}_j &:= \left(W_{r}((X_{i,k}')_{i\in S_j})\right)_{1\le k\le d},\quad\mbox{for}\quad 1 \le j\le B.
    \end{align*}
    While $\vartheta^*, \widetilde{\theta}_{j}\in\mathbb{R}^d, 1 \le j\le B$ cannot be computed from the observed data, we can consider the oracle version of the $\cosi{}$ confidence set as
    \[
    \widetilde{\mathrm{CI}}_{n,\alpha}^{\mathtt{cosi}\mbox{-}\mp} = \left\{\theta\in\mathbb{R}^d:\, \|\theta - \widetilde{\theta}_0\|_{\infty} \le \max_{1\le k\le d}(|\widetilde{\theta}_{0,k} - \widetilde{\theta}_{\cdot,k}|)_{(c^*_{B,\alpha} + 1)}\right\}.
    \]
    From Theorem~\ref{thm:coverage-of-cosi-m+-diameter-analysis}, we conclude
    \begin{equation}\label{eq:clean-sample-coverage-guarantee}
    \mathbb{P}(\mu\notin\widetilde{\mathrm{CI}}_{n,\alpha}^{\mathtt{cosi}\mbox{-}\mp}) \le \alpha\times\exp\left(2B\Delta^{\mathfrak{m}}\right),
    \end{equation}
    where
    \[
    \Delta^{\mathfrak{m}} = \max_{1\le k\le d}\,\frac{1}{B}\sum_{j=1}^B \left(\frac{1}{2} - \min\left\{\mathbb{P}(\widetilde{\theta}_{j,k} \ge \mu_k),\,\mathbb{P}(\widetilde{\theta}_{j,k} \le \mu_k)\right\}\right)_+.
    \]
    This median bias can be controlled using Lemma~\ref{lem:winsorized-mean-median-bias}.
    Observe now that if $x, y\in\mathbb{R}^m$ are such that $\sum_{\ell = 1}^m \mathbf{1}\{x_{\ell} \neq y_{\ell}\} \le q$, then $y_{(\ell - q)} \le x_{(\ell)} \le y_{(\ell + q)}$. This implies that
    \begin{equation}\label{eq:shift-of-quantiles}
    {W}_{r,q}^-(y) \le W_r(x) \le {W}_{r,q}^+(y).
    \end{equation}
    Applying this with $q = \eta_n n$ and $r$ as in~\eqref{eq:winsorization-rank}, from~\eqref{eq:cellwise-adversarial}, we get
    \begin{equation}\label{eq:clean-to-corrupted}
    \begin{split}
    \vartheta^-_k ~&\le~ \vartheta^*_k ~\le~ \vartheta^+_k,\\ 
    \widehat{\theta}_{j,k}^- ~&\le~ \widetilde{\theta}_{j,k} ~\le~ \widehat{\theta}_{j,k}^+,\quad\mbox{for all}\quad 1 \le j\le B, 1\le k\le d.
    \end{split}
    \end{equation}
    Hence, we conclude
    \[
    \widetilde{\mathrm{CI}}_{n,\alpha}^{\mathtt{cosi}\mbox{-}\mp} \subseteq \cosici{n}{\alpha}{\mathtt{m}\mbox{-}\mathtt{rob}+},
    \]
    and hence, the coverage guarantee~\eqref{eq:clean-sample-coverage-guarantee} continues to hold for the robust confidence set as well.

    For the radius analysis, we note by~\eqref{eq:shift-of-quantiles} that
    \[
    \cosici{n}{\alpha}{\mathtt{m}\mbox{-}\mathtt{rob}+} \subseteq \widetilde{C}_{\alpha},
    \]
    where
    \[
    \widetilde{C}_{\alpha} := \bigotimes_{k=1}^d \left[W_{r_0,2\eta_nn}^{-}((X_{i,k}')_{i\in S_0}) - \widetilde{R}_{\alpha},\, W_{r_0, 2\eta_nn}^{+}((X_{i,k}')_{i\in S_0}) + \widetilde{R}_{\alpha}\right],
    \]
    with $\widetilde{R}_{\alpha} = \max_{1\le k \le d}(R_{k,\cdot})_{(c^*_{B,\alpha}+1)}$, where
    \begin{align*}
    R_{k,j} &:= \max\{|W_{r_0, 2\eta_nn}^{+}((X_{i,k}')_{i\in S_0}) - W_{r,2\eta_nn}^{-}((X_{i,k}')_{i\in S_{j}})|,\,|W_{r_0, 2\eta_nn}^{-}((X_{i,k}')_{i\in S_0}) - W_{r,2\eta_nn}^{+}((X_{i,k}')_{i\in S_{j}})|\}.
    \end{align*}
    Note that $\widetilde{C}_{\alpha}$ is not computable from observed data. Nevertheless, we have
    \begin{align*}
    &\sup_{\theta\in\cosici{n}{\alpha}{\mathtt{m}\mbox{-}\mathtt{rob}+}}\,\|\theta - \mu\|_{\infty} \\
    &\le \sup_{\theta\in\widetilde{C}_{\alpha}}\,\|\theta - \mu\|_{\infty}\\
    &\le 2\max_{1\le k\le d}\max\{|W_{r_0,2\eta_nn}^{-}((X_{i,k}')_{i\in S_0}) - \mu_k|,\, |W_{r_0, 2\eta_nn}^{+}((X_{i,k}')_{i\in S_0}) - \mu_k|\}\\
    &\quad+ \max_{1\le k\le d}\left(\max\{|\mu_k - W_{r,2\eta_nn}^{-}((X_{i,k}')_{i\in S_{\cdot}})|,\,|\mu_k - W_{r,2\eta_nn}^{+}((X_{i,k}')_{i\in S_{\cdot}})|\}\right)_{(c^*_{B,\alpha}+1)}\\
    &= \mathbf{I} + \mathbf{II}.
    \end{align*}
    Set $Y_{i,k} = X_{i,k}' - \mu_{k}$ for $1\le i\le n, 1\le k\le d$. From Lemma~\ref{lem:some-ineq-for-winsorized-means}, we have
    \begin{align*}
        W_{r_0,2\eta_nn}^{\pm}((X_{i,k}')_{i\in S_0}) - \mu_k &= W_{r_0,2\eta_nn}^{\pm}((Y_{i,k})_{i\in S_0})\\ 
        W_{r,2\eta_nn}^{\pm}((X_{i,k}')_{i\in S_{j}}) - \mu_k &= W_{r,2\eta_nn}^{\pm}((Y_{i,k})_{i\in S_j}).
    \end{align*}    
    From Theorem~\ref{thm:supremum-of-quantiles}, with probability at least $1 - \gamma$,
    \begin{equation}\label{eq:part-II-diameter-cellwise}
    \mathbf{II} \le \frac{4B(e^{4\log(1/\gamma)/(B-c^*_{B,\alpha} - 1)} - 1)}{\log(1/\gamma)}\mathfrak{R} + 2\sigma\sqrt{\frac{B(e^{4\log(1/\gamma)/(B-c^*_{B,\alpha} - 1)} - 1)}{\log(1/\gamma)}},
    \end{equation}
    where for independent Rademacher random variables $\varepsilon_1, \ldots, \varepsilon_B$,
    \begin{align*}
    \mathfrak{R} &:= \mathbb{E}\left[\max_{1\le k\le d}\,\frac{1}{B}\sum_{j=1}^B \varepsilon_j\max\{|W_{r,2\eta_nn}^{-}((Y_{i,k})_{i\in S_{j}})|,\,|W_{r,2\eta_nn}^{+}((Y_{i,k})_{i\in S_{j}})|\}\right],\\
    \sigma^2 &:= \max_{1\le k\le d}\,\frac{1}{B}\sum_{j=1}^B \mathbb{E}[\max\{|W_{r,2\eta_nn}^{-}((Y_{i,k})_{i\in S_{j}})|^2,\,|W_{r,2\eta_nn}^{+}((Y_{i,k})_{i\in S_{j}})|^2\}].
    \end{align*}
    To control $\sigma^2,$ we note from Lemma~\ref{lem:some-ineq-for-winsorized-means} (with $p = 2 + \delta$) that
    \begin{align*}
    &\left|W_{r,2\eta_nn}^{\pm}((Y_{i,k})_{i\in S_j}) - \frac{1}{|S_j|}\sum_{i\in S_j} Y_{i,k}\right|\\ 
    ~&\le~ \frac{1}{|S_j|}\left[\frac{r + 2\eta_nn - 1}{(r-2\eta_nn)^{1/(2 + \delta)}} + (r + 2\eta_nn - 1)^{(1+\delta)/(2 + \delta)}\right]\left(\sum_{i\in S_j} |Y_{i,k}|^{2 + \delta}\right)^{1/(2 + \delta)}.
    \end{align*}
    This implies that 
    \begin{align*}
        \sigma^2 &\le \max_{1\le k\le d}\,\frac{2}{B}\sum_{j=1}^B \mathbb{E}\left(\frac{1}{|S_j|}\sum_{i\in S_j} Y_{i,k}\right)^{2}\\
        &\quad+ \max_{1\le k\le d}\left[\frac{r + 2\eta_nn - 1}{(r-2\eta_nn)^{1/(2 + \delta)}} + (r + 2\eta_nn - 1)^{(1+\delta)/(2 + \delta)}\right]^2\frac{2}{B}\sum_{j=1}^B\frac{1}{|S_j|^2}\left(\sum_{i\in S_j}\mathbb{E}[|Y_{i,k}|^{2+\delta}]\right)^{2/(2 + \delta)}.
    \end{align*}
    The second term can be controlled using~\eqref{eq:moment-equivalence-cellwise} as
    \[
    \left(\frac{1}{|S_j|}\sum_{i\in S_j} \mathbb{E}[|Y_{i,k}|^{2+\delta}]\right)^{2/(2 + \delta)} \le \frac{L^{2/(2+\delta)}}{|S_j|}\sum_{i\in S_j} \mathbb{E}[|Y_{i,k}|^2].
    \]
    This implies
    \begin{align*}
    \frac{2}{B}\sum_{j=1}^B \frac{1}{|S_j|^2}\left(\sum_{i\in S_j}\mathbb{E}[|Y_{i,k}|^{2+\delta}]\right)^{2/(2 + \delta)} &\le \frac{2L^{2/(2+\delta)}}{B}\sum_{j=1}^B \frac{1}{|S_j|^{2-2/(2+\delta)}}\frac{1}{|S_j|}\sum_{i\in S_j}\mathbb{E}[|Y_{i,k}|^2]\\
    &\le \left(\frac{2B}{n}\right)^{(2+2\delta)/(2+\delta)}\frac{2L^{2/(2+\delta)}}{B}\sum_{j=1}^B \frac{1}{|S_j|}\sum_{i\in S_j}\mathbb{E}[|Y_{i,k}|^2]\\
    &\le \left(\frac{2B}{n}\right)^{(2+2\delta)/(2+\delta)}{2L^{2/(2+\delta)}}\mathfrak{v}^2.
    \end{align*}
    Therefore, we get
    \begin{align*}
    \sigma^2 &\le \left(\frac{4B}{n} + 2L^{2/(2+\delta)}\left(\frac{2B}{n}\right)^{(2+2\delta)/(2+\delta)}\left[\frac{r + 2\eta_nn - 1}{(r-2\eta_nn)^{1/(2 + \delta)}} + (r + 2\eta_nn - 1)^{(1+\delta)/(2 + \delta)}\right]^2\right)\mathfrak{v}^2\\
    &\le \left(\frac{2B}{n} + 8L^{2/(2+\delta)}\left(\frac{4B}{n}\right)^{(2+2\delta)/(2+\delta)}\left(\frac{(r+2\eta_nn - 1)}{(r-2\eta_nn)^{1/(2+\delta)}}\right)^2\right)\mathfrak{v}^2,
    \end{align*}
    where the last inequality follows from the fact that $r - 2\eta_nn \le r + 2\eta_nn - 1$. Observe now that with $r = \lceil\lambda\eta_nn + 1\rceil$ for $\lambda > 2$, we get
    \begin{align*}
    \frac{(r+2\eta_nn - 1)}{(r-2\eta_nn)^{1/(2+\delta)}} &\le \frac{(\lambda+2)\eta_nn + 1}{((\lambda-2)\eta_nn+1)^{1/(2 + \delta)}}\\ 
    &\le (\eta_nn)^{(1+\delta)/(2 + \delta)}(\lambda+3)/(\lambda - 2)^{1/(2 + \delta)}.
    \end{align*}
    This yields
    \begin{equation}\label{eq:bound-on-sigma^2}
        \sigma^2 \le \mathfrak{v}^2\left(\frac{2B}{n} + 4L\left(2B\eta_n\right)^{(2+2\delta)/(2+\delta)}\frac{(\lambda+3)^2}{(\lambda - 2)^{2/(2 + \delta)}}\right).
    \end{equation}
    To control $\mathfrak{R}$, we apply the final statement of Lemma~\ref{lem:some-ineq-for-winsorized-means} to get
    \begin{align*}
        \mathfrak{R} &\le \mathbb{E}\left[\max_{1\le k\le d}\frac{1}{B}\sum_{j=1}^B \varepsilon_j\frac{1}{|S_j|}\sum_{i\in S_j}\tau_M(Y_{i,k})\right]\\
        &\quad+ \frac{1}{B}\sum_{j=1}^B \frac{2(r+2\eta_nn-1)M}{|S_j|} + \frac{2r}{(r-2\eta_nn)} \mathbb{E}\left[\max_{1\le k\le d}\frac{1}{B}\sum_{j=1}^B\frac{1}{|S_j|}\sum_{i\in S_j}(|Y_{i,k}| - M)_+\right].
    \end{align*}
    By the Lipschitzness of $u\mapsto\tau_M(u)$ and the contraction principle~\citep[Lemma 8]{mohri2014learning} (and symmetrization), we get
    \[
    \mathbb{E}\left[\max_{1\le k\le d}\frac{1}{B}\sum_{j=1}^B \varepsilon_j\frac{1}{|S_j|}\sum_{i\in S_j}\{\tau_M(Y_{i,k}) - \mathbb{E}[\tau_M(Y_{i,k})]\}\right] \le 8\mathbb{E}\left[\max_{1\le k\le d}\,\left|\frac{1}{n}\sum_{i=1}^n Y_{i,k}\right|\right].
    \]
    Also, note that
    \begin{align*}
    \mathbb{E}\left[\max_{1\le k\le d}\,\frac{1}{B}\sum_{j=1}^B \varepsilon_j\frac{1}{|S_j|}\sum_{i\in S_j} \mathbb{E}[\tau_M(Y_{i,k})]\right] &= \mathbb{E}\left[\max_{1\le k\le d}\,\frac{1}{B}\sum_{j=1}^B \varepsilon_j\frac{1}{|S_j|}\sum_{i\in S_j} \mathbb{E}[(\tau_M(Y_{i,k}) - Y_{i,k})]\right]\\
    &\le \max_{1\le k\le d}\frac{1}{B}\sum_{j=1}^B \frac{1}{|S_j|}\sum_{i\in S_j} \mathbb{E}[(|Y_{i,k}| - M)_+].
    \end{align*}
    Moreover, from~\eqref{eq:split-sizes}, we have
    \[
    \frac{1}{B}\sum_{j=1}^B \frac{2(r+2\eta_nn-1)M}{|S_j|} \le \frac{4B(r+2\eta_nn-1)M}{n}.
    \]
    Finally, 
    \begin{align*}
        &\mathbb{E}\left[\max_{1\le k\le d}\frac{1}{B}\sum_{j=1}^B \frac{1}{|S_j|}\sum_{i\in S_j}(|Y_{i,k}| - M)_+\right]\\ &\le \mathbb{E}\left[\max_{1\le k\le d}\,\frac{1}{B}\sum_{j=1}^B \frac{1}{|S_j|}\sum_{i\in S_j} \{(|Y_{i,k}| - M)_+ - \mathbb{E}[(|Y_{i,k}| - M)_+]\}\right]\\
        &\quad+ \max_{1\le k\le d}\frac{1}{B}\sum_{j=1}^B \frac{1}{|S_j|}\sum_{i\in S_j}\mathbb{E}[(|Y_{i,k}| - M)_+]\\
        &\le 16\mathbb{E}\left[\max_{1\le k\le d}\,\left|\frac{1}{n}\sum_{i=1}^n Y_{i,k}\right|\right] + \max_{1\le k\le d}\,\frac{1}{B}\sum_{j=1}^B \frac{1}{|S_j|}\sum_{i\in S_j}\mathbb{E}[(|Y_{i,k}| - M)_+].
    \end{align*}
    To bound the last term, observe that
    \begin{equation}\label{eq:bound-bias-winsorization}
    \mathbb{E}[(|Y_{i,k}| - M)_+] \le \frac{(1+\delta)^{1+\delta}}{(2+\delta)^{2+\delta}}\frac{\mathbb{E}[|Y_{i,k}|^{2+\delta}]}{M^{1+\delta}}.
    \end{equation}
    This inequality follows by maximizing $(x-M)/x^{2+\delta}$ over all $x \ge M$. This yields by~\eqref{eq:moment-equivalence-cellwise} that
    \begin{align*}
        \frac{1}{B}\sum_{j=1}^B \frac{1}{|S_j|}\sum_{i\in S_j}\mathbb{E}[(|Y_{i,k}| - M)_+] &\le \frac{(1+\delta)^{1+\delta}}{(2 + \delta)^{2+\delta}}\frac{1}{M^{1+\delta}}\frac{L}{B}\sum_{j=1}^B \left(\frac{1}{|S_j|}\sum_{i\in S_j} \mathbb{E}[|Y_{i,k}|^{2}]\right)^{1+\delta/2}\\
        &\le \frac{(1+\delta)^{1+\delta}}{(2 + \delta)^{2+\delta}}\frac{L\mathfrak{v}^{2 + \delta}}{M^{1 + \delta}}.
    \end{align*}
    This implies by contraction that
    \begin{align*}
        &\mathbb{E}\left[\max_{1\le k\le d}\frac{1}{B}\sum_{j=1}^B \frac{1}{|S_j|}\sum_{i\in S_j}(|Y_{i,k}| - M)_+\right]\\ 
       &\le 8\mathbb{E}\left[\max_{1\le k\le d}\,\left|\frac{1}{n}\sum_{i=1}^n Y_{i,k}\right|\right] + \frac{(1+\delta)^{1+\delta}}{(2 + \delta)^{2+\delta}}\frac{L\mathfrak{v}^{2 + \delta}}{M^{1 + \delta}}.
    \end{align*}
    Combining the preceding inequalities, for every $M>0$,
    \begin{align*}
    \mathfrak{R}
    &\le\left(8+\frac{16r}{r-2\eta_nn}\right)\mathfrak{E}_d
      +\frac{4B(r+2\eta_nn-1)M}{n}\\
    &\quad+\left(1+\frac{2r}{r-2\eta_nn}\right)
    \frac{(1+\delta)^{1+\delta}}{(2+\delta)^{2+\delta}}
    \frac{L\mathfrak{v}^{2+\delta}}{M^{1+\delta}}.
    \end{align*}
    Taking $M=L^{1/(2+\delta)}\mathfrak{v}(B\eta_n)^{-1/(2+\delta)}$
    and using $r/(r-2\eta_nn)\le\lambda/(\lambda-2)$ and
    $r+2\eta_nn-1\le(\lambda+3)\eta_nn$ gives
    \begin{equation}\label{eq:final-bound-R-cellwise}
    \mathfrak{R}\le C\left[\mathfrak{E}_d+
    \mathfrak{v}(B\eta_n)^{(1+\delta)/(2+\delta)}\right],
    \end{equation}
    for a constant $C$ depending only on $L, \lambda, \delta$.
    Combining~\eqref{eq:bound-on-sigma^2} and~\eqref{eq:final-bound-R-cellwise} with~\eqref{eq:part-II-diameter-cellwise} controls $\mathbf{II}$.

    Finally, to bound $\mathbf{I}$, one could apply the final inequality of Lemma~\ref{lem:some-ineq-for-winsorized-means} directly with arbitrary $M$ and optimize over $M$ but this is sub-optimal because $\max_{1\le k\le d} \sum_{i\in S_0} (|Y_{i,k}| - M)_+$ may not concentrate well around its expectation. For this reason, we choose a deterministic $M>0$ such that, with high probability,
    \[
    \max_{1\le k\le d}\sum_{i\in S_0}\mathbf{1}\{|Y_{i,k}|>M\}\le r_0-2\eta_nn-1.
    \]
    On this event, Lemma~\ref{lem:some-ineq-for-winsorized-means} implies
    \[
    \max_{1\le k\le d}\max_{\pm}
    \left|W_{r_0,2\eta_nn}^{\pm}((Y_{i,k})_{i\in S_0})
    -\frac{1}{|S_0|}\sum_{i\in S_0}\tau_M(Y_{i,k})\right|
    \le\frac{2(r_0+2\eta_nn-1)M}{|S_0|}.
    \]
    Symmetrization and contraction principles give
    \begin{align*}
    &\mathbb{E}\left[\max_{1\le k\le d}\left|
    \frac{1}{|S_0|}\sum_{i\in S_0}
    \{\tau_M(Y_{i,k})-\mathbb{E}[\tau_M(Y_{i,k})]\}\right|\right]\\
    &\le4\mathbb{E}\left[\max_{1\le k\le d}\left|
    \frac{1}{|S_0|}\sum_{i\in S_0}Y_{i,k}\right|\right]
    \le\frac{4n}{|S_0|}\mathfrak{E}_d.
    \end{align*}
    Additionally,~\eqref{eq:bound-bias-winsorization} and~\eqref{eq:moment-equivalence-cellwise} imply
    \begin{align*}
    \max_{1\le k\le d}\left|
    \frac{1}{|S_0|}\sum_{i\in S_0}\mathbb{E}[\tau_M(Y_{i,k})]\right|
    &\le\frac{(1+\delta)^{1+\delta}}{(2+\delta)^{2+\delta}}
    \frac{L\mathfrak{v}^{2+\delta}}{M^{1+\delta}}.
    \end{align*}
    From Theorem 1.1(c) of~\cite{KleinRio2005}, with probability at least $1-\gamma/3$,
    \begin{equation}\label{eq:prob-bound-max-winsorized-popu}
    \begin{split}
    &\max_{1\le k\le d}\left|
    \frac{1}{|S_0|}\sum_{i\in S_0}
    \{\tau_M(Y_{i,k})-\mathbb{E}[\tau_M(Y_{i,k})]\}\right|\\
    &\quad\le\frac{8n}{|S_0|}\mathfrak{E}_d
    +\mathfrak{v}\sqrt{\frac{2\log(3/\gamma)}{|S_0|}}
    +\frac{8M\log(3/\gamma)}{|S_0|}.
    \end{split}
    \end{equation}

    Finding $M$ to satisfy $\max_{1\le k\le d}\sum_{i\in S_0}\mathbf{1}\{|Y_{i,k}| > M\} \le r_0 - 2\eta_nn - 1$ can be done using the techniques in the proof of Theorem~\ref{thm:supremum-of-quantiles}. Following the proof that led to~\eqref{eq:expectation-sufficient-condition-bound-2}, we conclude that an $M$ such that
    \begin{equation}\label{eq:expectation-requirement}
    \mathbb{E}\left[\max_{1\le k\le d}\,\frac{1}{|S_0|}\sum_{i\in S_0} \mathbf{1}\{|Y_{i,k}| > M\}\right] \le \frac{\log(1/\gamma')}{|S_0|(\exp(4\log(1/\gamma')/(r_0 - 2\eta_nn - 1)) - 1)},    
    \end{equation}
    satisfies with probability at least $1 - \gamma'$,
    \[
    \max_{1\le k\le d}\,\frac{1}{|S_0|}\sum_{i\in S_0}\,\mathbf{1}\{|Y_{i,k}| > M\} \le \frac{r_0 - 2\eta_nn - 1}{|S_0|}. 
    \]
    Taking $\log(1/\gamma') = r_0 - 2\eta_nn - 1$ implies that we need to choose $M > 0$ such that
    \begin{equation}\label{eq:expectation-requirement-final}
    \mathbb{E}\left[\max_{1\le k\le d}\,\frac{1}{|S_0|}\sum_{i\in S_0} \mathbf{1}\{|Y_{i,k}| > M\}\right] \le \frac{r_0 - 2\eta_nn - 1}{|S_0|(e^4 - 1)}.
    \end{equation}
    The smoothing argument used in~\eqref{eq:expectation-bound}, with the $2+\delta$ moment in place of the second moment (in inequality (a) of~\eqref{eq:expectation-bound}), gives
    \begin{align*}
    &\mathbb{E}\left[\max_{1\le k\le d}\frac{1}{|S_0|}
    \sum_{i\in S_0}\mathbf{1}\{|Y_{i,k}|>M\}\right]\\
    &\le\frac{8n\mathfrak{E}_d}{|S_0|M}
    +\frac{2^{2+\delta}}{M^{2+\delta}}
    \max_{1\le k\le d}\frac{1}{|S_0|}
    \sum_{i\in S_0}\mathbb{E}[|Y_{i,k}|^{2+\delta}]\\
    &\le\frac{8n\mathfrak{E}_d}{|S_0|M}
    +\frac{2^{2+\delta}L\mathfrak{v}^{2+\delta}}{M^{2+\delta}}.
    \end{align*}
    Hence, a sufficient deterministic choice is
    \begin{equation}\label{eq:choice-of-M-winsorization}
    \begin{split}
    M&:=\frac{8(e^4-1)n\mathfrak{E}_d}{r_0-2\eta_nn-1} + 2\mathfrak{v}\left(
    \frac{(e^4-1)L|S_0|}{r_0-2\eta_nn-1}
    \right)^{1/(2+\delta)},
    \end{split}
    \end{equation}
    which satisfies~\eqref{eq:expectation-requirement-final}. Since $h=1+\log(3/\gamma)$ and $r_0=\lceil\lambda\eta_nn+h\rceil$,
    \begin{align*}
    \log(1/\gamma') = r_0-2\eta_nn-1&\ge(\lambda-2)\eta_nn+\log(3/\gamma)
    \ge\log(3/\gamma),\\
    r_0+2\eta_nn-1+\log(3/\gamma)
    &\le(\lambda+2)\eta_nn+2\log(3/\gamma)+1.
    \end{align*}
    Therefore, with probability at least $1-2\gamma/3$,
    \begin{align*}
    \frac{\mathbf I}{2}
    &\le\frac{8n}{|S_0|}\mathfrak{E}_d
    +\mathfrak{v}\sqrt{\frac{2\log(3/\gamma)}{|S_0|}}\\
    &\quad+\frac{(1+\delta)^{1+\delta}}{(2+\delta)^{2+\delta}}
    \frac{L\mathfrak{v}^{2+\delta}}{M^{1+\delta}}
    +\frac{8\log(3/\gamma)+2(r_0+2\eta_nn-1)}{|S_0|}M\\
    &\le C\Bigg[
    \left(1+\frac{r_0+2\eta_nn-1+\log(3/\gamma)}
    {r_0-2\eta_nn-1}\right)\mathfrak{E}_d
    +\mathfrak{v}\sqrt{\frac{\log(3/\gamma)}{n}}\\
    &\hspace{18mm}+\mathfrak{v}
    \frac{r_0+2\eta_nn-1+\log(3/\gamma)}
    {n^{(1+\delta)/(2+\delta)}
    (r_0-2\eta_nn-1)^{1/(2+\delta)}}\Bigg]\\
    &\le C\left[\mathfrak{E}_d
    +\mathfrak{v}\sqrt{\frac{\log(3/\gamma)}{n}}
    +\mathfrak{v}\left(\eta_n+\frac{\log(3/\gamma)}{n}\right)^{(1+\delta)/(2+\delta)}\right]\\
    &\le C\left[\mathfrak{E}_d
    +\mathfrak{v}\sqrt{\frac{\log(3/\gamma)}{n}}
    +\mathfrak{v}\eta_n^{(1+\delta)/(2+\delta)}\right].
    \end{align*}
    Here, we used $|S_0|\ge n/4$ and $\log(3/\gamma)\le r_0-2\eta_nn-1\le n$.

       Finally, the elementary inequalities $e^x-1\ge x$ and $e^x-1\ge x^2/2$ for $x\ge0$ imply
    \begin{align*}
    \frac{B(e^{4\log(3/\gamma)/(B-c^*_{B,\alpha}-1)}-1)}
    {\log(3/\gamma)}&\ge4,\\
    \frac{B^2(e^{4\log(3/\gamma)/(B-c^*_{B,\alpha}-1)}-1)}
    {\log(3/\gamma)}&\ge8\log(3/\gamma).
    \end{align*}
    Also, since $B\ge1$,
    \[
    \eta_n^{(1+\delta)/(2+\delta)}
    \le(B\eta_n)^{(1+\delta)/(2+\delta)}.
    \]
    Combining the bounds for $\mathbf I$ and $\mathbf{II}$, and taking a union bound over the three exceptional events, we conclude that with probability at least $1-\gamma$,
    \begin{align*}
    &\frac{1}{C}\sup_{\theta\in\cosici{n}{\alpha}{\mathtt{m}\mbox{-}\mathtt{rob}+}}\,\|\theta-\mu\|_\infty\\
    &\le\mathfrak{E}_d+
    \mathfrak{v}\sqrt{\frac{\log(3/\gamma)}{n}}
    +\mathfrak{v}\eta_n^{(1+\delta)/(2+\delta)}\\
    &\quad+\frac{B(e^{4\log(3/\gamma)/(B-c^*_{B,\alpha}-1)}-1)}
    {\log(3/\gamma)}
    \left[\mathfrak{E}_d+
    \mathfrak{v}(B\eta_n)^{(1+\delta)/(2+\delta)}\right]\\
    &\quad+\mathfrak{v}
    \sqrt{\frac{B(e^{4\log(3/\gamma)/(B-c^*_{B,\alpha}-1)}-1)}
    {\log(3/\gamma)}}
    \left[\sqrt{\frac Bn}+(B\eta_n)^{(1+\delta)/(2+\delta)}\right].
    \end{align*}
    The preceding elementary inequalities absorb the pilot terms and the square-root multiplier on the contamination term. Increasing $C$ if necessary, we obtain
    \begin{align*}
    &\frac{1}{C}\sup_{\theta\in\cosici{n}{\alpha}{\mathtt{m}\mbox{-}\mathtt{rob}+}}\,\|\theta-\mu\|_\infty\\
    &\le
    \frac{B(e^{4\log(3/\gamma)/(B-c^*_{B,\alpha}-1)}-1)}
    {\log(3/\gamma)}
    \left[\mathfrak{E}_d+
    \mathfrak{v}(B\eta_n)^{(1+\delta)/(2+\delta)}\right]\\
    &\quad+\frac{B\mathfrak{v}}{\sqrt n}
    \sqrt{\frac{e^{4\log(3/\gamma)/(B-c^*_{B,\alpha}-1)}-1}
    {\log(3/\gamma)}}.
    \end{align*}
    This completest the proof.
    
\section{Proof of Theorem~\ref{thm:metric-space-mean}}\label{appsec:proof-of-thm-metric-space-mean}
    For coverage, note that from Proposition~\ref{prop:poisson-binomial-to-binomial},
    \[
    \mathbb{P}(\mu\notin\cosicit{n}{\alpha}{\mp}\big|\vartheta) \le \alpha\times\exp(2B\Delta^{\mathfrak{m}}),
    \]
    where
    \[
    \Delta^{\mathfrak{m}} = \frac{1}{B}\sum_{j=1}^B \left(\frac{1}{2} - \mathbb{P}\left(\sum_{i\in S_j} \langle \log_{\mu}(X_i), \log_{\mu}(\vartheta)\rangle_{\mu} \le 0\big|\vartheta\right)\right)_+.
    \]
    Note from~\eqref{eq:mean-zero-Frechet-mean}, the summands $\langle \log_{\mu}(X_i), \log_{\mu}(\vartheta)\rangle_{\mu}, i\notin S_0$ are mean zero conditional on $\vartheta$. 
    From the univariate Berry--Esseen bound~\citep[Ineq. (1.5) with Corollary 2]{korolev2017bounds}, we obtain
    \[
    \Delta^{\mathfrak{m}} \le \frac{1}{B}\sum_{j=1}^B 1.87\frac{\sum_{i\in S_j}\mathbb{E}[|\langle \log_{\mu}(X_i), \log_{\mu}(\vartheta)\rangle_{\mu}|^{2 + \delta}]}{(\sum_{i\in S_j}\mathbb{E}[|\langle \log_{\mu}(X_i), \log_{\mu}(\vartheta)\rangle_{\mu}|^2])^{1 + \delta/2}} \le \frac{1}{B}\sum_{j=1}^B\frac{1.87L}{|S_j|^{\delta/2}}.
    \]
    Hence, from~\eqref{eq:split-sizes}, we conclude
    \[
    \mathbb{P}(\mu\notin\cosicit{n}{\alpha}{\mp}) \le \alpha\times\exp(5.3B^{1 + \delta/2}L/n^{\delta/2}).
    \]
    To prove the coverage guarantee for $\cosici{n}{\alpha}{\mp}$, observe that, by~\eqref{eq:tangent-cone-bracket} and non-negativity of $x\mapsto \kappa_{\mu}^{\vartheta}(x)$ for $x\in\{X_i, i\notin S_0\}$,
    \[
    \langle \log_{\mu}x,\,\log_{\mu}\vartheta\rangle_{\mu} ~=~ \frac{d^2(x, \mu) - d^2(x, \vartheta) + d^2(\mu, \vartheta)\kappa_{\mu}^{\vartheta}(x)}{2} \ge \frac{d^2(x, \mu) - d^2(x, \vartheta)}{2}.
    \]
    Therefore, 
    \[
    S_j(\mu) \ge R_j(\mu)\quad\mbox{for all}\quad 1\le j\le B.    
    \]
    Hence, $\max\{S_j(\mu), R_j(\mu)\} = S_j(\mu)$ and
    \begin{align*}
        \mathbb{P}(\mu\notin\cosici{n}{\alpha}{\mp}) &= \mathbb{P}\left(\sum_{j=1}^B \mathbf{1}\{\max\{S_j(\mu), R_j(\mu)\} \le 0\} < B - c^*_{B,\alpha}\right)\\
        &= \mathbb{P}\left(\sum_{j=1}^B \mathbf{1}\{S_j(\mu) \le 0\} < B - c^*_{B,\alpha}\right)\\
        &= \mathbb{P}(\mu\notin\cosicit{n}{\alpha}{\mp}).
    \end{align*}

    For the radius analysis, we first control the radius of $\cosicit{n}{\alpha}{\mp}$ under the assumption of non-negativity of $(\theta, x, y)\mapsto \kappa_{\theta}^y(x)$ over $\mathcal{X}^3$, and then we show that without this assumption, the radius of $\cosici{n}{\alpha}{\mp}$ can also be controlled by the same quantity. Note from~\eqref{eq:tangent-cone-bracket} that
    \begin{align*}
    \langle \log_{\theta}x,\, \log_{\theta}y\rangle_{\theta} &= \frac{d^2(x, \theta) - d^2(x, y) + d^2(\theta, y)\kappa_{\theta}^y(x)}{2},\\
    \langle \log_{\mu}x,\,\log_{\mu}\theta\rangle_{\mu} &= \frac{d^2(x, \mu) - d^2(x, \theta) + d^2(\mu, \theta)\kappa_{\mu}^{\theta}(x)}{2}.
    \end{align*}
    This yields
    \begin{equation}\label{eq:sum-of-inner-products-metric-space}
    \begin{split}
        &\langle \log_{\theta}x,\, \log_{\theta}y\rangle_{\theta} + \langle \log_{\mu}x,\,\log_{\mu}\theta\rangle_{\mu}\\
        &\quad= \frac{d^2(x, \mu) - d^2(x, y) + d^2(\theta, y)\kappa_{\theta}^y(x) + d^2(\mu, \theta)\kappa_{\mu}^{\theta}(x)}{2}. 
    \end{split}
    \end{equation}
    Set
    \[
    C_j := \left\{\theta:\,\sum_{i\in S_j} \langle \log_{\theta}(X_i), \log_{\theta}(\vartheta)\rangle_{\theta} \le 0\right\}.
    \]
    For any $\theta\in C_j$, we have
    \[
    \sum_{i\in S_j} \langle \log_{\theta}(X_i), \log_{\theta}(\vartheta)\rangle_{\theta} \le 0.
    \]
    Adding $\langle \log_{\mu}(X_i), \log_{\mu}(\theta)\rangle_{\mu}$ on both sides and applying~\eqref{eq:sum-of-inner-products-metric-space} (with $x = X_i, y = \vartheta, \theta = \theta$), we get for any $\theta\in C_j$,
    \[
    \sum_{i\in S_j}\frac{d^2(X_i, \mu) - d^2(X_i, \vartheta) + d^2(\theta, \vartheta)\kappa_{\theta}^{\vartheta}(X_i) + d^2(\mu, \theta)\kappa_{\mu}^{\theta}(X_i)}{2} \le \sum_{i\in S_j} \langle \log_{\mu}(X_i),\,\log_{\mu}(\theta)\rangle_{\mu}.
    \]
    Because $\kappa_{\theta}^y(x) \ge 0$ for all $x, y, \theta\in\mathcal{X}$, this implies
    \[
    \frac{d^2(\mu,\theta)}{2}\sum_{i\in S_j} \kappa_{\mu}^{\theta}(X_i) ~\le~ \sum_{i\in S_j} \langle \log_{\mu}(X_i), \log_{\mu}(\theta)\rangle_{\mu} + \frac{1}{2}\sum_{i\in S_j} [d^2(X_i, \vartheta) - d^2(X_i, \mu)].
    \]
    Because $\kappa_{\mu}^y(x) \ge \kappa_{\min} > 0$, this yields
    \[
    \frac{\kappa_{\min}}{2}d^2(\mu, \theta) \le \frac{1}{|S_j|}\sum_{i\in S_j} \langle \log_{\mu}(X_i), \log_{\mu}(\theta)\rangle_{\mu} + \frac{1}{2|S_j|}\sum_{i\in S_j} [d^2(X_i, \vartheta) - d^2(X_i, \mu)].
    \]
    Set, for $1\le j\le B$,
    \[
    f_1(S_j; \theta) := \frac{1}{|S_j|}\sum_{i\in S_j} \frac{\langle \log_{\mu}(X_i), \log_{\mu}(\theta)\rangle_{\mu}}{d(\mu, \theta)}\quad\mbox{and}\quad f_2(S_j) := \frac{1}{2|S_j|}\sum_{i\in S_j} [d^2(X_i, \vartheta) - d^2(X_i, \mu)]. 
    \]
    Then $\theta\in C_j$ implies
    \[
    \frac{\kappa_{\min}}{2}d^2(\mu, \theta) \le d(\mu, \theta)(f_1(S_j; \theta))_+ + (f_2(S_j))_+.    
    \]
    Solving the quadratic inequality, this further yields
    \begin{equation}\label{eq:superset-of-C-j}
    d(\mu, \theta) \le \frac{2(f_1(S_j; \theta))_+}{\kappa_{\min}} + \sqrt{\frac{2(f_2(S_j))_+}{\kappa_{\min}}}\quad\mbox{for all}\quad \theta\in C_j.
    \end{equation}
    Note that $\theta\in\cosicit{n}{\alpha}{\mp}$ if and only if there exists a set $\mathcal{J}_{\theta}\subseteq \{1, 2, \ldots, B\}$ with $|\mathcal{J}_{\theta}| \ge B - c^*_{B,\alpha}$ such that 
    \[
    \theta\in C_j\quad\mbox{for all}\quad j\in\mathcal{J}_{\theta}. 
    \]
    This implies that
    \[
    \theta\in\cosicit{n}{\alpha}{\mp}\quad\Rightarrow\quad d(\mu, \theta) \le \left(\frac{2(f_1(S_{\cdot}; \theta))_+}{\kappa_{\min}} + \sqrt{\frac{2(f_2(S_{\cdot}))_+}{\kappa_{\min}}}\right)_{(c^*_{B,\alpha}+1)},
    \]
    and hence,
    \begin{equation}\label{eq:bound-on-cosi-tilde-metric-space}
    \sup_{\theta\in\cosicit{n}{\alpha}{\mp}}\,d(\theta,\mu) 
    \le \mathcal{R}_{n,\alpha} := \sup_{\theta\in\mathcal{X}}\,\left(\frac{2(f_1(S_{\cdot}; \theta))_+}{\kappa_{\min}} + \sqrt{\frac{2(f_2(S_{\cdot}))_+}{\kappa_{\min}}}\right)_{(c^*_{B,\alpha}+1)}.
    \end{equation}
    (We assume that the supremum on the right can be written as a supremum over countable dense subset.)
    Before controlling the right hand side of~\eqref{eq:bound-on-cosi-tilde-metric-space}, we show that without assuming non-negativity of $(\theta, x, y)\mapsto\kappa_{\theta}^y(x)$, $\cosici{n}{\alpha}{\mp}$ satisfies the same bound. Note that $\theta\in\cosici{n}{\alpha}{\mp}$ if and only if there exists $\mathcal{J}_{\theta}\subseteq\{1, 2, \ldots, B\}$ with $|\mathcal{J}_{\theta}| \ge B - c^*_{B,\alpha}$ such that for all $j\in\mathcal{J}_{\theta}$,
    \[
    \sum_{i\in S_j}\,\langle\log_{\theta}(X_i),\log_{\theta}(\vartheta)\rangle_{\theta} \le 0\quad\mbox{and}\quad \sum_{i\in S_j} [d^2(X_i, \theta) - d^2(X_i, \vartheta)] \le 0.
    \]
    From~\eqref{eq:tangent-cone-bracket}, the second inequality is equivalent to
    \begin{align*}
    &\frac{1}{2}\sum_{i\in S_j} [d^2(X_i, \theta) - d^2(X_i, \mu)] \le \frac{1}{2}\sum_{i\in S_j} [d^2(X_i, \vartheta) - d^2(X_i, \mu)]\\
    &\equiv\quad \frac{1}{2}\sum_{i\in S_j}\kappa_{\mu}^{\theta}(X_i)d^2(\mu, \theta) - \sum_{i\in S_j}\,\langle\log_{\mu}(X_i), \log_{\mu}(\theta)\rangle_{\mu} \le \frac{1}{2}\sum_{i\in S_j} [d^2(X_i, \vartheta) - d^2(X_i, \mu)].
    \end{align*}
    Hence, $\theta\in\cosici{n}{\alpha}{\mp}$ if and only if for all $j\in\mathcal{J}_{\theta}$,
    \[
    d^2(\theta, \mu) \le d(\theta,\mu)\frac{2}{\kappa_{\min}}\sum_{i\in S_j}\frac{\langle\log_{\mu}(X_i),\,\log_{\mu}(\vartheta)\rangle_{\mu}}{d(\theta,\mu)} + \frac{1}{\kappa_{\min}}\sum_{i\in S_j} [d^2(X_i, \vartheta) -d^2(X_i, \mu)].
    \]
    As before solving the quadratic inequality and taking the supreumum over all $\theta\in\mathcal{X}$, yields
    \[
    \sup_{\theta\in\cosici{n}{\alpha}{\mp}}\,d(\theta,\mu) \le \mathcal{R}_{n,\alpha}.
    \]
    Note that for this inequality and also~\eqref{eq:bound-on-cosi-tilde-metric-space}, it suffices to have
    \[
    \inf_{\theta\in\mathcal{X}}\frac{1}{|S_j|}\sum_{i\in S_j} \kappa_{\mu}^{\theta}(X_i) \ge \kappa_{\min} > 0.
    \]
    Hence, these radius bounds hold true on the event 
    \[
    \mathcal{E}_n := \left\{\frac{1}{|S_j|}\sum_{i\in S_j} \inf_{\theta\in\mathcal{X}}\,\kappa_{\mu}^{\theta}(X_i) \ge \kappa_{\min}\right\}.
    \]
    This proves the remark after Theorem~\ref{thm:metric-space-mean}. In the following, we control $\mathcal{R}_{n,\alpha}$ assuming $\mathcal{E}_n$ holds.
    
    Applying Theorem~\ref{thm:supremum-of-quantiles} (conditional on $\vartheta$), we get with probability at least $1 - \gamma$,
    \[
    \mathcal{R}_{n,\alpha} \le \frac{4B(e^{4\log(1/\gamma)/(B-c^*_{B,\alpha} - 1)} - 1)}{\log(1/\gamma)}\mathfrak{R}(\mathcal{X}) + 2\sigma\sqrt{\frac{B(e^{4\log(1/\gamma)/(B-c^*_{B,\alpha} - 1)} - 1)}{\log(1/\gamma)}},
    \]
    where
    \begin{align*}
        \mathfrak{R}(\mathcal{X}) &:= \mathbb{E}\left[\sup_{\theta\in\mathcal{X}}\,\frac{1}{B}\sum_{j=1}^B \varepsilon_j\left(\frac{2(f_1(S_{j}; \theta))_+}{\kappa_{\min}} + \sqrt{\frac{2(f_2(S_{j}))_+}{\kappa_{\min}}}\right)\big|\vartheta\right],\\
        \sigma^2 &:= \sup_{\theta\in\mathcal{X}}\, \frac{1}{B}\sum_{j=1}^B \mathbb{E}\left[\left(\frac{2(f_1(S_{j}; \theta))_+}{\kappa_{\min}} + \sqrt{\frac{2(f_2(S_{j}))_+}{\kappa_{\min}}}\right)^2\big|\vartheta\right].
    \end{align*}
    To control these quantities in terms of global variance and weak variance, we need Hilbert subspace at the Fr{\'e}chet mean. This is offered by Theorem 2.4 of~\cite{le2022fast}. Formally, let $P_i$ be the probability measure of $X_i$ and define the mixture distribution
    \[
    \overline{P}:=\frac{1}{n}\sum_{i=1}^n P_i.
    \]
    Since $\mu$ is a Fr{\'e}chet mean of every $P_i$, it is also a Fr{\'e}chet mean of $\overline{P}$. Moreover,
    \[
    \operatorname{supp}(P_i)
    \subseteq
    \operatorname{supp}(\overline{P}),\quad\mbox{for}
    \quad 1\le i\le n.
    \]
    Theorem 2.4(3), applied to $\overline{P}$, therefore yields a Hilbert subspace $L_\mu\mathcal{X}\subseteq T_\mu\mathcal{X}$ such that
    \[
    \log_\mu\bigl(\operatorname{supp}(P_i)\bigr)
    \subseteq L_\mu\mathcal{X},
    \quad\mbox{for}\quad 1\le i\le n.
    \]
    Furthermore, applying the first-order optimality condition in Theorem 2.4(4) of~\cite{le2022fast} to each $P_i$ gives
    \[
    \mathbb{E}[\log_\mu(X_i)]=\log_{\mu}\mu =: o_\mu,
    \quad\mbox{for}\quad 1\le i\le n,
    \]
    where the expectations are understood as Pettis integrals in
    $L_\mu\mathcal{X}$. Consequently, $\log_\mu(X_i)$, $1\le i\le n$ may be regarded as independent, mean-zero random elements of the common Hilbert space $L_\mu\mathcal{X}$.
    Because $\mathbb{E}[\langle\log_{\mu}(X_i), \log_{\mu}(\theta)\rangle_{\mu}] = 0$, from~\citet[Lemma 8]{mohri2014learning} and the proof of Lemma 4.14 of~\cite{bartl2025robust}, we have
    \begin{align*}
    \mathfrak{R}(\mathcal{X}) &= \mathbb{E}\left[\sup_{\theta\in\mathcal{X}}\,\frac{1}{B}\sum_{j=1}^B \varepsilon_j\left(\frac{2(f_1(S_{j}; \theta))_+}{\kappa_{\min}}\right)\right]\\
    &\le \frac{2}{\kappa_{\min}}\mathbb{E}\left[\sup_{\theta\in\mathcal{X}}\,\frac{1}{B}\sum_{j=1}^B \varepsilon_jf_1(S_j; \theta)\right]\\
    &\le \frac{2}{\kappa_{\min}}\mathbb{E}\left[\sup_{\theta\in\mathcal{X}}\,\frac{1}{B}\sum_{j=1}^B \frac{\varepsilon_j}{|S_j|}\sum_{i\in S_j} \frac{\langle \log_{\mu}(X_i), \log_{\mu}(\theta)\rangle_{\mu}}{d(\mu, \theta)}\right]\\
    &\le \frac{2}{\kappa_{\min}}\left(\frac{1}{B^2}\sum_{j=1}^B \frac{1}{|S_j|^2}\sum_{i\in S_j}\mathbb{E}[d^2(X_i, \mu)]\right)^{1/2}\\
    &\le \frac{4}{\kappa_{\min}}\left(\frac{1}{n^2}\sum_{j=1}^B\sum_{i\in S_j} \mathbb{E}[d^2(X_i, \mu)]\right)^{1/2}\\
    &\le \frac{4}{\kappa_{\min}\sqrt{n}}\left(\frac{1}{n}\sum_{i=1}^n \mathbb{E}[d^2(X_i, \mu)]\right)^{1/2}.
    \end{align*}
    To control $\sigma^2$, observe by the mean zero property of ``inner products'' that
    \begin{equation}\label{eq:weak-variance-control-metric-space}
    \begin{split}
    \sigma^2 &\le \sup_{\theta\in\mathcal{X}}\,\frac{2}{B}\sum_{j=1}^B \mathbb{E}\left[\frac{4f_1^2(S_{j}; \theta)}{\kappa_{\min}^2} + \frac{2(f_2(S_j))_+}{\kappa_{\min}}\big|\vartheta\right]\\
    &\le \sup_{\theta\in\mathcal{X}}\,\frac{8}{B\kappa_{\min}^2}\sum_{j=1}^B \left(\frac{1}{|S_j|^2}\sum_{i\in S_j}\frac{\mathbb{E}[|\langle \log_{\mu}(X_i),\,\log_{\mu}(\theta)\rangle_{\mu}|^2]}{d^2(\mu, \theta)}\right) + \frac{4}{B\kappa_{\min}}\sum_{j=1}^B \mathbb{E}[(f_2(S_j))_+\big|\vartheta].
    \end{split}
    \end{equation}
    For the first term in~\eqref{eq:weak-variance-control-metric-space}, we have from~\eqref{eq:split-sizes} that
    \[
    \sup_{\theta\in\mathcal{X}}\,\frac{1}{B}\sum_{j=1}^B \left(\frac{1}{|S_j|^2}\sum_{i\in S_j}\frac{\mathbb{E}[|\langle \log_{\mu}(X_i),\,\log_{\mu}(\theta)\rangle_{\mu}|^2]}{d^2(\mu, \theta)}\right) \le \frac{4B\sigma_{w,n}^2}{n}.
    \]
    For the second term in~\eqref{eq:weak-variance-control-metric-space}, note from the definition of inner product~\eqref{eq:tangent-cone-bracket} that
    \[
    d^2(x, \vartheta) - d^2(x, \mu) = \kappa_{\mu}^{\vartheta}(x)d^2(\mu, \vartheta) - 2\langle\log_{\mu}(x),\,\log_{\mu}(\vartheta)\rangle_{\mu}\quad\mbox{for all}\quad x\in\mathcal{X}.
    \]
    This implies
    \[
    f_2(S_j) = d^2(\mu, \vartheta)\frac{1}{2|S_j|}\sum_{i\in S_j}\kappa_{\mu}^{\vartheta}(X_i) - \frac{1}{|S_j|}\sum_{i\in S_j}\, \langle\log_{\mu}(X_i),\,\log_{\mu}(\vartheta)\rangle_{\mu}\quad\mbox{for}\quad 1\le j\le B.
    \]
    By sub-additivity of $x\mapsto\max\{x, 0\}$ and non-negativity of $x\mapsto \kappa_{\mu}^{\vartheta}(x)$, we have
    \[
    (f_2(S_j))_+ \le d^2(\mu, \vartheta)\frac{1}{2|S_j|}\sum_{i\in S_j} \kappa_{\mu}^{\vartheta}(X_i) + \left(\frac{-1}{|S_j|}\sum_{i\in S_j}\,\langle\log_{\mu}(X_i),\,\log_{\mu}(\vartheta)\rangle_{\mu}\right)_+.
    \]
    Hence,
    \begin{align*}
    \frac{1}{B}\sum_{j=1}^B \mathbb{E}[(f_2(S_j))_+|\vartheta] &\le d^2(\mu, \vartheta)\frac{1}{B}\sum_{j=1}^B \frac{1}{2|S_j|}\sum_{i\in S_j} \mathbb{E}[\kappa_{\mu}^{\vartheta}(X_i)|\vartheta]\\ 
    &\quad+ \frac{1}{B}\sum_{j=1}^B \mathbb{E}\left[\left(\frac{-1}{|S_j|}\sum_{i\in S_j}\,\langle\log_{\mu}(X_i),\,\log_{\mu}(\vartheta)\rangle_{\mu}\right)_+|\vartheta\right]\\
    &\le \frac{d^2(\mu, \vartheta)}{2}e_{\kappa,n}\\ 
    &\quad+ \frac{1}{2B}\sum_{j=1}^B \left(\frac{1}{|S_j|^2}\sum_{i\in S_j} \mathbb{E}[|\langle\log_{\mu}(X_i), \log_{\mu}(\vartheta)\rangle_{\mu}|^2|\vartheta]\right)^{1/2}\\
    &\le \frac{d^2(\mu, \vartheta)e_{\kappa,n}}{2} + d(\mu,\vartheta)\sigma_{w,n}\sqrt{\frac{B}{n}}.
    \end{align*}
    The second inequality here follows from $E[(W)_+] = (1/2)\mathbb{E}[|W|] \le (1/2)\sqrt{\mathbb{E}[W^2]}$ for a mean zero random variable $W$.
    This implies that
    \begin{align*}
    \sigma^2 &\le \frac{32B\sigma_{w,n}^2}{n\kappa_{\min}^2} + \frac{2d^2(\mu, \vartheta)e_{\kappa,n}}{\kappa_{\min}} + \frac{4d(\mu,\vartheta)\sigma_{w,n}}{\kappa_{\min}}\sqrt{\frac{B}{n}}\\
    &\le \frac{34B\sigma^2_{w,n}}{n\kappa_{\min}^2} + 2d^2(\mu,\vartheta)\left(e_{\kappa,n}/\kappa_{\min} + 1\right).
    \end{align*}
    Therefore, with probability at least $1 - \gamma$,
    \begin{align*}
    \mathcal{R}_{n,\alpha}
    &\le~ \frac{16B(e^{4\log(1/\gamma)/(B-c^*_{B,\alpha} - 1)} - 1)}{\log(1/\gamma)}\sqrt{\frac{V_{2,n}}{n\kappa_{\min}^2}}\\ 
    &\quad+ \sqrt{\frac{B(e^{4\log(1/\gamma)/(B-c^*_{B,\alpha} - 1)} - 1)}{\log(1/\gamma)}}\left(20\sqrt{\frac{B\sigma_{w,n}^2}{n\kappa_{\min}^2}} + 2d(\mu,\vartheta)(e_{\kappa,n}/\kappa_{\min}+1)^{1/2}\right).
    \end{align*}
    This completes the proof.
    
    \section{Distance Characterization for Central and Angular Symmetry}\label{appsec:proof-dist-charact-symmetry}
    \begin{proposition}\label{prop:dist-charact-symmetry}
    Suppose $W$ is a random variable in an arbitrary measurable space, and $\mathfrak{P}$ is the scale-invariant property of central symmetry, then setting $\mathcal{C}$ as the set of all Borel measurable sets, we have
    \[
    \mathrm{dist}_{\mathcal{C}}(W,\,\mathfrak{P}) ~=~ \frac{1}{2}\mathrm{TV}(W,\,-W),
    \]
    Alternatively, if $\mathfrak{P}$ is the scale-invariant property of angular symmetry, then
    \[
    \mathrm{dist}_{\mathcal{C}}(W,\,\mathfrak{P}) ~=~ \frac{1}{2}\mathrm{TV}\left(\frac{W}{\|W\|},\,-\frac{W}{\|W\|}\right).
    \]
\end{proposition}
    \begin{proof}[Proof of Proposition~\ref{prop:dist-charact-symmetry}]
    From the definition of $\mathcal{C}$, we have
    \[
    \mbox{dist}_{\mathcal{C}}(W,\,\mathfrak{P}) = \inf_{L\in\mathcal{Q}_{\mathfrak{P}}}\, \mbox{TV}(W,\, L).
    \]
    Note that for any $L$ that is centrally symmetric (i.e., $L \overset{d}{=} -L$), we have
    \begin{align*}
    \mbox{TV}(W,\, -W) &\le \mbox{TV}(W,\, L) + \mbox{TV}(-W,\,L)\\
    &= \mbox{TV}(W,\, L) + \mbox{TV}(-W,\,-L)\\
    &= 2\mbox{TV}(W,\,L).
    \end{align*}
    Therefore, $\mbox{dist}_{\mathcal{C}}(W,\,\mathfrak{P}) \ge (1/2)\mbox{TV}(W,\,-W)$. To prove the upper bound, define
    \[
    \mu_W(A) = \mathbb{P}(W \in A),\quad\mbox{and}\quad \begin{cases}
    \nu(A) = \min\{\mu_W(A),\,\mu_{W}(-A)\},\\
    \sigma_+(A) = \mu_{W}(A) - \nu(A).
    \end{cases}
    \]
    Clearly, $\sigma_+(A) \ge 0$ for all $A$ and $\nu(\mathcal{X}) \le \mu_{W}(\mathcal{X}) = 1$. The measure $\nu$ is the largest centrally symmetric distribution below the given distribution of $W$, satisfying $\nu(A) = \nu(-A)$ for all $A$. Setting $\kappa = \nu(\mathcal{X})$. Let $X$ be a random variable with probability measure $\nu(\cdot)/\kappa$ and $X'$ be a random variable with probability measure $\sigma_+(\cdot)/(1-\kappa)$. Set $U$ and $U'$ as two IID copies of a standard uniform random variable. Define random variable $W'$ as follows:
    \[
    \begin{cases}
    W' = W = X, &\mbox{if }U \le \kappa,\\
    W = X', \quad\mbox{and}\quad W' = \mbox{sign}(2U' - 1)X', &\mbox{if }U > \kappa.
    \end{cases}
    \]
    Note that
    \begin{align*}
    \mathbb{P}(W \in A) &= \kappa\mathbb{P}(X \in A) + (1-\kappa)\mathbb{P}(X' \in A)\\ 
    &= \kappa \frac{\nu(A)}{\kappa} + (1-\kappa) \frac{\sigma_+(A)}{1-\kappa} = \nu(A) + \sigma_+(A) = \mu_W(A). 
    \end{align*}
    Moreover, 
    \begin{align*}
        \mathbb{P}(W'\in A) &= \kappa\mathbb{P}(X\in A) + \frac{1-\kappa}{2}\mathbb{P}(-X'\in A) + \frac{1-\kappa}{2}\mathbb{P}(X' \in A)\\
        &= \kappa\nu(A) + \frac{1}{2}\sigma_+(-A) + \frac{1}{2}\sigma_+(A).
    \end{align*}
    This implies that $\mathbb{P}(W'\in A) = \mathbb{P}(W'\in -A)$ and hence, $W'$ is centrally symmetric. It also follows that
    \[
    \mathbb{P}(W\in A) - \mathbb{P}(W'\in A) = \frac{1}{2}(\sigma_+(A) - \sigma_+(-A)) = \frac{1}{2}(\mu_W(A) - \mu_W(-A)).
    \]
    Hence, 
    \[
    \mbox{TV}(W,\, W') = \frac{1}{2}\mbox{TV}(W,\,-W),
    \]
    which proves the result.

    To prove the result when $\mathfrak{P}$ is the property of angular symmetry, fix any angularly symmetric random variable $L\in \mathcal{Q}_{\mathfrak{P}}$.
    The TV distance only decreases by transforming random variables and hence,
    \[
    \mbox{TV}(W,\,L) \ge \mbox{TV}\left(\frac{W}{\|W\|},\, \frac{L}{\|L\|}\right).
    \]
    Because $L/\|L\|$ is centrally symmetric, following the previous argument for central symmetry, we conclude
    \[
    \mbox{TV}\left(\frac{W}{\|W\|},\, \frac{L}{\|L\|}\right) \ge \frac{1}{2}\mbox{TV}\left(\frac{W}{\|W\|},\,-\frac{W}{\|W\|}\right)\quad\Rightarrow\quad \mbox{dist}_{\mathcal{C}}(W,\,\mathfrak{P}) \ge \frac{1}{2}\mbox{TV}\left(\frac{W}{\|W\|},\,-\frac{W}{\|W\|}\right).
    \]
    To prove the other direction, define for any $r > 0$
    \[
    \mu_r(A) = \mathbb{P}(W/\|W\| \in A\big|\|W\| = r),\quad\mbox{and}\quad\begin{cases}\nu_r(A) = \min\{\mu_r(A), \mu_r(-A)\},\\
    \sigma_r(A) = \mu_r(A) - \nu_r(A).\end{cases}
    \]
    Generate $W$ and $W'$ as follows: Let $U$ and $U'$ be two IID copies of the standard uniform random variable.
    \begin{enumerate}
        \item Generate $R$ from the distribution of $\|W\|$. Let the realized value of $R$ be $r$.
        \item If $U \le \kappa$, generate $S = S' = X$ from the probability measure $\nu_r(\cdot)/\nu_r(\{x\in\mathcal{X}:\,\|x\| = 1\})$.
        \item If $U > \kappa$, generate $X'$ from $\sigma_r(\cdot)/\sigma_r(\{x\in\mathcal{X}:\,\|x\| = 1\})$ and set $S = X'$, $S' = \mbox{sign}(2U' - 1)X'$.
        \item Set $W = RS$ and $W' = RS'$.
    \end{enumerate}
    Clearly, $W'$ is angularly symmetric, and
    \[
    \mathbb{P}(W \neq W') = \mathbb{P}(S \neq S') = \frac{1}{2}\mbox{TV}(S,\, S') = \frac{1}{2}\mbox{TV}\left(\frac{W}{\|W\|},\, -\frac{W}{\|W\|}\right).
    \]
    This proves the result.
    \end{proof}

\section{Proof of Theorem~\ref{thm:beta-dependent-COSI-validity}}\label{appsec:proof-of-beta-dependent-COSI-validity}
    We shall first prove that $\widetilde{\theta}_j, 1\le j\le B$ can be defined in such a way that
    \begin{equation}\label{eq:TV-beta-mixing}
    \mbox{TV}((\widehat{\theta}_1,\ldots,\widehat{\theta}_B),\, (\widetilde{\theta}_1, \ldots, \widetilde{\theta}_B)) \le (B-1)\beta(k_n). 
    \end{equation}
    The result then follows from Theorem~\ref{thm:finite-sample-validity} by noting that
    \begin{align*}
    \mathbb{P}_{P^n}\left(\theta(P^n)\notin\cosicit{n}{\alpha}{dep\mbox{-}a}\right) &= \mathbb{E}_{P^n}\left[\phi_{\mathfrak{P}}(\{\widehat{\theta}_j - \theta(P^n)\}_{j=1}^B)\right]\\
    &\le \mathbb{E}_{P^n}\left[\phi_{\mathfrak{P}}(\{\widetilde{\theta}_j - \theta(P^n)\}_{j=1}^B)\right] + \mbox{TV}((\widehat{\theta}_1,\ldots,\widehat{\theta}_B),\, (\widetilde{\theta}_1, \ldots, \widetilde{\theta}_B))\\
    &\le\sup_{P^n\in\mathcal{P}^n}\mathbb{P}_{P^n}\left(\theta(P^n) \notin \cosicit{n}{\alpha}{dep\mbox{-}a}\right) + (B-1)\beta(k_n).
    \end{align*}
    Similar calculations imply the result for $\cosi{dep\mbox{-}m}$ confidence set.

    Inequality~\eqref{eq:TV-beta-mixing} is a standard result which can be proved by an induction on $B$ as in Lemma 2 of~\cite{eberlein1984weak} along with Berbee's lemma. For $B = 2$, from Lemma 5.1 of~\cite{rio2017asymptotic}, inequality~\eqref{eq:TV-beta-mixing} follows. For $B\ge 2$, the induction argument as in Lemma 2 of~\cite{eberlein1984weak} implies~\eqref{eq:TV-beta-mixing}.

\section{Proof of Theorem~\ref{thm:HulC-strong-mixing}}\label{appsec:proof-of-Theorem-HulC-strong-mixing}
    By definition, 
    \begin{align*}
    &\left\{\theta(P)\notin\widehat{\mathrm{CI}}_{n,\alpha}^{\mathtt{HulC}\mbox{-}\mathtt{dep}}\right\}\\ 
    &\subseteq \left\{\sum_{j=1}^B \mathbf{1}\{\widehat{\theta}_j \le \theta(P^n)\} < \lfloor B/2\rfloor - c_{B,\alpha}\right\}\bigcup\left\{\sum_{j=1}^B \mathbf{1}\{\widehat{\theta}_j \ge \theta(P^n)\} < \lfloor B/2\rfloor - c_{B,\alpha}\right\}.
    \end{align*}
    We now apply Theorem 2 of~\cite{bradley1983approximation}. Define
    \[
    W_1 = \sum_{j=1}^{B-1}\begin{pmatrix} \mathbf{1}\{\widehat{\theta}_j \le \theta(P^n)\}\\\mathbf{1}\{\widehat{\theta}_j \ge \theta(P^n)\}\end{pmatrix},\quad\mbox{and}\quad W(\overline{\theta}) = \begin{pmatrix} \mathbf{1}\{\overline{\theta} \le \theta(P^n)\}\\\mathbf{1}\{\overline{\theta} \ge \theta(P^n)\}\end{pmatrix}.
    \]
    With the partition $\mathcal{H} = \{(-\infty,\,\theta(P^n)),\,\{\theta(P^n)\},\, (\theta(P^n),\,\infty)\}$ of $\mathbb{R}$, Theorem 2 of~\cite{bradley1983approximation} implies that we can construct $\widetilde{\theta}_{B} \overset{d}{=} \widehat{\theta}_B$ independent of $W_1$, and $\mathbb{P}(W(\widehat{\theta}_B)\neq W(\widetilde{\theta}_B)) \le 3\alpha(k_n)$. (Here, we use 3 instead of $\sqrt{24}$, which is valid given the comments after Theorem 1 of~\cite{bradley1983approximation}. For our partition, $N \le \sqrt{8N}$.)
    Therefore, setting $e_1 = (1, 0)^{\top}$ and $e_2 = (0, 1)^{\top}$, we get
    \begin{align*}
        &\mathbb{P}_{P^n}(\theta(P^n) \notin \widehat{\mathrm{CI}}_{n,\alpha}^{\mathtt{HulC}\mbox{-}\mathtt{dep}})\\
        &\le \mathbb{P}\left(\left\{\max_{l \in \{1, 2\}}\,e_l^{\top}W_1 + e_l^{\top}W(\widehat{\theta}_B) < \lfloor B/2\rfloor - c_{B,\alpha}\right\}\cap\{W(\widehat{\theta}_B) = W(\widetilde{\theta}_B)\}\right)
        + \mathbb{P}(W(\widehat{\theta}_B) \neq W(\widetilde{\theta}_B))\\
        &\le \mathbb{P}\left(\left\{\max_{l \in \{1, 2\}}\,e_l^{\top}W_1 + e_l^{\top}W(\widetilde{\theta}_B) < \lfloor B/2\rfloor - c_{B,\alpha}\right\}\right)
        + 3\alpha(k_n).
    \end{align*}
    Repeating this procedure, we get mutually independent estimators $\widetilde{\theta}_j, 1\le j\le B$ such that $\widetilde{\theta}_j \overset{d}{=} \widehat{\theta}_j$ and 
    \begin{align*}
        &\mathbb{P}_{P^n}\left(\theta(P^n)\notin\widehat{\mathrm{CI}}_{n,\alpha}^{\mathtt{HulC}\mbox{-}\mathtt{dep}}\right)\\
        &\le \mathbb{P}_{P^n}\left(\left\{\max_{l\in\{1, 2\}}\sum_{j=1}^B e_l^{\top}W(\widetilde{\theta}_j) < \lfloor B/2\rfloor - c_{B,\alpha}\right\}\right) + 3(B-1)\alpha(k_n).
    \end{align*}
    The first probability on the right is the miscoverage of the HulC interval constructed for independent data with $B$ splits. Hence, the result follows from Proposition~\ref{prop:binomial-probabilities}.

\section{Proof of Theorem~\ref{thm:coverage-of-LM}}\label{appsec:proof-of-coverage-of-LM}
    
    Let $U_1, \ldots, U_B$ be independent standard uniform random variables and set $R_j = \mathbf{1}\{U_j \le p_j(\vartheta)\}$ so that $R_1, \ldots, R_B$ are mutually independent Bernoulli random variables conditional on $\vartheta$ and 
    \begin{align*}
    \mathbb{P}_P(\theta(P)\notin\widehat{\mathrm{CI}}_{n,\alpha}^{\mathtt{LM}}) &= \mathbb{P}\left(\sum_{i=1}^B R_j \le \frac{B}{2}\right)\\
    &= \mathbb{E}\left[\mathbb{P}\left(\sum_{j=1}^B R_j \le \frac{B}{2}\middle|\overline{P}(\vartheta)\right)\right]\\
    &\le \mathbb{P}(\overline{P}(\vartheta) < 1/2) + \mathbb{E}\left[\mathbb{P}\left(\sum_{j=1}^B R_j \le \frac{B}{2}\middle|\overline{P}(\vartheta)\right)\mathbf{1}\{\overline{P}(\vartheta) \ge 1/2\}\right].
    \end{align*}
    By the Hoeffding comparison theorem for Poisson-Binomial distribution~\citep[Theorem 2.1]{tang2023poisson}, we conclude that
    \begin{align*}
    \mathbb{P}\left(\sum_{j=1}^B R_j \le \frac{B}{2} \middle| \overline{P}(\vartheta)\right)
    \mathbf{1}\{\overline{P}(\vartheta) \ge 1/2\} &\le \mathbb{P}\left(\mbox{Binom}(B, \overline{P}(\vartheta)) \le B/2\middle|\overline{P}(\vartheta)\right)\mathbf{1}\{\overline{P}(\vartheta) \ge 1/2\}\\
    &\le \mathbb{P}(\mbox{Binom}(B, \max\{1/2, \overline{P}(\vartheta)\}) \le B/2|\overline{P}(\vartheta)).
    \end{align*}
    Taking expectation on both sides, completes the proof.
\section{Proof of Corollary~\ref{cor:coverage-of-LM-asym-normality}}\label{appsec:proof-coverage-LM-asymp-normality}
    From the definition of $\Delta$, we get that
    \begin{equation}\label{eq:implication-of-Delta}
    \left|\overline{P}(\vartheta) - \left(1 - \Phi\left(-\frac{r_m\|\theta(P) - \vartheta\|^2}{2\|\theta(P) - \vartheta\|_{\Sigma}}\right)\right)\right| \le \Delta.
    \end{equation}
    Set
    \[
    \|\lambda\|_{\Sigma} = \sqrt{\langle\lambda,\,\Sigma\lambda\rangle},\quad\mbox{and}\quad\|\Sigma\|_{\mathrm{op}}^{1/2} = \sup_{\lambda\in\mathbb{H}}\, \frac{\|\lambda\|_{\Sigma}}{\|\lambda\|},
    \]
    and for $\varepsilon > 0$, let $\lambda_{\varepsilon}\in\mathbb{H}$ be such that
    \[
    \|\Sigma\|_{\mathrm{op}}^{1/2}(1 - \varepsilon) \le \frac{\|\lambda_{\varepsilon}\|_{\Sigma}}{\|\lambda_{\varepsilon}\|} \le \|\Sigma\|_{\mathrm{op}}^{1/2}.
    \]
    Then, we conclude for any $\theta\in\mathbb{H}$
    \[
    \frac{\|\theta - \vartheta\|^2}{\|\theta - \vartheta\|_{\Sigma}} \ge \frac{\|\theta - \vartheta\|^2}{\|\Sigma\|_{\mathrm{op}}^{1/2}\|\theta - \vartheta\|} \ge \frac{|\langle \lambda_{\varepsilon}, \theta - \vartheta\rangle|}{\|\lambda_{\varepsilon}\|\|\Sigma\|_{\mathrm{op}}^{1/2}} \ge (1-\varepsilon)\frac{|\langle \lambda_{\varepsilon}, \theta - \vartheta\rangle|}{\|\lambda_{\varepsilon}\|_{\Sigma}},
    \]
    and hence, (using $1 - \Phi(-a) = \Phi(a)$),
    \[
    \overline{P}(\vartheta) ~\ge~ \Phi\left(\frac{(1-\varepsilon)}{2}\frac{r_m|\langle\lambda_{\varepsilon},\,\theta(P) - \vartheta\rangle|}{\|\lambda_{\varepsilon}\|_{\Sigma}}\right) ~-~ \Delta.
    \]
    This along with~\eqref{eq:implication-of-Delta} implies the first result.
    
    For the second result, note from the definition of $\Delta'$ that
    \begin{align*}
    \mathbb{P}_P(\overline{P}(\vartheta) < 1/2) &\le \mathbb{P}\left(\frac{(1-\varepsilon)}{2}\frac{r_m|\langle\lambda_{\varepsilon},\theta(P)-\vartheta\rangle|}{\|\lambda_{\varepsilon}\|_{\Sigma}} \le \Phi^{-1}(1/2 + \Delta)\right)
    \\
    &\le \mathbb{P}\left(\frac{(1-\varepsilon)}{2}\frac{r_m|Z|}{r_{km}} \le \Phi^{-1}(1/2 + \Delta)\right) + 2\Delta'.
    \end{align*}
    Using the fact that $\mathbb{P}(|Z| \le cx) \le \max\{1,c\}\mathbb{P}(|Z| \le x)$, and letting $\varepsilon\downarrow0$, we conclude
    \begin{equation}\label{eq:first-term-coverage-LM}
       \mathbb{P}_P(\overline{P}(\vartheta) < 1/2) \le \max\left\{1, \frac{2r_{km}}{r_m}\right\}\Delta + 2\Delta'.
    \end{equation}
    Additionally, we have
    \begin{align*}
        &\mathbb{P}(\mbox{Binom}(B, \max\{1/2, \overline{P}(\vartheta)\}) \le B/2)\\
        &\le \mathbb{P}\left(\mbox{Binom}\left(B, \max\left\{\frac{1}{2},\, \Phi\left(\frac{(1-\varepsilon)}{2}\frac{r_m|\langle\lambda_{\varepsilon},\,\theta(P)-\vartheta\rangle|}{\|\lambda_{\varepsilon}\|_{\Sigma}}\right) - \Delta\right\}\right) \le \frac{B}{2}\right).
    \end{align*}
    It is easy to see that
    \begin{align*}
    &\mathrm{KS}\left(\max\left\{\frac{1}{2},\, \Phi\left(\frac{(1-\varepsilon)}{2}\frac{r_m|\langle\lambda_{\varepsilon},\,\theta(P)-\vartheta\rangle|}{\|\lambda_{\varepsilon}\|_{\Sigma}}\right) - \Delta\right\},\, \max\left\{\frac{1}{2},\, \Phi\left(\frac{(1-\varepsilon)}{2}\frac{r_m|Z|}{r_{km}}\right) - \Delta\right\}\right)\\ 
    ~&\le~ 2\sup_{\lambda\in\mathbb{H}}\mathrm{KS}\left(\frac{\langle\lambda,\,\theta(P) - \vartheta\rangle}{\|\lambda\|_{\Sigma}},\, Z\right) = 2\Delta'. 
    \end{align*}
    Hence, from Eqns. (2.1)--(2.4) of~\cite{hall1986number}, we get
    \begin{align*}
        &\mathbb{P}(\mbox{Binom}(B, \max\{1/2, \overline{P}(\vartheta)\}) \le B/2)\\
        &\le \mathbb{P}\left(\mbox{Binom}\left(B, \max\left\{\frac{1}{2},\, \Phi\left(\frac{(1-\varepsilon)}{2}\frac{r_m|Z|}{r_{km}}\right) - \Delta\right\}\right) \le \frac{B}{2}\right) + 2\Delta'.
    \end{align*}
    Because 
    \[
    |\mathbb{P}(\mbox{Binom}(B, p) \le B/2) - \mathbb{P}(\mbox{Binom}(B, q) \le B/2)| \le 2\sqrt{B}|p - q|,
    \]
    letting $\varepsilon \downarrow 0$, we get
    \begin{align*}
        &\mathbb{P}(\mbox{Binom}(B, \max\{1/2, \overline{P}(\vartheta)\}) \le B/2)\\
        &\le \mathbb{P}\left(\mbox{Binom}\left(B, \Phi\left(\frac{r_m|Z|}{2r_{km}}\right)\right) \le \frac{B}{2}\right) + 2\sqrt{B}\Delta + 2\Delta'.
    \end{align*}
    Combined with~\eqref{eq:first-term-coverage-LM}, Theorem~\ref{thm:coverage-of-LM} now implies the result.
\end{document}